# The State of AI Ethics
## Volume 6

February 2022

MAIEI

This report was prepared by the **Montreal AI Ethics Institute (MAIEI)** — an international non-profit organization democratizing AI ethics literacy. **Learn more on our website** or subscribe to our weekly newsletter **The AI Ethics Brief**.

This work is licensed open-access under a **Creative Commons Attribution 4.0 International License**.

Primary contact for the report: **Abhishek Gupta (abhishek@montrealethics.ai)**

**Full team behind the report:**

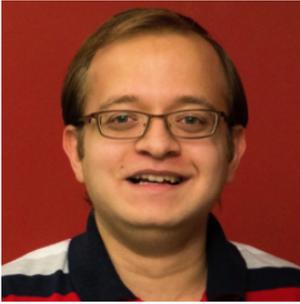

Abhishek Gupta
FOUNDER, PRINCIPAL RESEARCHER, AND DIRECTOR

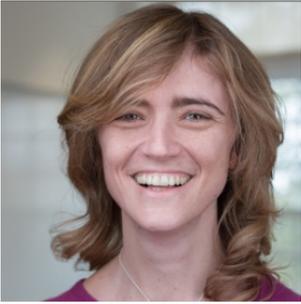

Marianna Ganapini, PhD
FACULTY DIRECTOR

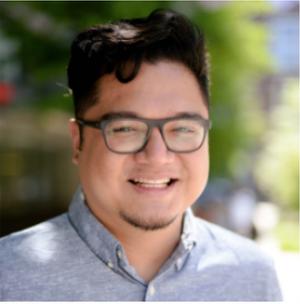

Renjie Butalid
CO-FOUNDER AND DIRECTOR

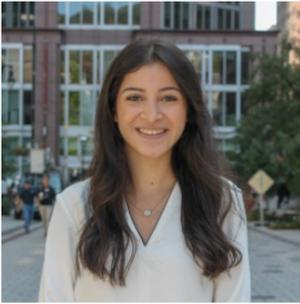

Masa Sweidan
BUSINESS DEVELOPMENT MANAGER

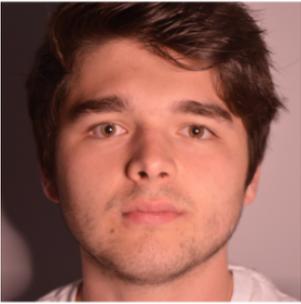

Connor Wright
PARTNERSHIPS MANAGER

**Special thanks to our contributor network:**

Ramya Srinivasan, Jonas Schuett, Jimmy Huang, Robert de Neufville, Natalie Klym, Andrea Pedeferri, Andrea Owe, Nga Than, Khoa Lam, Angshuman Kaushik, Avantika Bhandari, Sarah P. Grant, Anne Boily, Philippe Dambly, Axel Beelen, Laird Gallaghar, Ravit Dotan, Sean McGregor, and Azfar Adib.

# Table of Contents

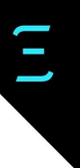

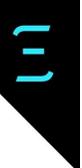

**\*Note:** The original sources are linked under the title of each piece. The work in the following pages combines summaries of the original material supplemented with insights from MAIEI research staff, unless otherwise indicated.

# Founder's Note

Welcome to 2022! (if you're reading this report at the time of the release, or hello to you in the future!) 2021 was a year that showed us a lot of things. We came together as society to fight one of the toughest battles of our time: the COVID-19 pandemic that reshaped our world and society. Operation Warp Speed helped us get vaccines and distribution around the world worked in overdrive to bring elevated safety against this pandemic. Yet, so much more needs to be done if we're to beat this back and go back to the world as it was before. At the same time, technological progress didn't pause. In most cases, it accelerated. This warrants continued examination to ensure that this technology doesn't insert itself into our society and lives in a way that is discordant with our values.

Now in its sixth cycle, this edition of the [State of AI Ethics Report](#) comes to you with a wide array of topics and contributions from leading lights in the field. For the first time, we have a Spanish text contribution in the report in our endeavor to produce multilingual content for the community to consume. We've added a new chapter on **Trends** that highlights subtle and not so subtle changes taking place in the AI ethics landscape. This one is a must-read for anyone who is planning on bring AI ethics meaningfully into their organizations, or pursuing research and looking for ideas on which areas to make an impact in.

As always, we've got our **What we're thinking** section that brings to you original contributions and essays diving into areas like **How to build an AI ethics team at your organization?** to other subjects like **Constructing and Deconstructing Gender with AI-generated art**. We cover other ideas in this chapter detailing developments around the world such as changes in talent, funding, and ethics as AI development picks up in Vietnam. In-depth interviews with industry experts to understand what it takes to bring AI ethics effectively into an organization's practices are supplemented by interviews with educators who are working hard to bring AI ethics education into classrooms around the world. They yield insights for anyone interested in either building training programs at a corporation or those who have students coming to them with questions about AI and want to have structured courses to guide them on this journey of building Responsible AI systems. A few other pieces make sure though that the discussions don't just focus on principles but also practical advice such as **The Proliferation of AI Ethics Principles: What's Next?** that analyzes what the gaps are between principles and practice today.

Another new addition to the report is the **Analysis of the AI Ecosystem** chapter that has the goal of taking a meta-level approach to understanding the dynamics at play in the field, including pieces like **The Values Encoded in Machine Learning Research** and **Putting AI ethics** **to work: are the tools fit for purpose?** We also cover various AI regulations that are in development around the world, looking at those coming out of the EU, US, NATO, UK, and UNESCO.

**Privacy**, **Bias**, and **Social Media and Problematic Information** also have a presence as chapters in this report. They continue to remain significant areas within Responsible AI and cover a lot of ground (though it is impossible to be exhaustive even in a ~300 page report). We've worked hard to curate those pieces that go beyond what is most often covered and pieces that we thought deserved a bit more attention for their particular lens on the issues in each of these domains.

Another new addition to this report which builds on our push towards moving from principles to practice is the chapter on **AI Design and Governance** which has the goal of dissecting the entire ecosystem around AI and the AI lifecycle itself to gain a very deep understanding of the choices and decisions that lead to some of the ethical issues that arise in AI. It constitutes about one-sixth of the report and is definitely something that I would encourage you to read in its entirety to gain some new perspectives on how we can actualize Responsible AI.

Given all the regulations coming out, the **Laws and Regulations** chapter provides a dedicated space to discuss the changes that are taking place in this landscape and will certainly provide you with markers on what to watch out for in 2022 and beyond as lawmakers and governments around the world scramble into action to regulate the relentless march of AI development and deployment.

And finally, as always we have our much-enjoyed **Outside the Boxes** chapter that captures eclectic developments in the field, things that might evolve into their own subfields as the years roll by. From covering things like Ubuntu ethics to animism and Rinri, you get the chance to zoom out and see the unbelievable ways that AI is impacting and transforming our society.

I hope that you will enjoy this edition of the report as much as we've enjoyed putting it together. We encourage you to share it with colleagues and friends, and those who are interested in getting an in-depth and broad understanding of the field. My recommendation for those who are wondering on how to work through almost ~300 pages of this report is this: grab a beverage of choice, and flip over to the **Table of Contents** and click through on a title that catches your eye and go on from there. If you're already familiar with the domain of AI ethics, I recommend starting with the **Trends** chapter and for those who are new and looking to get started in the domain, I encourage you to begin with **"Welcome to AI"**.

If this is your first time reading our reports, please don't hesitate in reaching out to us to let us know how we're doing! If you've been with us on this fantastic journey before, we'd be delighted to learn more about what brings you back to the report, what are your favorite parts and how we can improve. Thank you for the trust you place in us to bring you the latest in research and reporting in the domain of AI ethics.

For now, please enjoy the pages ahead, and I will see you again at the end of the report in the Closing Remarks!

---

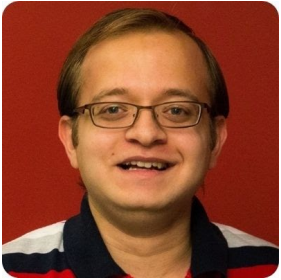

**Abhishek Gupta (@atg_abhishek)**
Founder, Director, & Principal Researcher
Montreal AI Ethics Institute

Abhishek Gupta is the Founder, Director, and Principal Researcher at the Montreal AI Ethics Institute. He is a Machine Learning Engineer at Microsoft, where he serves on the CSE Responsible AI Board. He also serves as the Chair of the Standards Working Group at the Green Software Foundation.

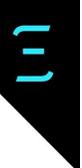

# 1. What we're thinking by MAIEI Staff and Community

**Introduction** by Abhishek Gupta, Founder and Principal Researcher, Montreal AI Ethics Institute

This section is geared towards showcasing ideas that our team and close network of collaborators has on the field of AI ethics with a focus on things that remain currently underexplored and unexamined.

If you are excited about what it takes to operationalize AI ethics in practice and how to effectively govern the development of AI systems, then you will find a lot to take away from the first piece on "**Patterns of Practice will be fundamental to the success of AI governance.**" But, such an effort does require robust supporting infrastructure, and it all starts with people. The next article on "**How to build an AI ethics team at your organization**" gives insights into a few actions like getting leadership buy-in and empowering people to make necessary changes along with aligning these principles with organizational values offer some concrete advice on a way forward.

A recurring question that we often get asked at the institute is how to increase literacy in AI ethics, which is an apt question given that our mission is to "**Democratize AI Ethics Literacy**." The next article from Marianna Ganapini, our Faculty Director, dives into a conversation with Chris McLean from Avanade who shares some underexplored areas in tech ethics today and what we can do to better cover them in curricula and elsewhere.

We also get a chance to talk about the role that AI-generated art can play in constructing and deconstructing gender. MAIEI's close collaborator Jimmy Huang dives into a conversation with artist Jake Elwes to understand how queerness and latent spaces come together and highlight some of the work that Elwes has done in this space to cross-pollinate ideas between the worlds of machine learning and art.

Our resident foodie went on out on a deep exploration of what it might be like to have a Michelin-star quality meal made by an AI system. Masa Sweidan, our Business Development Manager, talks about what role AI can play in the kitchen, as a tool for discovery and inspiration when it comes to picking flavors and more.

Another close collaborator of the institute Natalie Klym dives into wide-ranging, in-depth conversations with veterans of the technology and internet domains to pick their brains on ideas such as what we might learn from the internet experience and its journey to what it is today to guide our thoughts in addressing the ethics challenges in the domain of AI. She speaks to David Clark, a senior research scientist at the esteemed MIT CSAIL to bring us nuggets that I'm sure you will ruminate on long after perusing these pages. In another conversation, this time with Domhnaill Hernon who leads the Cognitive Human Experience at EY, we learn about how fusing art and engineering could lead us to a more humane tech future.

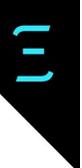

A lot of conversations in the domain of AI ethics are still quite Western-centric but there is a lot more out there and happening in other parts of the world. Two of my frequent collaborators, Nga Than and Khoa Lam, come together to share with us insights into the state of funding, talent, and ethics in the AI ecosystem in Vietnam. A must-read piece for those who want to break out of the more frequently covered areas and gain an outside perspective on how other countries are engaging with AI. This is followed by a piece by Philippe Dambly and Axel Beelen who explain the impacts of the first EU regulation that is geared towards the insurance industry.

An exclusive piece by my collaborator Ravit Dotan dives into the ever-growing landscape of AI ethics principles and what organizations should do in the face of so much information. It offers advice on how organizations can better navigate this space and highlights some limitations in the search for finding unifying principles across the globe. In "**Representation and Imagination for Preventing AI Harms**", Sean McGregor details the work he has done for Partnership on AI on the AI Incident Database that seeks to provide a public repository of ethical issues that have been documented in the real-world use of AI systems. The aspiration is that having a centralized place from which we can draw lessons will help to reduce those harms in the future as people can learn from prior mistakes.

Azfar Adib then dives into a piece exploring how the industry for age-verification is being transformed by the use of AI and what that means for us in the future as we go out to our favorite clubs and bars (whenever they open up!) In a follow-up piece, Adib explores what it might mean if we had robot co-workers and if we might need new policies for their governance or we would apply the same ones that are applied to human workers.

Even as I wrote the text above and read it again, it was shocking to see that the diversity of areas being covered has grown significantly over the past few years in exploring the impacts that AI is going to have on society. This is a testament to how versatile AI is as a piece of technology but also a warning sign that it will infiltrate all parts of our lives.

We are the ones who need to play the role of a sentinel, carefully guarding all that we hold sacred so that **we are the ones who shape AI systems to help us build a better world rather than letting the invisible hand of AI shape ours without our consent**. As they say, the power lies in our hands! I hope you enjoy this chapter and it gives you a lot to think about the AI systems that surround you all around.

---

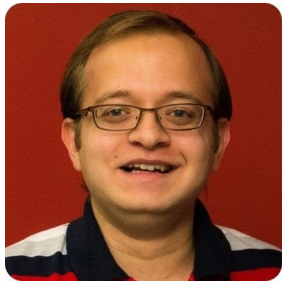

**Abhishek Gupta (@atg_abhishek)**
Founder, Director, & Principal Researcher
Montreal AI Ethics Institute

Abhishek Gupta is the Founder, Director, and Principal Researcher at the Montreal AI Ethics Institute. He is a Machine Learning Engineer at Microsoft, where he serves on the CSE Responsible AI Board. He also serves as the Chair of the Standards Working Group at the Green Software Foundation.

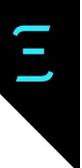

# From the Founder's Desk

## Patterns of practice will be fundamental to the success of AI governance

**[Original article by Abhishek Gupta]**

AI governance has certainly gained steam in 2020 with a lot of calls to action that have leveraged expertise in both the legal and technical fields to propose frameworks to govern both the development and deployment of AI systems. There are a lot of commonalities in these initiatives, with most of them focusing on areas of transparency, accountability, bias, privacy, non-discrimination, and other generally agreed upon values from the over 100 sets of principles in AI ethics, with most having at least some component focused on AI governance.

There has been a noticeable movement from 2019 when AI governance was a topic of discussion where people talked about abstract ideas and 2020 saw much more of a push to actually put those ideas into practice. Yet, as much as we saw movement, there were still some shortcomings that hindered the deployment of these governance mechanisms. In particular, 2020 was a year where we saw hasty roll-outs of these systems in tracking face mask compliance[1], grading students[2], handing out unemployment benefits[3], and more. So, what could we have done better?

As I have detailed in my work titled *Green Lighting ML: Confidentiality, Integrity, and Availability of Machine Learning Systems in Deployment*[4] that I presented with my co-author Erick Galinkin at several conferences in 2020 including ICML, what we have seen is that there is little focus on the practical manifestation of these ideas. Specifically, there is a missing focus on the needs and patterns of practice of designers and developers on the ground who will have at least partial responsibility in operationalizing these ideas. This is not to say that government mandates and management of the organization is not going to play an important role in how AI systems are governed. Quite the contrary. It is in fact essential that we consider the measures I am going to recommend as a supplement to the others, especially as they will help to bolster the efficacy of any other organizational-scale mechanisms that are applied in AI governance.

---

[1] https://www.theverge.com/2020/5/7/21250357/france-masks-public-transport-mandatory-ai-surveillance-camera-software

[2] https://www.wired.co.uk/article/gcse-results-alevels-algorithm-explained

[3] https://www.usnews.com/news/best-states/articles/2020-02-14/ai-algorithms-intended-to-detect-welfare-fraud-often-punish-the-poor-instead

[4] Gupta, A., & Galinkin, E. (2020). Green Lighting ML: Confidentiality, Integrity, and Availability of Machine Learning Systems in Deployment. arXiv preprint arXiv:2007.04693.

From a practitioner's perspective, there are numerous challenges that one faces when they encounter abstract principles coupled with business pressures and deadlines to deliver products and services on time and with high quality. It is at these points that there is a breakdown in the actual operationalization of the AI governance mechanisms which needs to be fixed. From my experience, the first method that helps to mitigate this issue is to strive to incorporate pieces of the governance requirements within existing workflows of designers and developers rather than jumping to create net new mechanisms. The benefit of doing so is that there is lower friction in the acceptance of these new requirements and they are also quicker to deploy and then gather evidence to see if they are effective or not. Armed with this evidence, one can make a stronger case for their incorporation at a wider level.

Second, and perhaps the most important aspect of creating AI governance solutions is to include the practitioners in the process of developing these mechanisms. The requirement there is two-fold: one, you are able to surface the exact places where the AI governance solutions might fail when they are asked to be implemented in practice based on the experience of the practitioners and two, you also build trust with those practitioners so that they are not only aware of what will be asked of them but given that they are active contributors, they will have a strong sense of ownership and desire to see this succeed.

Thus, keeping in mind these patterns of practice will be crucial if we are to actually move forward in putting AI governance to work rather than spend another precious few months and years debating on the abstract ideas. The time for action is now and it starts by paying attention to how these systems are actually designed and developed in practice.

### How to build an AI ethics team at your organization?
[Original article by Abhishek Gupta]

So you're working on AI systems and are interested in Responsible AI? Have you run into challenges in making this a reality? Many articles mention a transition from principles to practice but end up falling flat when you try to implement them in practice. So what's missing? Here are some ideas that I think will help you take the first step in making it a reality.

**Get leadership buy-in**
Yes, this is important! Why? Well, different units within your organization have different incentives and goals that they are working towards. Achieving Responsible AI in practice requires coordination across different units. The leadership team can help provide a unifying mandate to bring together different units to achieve this goal.

More so, they can act as a central point of dissemination of the "why" behind pursuing Responsible AI at your organization. They have the authority to create policy and drive change en masse that can make on-the-ground work easier and more effective. Especially in cases where you face reluctance from colleagues, a clear message from leadership provides a North Star for everyone.

Finally, leadership plays an essential role in providing you with necessary resources and "air-cover" to experiment with tools and techniques as we (the research and practitioner community) figure out practical solutions to some very complex challenges in the field of AI ethics.

**Setup feedback mechanisms**
As a complementary point to the above recommendation, we should also make it easy for on-the-ground practitioners to provide feedback on the tools, techniques, and processes that work well and those that don't. This is critical when you have a large organization with many teams working on very different product and service offerings. The guidelines and mandates coming top-down can suffer from a lack of context and nuance, which only gets clearer closer to the place of operation.

Effective feedback mechanisms have two qualities: they are easy to file and have transparency on which of the pieces of feedback have been acted upon. Many places fail on the second aspect, without which the entire exercise of feedback solicitation becomes fruitless. This also disincentivizes employees from sharing feedback in the first place and makes them lose trust in the process. Sharing results of the pieces of feedback that have been acted upon (which will often be visible through changes in the tools, techniques, and processes) and, more importantly, those that haven't been acted upon along with a reason on why they have not been acted upon will evoke higher levels of trust from the employees in the organization.

**Empower people to make decisions**
Often, those closest to the problems and building solutions to address those problems have highly contextual insights. We can leverage these insights by empowering those people to make decisions. This empowerment is important because it ensures that people feel a greater sense of ownership in the solutions that they are building. They become more capable of solving problems that really matter to their users and customers.

A hierarchical organization can help to align different teams together towards a common vision. Still, when complemented with the bottom-up approach of generating solutions and empowering the staff to act on those solutions, we increase the likelihood of achieving our Responsible AI goals.

A practical approach to building up this empowerment is to start allocating small amounts of direct responsibility on amending product and service offerings and increasing the scope of that responsibility over time as people demonstrate aptitude and skill for it. More so, proactively supplementing this on-the-job experience with training programs that promote decision-making regarding Responsible AI will make this approach successful.

**Align with organization values**
One of the core places of dissonance occurs when AI ethics is framed in a context aligned to the organization's mission and values. Drawing a clear connection between them helps boost uptake and leverage other evaluative instruments (like performance reviews) and policies within the organization.

It also helps percolate the idea of responsible AI as a key function of every person's job role in the organization that helps with organic integration of these responsibilities in the existing job roles and making it easy to create new job roles in the organization that are tasked with implementing AI ethics within the organization.

[Make RAI the norm rather than the exception](#)
Just as Microsoft has invested years of effort in tooling and processes to make accessibility a first-class citizen in their products and services, making Responsible AI the norm rather than the exception should be our North Star.

If, through investments, we can make the implementation of these ideas a default action and easy action, then not only will we get higher traction, but it will also disincentivize developers from doing anything other than the "right" thing. Yes, the last part is a little bit aspirational, but it isn't unrealistic!

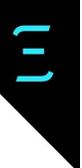

# Office Hours

**Exploring the under-explored areas in teaching tech ethics today**

**[Original article by Marianna Ganapini]**

Chris McClean shares his experience as the global lead for digital ethics at Avanade, and we are excited to learn more about how it trains tech and business professionals to recognize the most pressing ethical challenges. And as always, please get in touch if you want to share your opinions and insights on this fast-developing field.

**What is your background? What courses do (or did) you teach connected to Tech Ethics, and who's your audience (e.g., undergrads, professionals)?**
I am the Global Lead for Digital Ethics at Avanade, a 40,000-employee technology consulting and advisory firm. A substantial part of my role includes training our tech and business employees worldwide about how best to recognize and address ethical issues that arise in the technology we design, develop, deploy, and operate. I also offer Digital Ethics training, assessments, and program design for our clients (technology and business executives) as part of a broad advisory practice.

**What kind of content do you teach? What topics do you cover? What types of readings do you usually assign?**
I teach general concepts and trends in Digital Ethics, which covers a wide range of ways technology impacts individuals (such as privacy, accessibility, financial health and opportunity, mental well-being, personal dignity, and legal status), society (such as health care, education, the economy, criminal justice, and law enforcement), and the environment (such as energy use, material use, waste, pollution, and impact on biodiversity). I also cover a wide range of ethical controls, such as values alignment, ethical testing, security, resilience, monitoring, oversight, recourse, and accountability. I usually distill academic research for my audience given the amount of time such reading might take, and I rely heavily on real-world cases of ethics done well or done poorly.

**What are some teaching techniques you have employed that have worked particularly well? For Tech Ethics, what kind of approach to teaching do you recommend?**
I found it's especially helpful to run audiences through scenario analysis, especially if we can use real case examples. I've also run workshops that include a detailed assessment of a technical product or project using our Digital Ethics Assessment Framework.

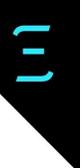

**In your opinion, what are some of the things missing in the way Tech Ethics is currently taught? For instance, are there topics that are not covered enough (or at all)? What could be done to improve this field?**

It's hard to say, as I don't have much visibility into all the different ways people are teaching these topics. However, given what we're seeing in the industry, it seems like we're spending a good deal of time on data ethics/privacy and responsible AI (which are critically important) but not enough time on the mental health, personal dignity, and environmental impacts of technology. I also don't see enough emphasis on how to incorporate ethical practices into various professional disciplines, like design, engineering, marketing, or audit.

**How do you see the Tech Ethics Curriculum landscape evolve in the next 5 years? What are the changes you see happening?**

I'm encouraged to see how much more often Tech Ethics is taught as part of general computer science and data science curricula. I'm hopeful that this trend will carry into business curricula as well, just as we've seen topics like sustainability and corporate responsibility become more popular. Ideally, I think our ethics-related education needs to include perspectives from economics, sociology, and even marketing to show that taking ethics seriously can positively impact business and social performance.

**Is there anything else you'd like to add?**

We should look carefully at the value of having stand-alone ethics training versus embedding ethics consideration into other aspects of training. As a disparate subject, it's very easy to compartmentalize ethics as something that's done occasionally, possibly by other people. But if it's incorporated as a standard element of other courses, it's easier to see that considering and addressing ethics is everyone's job throughout the entire tech lifecycle.

**Bio of interviewee:**

As the global lead for digital ethics at Avanade, Chris McClean is responsible for driving the company's digital ethics fluency and internal change and advising clients on their digital ethics journey. Prior to Avanade, Chris spent 12 years at Forrester Research, leading the company's analysis and advisory for risk management, compliance, corporate values, and ethics. Chris earned his MS in Business Ethics and Compliance in 2010 and BS in Business with a Marketing emphasis in 2001.

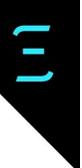

# AI Application Spotlight

**Jake Elwes: Constructing and Deconstructing Gender with AI-Generated Art**

**[Original article by Jimmy Huang]**

"The idea behind latent space is that there's this continuous space between the classes. You have these multi-dimensional vectors which relate everything it [the artificial intelligence] has learned about, say, a female face as well as everything it has learned about a male face, and there's this continuous space in between. It doesn't actually have those gendered binaries anymore – it's a continuation, and with unsupervised learning it doesn't even have the gendered labels…"

In the burgeoning field of artificial intelligence (AI) ethics, researchers at the Montreal AI Ethics Institute have been analyzing how AI applications frequently learn discriminatory behaviour from being fed biased training datasets. This could be, for example, from a lack of inclusion in the training set resulting in an application's inability to detect the faces of minorities [1] to an overinclusion within other sets for the express purpose of surveilling certain minority groups. [2]

There are also statistically significant, yet, barely perceptible biases we can only uncover through careful research such as when using historical US mortgage data to predict creditworthiness. Using standard logistic regression and Random Forest models, Fuster et al.'s CEPR discussion paper concludes: "minority groups appear to lose, in terms of the distribution of predicted default propensities, and in our counterfactual evaluation, in terms of equilibrium rates…" [3]

All this is to say that without guidance from ethics, by the very nature of training sets requiring bias to perform, discriminatory behaviour will not only persist but increase in ubiquity enhanced with modern, far-reaching technology.

Enter London-based artist, Jake Elwes. Carrying a strikingly warm presence, Elwes takes a seat across from me at Ditto Coffee in Shoreditch for the interview.

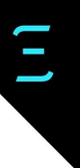

For the past half-decade, Elwes has been using various machine learning techniques to generate media art that compels us to consider our place in society giving a unique perspective in viewing human identity through the lens of modern technology.

Elwes' latest venture, The Zizi Project, is an ongoing series of works applying diverse drag and gender fluid identities as training sets for positive AI outcomes. The project started with the "Zizi – Queering the Dataset" installation in 2019 where an AI program attempts to constantly generate, shift, and regenerate non-binary faces in a work that celebrates difference and ambiguity. In 2020, Elwes produced the "Zizi & Me" installation, a double act between London drag queen Me [4] and a deep fake (AI) as well as "The Zizi Show" [5], a deep fake drag cabaret featuring a number of acts.

Elwes has taken a look at all emerging Generative Adversarial Networks (GAN) techniques and wonders 'how can we use this as a performance tool?'. "The Zizi Project" explores the effects of technology on gender identity through performance and in the process of creating the show, a variety of ethical topics are brought to light within the confines of a safe environment. Elwes leans forward over the table between us and, with passion, explains the discourse within both the drag and transgender communities around data consent, namely, how an individual's image may be used as well as what the use would be for.

On the data consent side, Elwes ensures that the performers who contributed visual content to "The Zizi Show" and "Zizi & Me" will have control over their image. They can retract their likeness from training sets and have derived performances from their likeness taken down. However, a much more interesting concern is brought forth by whether the inclusion of queer identities in training datasets have inherent issues. Some may posit that since we live in a technology-driven world, real harm could arise due to, for example, doctors not having the proper data points to adequately come up with treatments for transgender physiologies. On the other side, Elwes explains, "there is a real pride to being queer and having this otherness" and, historically speaking, marginalized communities are right to be wary of how changing technological and societal landscapes affects them. Given this pride in otherness, some members of the queer community feel hesitant to be included in training datasets or in having their identities assimilated to an extent.

Elwes aims to honour underrepresented and historically marginalized non-binary groups while also creating a uniquely charming cabaret show. In creating the show, from a technological standpoint, Elwes was largely inspired by the idea behind latent space.

*"There's a queerness to latent space"*

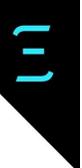

In simple terms, latent space is this hidden world, opaque to human intuition, of compressed data where similar features are mapped closer together.

Data is only useful insofar as there is bias in the set. Without bias, data would either be completely random or, on the other side, uniform and therefore largely useless. Machine learning applications find similarities in features by first compressing data into latent space, a vector space represented mathematically, and then in grouping similar data points closer together according to meaningful features.

There's ambiguity and nigh-infinite spectrums of data groupings that may be applied in a variety of contexts hidden within latent space. In this way, the vague, unlabelable inner-mechanism of how deep learning works have profound parallels to the gender fluidity of non-binary identities. It is especially fascinating how a non-binary group of identities is, in turn, grouped within latent space on a previously undefined spectrum.

Elwes' works aim, in part, to deconstruct gender and then reconstruct the features in an ever-transitory state. The gender-fluid appearances are distilled into groupings hidden within latent space and then constantly reconstructed becoming an evolving spectrum of an input set that is already a spectrum of gender identities.

In a world where we're constantly inundated with articles on the negative effects that AI applications may have on society at-large, seeing a positive outcome for a historically marginalized group, if only for cultural and artistic insight, is a breath of fresh air. Elwes works at the frontier of this innovative space, using emerging generative adversarial networks and deep fake techniques as they're discovered to create thoughtful, ethical art.

**References**
[1] https://news.mit.edu/2018/study-finds-gender-skin-type-bias-artificial-intelligence-systems-0212

[2] https://www.nature.com/articles/d41586-020-03187-3

[3] https://papers.ssrn.com/sol3/papers.cfm?abstract_id=3072038

[4] https://www.instagram.com/methedragqueen/?hl=en

[5] https://zizi.ai/

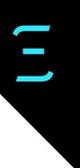

## Will an Artificial Intellichef be Cooking Your Next Meal at a Michelin Star Restaurant?

**[Original article by Masa Sweidan]**

The use of AI in the food industry has been growing over the past few years with applications such as robotics, kiosks, chatbots and recommendation engines. According to Mordor Intelligence, AI in the food and beverage market was valued at $3.07 billion in 2020 and is expected to reach $29.94 billion by 2026. A prime example of this progress is Miso Robotics' autonomous robotic kitchen assistant, Flippy, which was introduced in 2018 when it was able to grill 150 burgers per hour. Three years (and a pandemic later) and it can now cook 19 different foods including burgers, chicken wings and onion rings while keeping track of cooking times and temperatures. Not only will this machine reduce labor costs, but it will also improve the quality of food while providing deep insight into oil usage and product counts.

With various innovative solutions like Flippy popping up on the market, the benefits of AI in the restaurant industry all seem to stem from its efficiency and precision. When working properly, AI can increase savings and improve food safety, which is especially important as we navigate this COVID-19 era. Although these aspects are vital to the success of any restaurant, the true magic happens inside the kitchen where chefs cook delicious meals that often reflect unique social, cultural, and environmental influences. With this in mind, does AI have a place in the kitchen to support the creative process of professional chefs who have dedicated their life to learning the techniques and intricacies of high-quality cooking?

It is difficult to imagine a world where a machine prepares an entire meal from start to finish, because there is no doubt that the preparation and consumption of a delicately assembled meal is inherently linked to unique human experiences. How will a machine be able to understand and, more importantly, communicate a story through a particular spice that may remind you of your grandma's specialty dish or the smell of a dessert that transports you back to your favorite childhood memories?

At the tail end of 2020, Sony launched their Gastronomy Flagship Project to explore the potential of new technologies like AI and Robotics through interviews with chefs and professionals in the industry. This endeavor consists of the "research and development of an AI application for new recipe creation, a robotics solution that can assist chefs in their cooking process, and a community co-creation initiative that will serve as a foundation to these activities." Perhaps the most interesting aspect of this project is the focus on using technology to achieve even greater creativity, rather than replacing the more repetitive tasks in the kitchen.

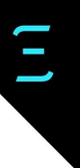

Surprisingly, the sentiments of chefs towards the use of AI in the world of gastronomy seem to be quite optimistic. By shifting the narrative to emphasize the fact that cooking is actually both art and science, the potential of technology becomes more apparent. Jordi Roca, a co-owner of 3 Michelin star restaurants, said it perfectly: "AI has been used to evolve musical compositions. My thinking is if it can be done with music, it can be done with flavors, because at the end of the day it consists of harmonizing a score or an aromatic chord."

This is the key. At its core, the process of cooking and baking is an applied science, because the building blocks of all food are large biological molecules such as proteins, carbohydrates, and fats. Put simply, the structure of a molecule defines how it functions in a cell and how a food may taste or react when being prepared. However, the artistic element comes into play with the creativity and emotion that is involved throughout the whole process. The tricky part is establishing the balance between using AI to optimize the molecular gastronomy of a dish, yet leaving room for the human chefs to express their imagination during the preparation of that meal. IBM spotted this opportunity and decided to develop Chef Watson.

Through computational creativity, Chef Watson can create recipes that suggest ingredient combinations and styles of cooking that humans would never have considered, due to its ability to analyze large data sets. After being fed 10,000 recipes from Bon Appetit's archives, it used natural language processing to learn the underlying logic of how ingredients were combined. This is particularly useful in the context of gastronomy because even the best professional chefs can only reason about pairing three ingredients, whereas Chef Watson can examine up to nine ingredient combinations. The power of this machine stems from its ability to model both the chemistry of the ingredients and the human perception of flavor.

It should be noted that this particular invention has not received much media attention or coverage since 2015, but it does support the vision that the future of gastronomy can integrate technology to achieve emotion. Josep Roca, co-owner and sommelier of El Celler de Can Roca, which was ranked the best restaurant in the world in 2015, takes it a step further and believes that AI can be used to create more personalized dining experiences for each individual. By using inputs about a customer's origin and preferences, they could be provided with a tailored menu that transports them back to their favorite memories.

These novel ideas that incorporate AI in the professional kitchen are surely exciting, but they also beg the question: will certain cuisines be misrepresented or completely left behind? Moreover, culinary traditions and recipes are typically passed down from generation to generation and are not necessarily meant to be perfect. Therefore, the most potential seems to lie in developing AI for the discovery and inspiration of flavors, rather than the actual cooking of a meal. Although it is unlikely that an "artificial intellichef" will be preparing your order at a fine dining restaurant in the near future, the developments in this space point towards a new reality where AI can assist with the recipe creation, but humans still have the final, artistic touch.

Credits to my dear friend and colleague, Connor Wright, for coming up with the clever term: "Artificial Intellichef!"

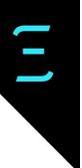

## Permission to be uncertain

**The Technologists are Not in Control: What the Internet Experience Can Teach us about AI Ethics and Responsibility**

**[Original article by Natalie Klym]**

**Interview with David Clark, Senior Research Scientist, MIT Computer Science & Artificial Intelligence Lab**

Artificial intelligence has recently emerged from its most recent winter. Many technical researchers are now facing a moral dilemma as they watch their work find its way out of the lab and into our lives in ways they had not intended or imagined but more importantly, in ways they find objectionable.

The atomic bomb is a classic example that many commentators on contemporary technologies refer to when discussing ethics and responsibility. But a more recent and relevant example that I would like to draw lessons from is the Internet–a foundational technology that has reached maturity and is fully embedded in society.

My focus is not on the specific social issues per se, e.g., net neutrality or universal access, rather, my goal is to provide a glimpse into some of the dynamics associated with the Internet's transition from lab to market as experienced by one prominent member of the research community, Dr. David Clark, Senior Research Scientist at MIT's Computer Science and Artificial Intelligence Lab (CSAIL).

Clark has been involved with the development of the Internet since the 1970s. He served as Chief Protocol Architect and chaired the Internet Activities Board throughout most of the 80s, and more recently worked on several NSF-sponsored projects on next generation Internet architecture. In his 2019 book, Designing an Internet, Clark looks at how multiple technical, economic, and social requirements shaped and continue to shape the character of the Internet.

In discussing his lifelong work, Clark makes an arresting statement: "The technologists are not in control of the future of technology." In this interview, I explore the significance of those words and how they can inform today's discussions on AI ethics and responsibility.

**You describe your observation that "the technologists are not in control" as a revelation that came to you during the commercialization phase of the Internet. Can you describe this moment and why it was revelatory?**

The goals of the research community in the 1970s and 1980s were purely technical. In the 70s, we were just trying to get the protocols to work. In the 1980s the challenge was scale. We went from a goal of connecting about 100,000 institutional computers to millions of personal computers, and now of course, we're looking at billions of connected devices including cell phones and sensors of all kinds.

Commercialization of the Internet began in the 1990s. During this period, it went through a rapid transition from being an infrastructure run by the US government to a service provided by the private sector.

It was an interesting as well as surprising time since we didn't know what a commercial Internet would look like. We had never thought about it that way. As the source of investment changed, so did the drivers and the goals of the research, which were now being led by industry. All of a sudden, a new set of factors emerged, things like profit-seeking, competition, etc.

The example that first brought this home to me concerned QoS, or "quality of service" controls, which enable the prioritization of packets. Our goal in designing these controls was to make time-sensitive applications like real time voice and games work better, and the controls did that. We initially saw that as a technical enhancement. However, it's not difficult to understand that in a commercial context packet prioritization has everything to do with industry competition and therefore money. In the early days of online phone and video services it was difficult for providers of Voice over IP and IPTV to compete with the telco and cableco's proprietary services because the quality of transmission over the public Internet was still relatively poor at the time and there were no QoS capabilities. So, if you were an ISP offering traditional phone and TV services, why would you build capabilities in your IP network that would offer a means for these new entrants to compete with you? As one ISP executive said to me, "why should I spend money on QoS so that Bill Gates can make money selling Internet telephony?"

The point is that what I had considered packet-routing protocols for decades were in effect money-routing protocols. This was pointed out to me by an economist who said: "the Internet is about routing money; routing packets is a side effect, and you screwed up the money-routing protocols." In my defense, I replied, "I didn't design any money-routing protocols," and his response was, "that's what I said." We were joking, but the point was real.

That the technical design of the Internet has implications for industry dynamics is obvious to economists and business people, and it's obvious to me now, but until we, as technologists, were compelled to solve industry problems, none of this was obvious. We did not understand at the time that we were now engineering both a technology and an industry structure that determined who had economic power.

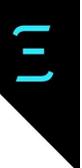

**So, you're saying that industry, or the private sector, is driving the future?**
Societal concerns have become more important in the last decade. We now have a fundamental tussle between the objectives of the private sector–concerned with things like commercialization and profitability–and those of the public sector. I worked with kc claffy, who is the Director of the Center for Applied Internet Data Analysis at the University of California, on research that explored the societal aspirations for the future of the Internet. We collected statements from a variety of stakeholders including governments and public interest groups and cataloged them into a list–things like reach, ubiquity, trustworthiness. These social aspirations can be in direct conflict with private sector goals. And it's this tension that shapes the future of the Internet.

**You changed your focus to the social implications of the Internet and policy matters in the 2000s. Would it be accurate to say this was the result of realizing you were now also engineering a social structure?**
Yes and no. There's a difference between the Internet itself—the package carriage system—and the applications that run on top of it, like the Web or Facebook, for example. These days, when people talk about the Internet, they are often talking about the latter.

So, yes, I have been more focused on the social implications of the Internet in the last 2 decades, but in terms of engineering a social structure, this stems from the application space, as opposed to the network itself.

And within the application space, it's the ad-driven business model that I most have issues with. This model creates all kinds of incentives that have negative consequences for society. Facebook, for example, is designed to be addictive. They want to keep you on their site so they can show you lots of ads, so they manipulate the experience to make it "sticky," and influence your personal behavior with all kinds of tactics. It's an incredibly distorted space.

**The 2020 documentary film, The Social Dilemma, and Shoshana Zuboff's 2019 book The Age of Surveillance Capitalism explore this distortion. And a few years ago, in 2014, Ethan Zuckerman (formerly at the MIT Media Lab and now at U. of Western Mass.) wrote a public apology for designing the pop-up ad back in 1997, declaring advertising in general as "the Internet's original sin." Do you feel personally responsible for the things you think are bad about today's Internet?**
We designed the Internet (the packet carriage system) for generality; that was its strength. I don't think there's any way I could have built an Internet that would allow for generality and at the same time preclude bad behavior at the application layer. Maybe there was a fork in the road where someone could have pushed things in a different direction—but not at the packet level.

**What are your thoughts on the moral dilemmas facing AI researchers today?**
I would say AI is probably more like the packet carriage layer in that it's a basic technology, with applications that can take many forms. And it's difficult if not impossible to preclude bad behavior or only allow for good behavior, nor is it easy to define what "good" vs "bad" behavior is. Stephen Wolff, who ran NSFNET in the 80s, said back then that every behavior we see in the real world is going to manifest in cyberspace including behaviors that we find unwelcome and offensive.

**The sociologist and Internet historian, Manuel Castells, has said that the Internet is the mirror of society. It is neither good nor bad, nor is it neutral, his point being that its uses are socially determined.**
I agree; the Internet evolved to defy both the original utopian and dystopian visions. The more general question regarding the moral responsibility of scientists has been debated over and over again. My view is that technology can be used in so many different ways, and rarely do we understand the future implications of what we have made, even if we have a clear sense of intended uses. A creative person can and will come along and use it in ways we never imagined.

The history of technology is full of stories of unintended consequences, whether good, bad, or simply frivolous. GPS is an interesting case. The early research papers stressed military applications of GPS because the researchers wanted the military to fund its development. Those who understood some of the broader societal benefits were afraid that if they put too much emphasis on these, the military would not pay for it. So, GPS emerged as a military technology. But today, we all have access to maps and directions, and my children have not had the experience of being lost. That is a good thing. But the negative consequences include things like neighborhood traffic congestion that results from traffic apps sending cars down residential streets, and global tracking of everyone's location. These outcomes are not something that could have been foreseen, nor is it at the level of something like autonomous weapons, even though it can lead to fatal accidents. But these are negative consequences that no one could have predicted at the time.

I think we should teach scientists to think through the social consequences of what they're doing, but I don't know if we can put a burden on researchers that says they have some obligation or responsibility to predict all the consequences and then try to embed mechanisms to prevent harm. I just don't think the world works that way.

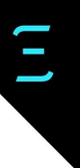

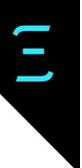

**The cryptographer Phillip Rogaway makes a strong case for computer scientists taking responsibility for their work, arguing that their work is political in nature.**
I'm very sympathetic to what Rogaway says, but it's one thing to conclude that encryption is going to shift power balances, and another to then design the technology to preclude bad behaviors by its users.

There should definitely be a sense of awareness during the more abstract exploration phase, but it's not until you get closer to a specific application that you need to think through the ethical implications. You're going to have to rectify problems as they emerge in each context, on a case by case basis.

**Carly Kind from the Ada Lovelace Institute in the UK refers to an emerging "third wave" of ethical AI that addresses specific use cases framed as social problems as opposed to philosophical concepts (the first wave) or narrowly-defined technical issues focused on algorithmic bias (the second wave). As we enter this third wave, it's clear that we need many voices at the table, but understanding each other and integrating multiple perspectives isn't always easy or straightforward. How have you addressed this challenge?**
When I began to realize that the Internet was no longer a purely technical problem I stopped running a purely technical research group at MIT. I started by hiring an economist and have also hired political scientists and collaborated with philosophers like Helen Nissanbaum. The last project you and I worked on regarding convergence at the application layer integrated ideas from media studies and other social sciences. Taking a multidisciplinary or interdisciplinary approach is key to understanding and shaping innovation in a way that benefits society.

I actually first started thinking in multidisciplinary terms much earlier, in the late 1980s. I got involved with the Computer Science and Telecommunications Board at the National Academies, which is an organization chartered by the US government to advise them in areas where a multi-stakeholder assessment of a problem is needed. I chaired an early study on computer security that was published in 1991. And I got a lot out of it, but most significantly, my experience with the National Academies taught me the importance of having conversations with people who were not like me—economists, social scientists, artists, lawyers, regulators. I chaired the board for 8 years, learning what happens when you get people with very different points of view and stakeholder biases together to produce something coherent. And so, as I moved forward with my technical research, I carried with me experiences and expertise that most pure technologists did not have.

**This third wave of ethical AI is intersecting with the long-awaited regulation of big tech. What are your thoughts on regulation?**

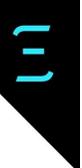

In terms of AI, the government has awakened to social implications a lot earlier in the lifecycle of the technology compared to the Internet. With the Internet, it took about 15 years after commercialization for the government to really wake up, so sometime around 2010.

There is always tremendous resistance from the private sector around regulation because they need to compete, which can mean doing things that harm society. When it comes to matters of public interest, as opposed to anti-competitive monopolistic practices, if one actor tries to be "good," they will lose. But if you impose regulation on everyone, it levels the playing field. The financial services sector is an example. It is heavily regulated—not to address monopolistic practices, but for public interest reasons. It adds to costs, and can stifle innovation, but it affects all players equally.

It can take a while for governments to figure out how to be effective. They may, for example, impose regulation on the wrong players for the wrong reasons. In the case of the Internet, we've seen governments impose regulation on the ISPs regarding objectionable content, like child pornography or terrorist activities, rather than the application providers. The rationale is that it's easier for the application providers to escape regulation by relocating operations to foreign countries. So, the ISPs are an easier target. But they are not necessarily the right target, or an effective one.

**You have referred to an "abstract" exploration phase of research and "basic" vs applied technology in a way that suggests more neutral phases in the overall research and innovation process. But do you think that the relationship between basic and applied research; between academic and industry research; and the path from discovery to invention to innovation in general, has changed over the years? How so? Are universities doing less basic, curiosity-driven research as collaborative innovation increases?**
There has been a large growth in computer science (CS) research, and the balance has certainly shifted toward more applied research—closer to commercialization. Our government is pushing investment to drive innovation, make our country more competitive, and so on. And innovation is not the only driver of this shift. As our field matures, some of the basic questions get answered. It is important to remember that in the late 1960s and early 1970s, when the early concepts behind the Internet were emerging, this was totally a venture into the unknown. But there are still folks who look further into the future. I am actually not sure I buy the distinction between basic and applied research. Those may not be the right divisions. Some research is more speculative, more driven by curiosity—a sense of exploration. If you are looking for a venture today that is going into the great unknown, crypto-currency comes to mind. Speculate on good and bad consequences of that idea, and how it will (or may) be shaped by various forces.

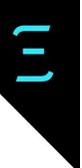

**Is there anything you would have done differently when designing the Internet?**
We designed the Internet for generality, that was its strength. The whole idea was that you could build anything on top of that platform. Of course, I hoped that smart people would come in and build really cool, useful things. I also expected that people would come in and build frivolous things, which is fine, and I always knew that people and organizations would eventually do wicked things on the Internet.

I love the writings of Terry Pratchet. He writes social satire cast as science fiction. His view of the world is that life is about performing a series of experiments that reveal how people really are. I see the Internet as such an experiment, and what we've discovered is that much of the world is evil, but I guess we knew that already.

## Fusing Art and Engineering for a more Humane Tech Future

[Original article by Natalie Klym]

**Interview with Domhnaill Hernon, Global Lead of Cognitive Human Enterprise at EY and former Head of Experiments in Arts and Technology (E.A.T.) at Nokia Bell Labs**
Marshall McLuhan believed that artists were the best probes into the future of technology because they lived on the frontiers. They were the most likely to take technology in directions beyond the intentions of the scientists and engineers. But according to Domhnaill Hernon, artists don't just think outside the box in terms of features and applications, their most important contribution to tech innovation is the ability to create a much needed human-centric vision of the future.

**Domhnaill, you just ended a 5-year term leading Bell Labs' Experiments in Arts and Technology program, one of the handful of corporate programs in the U.S. that integrated the arts with R & D. And now you are creating a new initiative at EY (Ernst & Young) called the Cognitive Human Enterprise. Can you tell us more about the work you do with artists and how it brought you to EY?**
I was asked to lead a new initiative at EY to show the potential of fusing art/creativity and technology to create the most cognitively diverse organizations possible. The role builds directly on what I had achieved at Bell Labs' Experiments in Arts and Technology program and supports EY's commitment to what they call Humans@Center.

My unique approach is to leverage the significant differences between the world of technology/business and art/creativity. This is a lofty goal but I truly believe that the future of human-centered innovation lies at the intersection of art and technology.

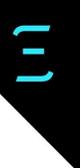

**Can you tell us more about the history of the E.A.T. program and how you ended up there?**
Bell Labs had been bringing engineering science and the arts together since its inception in 1925, all the way up until about the early 1980s. One of the standout moments from that period was the creation of a global not-for-profit organization, called E.A.T., which stands for Experiments in Art and Technology, in the 1960s. It emerged out of a series of art-performance events called 9 Evenings: Theatre and Engineering held in 1966. These events comprised collaborations between Bell Labs' engineers and several prominent artists of the time including experimental music composer John Cage, the abstract expressionist painter Robert Rauschenberg, dancer and choreographer Yvonne Rainer, and many others. But from the 1980s until about 2016 the fusion of art and engineering was largely non-existent at Bell Labs.

**What happened in the 80s? Why did the E.A.T. program end?**
There were major changes in U.S. socioeconomic policy that changed how industry in general was being regulated and how research was being funded, and a lot of other shifts including things like how employees were treated. There was pressure on corporations investing in what many perceived as frivolous artistic activity–and R&D in general–to reduce funding to those programs.

**And then what happened in 2016?**
I moved in late 2015 from Bell Labs in Ireland to the HQ of Bell Labs research in New Jersey. Soon after I arrived it was the 50th anniversary of E.A.T.'s incarnation. Several of us at the leadership level got invited to several celebratory events in New York City that were essentially engineers + artist meetups. Through those events we learned about the history of Bell Labs' work with artists and I realized that we, as an institution, had forgotten about that part of our history. From that point onward, we learned more and more about the critical value of fusing art and engineering, and the immensely significant role that Bell Labs had played.

And at the same time, but separately, we were having internal conversations about what was missing from our research strategy and what we wanted from new talent and our organizational culture. So, these were the conversations we had during the day, and then in the evenings, we attended the artist meetups commemorating the E.A.T. program.

Every one of the interactions blew my mind. I realized that the artists had a completely different perspective on everything — from the intersection of technology with society to life in general. More specifically, the role that humans play in technology development was at the center of every answer they provided to my questions. This was impressive because, as an engineer, I had not been trained that way and I could not believe that I was so blind to this perspective. It struck me that we needed an organizational culture that emphasized a more human-focused approach to innovation.

So I decided to establish a new initiative, based on the original E.A.T. program I had just learned about, but focused on modern day needs.

**So, how was the new program different from the original one?**
The original program was a not-for-profit entity, separate from Bell Labs. It grew organically out of the interpersonal connections between the artists and engineers, whereas the new initiative was a sanctioned, funded, internal initiative that evolved to become its own research lab within Bell Labs. It was designed more purposefully, based on what we learned from the original program.

The context was also very different. Back in the 60s and 70s digital technology was very new. Artists had a lot of creative ideas but didn't have the technological knowhow to manifest them. They leveraged Bell Labs' engineers and scientists expertise to make the technology required to enable their creative ideas. That is not the case today as many multimedia artists are technologically gifted.

There's also just that much more user-friendly consumer technology today, but it wasn't always the case. Dan Richter, who played the ape, Moonwatcher, in the opening scene of Stanley Kubrick's iconic AI film, 2001: A Space Odyssey, was a guest on the seminar series I run at the University of Toronto's BMO Lab, which focuses on the relationship between AI and art. He talked about how much new technology had to be created to enable Kubrick's vision. That was back in the early 1960s.

Yes, in fact, Artur C. Clarke (the author of the novel) spent a lot of time at Bell Labs. There was a major relationship between Bell Labs and the production of that film. They developed futuristic props such as the video phone, and Max Matthews, who's considered the godfather of computer music, inspired some of the music in 2001 such as HAL singing "Daisy Bell" towards the end of the film.

**How does the E.A.T. program benefit the engineers?**
Today, the E.A.T. artists are more technologically savvy and the program is designed to be more mutually beneficial. When we pair artists with engineers and scientists, the artists, as before, get access to tech they wouldn't otherwise, but in the new program they infuse their human focus deep into the R & D community. In other words, the artists are there to enlighten the engineers. And by that I mean, STEM practitioners are very well trained in the scientific methods but that blinds us to other ways of thinking and problem solving. We wanted the artists to infuse R & D with an ethos of humanizing technology. We wanted the engineers and scientists to always have in their minds the human aspect of technology; to question how this technology might do good or harm to society, and how they might design out the ability to do harm in the earliest stages of a research project. We also wanted to expose our R&D community to new forms of creativity.

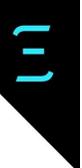

In my previous interview with MIT's David Clark, one of the early Internet pioneers, he emphasizes how difficult it is to predict the outcomes of technology–to "design out" those possibilities as you put it.

I'm not saying it's easy. It's very difficult, but at the very least, technologists, engineers, scientists, researchers should be asking those questions, they should be aware of that human element. That awareness isn't part of how engineers are educated or expected by their employers to create value in the marketplace. Whereas artists have an inherent way of keeping the human in mind, first and foremost. I wanted to integrate their way of thinking into our R & D community in a deeply purposeful way. That way, we could drive real cultural change around this foundational concept of humanizing technology. I see this type of holistic approach as the driver of human-centered innovation.

That might be one of the great untapped potentials of fusing art and technology–the ability to sense and create a human-centric vision for the future.

**Your point about creating value in the marketplace is interesting and makes me question whether it's the technology and technologists that need to be "humanized" or business and the executives managing firms. In other words is tech the problem or is it the tech industry?**
I don't think the problems in society can be blamed on technology directly. Technology is just a tool that is designed and used by humans in various ways. I also don't think it's fair to say it's purely a business problem either. I think every aspect of the chain needs redefining and all elements of the chain need to work together in tandem. Much of my work is about getting to the core of where the tensions reside, and fundamentally, it's about adding the human element to the design of technology and a human element to how businesses leverage the skills of engineers to create value and a human element to how businesses push technology out into the market.

We've evolved to a point where we largely rely on markets and we develop technology to survive and thrive as humans. A core part of the human condition is that we're going to develop systems and paradigms and tools that are developed by humans for humans and they will have an impact on society.

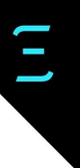

My main issue today when I look across the chain is with the education and training aspect of science and engineering. When you're studying or working in technology, all problems are technology problems and all solutions are technology solutions. It was really an eye opening experience to work with artists. It made me realize the trap I had fallen into and that I was blind to the other lenses through which you can view the world and solve problems. There's a lack of connection to the humanities, a gap. However, making that connection in an impactful way is not easy. The E.A.T. program, as I said earlier, was about making purposeful connections and really bringing the best of both worlds together.

**You're reminding me of my experience working with a research group at the MIT Media Lab that made the integration of art as one of its goals. But it was a vaguely defined objective and the project leaders, neither of whom were artists, didn't know exactly what it meant or how to do it, and were very open to suggestions. I appreciated their honesty, because it isn't easy, as you say, to make the connection in an impactful way.**
There is a real lack of understanding of how to bring these different ways of thinking together. It's very hard work. I still see a lot of efforts that are quite superficial, what I would call a "check the box" exercise. And I see a lot of efforts that are random–an artist is randomly selected and paired with a randomly selected engineer and they are put together randomly in some common space for a short period of time. In that model if anything good was to come out of the interaction it would be just fluke. These initiatives need to be thought through purposefully and strategically and executed with precision within the bounds of what you have control over.

**I also encountered an attitude from some of the engineers I worked with over the years that art, or any of the social sciences for that matter, was somehow inferior or insubstantial. The word "fluff" was used on many occasions to describe these disciplines.**
Yes, and when I started the Bell Labs program I had to think through all the ways in which the program could be killed, given that kind of attitude.

But even when there's a lot of goodwill, and good intention, there's still not a lot of good execution. I found there were two main approaches to art and tech fusion. One was extremely transactional. A company would bring in an artist for a couple of weeks and say, here's our new product, do something cool with it. But then that was it. The impact was short term and superficial, driven primarily by communications and branding goals.

Then there was the completely ad hoc approach where someone in the organization would say, oh, we need to bring artists in, and they would randomly select an artist and likewise a random group of employees who would engage with the artist. They would put them in a space together and think something would just emerge and that the organization would suddenly become more creative.

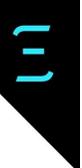

By contrast, I designed the modern E.A.T. program more strategically and more purposefully, with ways to measure impact. And again, I designed around the modes of failure I was aware of and I applied the concept of a pre-mortem to my analysis and design of the new initiative. I spent a lot of time getting to know artists, their personalities, their openness to collaboration and their technological capabilities. And the same thing for the scientists and engineers, so I could make the right match. I also had to factor in deliverables and schedules of those engineers and the perception of their management chain so that we covered all dimensions of success as much as possible.

I had to think it through as much as I would an actual technical product that would go to market. Also very important is the fact that I viewed it as a major cultural change initiative, which are known to have a high failure rate.

**What were some of the early proof of value experiences you had, and how did they evolve over time?**
In some cases, the proof of value was an exceptionally insightful conversation that completely changed our perspective on technology and informed a new research direction.

From there we developed whole new classes of technology–not just out of the conversations with artists but also out of the collaborations where artists were using our technology in very different ways.

**Technology is often used differently than how its inventors intended. In cases where the technology in question is a creative tool, you get some amazing stories. The electric guitar, for example, was a technical solution to the very practical problem of amplification, but Jimi Hendrix and other musicians created a whole new sound. Stevie Wonder did the same thing with the synthesizer, turning technical and gimmicky sounds into a whole new artistic practice. What were some of the artist-driven consequences you saw at Bell Labs?**
One of the earliest examples was in the area of wearables. We had asked, what's the next communication device after the smartphone? This was around 2016. We were looking 10 years out. You had to assume the smartphone didn't exist anymore. We started from a technological research perspective that led to ideas of disaggregating smart phone functionality so that we could communicate, control and sense the world around us in new ways. Our earliest designs and prototypes were very utilitarian and clearly designed with technology at the center. Then we brought in artists and approached the question from completely new angles.

One of the first artists we collaborated with in the modern E.A.T. era was Jeff Thompson. He pointed out to us that, even at that time in 2016, we were all spending an order of magnitude more time on our smartphones than we were with the people we most loved in the world! This was an eye opening observation that helped us completely rethink the design and development of these new wearable concepts to be more human centric. We designed a wearable for your arm and one for your head–the Sleeve and the Eyebud–that worked in combination in much more intuitive and non-intrusive ways and removed the need to keep looking at your smartphone. So our initial conversations focused on the problem from a technological perspective (solving the biggest tech challenges in creating wearables), but the solution we ended up with came directly out of our artistic collaborations and showed you could sense and control the world around you in much more human centric ways, using the more natural forms of your body and leveraging the technology in a symbiotic way.

**What about AI?**
I think there are two main popular narratives surrounding Machine Learning (ML) and Artificial Intelligence (AI) at the moment and both stem from different interpretations in the value of automating "mundane" tasks.

In one argument people talk a lot about how AI can be used in industry to enhance efficiency/productivity through automation of the mundane and the popular assumption is that this approach will lead to job losses. I think this is probably a reasonable current assumption since very few in industry or academia that are researching and developing these AI tools have provided a strong counter argument. It is clear that current business imperatives are based on cost savings and margin increases and AI has the potential to benefit companies across all industries in that regard.

The second argument is that automation will free up people's time and then they can be more creative, productive and strategic with that time and create more value. The difficulty with this argument is that people can't just become more creative/productive/strategic–we need to develop tools that will help them on that journey.

So either way, we have a gap between the benefits that AI can provide and the narrative surrounding AI and its use in industry. We need to figure out a way to free up people's time from the mundane tasks and help them be more creative and productive with that time to create more value. The value they create needs to be more than the savings created through the potential of job reductions.

I'm also very interested in counteracting the dystopian narratives around AI. These negative stories are typically based on a fundamental lack of understanding of the technology and the lack of understanding on the potential for the technology to enhance human creativity and potential.

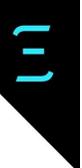

For example, at Bell Labs we wanted to showcase instead, the potential for AI to enhance human creativity. One project is "We Speak Music" and features the beatboxer Reeps One. We trained ML algorithms on his voice as he was beatboxing to the point where they started generating sounds and techniques that he had never created in his life, yet the AI voice kind of sounded like him.

Prior to this experiment, Reeps One felt that he had pushed the capability of his voice to the absolute limit. He didn't think there was anything else he could do to augment his voice and had started branching into other areas of art to satisfy his creativity and curiosity. But through this experiment we gave him what we called a "second self", an AI digital beatboxing twin, for him to collaborate with and according to him this enabled him to "level up" his voice and he is now creating new sounds and techniques and composing and performing in new ways.

Think about that–we took one of the best beatboxers that ever lived and one of the most creative people I've had the pleasure of working with and we helped him be more creative by creating an AI digital twin of/for him to collaborate with. Can you imagine the potential for AI to enhance the creativity of all people if it was developed right and for everyone?

**The seminar that I run at U of T's BMO Lab questions the role of technology in the creative process and there's definitely a tension around the question of how much tech is too much? At what point is it no longer human creativity? Is that a good thing or a bad thing? Has anyone ever viewed the idea of an AI-based digital twin as a dystopian narrative?**
Never. I've never heard anyone even question the experiment from that perspective. What was important to me was developing AI in a way that involves actual humans, that took embodied cognition or embodied intelligence into consideration and where the technology was in service to our humanity and not viewed as a replacement.

I believe the reason this question didn't arise out of our work is because we collaborated with this intent from day 1. The whole point of the collaboration was to dispel this sentiment.

**What's interesting to me about the work of Reeps One is that it's not about automatically producing a piece of music "in the style of," like a deep fake, it's about an actual collaboration between a human and machine. Can you say a little more about embodied intelligence?**
I think there is a lot of work to be done to dispel some of the myths and assumptions in AI today. For example, this notion that AI will supersede human intelligence is nonsensical to me because the way AI is developed today is based on a flawed understanding of human intelligence.

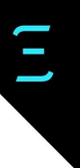

The neural net (machine learning and deep learning) type architectures of today are based on the mathematical models that AI pioneers created 30-40 years ago based on how they thought the human brain worked. We know now that the model is more of a metaphor and not how the brain actually works today based on neuroscience; however, the model is simple and pervasive and won't change anytime soon.

The problem with the neural net model of human intelligence is the aspect of disembodiment — the absence of a human body. The human brain on its own has no intelligence, cognition, creativity or consciousness. It has to be connected to the human body. The brain is a computational pattern recognition engine that requires sensory inputs from your physical body. Without the body, the brain is nothing.

But because of the flawed foundations of AI, we have this equally flawed idea of intelligence that encourages us to imagine we can replicate or supersede human intelligence, which has all sorts of practical and ethical implications.

I have no doubt that we are creating a new type of intelligence, which may be able to do things humans can't, but it's not going to be more intelligent in the way humans are intelligent. It's going to be different.

**What do you hope to achieve in your new role at EY?**
One of the powerful lessons I learn everyday working with artists is to remember that we are human, remember what is special about humanity and keep that front and center when developing technology. This is something that I am very excited about with my new role at EY.

EY have invested in diverse communities for decades. For example, they set up more than 10 global neurodiverse centers of excellence and hired hundreds of people from that community showing the world the immense value that people with different experience and skills can bring to any organization.

I co-founded a new initiative called the Cognitive Human Enterprise. The objective is to solve global-scale human and business challenges by investing in massively interdisciplinary collaboration and full-spectrum diversity to create the most cognitively diverse organizations possible. One aspect of accelerating towards that cognitive diversity is to leverage the benefits of fusing art and technology. Given EY's commitment to Humans@Center I am excited to see how far we can take this and deliver on human-centered innovation.

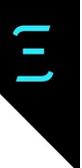

# Sociology of AI Ethics

**Challenges of AI Development in Vietnam: Funding, Talent and Ethics**

[Original article by Nga Than and Khoa Lam]

Vietnam in 2020 overtook Singapore's gross domestic product (GDP), and became the third largest economy in ASEAN, the Association of Southeast Asian Nations. Immediately after the new national leadership was elected at the Communist Party of Vietnam's Congress in January 2021, President Nguyen Xuan Phuc signed an important document entitled National Strategy on R&D and Application of Artificial Intelligence, or the Strategy Document. The 14-page document outlines plans and initiatives for Vietnam to "promote research, development and application of AI, making it an important technology of Vietnam in the Fourth Industrial Revolution." Vietnam aims to become "a center for innovation, development of AI solutions and applications in ASEAN and over the world" by 2030.

With ambitious goals, the strategy document provides some directions to where Vietnam should go in the next decade. It shows that it follows China's and other Asian countries' footsteps in becoming a techno-developmental state which takes advantage of technological changes for economic developments. While outlining what 16 ministries and the Vietnam Academy of Science and Technology need to do in the next 10 years, the document does not show how other players such as startup founders, civil society, and beneficiaries of AI, common users in Vietnam's AI economy should do. It also has no mention of the role of AI ethics in this development. Without any consideration to important ethical issues such as privacy and surveillance, bias and discrimination, and the role of human judgment, AI development in the country might only benefit a small group of people, and possibly bring harms to others.

In this op-ed we examine three key issues regarding AI development that any country would have to tackle when joining the AI global race: Funding, Talent and Ethics.

**Funding**
AI developments need a large amount of funding coming from a variety of sources such as international venture capital firms, local venture capitalists, government fundings, or companies' own profits. Funding of AI development in Vietnam is lagging behind other Southeast Asian countries. In 2019, Vietnam's AI investment per capita was under $1, while the Southeast Asian leader Singapore has $68 worth of AI investment per capita. Venture capital investment suffered in the first half of 2020 due to COVID-19. However with the government's assistance, there has been some sign of improvement regarding funding in the near future. At the Vietnam Venture Summit 2020, both foreign and domestic investors pledged to invest $800 millions in Vietnam's startup ecosystem. According to Crunchbase, currently, there are 155 venture capital investors with investments in the country.

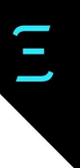

Tech startups received the most investment funding especially in e-commerce, fintech, and AI. The government also provided state funding at the national and city level to encourage entrepreneurship. As a result, the startup ecosystem in cities like Ho Chi Minh City and Hanoi thrived in 2020, before the fourth wave of COVID-19 hit the country in April 2021.

The strategy document outlines the role of the Ministry of Planning and Investment to "to attract venture capital funds to innovative AI start-ups in Viet Nam." The question remains open as to what the plans to bring international capital for domestic technological development are, which specific areas of AI should be the main areas of investment, how would the capital be distributed, and will there be any accountability mechanisms, and who are these entities enforcing accountability?

**Businesses**
The development of AI in Vietnam has been driven primarily by private businesses. The strategy document outlines a push towards digitization and industry 4.0 to create incentives for businesses to become more aware of the potential of data science and AI. Vietnamese companies are still in the early stages of development. Only a few large corporations are prominent in the AI space, notably FPT, Vingroup, Zalo, who have the resources to invest in the research, development, and deployment of AI.

From our conversations with professionals in the space, smaller companies run into a key challenge: product-market fit. To what extent is the Vietnamese public willing to adopt new AI solutions as opposed to existing solutions? As Nam Nguyen, the CTO of an ecommerce company in Ho Chi Minh City, puts it: "If it takes a lot of money to invest in AI, but its economic benefits are not yet significant. Businesses in Vietnam will not jump on this AI bandwagon. Only big companies with extra capital can be in this AI playing field." This problem is also prevalent in countries where AI development is more mature. Many companies in the US, for example, are still struggling to scale AI solutions where AI was developed prior to finding customers who are willing to adopt it. Vietnamese companies also have to compete against foreign or imported AI solutions, and the lack of venture capital investment from both domestic and foreign funds. Future strategy documents should address these particular issues in detail.

**Talent Pool**
There is no shortage of technical talent in Vietnam. However, AI education is relatively new in Vietnam. Most of the tech workforce are still working in outsourcing. The talent pool is young and specialized: young because the majority of the talent pool is IT graduates, working data scientists, or software engineers with few years of experience, and specialized because there is a strong affinity to acquire a technical skill set in niche machine-learning areas (e.g., deep learning, GANs, reinforcement learning)—as opposed to a more general product or project management skill set.

Skilled talent often looks for professional opportunities abroad, where salaries would be drastically higher. Furthermore, these opportunities would enable them to actively participate in the research, development, and deployment of state-of-the-art AI technologies in more AI-mature countries.

Given this landscape, there are challenging conditions to effectively retain talent in Vietnam:

- Salaries have to be competitive, compared to both regional (i.e., Southeast Asia) and global markets.
- There have to be professional development opportunities for talent (e.g., courses, international conferences, etc.) where they can keep up-to-date with the latest trends and practices in AI development.

As Tuan Anh, research scientist at VinAI, claims: "We need to attract Vietnamese scientists back to Vietnam. The key issue is still the salary. It's difficult for a Vietnamese-based company to compete with Google, DeepMind, Microsoft when it comes to salary."

It is worth mentioning that there is also a language barrier to learning AI. As AI education material is predominantly in English, it is crucial to enable young talent with the necessary language learning support in addition to a more technical education in AI. "Students in special programs have English curricula. However, it only accepts 50-60 students per year," says Khoat Than, a professor at Hanoi University of Science & Technology.

**Public Perception of AI and the Missing Ethics Conversation**
In Vietnam, AI is viewed overwhelmingly positively. It is regarded as a catalytic force for economic and technological advancement. In the public mind, the concept of what AI is, how it is used, and who it affects are not as clear. Due to the push towards digitization and industry 4.0, the Vietnamese may see AI only as a tool reserved for industries, where some implementation of natural language processing and computer vision are used to further business objectives. However, these cases are only among a plethora of AI applications that the public have already been using in their everyday life. It might not be immediately obvious that the routes that Grab drivers use to navigate the heterogeneous street network in Saigon are selected by an algorithm, or that the discounted products they see as they log onto e-commerce websites such as Shopee or Tiki may be recommended to them by an algorithm.

This acute awareness is essential because it expands the public's perspective on the role AI plays in benefiting or harming their lives. Amidst the COVID-19 pandemic, "rice ATMs," automatic rice dispensing machines, were invented and deployed in many cities to provide rice both contactless and free-of-charge to low-income communities. What is often left out in the reports of this story is that facial recognition was also used to ensure compliance with the authorities. This critical emphasis on AI involvement is the first step in shaping the conversation around AI and its impacts in Vietnam as a part of the much larger global discourse. The public needs to start having the many necessary conversations about AI around privacy, trust, bias, cybersecurity, and ethics, as well as the nuances, risks, and trade-offs of these aspects (e.g., privacy paradox).

Not only is AI ethics absent from media and public policy discussions, it's also missing in engineering education. Khoat Than, a professor at Hanoi University of Science & Technology notes: "AI ethics at the college level is lacking for engineering students. What students learn at universities are still ethics in computer science." Colleges and universities should invest in not only learning from the learning and teaching of this curriculum, adopting terminologies from the global discourse, they should also invest in doing research, particularly social science that examines societal impacts of technology in Vietnam.

At the governmental level, Vietnam can look to other Asian countries which have drafted national strategy documents that created a framework to make AI "for all." One example is the Responsible AI for All Strategy Document, recently published by Niti Aayog, a premier think-tank by the Indian government. It outlines potential ethical issues that AI would create, and that many of those issues need new legal frameworks that different governmental bodies need to work together to address.

**Conclusion**

Vietnam has entered the early phase of AI development, the strategy document is by no means the last that the government would produce. We recommend the new leadership to consider other aspects of AI development including ethical considerations, legal frameworks, as well as creating partnerships with investors, civil society, and common users to create frameworks to address ethical problems that are native to Vietnamese society. Vietnam should be in conversation with global AI technologists and ethicists as AI development is truly a global phenomenon.

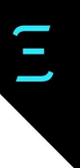

# Other

## Analysis of the "Artificial Intelligence governance principles: towards ethical and trustworthy artificial intelligence in the European insurance sector"

[Original document by EIOPA's Consultative Expert Group on Digital Ethics in insurance]
[Analysis by Philippe Dambly (Senior Lecturer at University of Liège) and Axel Beelen (Legal Consultant specialized in data protection and AI)]

**Overview**: After the 2020 White Paper on Artificial Intelligence and the Proposal for a new regulation on AI of 21 April 2021 published by the European Commission in April 2021, the European Insurance and Occupational Pensions Authority (« EIOPA ») published, on 18 June 2021, a report towards ethical and trustworthy artificial intelligence in the European insurance sector. This is the first AI EU regulation of insurance. The report is the result of the intensive work of EIOPA's Consultative Expert Group on Digital Ethics in insurance. The document aims in particular to help insurance companies when they implement AI applications/systems. The measures proposed in this document are risk-based and cover the entire lifecycle of an AI application.

**Objectives of the report**
The report begins by first identifying the legal framework currently applied to AI in the insurance sector in the EU. Existing legislation should indeed form the basis of any AI governance framework, but the different pieces of legislation need to be applied in a systematic manner and require unpacking to assist organizations understand what they mean in the context of AI. Furthermore, an ethical use of data and digital technologies implies a more extensive approach than merely complying with legal provisions and needs to take into consideration the provision of public goods to society as part of the corporate social responsibility of firms. The existing framework includes, in particular, the 2009 Solvency II Directive, the 2016 IDD Directive, the General Data Protection Regulation ("GDPR") and the 2002 ePrivacy Directive. Good to know is that the EIOPA report uses the definition of AI included in the Proposal for a regulation recently published by the European Commission.

**Six Key Principles**
The 6 key principles identified by the report, along with guidance for insurance companies on how to put them into practice throughout the AI system lifecycle for different applications, are:

1. the principle of proportionality;

2. the principle of fairness and non-discrimination;

3. the principle of transparency and explainability;

4. the principle of human oversight;

5. the principle of data governance of record keeping and

6. the principle of Robustness and Performance.

The high-level principles are accompanied by additional guidance for insurance firms on how to implement them in practice throughout the AI system's lifecycle. For example, in order to implement the principle of proportionality, the report develops an AI use case impact assessment which could help insurance firms understand the potential outcome of AI use cases and subsequently, determine in a proportionate manner the "mix" of governance measures necessary to implement ethical and trustworthy AI systems within their organizations.

With regards to the use of AI in insurance pricing and underwriting, the report includes guidance on how to assess the appropriateness and necessity of rating factors, noting that correlation does not imply causation. From a transparency and explainability perspective, consumers should be provided with counterfactual explanations, i.e. they should be informed about the main rating factors that affect their premium to promote trust and enable them to adopt informed decisions.

Each of the principles is analyzed in the light of the principle of ethics, a transversal principle in AI. The report focuses on private insurance (life, health and non-life insurance). The analysis of the six principles by EIOPA experts is very rich and complemented by multiple graphs and summary tables. The possible issues of big data and AI in social insurance should indeed be analyzed separately. The report considers each principle on the one hand in its generality and then on the other hand deepens it through two or three specific applications of the insurance sector (such as pricing and underwriting, claims management and fraud detection).

Against this background, several initiatives have proliferated in recent years at international, European and national level aiming to promote an ethical and trustworthy AI in our society. EIOPA also recognizes that AI is an evolving technology with an ever-growing number of applications and continuous and in-depth research. This is particularly the case in the areas of transparency and explainability, as well as in the areas of active fairness and non-discrimination principles. As these areas of application and research evolve, EIOPA warns that the recommendations included in the report may therefore need to be revised in the future.

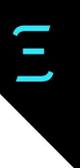

**Legal value of the report?**
But what is the legal value of this report? Do European insurance companies that want to introduce AI processes into their systems (whether opaque or not) have already to take into account its many recommendations on the basis of the well-known principle "satisfy or justify"? Do they have to justify themselves if they want to derogate from it?

The report states, on its page 2, that it was written by members of EIOPA's Consultative Expert Group on Digital Ethics in insurance.

The European regulator created this working group in 2019 to support its work. This group of experts was created in order to help the regulator in its activities, but their views are purely advisory. It will therefore be necessary to wait for the governing bodies of EIOPA to decide on the report to know whether its content could become mandatory for insurance companies or not.

However, given the excellence of the writing and the importance of the subject, there is no doubt that EIOPA will soon approve this document more formally. It is therefore of utmost importance that insurance companies (and others) take note of it and start to implement the 6 analyzed principles that we have just summarized for you.

## The Proliferation of AI Ethics Principles: What's Next?

**[Original article by Ravit Dotan]**

With the rise of AI and the recognition of its impacts on people and the environment, more and more organizations formulate principles for the development of ethical AI systems. There are now dozens of documents containing hundreds of principles, written by governments, corporations, non-profits, and academics. This proliferation of principles presents challenges. For example, should organizations continue to produce new principles, or should they endorse existing ones? If organizations are to endorse existing principles, which ones? And which of the principles should inform regulation?

In the face of the proliferation of AI ethics principles, it is natural to seek a core set of principles or unifying themes. The hope might be that the core set of principles would save organizations from reinventing the wheel, prevent them from cherry-picking principles, be used for regulation, etc. In the last few years, several teams of researchers have set out to articulate such a set of core AI ethics principles.

These overviews of AI ethics principles illuminate the landscape. In addition, they highlight the limitations of the search for unifying themes. They help us see that it is unlikely that a unique set of core principles will be found. And that, even if it is found, universally applying it runs the risk of exacerbating power imbalances.

**Five overviews of AI ethics principles**
Let's start with reviewing five studies that overview the landscape of AI ethics principles. What is their methodology? And what unifying themes do they identify?

**1. The Global Landscape of AI Ethics Guidelines, by Anna Jobin, Marcello Lenca, and Effy Vayena**
Jobin et al. conducted an extensive search and identified 84 papers producing AI ethics principles. The inclusion criteria were as follows: (i) The paper is written in English, German, French, Italian, or Greek. (ii) The paper was issued by an institutional entity. (iii) The paper refers to AI ancillary notions explicitly in its title or description. And (iv) the paper expresses a moral preference for a defined course of action.

The team used manual coding to identify unifying themes and came up with 11 of them: transparency (appeared in 87% of the documents), justice and Fairness (81%), non-maleficence (71%), responsibility (71%), Privacy (56%), beneficence (49%), freedom and autonomy (40%), Trust (33%), sustainability (17%), dignity (15%), and solidarity (7%).

While there is convergence on principles, Jobin et al. point out that there is divergence in how the principles are interpreted, why they are deemed important, and how they should be implemented.

**2. A Unified Framework of Five Principles for AI in Society, by Luciano Floridi and Josh Cowls**
Floridi and Cowl identify six high-profile and expert-driven AI ethics documents. The selection criteria were as follows: (i) The document was published no more than three years before the study. (ii) The document is highly relevant to AI and its impact on society as a whole. And (iii) the document is highly reputable, published by an authoritative and multi-stakeholder organization with at least national scope. In searching for unifying themes in AI ethics principles, the authors draw from the four ethical principles commonly used in bioethics: beneficence, non-maleficence, autonomy, and justice. They identify these same themes as unifying themes for AI ethics principles, and they add a fifth one: explicability.

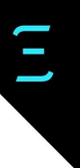

**3. Linking Artificial Intelligence Principles, by Yi Zeng, Enmeng Lu, and Cunqing Huangfu**

Zeng et al. collected 27 proposals of AI ethics principles and grouped them by background: (i) Academia, non-profits, and non-governmental organizations, (ii) government, and (iii) industry. The authors extracted principles from each text and tracked common themes using a keyword search. They started by choosing ten keywords as core terms: humanity, collaboration, share, fairness, transparency, privacy, security, safety, accountability, and AGI (artificial general intelligence). After identifying these core terms, Zeng et al. computationally expanded them, creating lists of related words and expressions. For example, the "accountability" theme was expanded to include "responsibility."  Zeng et al. then performed keyword searches for all the words on the lists, thereby discovering the frequency of appearance of each theme.

The team found that the prominence of each theme depends on the background of the document:

- **Corporations**: The top three themes are humanity, collaboration, fairness, transparency, safety. They mention privacy and security much less than the other institutions and mention AGI and collaboration much more.
- **Governments**: The top themes are privacy, security, humanity. They mention accountability much less than the other kinds of institutions.
- **Academia, non-profits, and non-government**: The top categories are humanity, privacy, accountability. They mention humanity much more than the other kinds of institutions.

**4. Principled Artificial Intelligence: Mapping Consensus in Ethical and Rights-based Approaches to Principles for AI, by Jessica Fjeld, Nele Achten, Hannah Hilligoss, Adam Christopher Nagy, and Madhulika Srikumar**

Fjeld et al. analyzed 36 documents. The selection criteria were as follows: (i) The document represents the views of an organization or institution. (ii) The document was authored by relatively senior staff. (iii) In multi-stakeholder documents, a breadth of experts were involved. (iv) The document was officially published. And (v) the document was written in English, Chinese, French, German, or Spanish.

The authors extracted ethical themes from these documents by manual coding, resulting in eight themes: fairness and non-discrimination (appeared in 100% of documents), privacy (97%), accountability (97%), transparency and explainability (94%), safety and security (81%), professional responsibility (78%), human control of technology (69%), and promotion of human values (69%).

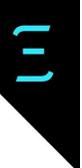

The paper recognizes that other teams of researchers may identify different themes. It also points out that, while there is a convergence on the themes, the principles are implemented differently in different documents.

**5. The Ethics of AI Ethics: An Evaluation of Guidelines, by Thilo Hagendorff**
Hagendorff analyzed 22 major ethical guidelines. The selection criteria were as follows: (i) The document was published no more than three years before the study. (ii) The document refers to more than a national context or has significant international influence. (iii) The document addresses AI ethics generally, not on specific aspects of AI. (iv) The principles are not corporate policies unless they have become well-known through media coverage.

Hagendorff identified eight themes: privacy protection (appeared in 82% of documents), fairness, non-discrimination, justice (82%), accountability (77%), transparency/openness (73%), safety, cyber-security (73%), common good, sustainability, well-being (73%), human oversight, control, auditing (54%), and solidarity, inclusion, social cohesion (50%).

Hagendorff also identified that most of the authors of the documents were men and that only one document included notes on the technical application of the principles, and even those were few and limited.

**Limitations of the search for unifying themes in AI ethics principles**
1. Is it likely to identify a unique set of core AI ethics principles?

As you can see, the different overviews resulted in different sets of unifying themes. Such differences are expected since the overviews differ on their choice of documents, methodology, and application of methodology.

What shall we do with the resulting multiplicity of unifying themes? One approach is to seek unifying themes in the proposed unifying themes. The hope might be to identify the "core" of the core AI ethics principles. However, it seems unlikely that such efforts will yield a unique set. We will once again need to ask: Which sets of unifying themes should be included? Which methodology should be chosen? And how should it be applied? Just as different overviews of AI ethics principles produced different unifying themes, it is likely that overviews of the overviews will produce different sets of "unifying unifying themes."

Therefore, finding a unique set of core AI ethics principles seems unlikely.

2. Suppose that a core set of principles were to be found, should it be universally adopted?

Even if a core set of AI ethics principles were to be found in the existing AI ethics principles, universally adopting it is problematic because of the lack of diversity in the perspectives that generated the principles.

To start, the vast majority of the existing AI ethics documents were written in North America and Europe, as some of the overviews highlight.

Moreover, even within the global north, the perspectives that are represented in the existing AI ethics documents are limited. As Hagendorff identified, the documents were written by men for the most part. We do not have statistics on the participation of other relevant identity categories, such as race, religion, and sexual orientation. However, the authors of the AI ethics documents are probably relatively homogenous along these axes as well.

Further, the voice of those impacted by AI systems is likely to be underrepresented. Zeng et al. suggest that AI ethics documents might reflect the interests and needs of the institutions that authored them. For example, Zeng et al. show that corporations mention privacy and security less than other types of institutions. The reason might be that privacy and security are sensitive topics for them. Similarly, governments mention accountability less, and academia, non-profits, and non-governmental organizations mention collaboration less. The reason might be that these are sensitive topics for them. Which institutions represent the interests and needs of the broader, global public impacted by AI systems? How influential are they in the production of AI ethics principles?

Given the lack of diversity in the perspectives involved in generating AI ethics principles, they seem to represent the preferences and interests of a selected few. If a core set of principles were to be found among them, it would represent these selected few as well. Therefore, universally adopting unifying themes found in the existing AI ethics principles would run the risk of subjugating broad populations to principles that were formulated by a small elite, thereby exacerbating existing power imbalances.

**What's next?**
Overviews of existing AI ethics principles help us see that it is unlikely that a core set of principles will be found and that, even if it were to be found, universally adopting runs the risk of exacerbating power imbalances. That brings us back to the questions with which we started. How do we navigate the proliferation of AI ethics principles? What should we use for regulation, for example? Should we seek to create new AI ethics principles which incorporate more perspectives? What if it doesn't result in a unique set of principles, only increasing the multiplicity of principles? Is it possible to develop approaches for AI ethics governance that don't rely on general AI ethics principles?

## Representation and Imagination for Preventing AI Harms

**[Original article by Sean McGregor]**

The AI Incident Database launched publicly in November 2020 by the Partnership on AI as a dashboard of AI harms realized in the real world. Inspired by similar databases in the aviation industry, its change thesis is derived from the Santayana aphorism, "Those who cannot remember the past are condemned to repeat it." As a new and rapidly expanding industry, AI lacks a formal history of its failures and harms were beginning to repeat. The AI Incident Database thus archives incidents detailing a passport image checker telling Asian people their eyes are closed, the gender biases of language models, and the death of a pedestrian from an autonomous car. Making these incidents discoverable to future AI developers reduces the likelihood of recurrence.

**What Have We Learned?**

Now with a large collection of AI incidents and a new incident taxonomy feature from the Center for Security and Emerging Technology, we have a sense of our history and two statistics are worth highlighting.

**CSET Harm Type Taxonomy**

First, the harm types seen in the real world are highly varied. Existing societal processes (e.g., formal lab tests and independent certification) are prepared to respond to just the 24 percent of incidents related to physical health and safety. While an autonomous car poses obvious safety challenges, the harms to social and political systems, psychology, and civil liberties represent more than half of the incidents recorded to date. These incidents are likely either failures of imagination, or failures of representation. Let's dial into "failures of imagination" with the observation that the majority of incidents are not distributed evenly across all demographics within the population.

**Uneven distribution of harms basis**

Of these "unevenly distributed" harms, 30 percent are distributed according to race and 19 percent according to sex. Many of these incidents could have been avoided without needing an example in the real world if the teams engineering the systems had more varied demographic identity.

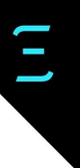

So is representation a panacea to the harms of intelligent systems? No. Even were it possible to have all identities represented, there will still be incidents proving the limits of our collective imagination. For these "failures of imagination", the AI Incident Database stands ready to ensure they can only happen once.

**What is next?**
If you compare the AI Incident Database to the Common Vulnerabilities and Exposures database and the US Aviation Accident Database both have extensive software, processes, community integrations, and authorities accumulated through decades of private and public investment. Comparatively, the AI Incident Database is only at the beginning of its work ensuring AI is more socially beneficial. Three thematic areas are particularly important for building on the early successes of the AIID in its current form. These include,

1) **Governance and Process.** The AIID operates within a space lacking established and broadly accepted definitions of the technologies, incident response processes, and community impacts. Regularizing these elements with an oversight body composed of subject matter experts in the space ensures quality work product and adoption across the corporate and governmental arenas.

2) **Expanding Technical Depth.** The AI Incident Database does not offer one canonical source of truth regarding AI incidents. Indeed, reasonable parties will have well-founded reasons for why an incident should be reported or classified differently. Consequently, the Database supports multiple perspectives on incidents both by ingesting multiple reports (to date, 1,199 authors from 547 publications), and by supporting multiple taxonomies for which the CSET taxonomy is an early example. The AIID taxonomies are flexible collections of classifications managed by expert individuals and organizations. The taxonomies are the means by which society collectively works to understand both individual incidents, as well as the population-level statistics for these classifications. Well structured and rigorously applied AI incident taxonomies have the ability to inform research and policy making priorities for safer AI, as well as help engineers understand the vulnerabilities and problems produced by increasingly complex intelligent systems.

The CSET taxonomy is a general taxonomy of AI incidents involving several stages of classification review and audit to ensure consistency across annotators. The intention behind the CSET taxonomy is to inform policy makers of impacts. Even with the success of the CSET taxonomy for policy makers, the AIID still lacks a rigorous technical taxonomy. Many technical classifications informing where AI is likely to produce future incidents are not currently captured. Identifying unsafe AI and motivating the development of safe AI requires technical classification.

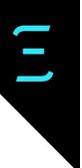

3) **Expand Database Breadth.** The AI Incident Database is built on a document database and a collection of serverless browser applications. This means that the database is highly extensible to new incident types and scalable to a very large number of incident reports. In short, the database architecture anticipates the need to record an increasing number of highly varied and complex AI incidents. While a large number of incidents currently in the database have been provided by the open source community, we know we are currently missing many incidents that should be included in the current criteria. This is one area where everyone has a role in the successful development of our collective perspective into AI incidents.

**How can you help?**
The AI Incident Database will not succeed without your input of incidents and analysis. When encountering an AI Incident in the world, we implore you to submit a new incident record to the database. We additionally ask that software engineers and researchers work with the codebase and dataset to engineer a future for humanity that maximally benefits from intelligent systems.

## Evolution in Age-Verification Applications: Can AI Open Some New Horizons?

**[Original article by Azfar Adib]**

Have you ever been asked to prove your age or verify your identity while you tried to buy any product or service? Many readers here may answer yes to this question. Whether it'd be buying a bottle of wine or signing up for our first driving lesson, age-verification requirements have existed for long. Similarly, a wide range of online applications now requires age-verification before providing service or content access to users. So, in this digital age, this has become a broad area of research and product development.

According to a report by MarketsandMarkets, the global market of identity verification is expected to grow from USD 7.6 billion in 2020 to USD 15.8 billion by 2025 [1]. In its recent directory, "Digital ID & Authentication Council of Canada" (DIACC) has enlisted 73 member companies providing a variety of digital authentication services in a continuously growing market [2].

Age verification is becoming crucial in various dimensions. An increasingly under-18 digital population has heightened pressure from regulatory and non-regulatory entities on service providers to implement more stringent assurances, so that children are kept safe online. Also, age verification is no longer just limited to segregation between adult and under-aged ones, it has other applications too. For instance, estimating the age of unidentified patients in hospital emergencies can be a crucial decision for physicians.

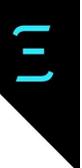

So accuracy and consistency of age verification tools are quite crucial. Equally important is to maintain their ethical standards; particularly privacy, bias-avoidance and data security. Advancement of AI is playing a key role here. Let us have a look that at some key transformation occurring in this arena:

**From paper-based to digital:**
Paper based identification documents (like- driving license, health card or other government issued ID documents) have been the most prevalent verification scheme during in-person transactions. For digital transactions and often during in-person transactions also, their paperless versions (picture of ID documents) are widely used.

The concurrent trend, particularly fueled by the COVID-19 pandemic, is inclining more towards the paperless version and beyond. In a survey carried out by Interac in August 2020 among adult Canadians, the majority of the respondents expressed hygiene concerns around physical ID. They were also concerned about keeping their identity data safe online, and felt it risky to take a picture of a physical ID [3]. Fraudulent activities are indeed becoming a major concern here for both users and service providers. This indicates the need for more secured and seamless digital identification schemes.

**Emergence of biometrics**
Biometric identification, being widely used across the globe, enables automated recognition of individuals through certain physiological characteristics like- facial image, fingerprint, iris, voice, gait, signature, heart signal, gait etc. A research by Juniper has predicted that 95 percent of mobile users will adopt biometrics for authentication by 2025. Interestingly, as this research showed, face biometrics expansion was not slowed down by increased face mask usage during COVID-19 pandemic [4].

Cutting-edge machine learning schemes have been playing a pivotal role in enhancing accuracy in biometrics. However, biometrics are a form of deterministic data, where people are identified by matching with their previously stored record, mostly through supervised algorithms. While facial recognition often claims high classification accuracy (over 90%), these outcomes may not be universal. Some research (like the 2018 "Gender Shades" project research carried out by MIT Media Lab and Microsoft Research) exposes increasing error rate of facial recognition across marginalized demographic groups, with the poorest accuracy consistently found in subjects who are female, Black, and 18-30 years old. Overcoming such bias remains a continuous endeavor for researchers-developers [5].

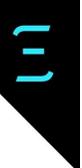

Despite increasing levels of implementation, some people still remain reluctant to use biometrics, especially when it involves their image capturing, voice recording or physical touch. So contactless, non-vocal and non-facial biometric schemes are also being developed as alternates. A very recent example is finger vein-based biometrics to verify a person's COVID-19 vaccination status, being developed by Hitachi and Kyushu University in Japan [6].

**Non-anonymous identification to anonymous identification**
As users are becoming more conscious regarding their data and privacy, they are preferring more controls over where their data is shared. In a July 2021 survey conducted by Liminal, a quarter of consumers stated that they actively avoided using fingerprint or facial biometrics on smartphones due to privacy concerns. Three-quarters of consumers desired to control and revoke access to their identity data any time [7].

Such demand from users puts forth the relatively novel concept of anonymous age verification. In existing technologies, age verification is actually a part of a holistic authentication scheme, where individuals get completely identified based on their previously stored credentials. The purpose of anonymous age verification is to estimate any person's age (or age range) instantly from certain biometrics data, without any prior info about them, thus avoiding their complete identification.

From technological and biological perspectives this remains a daunting task, but research on this is continuing, mostly through a combination of supervised and unsupervised algorithms. In a study jointly carried out in Michigan State University and Beihand University based on GAN (Generative Adversarial Network), researchers have estimated age progression through facial analysis with accuracies of over 99% [8]. Age estimation is also being attempted from ECG (Electrocardiogram), which has recently emerged as a promising biometric scheme [9].

As age-verification tools keep evolving, artificial intelligence will obviously continue its dominant role in this journey of continuous enhancement. Ethical considerations need to be at the core of this progress. And we may not be far away from a world where just an instant biometric signal will be enough to identify our age, without any document or prior data.

**References**
1. https://www.marketsandmarkets.com/Market-Reports/identity-verification-market-178660742.html

2. https://diacc.ca/membership/diacc-members/

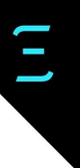

3. https://newsroom.interac.ca/core-principles-for-building-digital-identity-in-canada/

4. https://www.juniperresearch.com/whitepapers/how-to-maximise-mobile-payment-security

5. http://proceedings.mlr.press/v81/buolamwini18a/buolamwini18a.pdf

6. https://www.biometricupdate.com/202110/hitachi-trials-vein-biometrics-to-verify-vaccination-records

7. https://liminal.co/articles/bio-shock-is-biometric-avoidance-more-than-digital-luddism/

8. https://ieeexplore.ieee.org/document/8772083

9. https://www.nature.com/articles/s41467-021-25351-7

## Managing Human and Robots Together – Can That Be a Leadership Dilemma?

**[Original article by Azfar Adib]**

Have we all written poems sometime at some point in our lives? Some of us may have tried so, some may not. Poetry is mostly considered as a gifted talent. Now, can we expect robots to be poets? Well, that is quite possible now.

On 26 November 2021, Ai-Da (the world's first ultra-realistic humanoid robot artist) gave a live demonstration of her poetry [1]. The event took place at the University of Oxford's famous Ashmolean Museum, as part of an exhibition marking the 700th death anniversary of the great Italian poet Dante. Ai-Da produced poems there as an instant response to Dante's epic "Divine Comedy", which she consumed entirely, used her algorithms to analyze Dante's speech patterns, and then created her work utilizing her own word collection.

This is another solid example of the marvelousness of artificial intelligence. In the previous month also (October 2021), Ai-Da participated in the "Forever is Now ", an historic art exhibition in the Giza Pyramids , jointly arranged by UNESCO and the Government of Egypt. Prior to that event, one incident got quite highlighted. While entering into Egypt, , Ai-Da was seized by border agents who feared her robotics may have been hiding covert spy tools ! After 10 days she got released, thanks to continuous effort in diplomatic and other channels for that [2]. Egyptian border guards basically raised security concerns about modem and camera set inside Ai-Da. While the modem could be temporarily removed, removing the camera was not an option as those were crucial components for the robot's vision. "I can ditch the modems, but I can't really gouge her eyes out,"- this was the remark by Aidan Meller, creator of the robot Ai-Da [3].

While this was an unfortunate and discrete occurrence in a particular context, it does spark some interesting perspective. Can the same laws-regulations , which are used for humans, can be applied for robots? That will not be realistic, as we clearly find in the above example. Carrying a camera in a particular place may be prohibited for humans. However, a camera is a basic organ for a robot, without that it can not operate.

Let us consider a more familiar scenario to us, for instance a workplace. We are obviously habituated to human co-workers in workplaces. What will be the scenario if there are robot co-workers? In most organizations, there remains certain policies-procedures for human resources. Will we need separate policies for robots?

For leaders in the organization, it can be an interesting challenge. Effective leaders always try to engage and motivate people to bring out the best from them. Sometimes they need to lead in a diverse multicultural scenario, where also they may lead successfully following some basic values. But how can they adapt to a scenario when they will have both humans and robots in their workforce? Do they need to focus equally on the robots to facilitate performance, or can they just consider those as mechanical devices?Can they evaluate performance and provide feedback to the robots in the same manner they do for their human colleagues? In summary, can they manage humans and robots similarly and simultaneously?

We may need to wait sometime to find the answers to these questions. It might be a better approach to emphasize on how humans and robots can empower each other in a working environment. Few thoughts around that are mentioned below:

**Cross-training**
On-the-job learning is always an effective way to acquire practical working skills. It is actually a career-long process while we continuously learn from our own tasks,  from our colleagues-stakeholders along with traditional/ digital  learning mediums. Learning from robot co-workers can add new dimensions for the human workforce, and vice-versa. However, it also depends on the learning mode.  In a recent study carried out by a group of researchers from MIT and US Air Force, it was shown that AI agents  became frustrating teammates for human players  while playing a card game Hanabi.  The study also showed that human players preferred the classic and predictable rule-based AI systems over complex Reinforcement Learning (RL) based systems. So choosing the right mode of cross-learning between humans and robots is alway quite important.

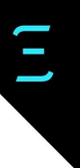

**Increased Productivity**

Several studies have demonstrated increased productivity in certain industries (particularly manufacturing) when humans and robot workers successfully collaborate. A study in MIT demonstrated 85% reduction in manufacturing idle time when people worked collaboratively with a human-aware robot, compared to when working in an all-human team [5]. Another study by "Advanced Robotics for Manufacturing" found that a collaborative approach cuts cycle time by almost two-thirds compared to a fully manual approach [6]. In many scenarios it is quite effective to deploy a hybrid mechanism of shared functionalities between humans and robots, rather than a fully manual or fully automated approach.

**Dealing With Unfavourable Circumstances**

Robots have been used for a long time to perform risky tasks in unfavourable scenarios, which humans can not attempt. Such examples include fire hazards, natural disasters (flood, earthquake, snowfall), nuclear hazards etc. During COVID-19 pandemic, robots often became essential workers by providing crucial support in cleaning-sanitization along with testing-screening procedures [7]. Last year a Canadian construction company deployed a robot dog in one of their sites in Montreal, which they declared as the first robot worker in the world being fully active on a daily basis [8]. So it will always remain quite natural for human workers to trust their robot colleagues to better deal with risky working conditions.

**Preferring Robots as Managers**

It may sound surprising enough, but some studies have shown that people have often preferred robots as their managers rather than human beings ! In a study carried out last year by Oracle and Workplace Intelligence, involving more than 12,000 people across 11 countries ,68% respondent say that during stress or anxiety they would prefer talking to a robot than their own manager. This opinion may not indicate robots as better managers for all. However, it does show that there remains significant improvement scope for managers (or "human managers") to support their employees, or there can be some better practices which can be learnt from robots in this regard.

Time will ultimately tell us whether managing humans and robots together may turn out as a dilemma for leaders. In any scenario, it always remains as a better option for leaders to combine the consistency, precision and speed of robots with the flexibility, creativity and emotional intelligence of human workers.

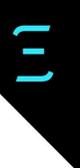

**References**
https://www.ai-darobot.com
https://www.cnn.com/2021/11/27/tech/ai-da-robot-intl-scli-gbr/index.html
https://www.bbc.com/news/world-us-canada-58993682
https://arxiv.org/pdf/2107.07630.pdf
https://dspace.mit.edu/handle/1721.1/63034
https://www.veobot.com/blog/2020/2/18/freemove-in-a-collaborative-palletizing-case-studypart-2
https://www.forbes.com/sites/saibala/2021/01/26/robots-have-become-an-essential-part-of-the-war-against-covid-19/?sh=1207f3ab5ef3
https://montrealgazette.com/news/local-news/spot-the-robot-makes-debut-on-quebec-construction-site
https://www.forbes.com/sites/tracybrower/2020/10/07/study-shows-people-prefer-robot-over-their-boss-6-ways-to-be-a-leader-people-prefer/?sh=445c661a45f4

## "Welcome to AI"; a talk given to the Montreal Integrity Network

**[Original article by Connor Wright]**

**Overview**: In a talk given to the Montreal Integrity Network, Connor Wright (Partnerships Manager) introduces the field of AI Ethics. From an AI demystifier to a facial recognition technology use case, AI is seen as a sword that we should wield, but only with proper training.

**Introduction**
In a talk given to the Montreal Integrity Network, I set about offering an overview of the AI Ethics field and the issues it contains. Stretching from defining AI, to doughnuts, to facial recognition technology (FRT) and current laws, I aimed to provide a fruitful introduction to the field. Like any good presentation, it all starts with some definitions.

**An AI demystifier**
I mentioned how AI is not just limited to the stereotypical view of killer terminator robots or anthropomorphic AI. Instead, depending on how you define AI (which is highly fluid), you can possess the technology in your very hand.

With some central themes of my talk emerging, I set about defining what AI is.

**How can AI learn?**

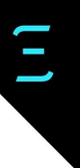

**Machine learning**

As a subset of AI, machine learning provides a more technical explanation of how an AI makes its predictions. Here, machine learning describes AI as algorithms, whereby a human sets the parameters and desirable features of the data that it will receive as input. For example, let's say we're designing an AI with the goal of identifying pictures of cats. I would set the parameters (like the 'rules of a game') for the AI to act towards this goal by identifying the desirable features (such as whiskers). The algorithm will be able to improve its predictions with the more input data (photos of cats) that I provide it.

**Deep learning**

As a subset of machine learning, deep learning algorithms set the desirable features themselves, unlike machine learning itself. Requiring large data sets and computing power, the algorithm goes about learning which features of the data it receives are conducive to achieving its goal.

For example, if we started a doughnut-ranking business, we could set the deep learning algorithm to try and discover the most popular Krispy Kreme doughnut in the world. The data supplied to it would contain every different type of doughnut sold in the world, and it would then set about identifying which features help it best to come to a decision. In this way, it would start eliminating all the doughnuts that Krispy Kreme doesn't supply, considering that popularity means how many are sold, etc.

Machine learning vs deep learningIn this way, we can come to a critical difference between machine learning and deep learning. Here, machine learning requires 'structured data' (data with labels set for the algorithm to learn and then use to identify objects pertaining to its goal). On the other hand, deep learning uses unstructured data (data without labels, which it then creates to identify the objects conducive to its goal). Here, machine learning needs to have it pointed out that cats have whiskers to recognize images of them, whereas deep learning creates its own label for whiskers to identify images as cats.

**AI in business**

**Chatbots**

I touched upon how chatbots (through their use of natural language processing which allows them to recognise words) are a straightforward way to leverage AI technology in business. They help free up resources to be dedicated elsewhere and can act as part of a 24/7 customer response service.

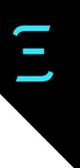

I then tackled the question of what makes a good chatbot? Part of the answer process is the right balance between anthropomorphism and truth. Here, a more 'natural' sounding conversation will help keep customers engaged (such as asking what the customer's name is), but this should not be taken to the point where the chatbot is confused with being a human. Furthermore, a decent knowledge of colloquialisms and the ability to adapt to typos are also key. At times, these two aspects may flumox the chatbot and that's alright, so long as a human agent can be introduced into the conversation quickly.

**The hiring process**
AI can also be used in business to help streamline the hiring process. Top businesses such as Hilton use AI to help deal with the thousands of applications they receive a day. For example, Hilton uses the end to end AI recruitment software of AllyO to help schedule final interview calls for call centre applications. AI in this capacity can again help dedicate finite resources to other tasks which require a more personal touch, instead of having to manually send thousands of emails and schedule thousands of calls. However, this is not without its problems.

**Issues and concerns**

**Problems with learning**
An unfortunately popular avenue for problems within AI is contained within the learning process of AI itself. I mentioned how with machine learning and deep learning, the sourcing of data is vital. Large data sets are required in order to better train the models being designed and create a more accurate product at the end of it. However, how this data is sourced can be problematic, with the consent of the 'producer' of the data (such as a Facebook user) not always being achieved.

**AI bias**
I took AI Bias to be the systematic prioritization of arbitrary characteristics in a model that leads to unfair outcomes. An AI is then biased if it makes decisions that favour or penalize certain groups for reasons that are not valid criteria for decision-making or for factors that are spuriously correlated with the outcome. For example, within predictive policing, an unrepresentative data set fed into the algorithm (such as featuring more criminal records of one race over another) would be more likely to predict a disproportionately higher likelihood to commit a crime for some races over others.

**AI fairness**
I commented on how algorithmic fairness is the principle that the outputs of an AI system should be uncorrelated with certain characteristics such as gender, race, or sexuality. There are many possible ways to consider a model fair. Common approaches include equal false positives across sensitive features, equal false negatives across sensitive characteristics, or minimising "worst group error", the algorithm's number of mistakes on the least represented group. Being able to best evaluate an AI's fairness is to know where and how it went wrong, preventing the proliferation of "black box" algorithms.

**Facial recognition technology (FRT) use case**

**What are the kinds of ethical issues involved in FRT? I was able to mention the following:**

**FRT needs specifics**
FRT does not like any "noise" present when it's trying to study photos (such as loads of different objects in the background). This doesn't bode too well for society, which is a busy place and isn't always posing for a clear photo.

**FRT is liable to bias**
What you give FRT, you get out. As with all AI, the dataset given to the algorithm or software is what it feeds off and learns from (like a newborn baby). For example, if we taught a newborn that the first letter of the alphabet is Z, it will continue to treat it as such. Likewise, if we present the FRT with a dataset that only comprises of white male faces, it will only be able to accurately identify white male faces. This could be a result of human error, or just a lack of awareness of what the database consists of.

**The level of trust**
It has become a 'malpractice' to question decisions made by technology. The technological mindset that has shaped our searches for solutions to problems, I argued, made it almost frowned upon to question the results of the technology. Technology is seen as something that, and with proof, is more accurate than humans can ever wish for. However, statistical accuracy is different to contextual accuracy and the human experience in general. In this sense, technology has definitely proven it can be trusted, but it must also warrant the trust.

**Difficulty with opting out:**
Just like with website cookies, where it's a lot easier just to "Accept all" or opt-in to whatever ad analysis they want to do to get rid of that annoying pop up, it's a lot easier just to consent to the use of FRT. If you do not opt-in, you are very much seen as an inconvenience.

How FRT affects our social behaviourA talk of mine would not be so without a little bit of philosophy. Here, I made sure to mention how the vigilance aspect of FRT could end up affecting how we conduct ourselves in social spaces. We could begin to become hyper aware of how we act, and whether this attends to the standards of the people behind the vigils. Here, we could now be treated as something 'to be monitored' and 'to be tracked' like with FRT being used in Myanmar to track protestors earlier this year.

**The need for industry ethics**
In this section, given the ethical issues at play, I highlighted the need for industry ethics. I mentioned Amazon's Rekognition moratorium in 2020, as well as IBM and Facebook cancelling their ventures into FRT.

**Current laws surrounding AI**
I made sure to give my audience a flavour of what AI regulation is currently active. I specifically mentioned the Bolstering Online Transparency Act in California, as well as the Illinois video analysis law. I also mentioned how age-old laws like the Civil Rights Act can still play a part in algorithmic design, making sure the technology can't discriminate on age, race, marital status etc.

**Between the lines**
My conclusion centres in how while AI is a sword to be wielded, it requires training to be used correctly. AI in the business world can serve as a great tool, but its potential issues are clear for all to see. However, through the study of the AI basics and the technology's current state, I believe that the right training can be provided in order to appropriately utilize AI.

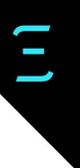

# 2. Analysis of the AI Ecosystem

**Introduction** by Connor Wright, Partnerships Manager, Montreal AI Ethics Institute

Such a question as the one above could generate hours of discussion yet only scratch the surface of the discipline. In recent years, AI Ethics has been moulded into its own field through increasing recognition of its importance and sheer complexity. Nevertheless, essential to note is while AI Ethics is now an established branch, it stretches into and requires input from its surrounding fields. As mentioned in my research summary on how [we cannot have AI Ethics without Ethics](#), the domain does not exist in a vacuum. It should be incorporated as a [standard practice](#) in different fields, rather than a stand-alone subject (rather than something to 'tick off the list'). Its independence requires interdependence on a diverse and multidisciplinary approach, not just on the individual level but also on the organisational.

The whole corporate involvement within AI Ethics is paramount. Establishing the link between those at the top of the tree and those closest to the AI product will be critical to transferring information. Open communication, here, can lead to honest feedback on how the technology is performing. Not only that but involving those closest to the application creates the [empowerment of employees](#), allowing them to take ownership of the solutions they are making. For example, should the company be involved in automated hiring, employees will be able to answer the question confidently, "why are you creating [automated hiring tools](#)?".

Furthermore, involving different business sections can avoid the [broken part fallacy](#), focussing on a broken part in a problem that you think, once fixed, resolves the issue. However, this dismisses the systemic nature of the problem. Hence, before looking for the broken part, we should ask ourselves how it got there. To be sure your decisions can combat this issue, academia and education form a crucial part of the AI Ethics identity.

We have explored how scenario analysis and real case studies are the best way to [educate](#) on the issues involved. Rather than treat AI Ethical issues as theoretical and intangible, basing them on practical considerations helps anchor the practice in the here and now. In this, academia serves as a rich resource from which to encounter the latest necessary concerns. Combating [racial issues](#), acknowledging [nonhuman considerations](#) and more are issues that can be brought to the fore through academia's involvement. What must now be considered is where or, better yet, to whom is the implementation of these considerations dedicated to?

In terms of their job title, an AI Ethicist is the person in charge of putting theory into practice. However, AI ethics isn't just for an ethicist (even if it's semantically similar). Instead, they have to act as engineers, data scientists, computer programmers, and more to best represent their view. To do so may involve [putting the brakes on](#) an initiative that is pressing and will make a company a healthy profit. You are, at times, the [only person saying no](#).

The timeframe to say no should certainly not be limited to purely the design stage. Instead, saying no can also involve decisions on products already present, such as [Facebook shutting down their use of facial recognition](#) amid privacy concerns, alongside [IBM's bias concerns](#). Monitoring the AI product in these early stages after deployment is crucial to mitigating potential problems, especially unintended consequences.

Such decisions often include thinking of [unintended consequences](#), a difficult task made even more so when some unintended consequences may actually be beneficial. For example, [the true visit of dolphins to the waters of Venice](#) in March 2021 thanks to the city's lockdown restrictions. In this way, thinking of all scenarios is hard and the field may be seen as a constant uphill push, but the key is to [work together](#).

As mentioned above, the AI Ethics space is incredibly encompassing and the benefit of including diverse perspectives is apparent for all to see. From our perspective here at MAIEI, nothing expresses this more than our [Learning Community 2021](#) Cohort. Featuring experts in minorities, disability and gender from the corporate, academic and political scene, participants engaged in topics from digital labour to emotional recognition. Alongside the geographical variation in the group, the discussions were both enlightening and rich in content, showing the true potential of an interdisciplinary discussion. If anyone ever asks for a one sentence answer to the question "what does AI Ethics involve?" interdisciplinarity is impossible to omit.

One of the many reasons for this, in my view, is diversity's way of tackling the difficult task if one person being able to consider all perspectives. It may be that the AI Ethicist is to assume the role of designer, programmer and so on, but this is still in desperate need of other peoples' inputs. As I mentioned in my TEDxYouth Talk in November 2020, AI Ethics is to involve allowing the public to become [authors of the AI story](#). That is to say, those directing AI's current trajectory cannot account for every single different experience, meaning without making our own unique experiences known the AI Ethics space will be worse off. It is the space's ability to be enriched by all those involved, no matter what qualification you possess, that makes it truly unique.

I hope the following chapter helps to deepen what I have already mentioned, as well as highlight the important role all can play in this constantly evolving space. It may be a lonely practice at times, but the benefits of working alongside each other to achieve a solution that benefits all is truly enticing. So, please join!

---

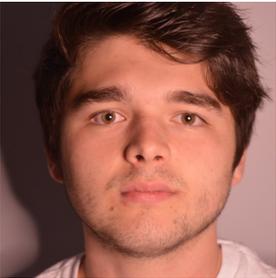

**Connor Wright ([@csi_wright](https://twitter.com/csi_wright))**
Partnerships Manager
Montreal AI Ethics Institute

Connor is the Partnerships Manager at the Montreal AI Ethics Institute and is currently pursuing a philosophy degree at the University of Exeter. He has featured on panels on the topics of facial recognition technology and post-pandemic education, while currently working on the relationship between anthropomorphism and AI. His main passion lies in the form of the cross-section between the Sub-Saharan African philosophy of Ubuntu and AI, stemming from his upbringing in South Africa.

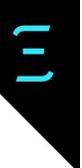

# Go Deep: Research Summaries

## The Values Encoded in Machine Learning Research

[Original paper by Abeba Birhane, Pratyusha Kalluri, Dallas Card, William Agnew, Ravit Dotan, Michelle Bao]
[Research Summary by Abhishek Gupta]

**Overview**: Machine learning is often portrayed as a value-neutral endeavor; even when that is not the exact position taken, it is implicit in how the research is carried out and how the results are communicated. This paper undertakes a qualitative analysis of the top 100 most cited papers from NeurIPS and ICML to uncover some of the most prominent values these papers espouse and how they shape the path forward.

**Introduction**
As we get a higher proliferation of AI in various aspects of our lives, critical scholars have raised concerns about the negative impacts of these systems on society. Yet, most technical papers published today pay little to no attention to the societal implications of their work. And this is despite emerging requirements like "Broader Impact Statements" that have become mandatory at several conferences. Through the manual analysis of 100 papers, this research surfaces trends that support this position and articulates that machine learning is not value-neutral. They annotate sentences in the papers using a custom schema, making open source annotated versions of the papers and their schema and code. The researchers used an inductive-deductive approach to capture the values that are represented in the papers. The researchers found that most of the technical papers focused on performance, generalizability, and building on past work to demonstrate continuity. There has also been a rising trend in the affiliations and funding sources for the authors of these papers to come from Big Tech and elite universities. Through these findings, the authors hope that technical research can become more self-reflective to achieve socially beneficial outcomes.

**Methodology**
The authors choose NeurIPS and ICML as the source of their papers because they have the highest impact (as quantified by the median h-5 index on Google Scholar), and conference submissions are a bellwether to judge where the field is headed and what areas of the field researchers care about and focus their efforts on as they form critical evaluative factors. Many of these papers are written to win approval from the community and the reviewers drawn from that community to achieve these goals. The annotation approach includes examining the content of the paper and creating a justificatory chain with a rating on the degree to which technical and societal problems serve as the motivation for the work. They also pay attention to the discussion of the negative impacts of the work as stated in those papers. The authors acknowledge that this methodology is limited because it is manual and can't be easily scaled. They justify that by pointing out that automated annotation would be limiting in that the categories would be pre-encoded and subtleties will be lost that, for the time being, only human reviewers would be able to pick up on.

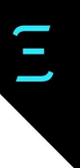

**Findings**

Values related to user rights and stated in ethical principles rarely occurred, if at all, in the papers. Other moral values like autonomy, justice, and respect for people were also noticeably absent. Most of the justifications provided for carrying out research point to the needs of the ML community with no relation to the societal impacts or problems that they are trying to solve. The negative potential of these works was also conspicuous by their absence. Though, some of those omissions might be the result of the taxonomy and awareness of the societal impacts of AI being more recently discussed, especially related to the analysis of the papers from 2008-09.

In terms of performance, the typical characterization for it is average performance over individual data points with equal weighting. This is a value-laden move as it deprioritizes those who are underrepresented in the datasets. In choosing the data itself, building on past work to show improvements on benchmarks is the dominant approach, but this presupposes a particular way of characterizing the world that might not be accurate, as demonstrated with many datasets codifying societal biases. The emphasis on large datasets also moves to centralize power because it shifts control over to those who can accumulate such data and then dictate what is included and what is not. The reliance on ground-truth labels in this case also codifies the assumptions that there is necessarily a correct ground-truth that is single-valued, which is not the case.

Representational harms arise from the excessive focus on generalization capabilities of the systems since it moves to disregard context and enforce a particular view of the world onto the rest of the incoming data in the interest of generalization. Efficiency is another value that is emphasized, but it is rarely discussed in the context of accessibility which could create more equity. Instead, it focuses on using fewer resources and scalability as the values that are the most important. The focus on novelty and building on previous work also entrenches existing positions further with limited critical examination of that prior work in the interest of continuity and demonstrating improvements on existing benchmarks rather than questioning if the benchmarks are representative in the first place.

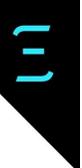

Finally, the increasing influence of Big Tech and elite universities on the state of research is another avenue through which ethical principles are being sidelined and a specific set of values, as highlighted above, are being pushed into the research and development of machine learning. The current trend of treating machine learning as neutral creates insulation for the field in terms of critiquing the values that it espouses, both implicitly and explicitly.

**Between the lines**
This meta-analysis of the state of affairs in machine learning research is a significant contribution to better understanding where we are headed. In particular, the authors' contribution of a data annotation schema and the set of annotated papers will be helpful for future analysis and research. Some developments that I'd like to see building on this work would be finding a way to scale this approach to have a more real-time analysis of where the field is headed and self-correct as we go along. A broader test of the inter-research agreement on the annotations would also be helpful. While the authors do indicate a high degree of inter-reviewer agreement through the Cohen Kappa coefficient, it would be interesting to see how that changes (if at all) when you get a broader set of people to take a look, especially those coming from a variety of fields (even though the authors themselves are quite diverse in the composition of this team).

## Combating Anti-Blackness in the AI Community

[Original paper by Devin Guillory]
[Research summary by Connor Wright]

**Overview**: Racism has the potential to establish itself in every corner of society, with the AI community being no different. With a mix of observations and advice, the paper harbours a need for change alongside the potential for the academic environment to manifest it. While some of the steps involved carry risk, the danger of not doing so is even greater.

**Introduction**
How badly does the AI community need diversity? What is academia's role in this process? Themes surrounding how those in the AI community can help to combat systemic racial injustice are lightly touched upon through an academic's lens. Here, acknowledging how racism permeates into every corner of our society is an important first step. However, further disparities and changes are still left unturned, with the AI community suffering as a result if nothing is done.

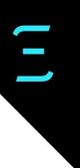

**Discrepancies in resources**
The AI field boasts some invisible barriers to entry between candidates of different ethnicities wanting access. These present themselves in 3 different categories:

**Physical discrepancies.** Disparities in resources, such as computers, are accentuated in a field that often requires large computing power to participate. Also included is the valuable asset of time.
**Social discrepancies.** Many AI jobs are now accessed through social networking and referrals. With there being gaps in physical resources, having access to the networking environments required varies hugely.
**The Measures used** – SAT systems have been seen to disproportionately disadvantage Black students.
In the admissions process, the disparity in terms of social and physical resources becomes even more apparent. Academia's role, and the potential problems it can propagate, become even more critical with its relation to the AI community.

**Academia as a well of information**
Academia and research are a direct feeder into the AI community, so any poisoning found in the field will propagate to other parts of society. In this way, academic faculties will have to wholeheartedly buy into the effort of combating the issues at their root. One way to do this is through feedback.

**The importance of feedback**
Any positive change will need to be grounded in information from those who have gone through the system. The experiences of those who have been discriminated against can provide crucial insights into how the system can change. Without such feedback, the same procedures and the same discrimination will continue to be present.

Given the need for change, the paper also offers views on what can be done. The first of which involves jumping into the unknown.

**Taking risks**
Starting to accept candidates with different applications to years gone by can be a first step toward combating any effects of systemic racism. Looking at other institutions from which students come from or prioritising different characteristics in successful candidates. Not emphasising the need for experience, a clear consequence of having lesser social and physical discrepancies, as much can be one way of doing this.

**Reflecting on your environment**
As a researcher, reflecting on which students you are mentoring can also bring up observations about the current diversity level in your environment. Furthermore, collaborating with different people than you usually can also help promote diversity by experiencing different ways of thinking.

**What diversity brings**
Such alternative views are not the only thing that diversity brings. Having a faculty with varied backgrounds can also allow students of similar experiences to better relate to their professors. Some students may feel that they can only talk about specific problems with professors of similar backgrounds, with such a presence bringing great comfort to the academic experience. However, this is not to say that underrepresented members are solely thought of in the value they add to an in-group institution. Instead, the benefits of diversity should be a consequence of the diverse professors' value.

**Between the lines**
The potential academia possesses to influence the proliferation of discriminatory practices in the AI community is extensive. Seen as the seed for the AI community, taking risks to effect change is a significant step for me. Nevertheless, any form of change will not be easy, especially if it involves self-reflection about your environment. However, not taking these steps could further drive any form of diversity away, which is simply a move that the AI community can no longer afford.

## Achieving a 'Good AI Society': Comparing the Aims and Progress of the EU and the US

[Original paper by Huw Roberts , Josh Cowls, Emmie Hine , Francesca Mazzi, Andreas Tsamados , Mariarosaria Taddeo, Luciano Floridi]
[Research Summary by Andrea Pedeferri]

**Overview**: Governments around the world are formulating different strategies to tackle the risks and the benefits of AI technologies. These strategies reflect the normative commitments highlighted in high-level documents such as the EU High-Level Expert Group on AI and the IEEE, among others. The paper "Achieving a 'Good AI Society': Comparing the Aims and Progress of the EU and the US compares strategies and progress made in the EU vs the US. The paper concludes by highlighting areas where improvement is still needed to reach a "Good AI Society" are "autonomous, interactive, and adaptive".

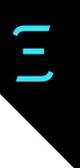

**Introduction**
Recently, we have focused on how designers should implement values in AI systems and how design choices can become more ethical. Now is the time to turn to the role of policymakers and governments in shaping strategies and regulations to tackle the risks and the benefits of AI technologies. The paper Achieving a 'Good AI Society': Comparing the Aims and Progress of the EU and the US compares the strategies and the progress made in the EU vs the US. The paper concludes by highlighting areas where improvement is still needed.

**Key Insights**
At MAIEI, we have looked at some recent research on AI governance in China. Similarly, the current paper gives us the chance to look at how AI-governance is shaping up in the US and in the EU. Why EU and US only? As the authors of the paper explain, "We chose to focus on the EU and US in particular because of their global influence over AI governance, which far exceeds other countries (excluding China). More substantively, the EU and the US make for an interesting comparative case study because of their often-touted political alignment over guiding values, such as representative democracy, the rule of law and freedom." Hence, the goal of the paper is to analyze those governments' "visions for the role of AI in society", and in particular how they intend to develop a 'Good AI Society'.

When making a comparative analysis of ethics-related issues, it is crucial to keep in mind that different societies and cultures may subscribe to different values and have a different understanding of what developing a 'good AI society' actually means. At the same time, the authors rightly point out that, "to consider no values as inherently 'good' is a form of extreme metaethical relativism (Brandt, 2001), according to which nothing of substance can ever be said justifiably about the respective merits of different visions." The authors' view on this is that we should adopt a form of "ethical pluralism". As they explain it, there are "many different valid visions of a 'Good AI Society', but […] each one needs to be underpinned by a set of values that are viewed at national and international levels as desirable. Such values are likely to include democracy, justice, privacy, the protection of human rights, and a commitment to environmental protection." Thus, while they want to avoid adopting ethical absolutism, the authors also voice the need to avoid the trap of ethical relativism.

**AI Governance in the European Union**
In particular since 2016, European countries have worked quite hard to find ways to regulate AI. They have put forward some high level requirements for a trustworthy AI (e.g. robustness, transparency). Most recently, the EU has released the draft Artificial Intelligence Act "which proposes a risk-based approach to regulating AI." As the authors explain, "the EU's long-term vision for a 'Good AI Society', including the mechanisms for achieving it, appears coherent. The vision for governing AI is underpinned by fundamental European values, including human dignity, privacy and democracy. […] The risk-based approach, which combines hard and soft law, aims to ensure that harms to people are minimized, while allowing for the business and societal benefits of these technologies."

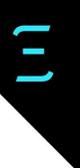

**However, this vision has some notable gaps:**

- No reference is made to the "contribution of training AI models to increased greenhouse gas emissions."
- Not enough to "support collective interests and social values" (e.g. no right to group-privacy)
- Not enough emphasis on "how to address systemic risk". The draft focuses on "the risk to individuals from specific systems" but does not really look at "the potential of AI to cause wider societal disruptions."
- No clear position on "the use of AI in the military domain."
- "The aim of boosting the EU's industrial capacity is hamstrung by the current funding of the EU AI ecosystem, which has been criticized as being inadequate when compared to the US's and China's"
- No clear path to tackle disparities among European countries: "Some Member States, typically in Western Europe, have developed AI strategies, yet this is mostly not the case in Eastern and Southern Europe".

The language around risk and risk-assessment in the draft is "vague and not-committal". "As a result, effective protection from high-risk systems will be largely reliant on interpretations by standards bodies and effective internal compliance by companies, which could lead to ineffective or unethical outcomes in practice."

**The US approach to AI**
In 2016, two broad US reports on AI were released: "Preparing for the Future of Artificial Intelligence' and the 'National Artificial Intelligence Research and Development Strategic Plan'. These and other documents released in the last few years focus mostly on making sure US leadership in AI is preserved while limiting regulatory overreach. When it comes to ensuring a 'Good AI Society', the documents focus on ethical principles such as privacy, fairness and transparency. These principles, however, do not translate into a real AI governance strategy and the tendency is to emphasize self-regulation by industry (as for instance, IBM's recent initiatives to ensure a trustworthy design and use of AI). The problem is that, as the authors point out, "the lack of specific regulatory measures and oversight can lead to practices such as ethics washing (introducing superficial measures), ethics shopping (choosing ethical frameworks that justify actions a posteriori) and ethics lobbying (exploiting digital ethics to delay regulatory measures)."

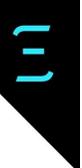

The US strategy is much more hand-on when it comes to international relations that concern the use and development of AI. For instance, the authors explain that "The American AI Initiative states the need to promote an international environment that opens markets for American AI industries, protects the US's technological advantage and ensures that international cooperation is consistent with 'American values'." This has translated into a clear effort to frame AI as a "a defense capability that is essential for maintaining technological, and therefore operational, superiority over the adversary." However, the overall assessment is that "US has not gone far enough in protecting its AI capacities, including its data sets and stopping the illicit transfer of technologies" (e.g. surveillance technology).

**Between the lines**
The paper concludes that when it comes to AI governance, "the EU's approach is ethically superior" as it strives to protect its citizens by implementing regulatory mechanisms. The US has mainly focused on making sure that "the governance of AI" is placed "in the hands of the private sector". What we have not seen discussed in the paper, though, is the role an independent auditing of AI systems could play in both the US and the EU. It would be important to see how and whether independent auditing in AI could be applied in the US' and/or the EU's regulatory systems, and what could be the advantages and disadvantages of doing so.

## Responsible Use of Technology: The IBM Case Study

[Original paper by World Economic Forum and the Markkula Center for Applied Ethics at Santa Clara University]
* Conflict of interest:  Marianna is currently collaborating with a research team at IBM led by Francesca Rossi
[Research Summary by Marianna Ganapini]

**Overview**: In recent summaries, we have stressed the fact that at times private companies have taken the lead in providing guidelines for the responsible use and development of AI technologies. The World Economic Forum and the Markkula Center for Applied Ethics at Santa Clara University are collaborating in surveying the work of these companies, and they recently have focused on IBM. In this summary, we will go through the main points of their most recent white paper, discussing the importance and novelty of the approach taken by IBM.

**Introduction**
Some tech companies have taken the lead in providing guidelines for the responsible use and development of AI technologies, especially where governments and public institutions are failing to establish clear guidelines and regulations. The World Economic Forum and the Markkula Center for Applied Ethics at Santa Clara University are collaborating in surveying the work of these companies, and in this paper, they have focused on IBM. In what follows, we will go through the main points of their recent white paper discussing the importance and novelty of the approach taken by IBM.

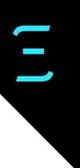

**Key Insights**

One of the key moments in IBM's development of AI Ethics strategy was the publication of IBM 5 key commitments to "accountability, compliance and ethics in the age of smart machines":

- Creating an IBM AI Ethics Board to "discuss, advise and guide (eventually govern) the ethical development and deployment of AI systems (by IBM and its clients)" — since 2019 it is co-chaired by Christina Montgomery (IBM Chief Privacy Officer) and Francesca Rossi (IBM Global Leader of AI Ethics)
- Designing a "company-wide educational curriculum on the ethical development of AI"
- Creating the IBM "Artificial Intelligence, Ethics and Society program": "a multidisciplinary research programme for the ongoing exploration of responsible development of AI systems aligned with the organization's values"
- Establishing an ongoing "participation in cross-industry, government and scientific initiatives and events on AI and ethics"
- Establishing a "regular, ongoing IBM-hosted engagement with a robust ecosystem of academics, researchers, policy-makers, non-governmental organizations (NGOs) and business leaders on the ethical implications of AI

How are these commitments being implemented in practice? To understand some of the recent key decisions of the IBM AI Ethics Board, we need to first zoom in on the fact that Trust & Trustworthiness are central concepts to the current IBM strategy, and they emerge out of 5 "pillars of trust": Explainability, Fairness, Robustness, Transparency, Privacy.

These are the key values that IBM pledges to follow in its design strategies, starting with the creation of ethics-sensitive technologies followed by close monitoring of downstream effects of the use of these technologies.

These pillars have been tackled at IBM by first developing some technical tools to ensure trust for their clients and for the public at large. Let's see what they are:

- **Explainability**: when AI is involved in a decision-making process, the reasons for the decisions are to be made available. IBM AI Explainability 360 toolkit aims at tackling some of the technical challenges of ensuring explainability

- **Fairness**: together with the IBM Cloud Pak for Data, the IBM AI Fairness 360 toolkit for detecting biases in AI is a tool that could help avoid discrimination and unequal treatment in the design of AI technologies.
- **Robustness**: to shield from adversarial attacks the IBM Adversarial Robustness 360 toolbox is a valuable defense tool
- **Transparency**: the IBM AI FactSheets 360 and the Uncertainty Quantification 360 toolkit are ways for AI developers to document key aspects of their models to ensure transparency
- **Privacy**: IBM pledges that "[o]nly necessary data should be collected, and consumers should have clear access to controls over how their data is being used."

Going beyond the technical tools, to operationalize those 5 pillars and ensure trust, IBM adopts an "ethics by design" approach. In our understanding, that should mean that the above values or pillars are embedded in the design of AI technology not only in the initial design phase but also in considering the downstream consequences and potential misuse of that technology. In some cases, that may require a company to re-design or change the technology altogether.

IBM seems committed to embedding values in this way, as shown, for instance, by their willingness to re-think the use and production of their facial recognition software. More specifically, according to the report, IBM is taking company-wide practical steps to implement its "ethics by design" approach. Some of these important steps are:

- Internal curriculum-development and repeated training activities to promote ethics sensitive design (IBM Garage)
- Fostering diversity, inclusion and equality in the workplace and at the HR level
- Stakeholders engagement with the goal of bringing together "AI corporations with civil society groups for conversations on the best practices for beneficial AI" (e.g. the collaboration with PAI)
- Stakeholders engagement through partnerships with universities (e.g. Notre Dame-IBM Tech Ethics Lab)
- Involvement in governmental discussion on AI (e.g., Francesca Rossi's involvement in the European Commission's High-Level Expert Group on AI)
- Promoting AI for social good (e.g. the Science for Social Good initiative; IBM signed the Vatican's Rome Call for AI Ethics in 2020).

These are some of the concrete initiatives taken by IBM to drive the company toward a more ethical and trustworthy design and use of AI. They can't do it alone, though: private companies can be fully trustworthy only if they are part of a broader value-sensitive environment that includes independent oversight organizations, a clear legislative framework, and an engaged and informed public.

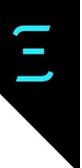

**Between The Lines**
IBM has taken the lead in setting the standards for private corporations' involvement in promoting AI Ethics, trying to learn from their past mistakes while looking for new ways to ensure a trustworthy AI for their clients and society at large. We hope to see more of that kind of engagement and commitment from the private sector going forward. More broadly, we believe that to reach a trustworthy AI we need to put more effort into the following:

- Precise and targeted government's AI regulations
- A private sector that genuinely committed to a trustworthy AI
- Independent oversight organizations & frameworks (e.g., Independent audit systems)
- Civic competence-promoting initiatives and organizations

## U.S.-EU Trade and Technology Council Inaugural Joint Statement – A look into what's in store for AI?

[Statements and Releases from The White House Briefing Room]
[Research Summary by Angshuman Kaushik]

**Overview**: This write-up focuses on the conclusions reached at the inaugural meeting of the U.S.-EU Trade and Technology Council ("TTC") held in Pittsburgh, Pennsylvania on September 29, 2021. It concerns only those aspects in the meeting that deal with the use of AI, its effects, and the areas of cooperation envisioned, going forward.

**Introduction**
The timing of the inaugural meeting of the TTC couldn't have been better, with the Facebook saga unfolding before the world. The reason I say this is because, the TTC's Inaugural Joint Statement discusses outcomes in respect of five key areas, one of them being development and implementation of AI systems that are trustworthy, and those that respect universal human rights. Set in motion by President Joe Biden, President of the European Commission Ursula von der Leyen and European Council President Charles Michel at the US-EU Summit in June 2021, TTC comprises 10 Working Groups, with AI falling within the Technology Standards Working Group. The importance of the TTC can be gauged by the fact that both the US and the EU have appointed some of their senior officials to spearhead it. The US side is led by Secretary of State Antony Blinken, US Trade Representative Katherine Tai and Secretary of Commerce Gina

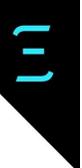

Raimondo. EU Commissioner for Competition Margrethe Vestager and Commissioner for Trade Valdis Dombrovskis are representing Brussels.

**Statement on AI**
Coming to the material contents of the Joint Statement, with regards to AI, it talks about the belief of both the sides in the potential of AI to bring about substantial benefits to their respective societies, and tackle various challenges. One significant aspect of the joint statement is the acknowledgement from both US and the EU of the risks associated with AI-enabled technologies that are either not developed and deployed responsibly, or misused. Further, they assert their willingness and intention to develop and implement trustworthy AI, and their commitment to a human-centered approach that buttresses shared democratic values and respects universal human rights. The key here is the choice of words i.e., trustworthy AI and human-centered approach. In fact, the EU has already shown the way to the world, by putting out the Proposal for a Regulation of the European Parliament and of the Council Laying Down Harmonized Rules on Artificial Intelligence and Amending Certain Union Legislative Acts on April 21, 2021 ("Artificial Intelligence Act"). The aforementioned Proposal delivers on the political commitment by President Ursula von der Leyen that the Commission would bring about legislation for a coordinated European approach on the human and ethical implications of AI. In pursuance of the above, the Commission published the White Paper on Artificial Intelligence – A European approach to excellence and trust on February 19, 2020. The White Paper sets out the policy routes on how to achieve the twin goals of promoting the uptake of AI and of addressing the risks associated with certain uses of such technology.

This proposal aims to implement the second goal for the development of an ecosystem of trust by proposing a legal framework for trustworthy AI. Moreover, both the US and the EU are the founding members of Global Partnership on AI, which brings together a group of like-minded partners seeking to support the responsible development of AI that is based on human rights and societal benefit. The joint statement also opposed and reflected its significant concern, regarding the social scoring systems deployed by authoritarian governments (without naming any country, in particular) with the aim to implement social control at scale. They reiterated that these systems pose threats to fundamental freedoms and the rule of law, which includes silencing speech, punishing peaceful assembly and unlawful surveillance. The statement also emphasized that the policy and regulatory measures should be based on and proportionate to the risks posed by the different uses of AI. Moreover, the US noted the European Commission's proposal for a risk-based regulatory framework for AI and the fact that, EU supports a number of research projects on trustworthy AI, as part of its AI strategy. The EU also noted the US government's development of an AI Risk Management Framework, as well as ongoing projects on trustworthy AI as part of the US National AI Initiative. Further, the joint statement reiterated the commitment of both the sides to work together to foster responsible stewardship of trustworthy AI and provide research-based methods to advance trustworthy approaches to AI that serve people in beneficial ways.

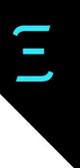

**Areas of cooperation**

The statement also mentions areas of cooperation between both the US and the EU. The objective is to translate shared common values into tangible action and cooperation for mutual benefit. It goes without saying that the commitment to the responsible stewardship of trustworthy AI seems to be on top of the agenda for both US and the EU, as they seek to develop a mutual understanding on the principles underlying 'trustworthy and responsible AI'. In this regard, they intend to discuss measurement and evaluation tools and activities to assess the technical requirements for trustworthy AI, concerning, for example, accuracy and bias mitigation. Further, they also expressed their desire to collaborate on projects furthering the development of 'trustworthy and responsible AI' to explore better use of machine learning and other AI techniques towards desirable impacts. The above quite clearly points toward the fact that, both sides are concerned about the damaging effects of certain algorithms on society. Both the US and EU also expressed their intention to explore cooperation on AI technologies designed to enhance privacy protections, in full compliance with their respective rules, coupled with additional areas of cooperation to be defined through dedicated exchanges. Further, they also stressed on upholding and implementing the OECD Recommendation on AI. They also intend to jointly undertake an economic study examining the impact of AI on the future of their workforces, with attention to outcomes in employment, wages and dispersion of labor market opportunities. They also expressed their willingness to inform approaches to AI consistent with an inclusive economic policy that ensures that the benefits of technological gains are broadly shared by workers.

**Between the lines**

As stated above, this meeting assumes a lot of significance, considering the developments taking place globally against the detrimental effects of AI on individuals in particular, and the society, in general. Undoubtedly, it is a herculean task to resolve the issue of bias and discrimination creeping into the AI systems, but someone has to make a start somewhere. Further, the aforesaid problem is exacerbated by interpretability and explainability issues associated with certain 'black box' algorithms. Governments around the world, either individually, or in a collective manner (as is the case here) can only enact policies and laws to enforce responsible and trustworthy AI, but self-regulation by the entities accountable for the development and deployment of AI is crucial. The latest example of government stepping in is the non-binding resolution passed by the European Parliament on October 6, 2021 calling for a ban on police use of facial recognition technology in public places and on predictive policing. It also called for a ban on private facial recognition databases in law enforcement and supported the European Commission's recommendations to put an end to social scoring systems. As far as self-regulation by the aforesaid entities are concerned, Francis Haugen (Facebook whistleblower), during her testimony before the Senate Subcommittee on Consumer Protection, Product Safety and Data Security within the Committee on Commerce, Science and Transportation stated – "they (she was referring to Facebook) have a 100% control over their algorithms", which means that companies who make use of AI, have 100% control over their algorithms. Therefore, it is imperative that both the government and the corporates need to join hands in order to arrest the further occurrence and spread of bias, discrimination, hatred and disintegration of the society caused by "toxic algorithms".

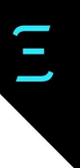

## Public Strategies for Artificial Intelligence: Which Value Drivers?

[Original paper by Gianluigi Viscusi, Anca Rusu, and Marie-Valentine Florin]
[Research Summary by Connor Wright]

**Overview**: Different nations are now catching on to the need of national AI strategies for the good of their futures. However, what really drives AI and whether this is in line with the current fundamental values at the heart of different nations is a different question.

**Introduction**
With the release of the UK Government's AI strategy, the question of what is driving its design comes to mind. AI has long been touted as a sure way to improve public service delivery and administration efforts. However, questioning what kind of values are currently driving government initiatives has not been too visited. Are the private and public spheres motivated by different values? Does humanity itself enter into the AI conversation? Is AI being treated in too much of an instrumental maner? I will now explore these three questions in turn.

**Do the private and public sectors differ?**
With different end goals and different audiences, the private and public sectors can differ on many aspects of governance. However, "accountability, expertise, reliability, efficiency, and effectiveness" (p.g. 2) were found to be held in common between the two spheres. Other aspects like "professionalism", "efficiency", "openness" and "inclusion" have been common to both as well (p.g. 2). However, I believe what can differentiate the two are the interpretations of the values listed. Efficiency will differ from business to business, especially in terms of what is deemed as the threshold. Moreover, "inclusion" (p.g. 2) and who is subjected to it can vary widely in terms of extent, depth and what inclusion entails. Being included in the AI governance could range from participating in its design or just occasionally being informed on the changes being made to an AI. Most of the time, a lack of inclusion is witnessed.

**A lack of focus on humanity**

Principles such as "transparency", "privacy" and "responsibility" are mentioned in AI governance strategies more so than "Human dignity" (pg. 4). AI is often touted in the media as the golden ticket towards greater 'prosperity' for humanity, but humanity's role in this prosperity is often left untouched. To illustrate, there were fewer risks and challenges identified in the AI strategies of each country than there were values. So, what are these values contributing towards if governments do not articulate the problem? From my reading of the report, AI is far more accepted as a tool than as a socially designed and sensitive technology.

**Instrumental normativity over social normativity**

Phrases in the shape of privacy, efficiency and transparency often are preferred by governments instead of words such as democracy. One way to express this substitution is by acknowledging the tension between improving administrative features and simultaneously focusing on societal issues. Resultantly, the values held at the core of constitutions and manifestos are often sidelined when thinking about AI, like with the lack of reference to democracy. Questions then arise of whether values serve as sound guiding principles for AI at all, or are they another example of a tokenistic gesture.

**Between the lines**

What is crucial for me to consider is how what is valued varies so differently across different nations. As explained in one of our event summaries on AI in different national contexts, the cultural interpretation of different values can vary widely. However, fragmentation is not always a bad thing if each country adheres appropriately to their given interpretation.

The next topic of debate comes from how the fundamental values that the entities in the report expounded were not concurrent with the values shown in the AI strategies. Committing to AI values beyond simply writing them down is a known struggle within the space. For me, writing down the value and what problem this value, if championed correctly, helps to prevent is one way of contextualizing and focussing efforts on the issue at hand. The more context and grounding in what AI is being deployed to do, the more the equally valuable social values we hold can appear.

## Artificial Intelligence: the global landscape of ethics guidelines

[Original paper by Anna Jobin, Marcello Ienca, Effy Vayena]
[Research Summary by Avantika Bhandari]

**Overview**: Many private companies, research institutions, and public sectors have formulated guidelines for ethical AI. But, what constitutes "ethical AI," and which ethical requirements, standards, and best practices are required for its realization. This paper investigates whether there is an emergence of a global agreement on these questions. Further, it analyzes the current corpus of principles and guidelines on ethical AI.

**Introduction**

There has been continuous and vigorous debate around AI technologies and their transformative impact on societies. While most studies establish that AI brings many advantages, they also underline numerous ethical, legal, and economic concerns primarily relating to human rights and freedoms. Then there are concerns that AI may "jeopardize jobs for human workers, be exploited by malicious actors, or inadvertently disseminate bias and thereby undermine fairness."

National and international organizations are looking at solutions to tackle the risks associated with the development of AI by developing ad hoc expert committees. Examples include: the High-Level Expert Group on Artificial Intelligence appointed by the European Commission, the Advisory Council on the Ethical Use of Artificial Intelligence and Data in Singapore, and the select committee on Artificial Intelligence of the United Kingdom (UK) House of Lords. Private companies like Google, and SAP have also released their principles and guidelines on AI. Professional associations and non-governmental organizations such as the Association of Computing Machinery (ACM), Access Now, and Amnesty International have come forward with their own recommendations. Active involvement of different stakeholders in issuing AI policies and guidelines proves the strong interest in shaping the ethics of AI in order to meet their respective priorities.

**The researchers pose the following questions:**
- Are these groups converging on what ethical AI should be, and the ethical principles that will determine the development of AI?
- And, if they diverge, then what are these differences, and can they be reconciled?

**Results**

The researchers conducted a review of the existing corpus of guidelines on ethical AI. The search identified 84 documents containing ethical principles or guidelines for AI.

- Data reveal a significant increase in the number of publications, with 88% having been released after 2016.
- Most documents were produced by private companies ( 22.6%) and governmental agencies respectively (21.4%), followed by academic and research institutions (10.7%),

- intergovernmental or supra-national organizations (9.5%), non-profit organizations, and professional associations/scientific societies ( 8.3% each), private sector alliances (4.8%), research alliances ( 1.2%), science foundations ( 1.2%), federations of worker unions (1.2%) and political parties (1.2%). Four documents were issued by initiatives belonging to more than one of the above categories and four more could not be classified at all (4.8% each).
- In terms of geographic distribution: a significant representation came from more economically developed countries (MEDC). The USA (23.8%) and the UK (16.7%) together account for more than a third of all ethical AI principles, followed by Japan (4.8%), Germany, France, and Finland (3.6% each).
- Ethical values and principles: Eleven (11) overarching ethical values and principles have emerged from the content analysis. These are by frequency of the number of sources in which they appeared: transparency, justice, and fairness, non-maleficence, responsibility, privacy, beneficence, freedom and autonomy, trust, dignity, sustainability, and solidarity.
- The researchers found that no single ethical principle was found common to the entire corpus of document, however, an emerging convergence was found around the following principles: transparency, justice and fairness, non-maleficence, responsibility, and privacy.

**Discussion**
- The proportion of documents issued by the public and private sectors indicate that ethical challenges of AI concern both the stakeholders. However, there is a notable divergence in the solutions proposed.
- Further, there seems to be an underrepresentation of geographic areas such as South and Central America, Africa, and Asia which insinuates that the international debate on AI may not be happening in equal measures. It seems that MEDC is shaping this debate, which raises concerns about "neglecting local knowledge, cultural pluralism and global fairness."
- There is an emergence of a cross-stakeholder convergence on promoting the ethical principles of transparency, justice, non-maleficence, responsibility, and privacy. However, the thematic analysis shows divergences in four (4) areas: 1) how ethical principles are interpreted, 2) why they are deemed important, 3) what issue, domain or actors they pertain to, and 4) how they should be implemented. It remains ambiguous as to which ethical principle should be prioritized, how the conflicts between the principles should be resolved, the enforcement mechanism on AI, and how institutions and researchers can comply with the resulting guidelines.

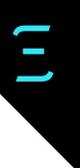

The research indicates an emerging consensus around the promotion of some ethical principles, however thematic analysis provides a complicated narrative as "there are critical differences in how these principles are interpreted as well as what requirements are considered to be necessary for their realization."

**Between the lines**
It seems that the different stakeholders seem to converge on the importance of transparency, responsibility non- non-maleficence, and privacy for the development and deployment of ethical AI. However, the researchers also call for underrepresented ethical principles such as solidarity, human dignity, sustainability that would most likely result in better articulation of the ethical landscape for AI. Moreover, it is high time there is a shift in focus from principle-formulation into actual practice. Finally, a global scheme for ethical AI should "balance the need for cross-national and cross-domain harmonization over the respect for cultural diversity and moral pluralism."

**NOTE**: The researchers acknowledge limitations in the study. First, the guidelines and soft-law documents are an example of gray literature, and thereby not indexed in conventional databases. Second, a language bias may have skewed the corpus towards English results. Finally, given the rapid frequency of publication, there is a possibility that new policies were published after the research was completed.

## UK's roadmap to AI supremacy: Is the 'AI War' heating up?

[[Document](#) from the UK Government]
[Research Summary by Angshuman Kaushik]

**Overview**: The UK's first National Artificial Intelligence Strategy was presented to the Parliament by Nadine Dorries, Secretary of State for Digital, Culture, Media and Sport by Command of Her Majesty on September 22, 2021. The highlight of the strategy is the highly ambitious ten-year plan 'to make Britain a global AI superpower'. Further, according to Dorries, 'this strategy will signal to the world the UK's intention to build the most pro-innovation regulatory environment in the world'.

**Introduction**
The celebrated Kai-Fu Lee in his seminal book, 'AI Super-Powers China, Silicon Valley, and the New World Order', observed, 'Harnessing the power of AI today – the "electricity" of the twenty-first century – requires four analogous inputs: abundant data, hungry entrepreneurs, AI scientists, and an AI-friendly policy environment'. The UK which has always been at the vanguard of AI – from Alan Turing to the present – seems to have taken the perspicacious words of 'the indisputable rock star of China's technology scene' earnestly, and has decided to follow them to the hilt, going forward. In fact, in the recent past, it has come up with a surfeit of Reports, Policy Papers etc., all touching upon the various facets of AI. One amongst them, is the House of Lords Select Committee Report on Artificial Intelligence published in 2018 titled 'AI in the UK: ready, willing and able?', followed by the House of Lords Liaison Committee Report called the 'AI in the UK: No Room for Complacency.', published in 2020. There are several other reports also. This strategy, which is the latest entrant into the scene, owes enormously to the AI Council's 16 recommendations to help the government develop a UK National AI Strategy. In fact, the strategy acknowledges the said fact by mentioning that the Council has played a central role in gathering evidence to inform its development. It further goes on to say that the government remains grateful to the AI Council for its continued leadership of the AI ecosystem. The strategy centers around three pillars with short, medium and long term timelines for achieving the delineated tasks.

**The Three Pillars**

**Pillar 1: Investing in the long-term needs of the AI ecosystem**
The first pillar focuses on investing in the long-term needs of the AI ecosystem. Some of the key ways in which the government intends to achieve the same is outlined below:

- continue to develop the brightest and the most diverse workforce, considering that UK suffers from AI skills gap;
- United Kingdom Research and Innovation (UKRI) will support the transformation of the UK's capability in AI by launching a National AI Research and Innovation (R&I) Programme;
- continue to use Official Development Assistance to support R&D partnerships with developing countries;
- publish a policy framework in autumn 2021, setting out its role in enabling better data availability in the wider economy;
- consult on the potential value of and options for a UK capability in digital twinning and wider 'cyber-physical infrastructure';
- continue to publish authoritative, open and machine-readable data on which AI models for both public and commercial benefit can be trained.
- the Office for AI will also work with teams across government to consider what valuable datasets government should purposefully incentivize, that will accelerate the development of AI applications;

- to better understand the UK's future AI computing requirements, the Office for AI and UKRI will evaluate the UK's computing capacity needs to support AI innovation, commercialization and deployment;
- continue to evaluate the state of funding specifically for innovative firms developing AI technologies across every region of the UK; and
- include provisions on emerging digital technologies, including AI, in the government's new trade deals.

**Pillar 2: Ensuring AI benefits all sectors and regions**
Significant modes to be pursued include:

- the Office for AI will publish research later this year into the drivers of AI adoption and diffusion;
- to stimulate the development and adoption of AI technologies in high-potential, low-AI maturity sectors the Office for AI and UKRI will launch a programme;
- the Office for AI will work closely with the Office for Science and Technology Strategy and government departments to understand the government's strategic goals and where AI can provide a catalytic contribution;
- through its leadership in international development and diplomacy, the government of UK will work to ensure that international collaboration can unlock the enormous potential of AI to accelerate progress on global challenges, from climate change to poverty;
- launch a draft National Strategy for AI in Health and Social Care in line with the National AI Strategy. This will set the direction for AI in health and social care up to 2030, and is expected to launch in early 2022; and
- publish the Defence AI Strategy, which will include the establishment of a new Defence AI Center.

**Pillar 3: Governing AI effectively**
Some of the tasks to be taken up are as stated below:

- the Office for AI will develop UK's national position on governing and regulating AI, which will be set out in a White Paper in early 2022;
- the government will continue to work with its partners around the world to shape international norms and standards relating to AI, including those developed by multilateral and multistakeholder bodies at global and regional level;
- to support the development of a mature AI assurance ecosystem, the CDEI is publishing an AI assurance roadmap;

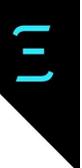

- the government is working with the Alan Turing Institute to update the guidance on AI Ethics and Safety in the Public Sector in order to provide the public servants with the most current information about the state of the art in responsible AI innovation;
- the Ministry of Defence has rigorous codes of conduct and regulation which uphold responsible AI use, and is working closely with the wider government on approaches to ensure clear alignment with the values and norms of the society;
- to ensure that citizens have confidence and trust in how data is being processed and analyzed to derive insights, the Central Digital and Data Office (CDDO) is conducting research with a view to developing a cross-government standard for algorithmic transparency in line with the commitment in the National Data Strategy; and
- the Office for AI will coordinate with cross-government processes to accurately assess long term AI safety and risks, which will include activities such as evaluating technical expertise in government and the value of research infrastructure.

**Next steps**

The present strategy talks about publication of a plan to execute the vision set out therein, in the near future. Further, mechanisms will be put in place to monitor and assess progress. The government also intends to publish a set of quantitative indicators, given the 'far-ranging' and 'hard-to-define' impacts AI will have on the economy and the society. These indicators will be published separately and at regular intervals to provide transparency and accountability. It is the Office for AI that will be responsible for overall delivery of strategy, monitoring progress and enabling its implementation across government, academia, industry and civil society.

**Between the lines**

On March 12, 2021 Oliver Dowden, the then Secretary of State for Digital, Culture, Media and Sport and the predecessor of Nadine Dorries announced the Government's Ten Tech Priorities. One of the priorities also included a commitment to publish a National AI Strategy (the present strategy). Dowden said 'Unleashing the power of AI is a top priority in our plan to be the most pro-tech government ever. The UK is already a world leader in this revolutionary technology and the new AI Strategy will help us seize its full potential – from creating new jobs and improving productivity to tackling climate change and delivering better public services'. Now, that the strategy is published, we will have to wait and watch the execution of the vision contained in it.

**Putting AI ethics to work: are the tools fit for purpose?**

[Original paper by Jacqui Ayling and Adriane Chapman]
[Research Summary by Ravit Dotan]

**Overview**: This paper maps the landscape of AI ethics tools: It develops a typology to classify AI ethics tools and analyzes the existing ones. In addition, the paper identifies two gaps. First, key stakeholders, including members of marginalized communities, under-participate in using AI ethics tools and their outputs. Second, there is a lack of tools for external auditing in AI ethics, which is a barrier to the accountability and trustworthiness of organizations that develop AI systems.

**Introduction**
As more and more AI ethics tools are developed, it becomes difficult to get a handle on the terrain. This paper addresses the challenge by mapping and analyzing the existing AI ethics tools (as of the end of 2020).

The authors conducted a thorough search and identified 169 AI ethics documents. Of those, 39 were found to include concrete AI ethics tools. Each of the 39 tools was classified using the following questions: (i) What sector are the document's authors from? And what sector are the users of the tools from? (ii) Which stakeholders would either use the tool or engage with the results? (iii) What type of tool is it? Which strategy does it employ? (iv) Were these tools for use internally, or did they have external elements? (v) In which stage in the AI production and use chain was the tool used? (vi) Was the tool appropriate for addressing the model, data, or both?

The paper presents statistics characterizing the tools using these questions. Among its findings, the paper uncovers that a wide stakeholder base, involving customers, the broader public, and the environment, is typically not a part of AI ethics evaluation processes. Moreover, the paper finds that almost all AI ethics tools are used internally, without external oversight. The authors emphasize that these characteristics stand in the way of accountability and trustworthiness of organizations that develop AI systems.

**The map of AI ethics tool landscape**
The paper divides AI ethics tools into three categories:

**Impact assessment tools**
Impact assessment is a fact-finding and evaluation process that precedes or accompanies the production of artifacts, systems, or research. Ex-ante assessments are used in the use case development and testing stages. Ex post assessments are used post-deployment, in the monitoring stage, to capture the impacts of the system. The most predominant tools for impact assessments in AI ethics are checklists and questionnaires.

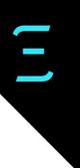

**Technical and design tools**
These tools are typically developed by the ML community. Some of them are computational, e.g., computationally identifying and mitigating bias. Others are design processes, e.g., workshop-style events for raising awareness in design teams or participatory design processes. These tools are used along the whole process and can facilitate impact assessment and auditing.

**Auditing tools**
An audit is an examination of evidence of a process or activity against some standards or metrics. To ensure transparency and to be able to place liabilities, the auditing process needs to be independent of the assessment process and from the day-to-day management of the auditee. AI ethics auditing tools are used in the late stages of the production process, when testing and monitoring the AI system. The focus of these tools is on appropriate documentation for verification and assurance. Checklists are also used for auditing, but less so.

**Some statistics**: The paper finds that tools for AI ethics are developed mostly by the private and academic sectors. However, the private and the public sectors are the ones that mostly use the tools. The paper also finds that more tools are developed for the early stages of the production process, namely the use case and design stages. Overall, AI ethics tools focus more on addressing models, as opposed to addressing data.

**A gap in stakeholder participation**
Typically, AI ethics tools are directly used by those who develop the AI system (e.g., development, delivery, quality assurance). The outputs of the AI ethics tools are typically used by decision-makers, such as elected officials and board members. There is typically little participation in the assessment and audit processes by traditionally marginalized groups, the users of the developed services, and vested interest stakeholders such as citizens, shareholders, and investors.

The paper emphasizes the relation between participation in AI ethics processes and power dynamics. The two are linked because participation has to do with who has the power to make decisions, who is invited to the table, and whose views and goals are prioritized. The paper recommends integrating a wider stakeholder base in AI ethics assessments and audits. It also recommends focusing the conversation on power relations rather than strictly on participation. Focusing on participation alone runs the risk of giving rise to "participation washing."

**A gap in auditing**
Nearly all the AI ethics tools are for internal self-assessment only. There are generally no requirements or processes for publishing the outputs externally. The authors emphasize that external oversight is required for the trustworthiness of organizations developing the systems. Without robust oversight, there is a risk that organizations that develop AI systems would fall into a "checklist mentality" and would settle for performative gestures that constitute "ethics washing."

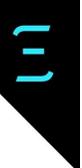

**Between the lines**
This paper gives us language to talk about the different AI ethics tools that are out there. In doing so, it helps in understanding the complex landscape of AI ethics. The identification of the participation and auditing gaps invites the reader to seek solutions.

One topic for further exploration is which strategies are appropriate for external oversight in the case of AI ethics. It might be tempting to think of auditing processes familiar from the financial and other sectors. However, in the case of AI ethics, the participation from a wider stakeholder base in external oversight seems especially important given that ethical evaluations involve the values and the perspectives available to the evaluator. Can sufficient participation be introduced into familiar auditing processes, and if so, how? Alternatively, would it be better to design different oversight procedures for AI ethics? If so, what should they look like?

## UNESCO's Recommendation on the Ethics of AI

[[Original document](#) by UNESCO]
[Research Summary by Angshuman Kaushik]

**Overview**: The Director-General of the United Nations Educational, Scientific and Cultural Organization (UNESCO) convened an Ad Hoc Expert Group (AHEG) for the preparation of a Draft Text of a Recommendation on the Ethics of Artificial Intelligence ("hereinafter the Recommendation") and submitted the draft text of the Recommendation to the special committee meeting of technical and legal experts, designated by Member States. The special committee meeting revised the draft Recommendation and approved the present text for submission to the General Conference at its 41st Session for adoption. Consequently, it was unanimously adopted by all its 193 Member States on 24.11.2021.

**Introduction**
The Recommendation addresses ethical issues related to AI to the extent that they are within UNESCO's mandate. Moreover, a significant feature of the Recommendation is that, it does not provide one single definition of AI, since such a definition would need to change over time, in accordance with technological developments. Rather, its ambition is to address those features of AI systems that are of central ethical relevance. Therefore, this Recommendation approaches

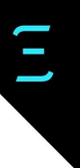

AI systems as systems which have the capacity to process data and information in a way that resembles intelligent behaviour, and typically includes aspects of reasoning, learning, perception, prediction, planning or control. Further, the aim of the Recommendation is to provide a basis to make AI systems work for the good of humanity and to prevent harm. To add to the above, it also aims at 'stimulating the peaceful use of AI systems'. The Recommendation probably refers to the use of AI systems in military warfare but, whatever that means, it needs elucidation.

**Core aspects of the recommendations**
The Recommendation states that the policy actions proposed in it are all directed at promoting trustworthiness in all the stages of the AI system life cycle. Its values and principles are outlined below;

**Values:-**
- Human rights and fundamental freedoms must be respected, protected and promoted throughout the life cycle of AI systems;
- All actors involved in the life cycle of AI systems must comply with laws, standards, practices etc., designed for environmental and ecosystem protection and restoration, and sustainable development;
- Respect, protection and promotion of diversity and inclusiveness should be ensured throughout the life cycle of AI systems, consistent with international law, including human rights law; and
- AI actors should play a participative and enabling role to ensure peaceful and just societies.

**Principles:-**
- The use of AI systems shall be governed by the principle of 'necessity and proportionality'. AI systems, in particular, should not be used for social scoring or mass surveillance purposes;
- Safe and secure AI systems shall be prioritized and any threat emanating from such systems shall be addressed to ensure human and environmental well-being;
- AI actors shall safeguard fairness and non-discrimination and also ensure that the benefits of AI technologies are available to all;
- The continuous assessment of the human, social, cultural, economic and environmental impact of AI technologies should be carried out to ascertain whether they are in conformity with the sustainable goals, such as, those currently identified in the United Nations Sustainable Development Goals (UNSDGs);
- Privacy shall be protected throughout the life cycle of the AI systems;

- Member States to ensure that it is always possible to attribute ethical and legal liability arising out of AI systems to humans. Further, as a rule, life and death decisions should not be ceded to AI systems;
- Efforts need to be made to enhance transparency and explainability of AI systems, including those having extra-territorial effect, to support democratic governance;
- Appropriate oversight, impact assessment, audit and due diligence mechanisms, including whistle-blowers' protection, should be developed to ensure accountability for AI systems;
- Public awareness and understanding of AI technologies should be promoted through open and accessible education, civic engagement, AI ethics training etc., so that people can take informed decisions regarding their use of AI systems and be protected from undue influence; and
- States shall be able to regulate the data generated within or passing through their territories, and take measures towards effective regulation of data in accordance with international law. Further, measures should be taken to allow for meaningful participation by marginalized groups.

**Areas of policy action**

The policy actions mentioned in the policy areas operationalize the values and principles set out in the Recommendation. It calls for member states to put in place effective measures, such as, policy frameworks and to ensure that stakeholders, such as private sector companies, academic and research institutions and civil society adhere to them by encouraging them to develop ethical impact assessment, due diligence tools etc., in line with guidance, including the United Nations Guiding Principles on Business and Human Rights. Listed below are the policy areas:

- Policy Area 1:- The Member States shall introduce frameworks for impact assessments, such as ethical impact assessments, to identify and assess benefits, concerns and risks of AI systems;
- Policy Area 2:- The Member States shall ensure that AI governance mechanisms are inclusive, transparent, multidisciplinary, multilateral and multi-stakeholder.
- Policy Area 3:– The Member States shall develop data governance strategies. Further, privacy shall be respected, protected and promoted throughout the life cycle of AI systems.
- Policy Area 4:– Both the Member States and transnational corporations shall prioritize AI ethics by including discussions on the topic in relevant international, intergovernmental and multi-stakeholder forums. Further, the Member States shall work to promote international collaboration on AI Research and innovation, particularly in the area of AI ethics.

- Policy Area 5:– The Member States and businesses shall assess the direct and indirect impact on the environment throughout the life cycle of an AI system. They shall also ensure compliance with environmental law, policies and practices by all the AI actors.
- Policy Area 6:– The Member States shall ensure that the potential of AI systems to contribute to achieving gender equality is fully maximized, and further, they must also ensure that the human rights and fundamental freedoms of girls and women, and their safety and integrity are not violated at any stage of an AI system life cycle.
- Policy Area 7:- The Member States are encouraged to incorporate AI systems, where appropriate, in the preservation, enrichment, understanding, promotion, management and accessibility of cultural heritage, including endangered languages as well as indigenous languages and knowledge.
- Policy Area 8:– The Member States shall work with international organizations, educational institutions and private and non-governmental entities to provide adequate AI literacy education to the public in order to empower people and reduce the digital divide and digital access inequalities resulting from the wide adoption of AI systems.
- Policy Area 9:– The Member States shall use AI systems to improve access to information and knowledge. This shall include support to researchers, academia, journalists, the general public and developers to enhance freedom of expression etc.
- Policy Area 10:– The Member States shall assess and address the impact of AI systems on labor markets.
- Policy Area 11:– The Member States shall endeavor to employ effective AI systems for improving human health and protecting the right to life, including mitigating disease outbreaks. Further, they shall implement policies to raise awareness about the anthropomorphization of AI technologies.

The Recommendation also directs the Member States to credibly and transparently monitor and evaluate policies, programmes and mechanisms related to ethics of AI, using a combination of quantitative and qualitative approaches, according to their specific conditions, governing structures and constitutional provisions. The Recommendation further directs that processes for monitoring and evaluation should ensure broad participation of all stakeholders, including, but not limited to, vulnerable people or people in vulnerable situations.

**Between the lines**
Although the Recommendation is voluntary and non-binding, it signifies 'consensus ad idem' amongst all the UNESCO Member States. However, there are suggestions which require elaboration. For instance, it is recommended that "Member States and business enterprises should implement appropriate measures to monitor all phases of an AI system life cycle". Now, guidance on the meaning and mode of operation of the term 'appropriate measures' is imperative. Another case in point is that the Recommendation states that the "Member States that acquire AI systems for human rights-sensitive use cases, such as ………..the independent judiciary system should provide mechanisms to monitor the social and economic impact of such systems by appropriate oversight authorities, including independent data protection authorities, sectoral oversight and public bodies responsible for oversight". A couple of questions that emerge here are firstly, can an independent judicial system be subjected to oversight by say, a data protection authority? Secondly, wouldn't subjecting a court to further oversight when it is already under the supervisory control of a superior court, lead to an issue of overlapping of jurisdiction? There are other key areas also that need to be clarified. Further, it is interesting to note that China being a member of UNESCO has adopted the Recommendation which follows its formulation of the AI Ethics Code. However, the US is not a signatory to the Recommendation as it is not an UNESCO Member State. Now, what needs to be seen is how the Member States incorporate and operationalize the various guidelines enshrined in the Recommendation in the future.

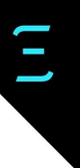

# Go Wide: Article Summaries (summarized by Abhishek Gupta)

## America's global leadership in human-centered AI can't come from industry alone

[Original article by The Hill]

**What happened**: Li, one of the people behind the famed ImageNet dataset and the co-Director of the Stanford HAI center talks about her work on the National AI Research Resource Task Force. As a part of this work, they are seeking to realize the vision of human-centered AI by democratizing access to AI systems and education to build capacity in the ecosystem for more people to build these systems. This will be supported through the construction of a National Research Cloud with the goal of making compute and data storage resources available to a wider set of people. There is vast inequity between those who are backed by large industrial and academic labs and those who are not in terms of the research work that they can carry out in the space, something that we have highlighted in our work here as well.

**Why it matters**: To achieve more responsible AI systems as well, widespread access to the necessary underlying infrastructure to run experiments and do research will be essential, quite in line with the mission of the Montreal AI Ethics Institute as well. More importantly, with dedicated resources being allocated by the federal government and a firm commitment from them to make the National Research Cloud a reality showcases a positive step in really making AI something that will empower a lot more people to build solutions for problems that are close to them.

**Between the lines**: What is particularly heartening is to see someone of Li's caliber and expertise being a part of the Task Force, especially given the deeply technical nature of the work that will be involved in making this a reality. Additionally, she is someone who is championing human-centered AI through her work at Stanford and elsewhere which hopefully will become a central tenet in the final structure that the National Research Cloud manifests in.

## Training self-driving cars for $1 an hour

[Original article by Rest of World]

**What happened**: The article highlights the abysmal rates that are paid out to workers who help to power the most lucrative and well-funded sub-industries within AI: self-driving vehicles. Given that the dominant paradigm for getting these systems to work effectively still requires large amounts of labeled data, it is not surprising that loads of money gets poured into building up datasets that can be used by companies to train their systems. Some interesting highlights from the article showcase how the demand for this has reshaped the crowdsourced work industry demanding much higher rates of accuracy from workers, supervision officers and checkers over the data labelers, and finally, additional distancing between those who provide these tasks and those who complete them.

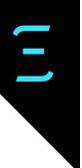

**Why it matters**: Fair compensation for work, especially for work that is crucial to the existence and continuation of the self-driving industry is the least that can be done, particularly when such companies are extremely well-funded. The platform called Remoteworks discussed in the article is owned by ScaleAI, a giant AI company valued at close to $7b. The AI engineers who develop models are typically based in places like SF while the workers who painstakingly construct the datasets to power these systems work in the Global South with none of the benefits offered to the employees of the organizations that contract out this work.

**Between the lines**: The evolving requirements put forth as the demands for dataset construction for these systems become more rigorous will herald a further reshaping of the crowdsourced work industry. One of the examples provided in the article talks about a new label category called "atmospherics" that requires labeling rain drops in an image so that the powerful cameras onboard the vehicle which capture those raindrops in their images don't mistake them for obstacles. The tasks are only going to become more tedious and will make the pace of such dataset construction unsustainable in the long run.

**How Data Brokers Sell Access to the Backbone of the Internet**

[Original article by Vice]

**What happened**: Netflow data is the data that tracks requests over the internet from one device to another. Piece enough of this together and you can learn about the patterns of communication of any individual or an organization. In this article, we learn more about Team Cymru, a firm that build products based on netflow data that it purchases from various internet service providers (ISPs) which are then sold on to other cybersecurity firms and organizations who want to perform analysis for intelligence, surveillance, and many other purposes.

**Why it matters**: This is important because it allows tracking even through virtual private networks (VPNs) stripping away anonymity on the internet even further than it already is. There is an inherent conflict with the collection of such data in the sense that on the one hand it is intrusive and strips away privacy but it also enables some great cybersecurity work that helps protect against virtual and physical threats. As with all other dual-purpose technologies, this one requires a thorough analysis of the pros and cons, especially the potential for misuse as the data stored with Team Cymru might fall into the hands of bad actors.

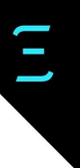

**Between the lines**: Something that caught my attention was how the Citizen Lab declined to comment on their use of a product from Team Cymru for a research report that they published. Given the strong upholding of rights and transparency that the Citizen Lab engages in, it seemed odd to not comment on the story. In addition, the Wyden requests for information to the Department of Defense for their purchase and use of internet metadata would also be interesting to examine to gain an understanding of the extent to which people's internet activities are monitored.

## How new regulation is driving the AI governance market

[Original article by VentureBeat]

**What happened**: The article highlights the trend in the current market towards a greater adoption of AI governance solutions, frameworks and tools, that will multiply the market value of such solutions to almost 10x the current amount over the next 6 years. This is being driven by incoming regulations, mostly from Europe with burgeoning efforts in the US, combined with increasing consumer savvy around data privacy and other harms like biases in algorithmic systems as they make purchase and use decisions for the various products and services around them.

**Why it matters**: As highlighted in a report from the Berkeley Center for Long-Term Cybersecurity, AI governance has undergone 3 major stages since 2016: development of high-level principles, consensus on those principle sets, and translating principles into practice. The trend observed here in terms of market value is just a natural extension of the final stage where a demand for solutions that can materialize the principles will be sought by organizations.

**Between the lines**: I believe that there is another era that we're entering with this trend which is going to involve immature solutions that claim to solve AI governance problems proliferating the market (the current phase), followed by a culling of players who aren't able to deliver on lofty promises, and finally an establishment of more mature companies that will cement their positions in various niches of the AI governance landscape selling more battle-tested solutions.

## How open-source software shapes AI policy

[Original article by Brookings]

**What happened**: This article dives into details about how the open-source software (OSS) ecosystem operates and the implications that it has on how policies are developed for the governance of AI systems. It identifies a gap in the current policy initiatives on examining the role of OSS in the power dynamics of the AI ecosystem. Notably, most policy discussions today focus on technology, data, talent, funding, but rarely do they look at how the OSS ecosystem impacts all of these factors. OSS provides benefits like speeding up AI adoption, bringing more transparency to the code bases used in products and services, and helps to accelerate fundamental advances in a lot of fields by making AI capabilities more accessible. But, this also has negative impacts in terms of the competitiveness in the market for AI solutions and sets standards implicitly, operating outside the purview of standards setting bodies that would typically help to bring counterweight to the development of the tools and methodologies in the domain.

**Why it matters**: The article highlights how the current ecosystem for AI frameworks is dominated by Google and Facebook through Tensorflow and PyTorch respectively. It is not a new phenomenon since both these companies have also published the popular Angular.js and React.js that dominate frontend web development frameworks. What is interesting on closer examination is that most of the core developers on Tensorflow and PyTorch still come from Google and Facebook giving them a much stronger implicit say in how the code develops in the future and thus potentially shaping also the standards that might follow since we would be locked into how these frameworks structure and operate.

**Between the lines**: OSS contributors need to be paid and the funding for that needs to come from somewhere. A lot of OSS projects end up being abandoned or suffer when there isn't adequate funding to compensate the contributors for their efforts and they choose to work on other things that help them pay their bills. If we talk about true democratization of tooling in OSS, we need to strongly consider whether we can reshape the structure of the ecosystem as it exists today towards something where there are perhaps external grants that are more widely available that allow anyone to sustainably contribute to such projects helping to bring more diversity to the contributors list. Until then, we at least do have access to such tooling benefitting from the investments made by corporate benefactors.

## AI industry, obsessed with speed, is loathe to consider the energy cost in latest MLPerf benchmark

**[Original article by ZDNet]**

**What happened**: In the latest MLPerf benchmark results, a benchmark that compares hardware performance in the domain of AI, with the rise in performance capabilities of the chips submitted, there was a notable drop in the reported values for energy consumption. The article posits that manufacturers are more interested in selling the high performance of their systems, and consider the energy efficiency as a secondary outcome, hence when asked to make tradeoffs, they choose to lean towards the former.

**Why it matters**: We've spoken about the environmental impact of AI, in our State of AI Ethics Report and in "The Imperative for Sustainable AI Systems", and information about the energy consumption of the physical infrastructure used to power AI applications is essential in guiding the actions of the practitioners in choosing an appropriate solution, something that we've highlighted in "The current state of affairs and a roadmap for effective carbon-accounting tooling in AI." Without that information, it is difficult to assess the energy efficiency of different systems without trying them all out, instrumenting them, and reporting the results, an exercise that manufacturers can do for a fraction of the cost.

**Between the lines**: With an overheated market for the very essential hardware that powers the booming AI market, it is understandable that manufacturers want to emphasize the performance of their hardware, rather than draw attention to the massive energy consumption of these chips. What lies in our control though is to demand that we be provided that information, and use our purchasing power to shape the market by rewarding those manufacturers who do, in essence setting a new status quo where reporting energy figures becomes the norm.

## Facebook Rolls Out News Feed Change That Blocks Watchdogs from Gathering Data

**[Original article by The Markup]**

**What happened**: In yet another blow to researchers who utilize data from Facebook to study its impacts on society, the platform has rolled out code changes by injecting superfluous elements into its website that make it even more difficult for research projects to operate and gather the necessary data that fuels their efforts to study, for example, how disinformation spreads on the platforms and biases in the kinds of advertisements shown to people of different demographic groups. The change has impacted the Citizen Browser project from The Markup, the team at the NYU Ad Observatory amongst others.

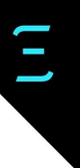

**Why it matters**: Not only are researchers affected, but screen readers, that rely on the HTML tags that have now been injected with junk code to foil these attempts, have seen performance problems making the site much less accessible for the visually impaired who rely on screen readers, and consequently the tags to navigate the website. This is not the first time that a change in the Facebook website code base has had an impact on accessibility. This violates some of the tenets of accessible web design all in the interest of decreasing transparent access to ad distribution on the platform and reducing the efficacy of ad blockers that people use.

**Between the lines**: One of the points highlighted in the article aptly sums up the current state of affairs: Facebook is working against researchers rather than with them, and this is only going to make problems worse. As pointed out by the article, there was also another instance this year when Facebook corrected previously supplied data about misinformation on the platform only after someone noticed a discrepancy in a report published by Facebook and the open data that they had made available, this potentially has impacts on years of research efforts. Moving away from an adversarial dynamic will be essential if we want to achieve the goal of having a healthier ecosystem.

### There's a Multibillion-Dollar Market for Your Phone's Location Data

**[Original article by [The Markup](#)]**

**What happened**: We all have tons of apps on our phones and for those of us who are more privacy-minded, we turn off location services. But, there exists a massive market for one's location data, and often some apps not only don't have the option to disable collecting location data, there are some who collect that data surreptitiously. And this market for location data is worth billions of dollars with many players who trade in billions of data points on millions of individuals selling that data for pennies. Often these are companies that we don't hear about very often. Data brokers are required to register in Vermont and California, but in many other places, they operate under different guises.

**Why it matters**: There have been many past cases where people have been identified for their sexual identity or religious beliefs based on location data that was obtained illicitly from such data brokers. The data brokers operate in a shadowy world, buying and selling from any and all sources, trading information to continue building richer datasets that unlock more information about each person. This happens through something called the mosaic effect where disparate pieces of information can be combined to fill in the blanks about our lives and make inferences about our identities and behaviors.

**Between the lines**: In a paper published in 2018, the role of data brokers was pointed out as being even more insidious in a world where biometric data about us also becomes more widely available through systems for DNA tests, facial recognition technology, and others. Without robust guarantees for the security of that data (which is always a challenge!), and without more stringent measures on how data brokers operate, we will continue to exacerbate risks for people in our society whose data can be weaponized against them. And this is a bigger problem in regimes where there are even fewer civil rights protections.

## Chinese AI gets ethical guidelines for the first time, aligning with Beijing's goal of reining in Big Tech

[Original article by the SCMP]

**What happened**: The Chinese Ministry of Science and Technology released more focused guidelines on AI ethics that place human control over the technology at its center. It has brought the broader Beijing AI Principles published earlier much more in line with the emphasis that the Chinese Government has placed on reigning in Big Tech. Some of the other values emphasized in the document include improving human well-being, promoting fairness and justice, protecting privacy and safety, and raising ethical literacy.

**Why it matters**: With the emphasis on human control, the guidelines set a strong example in terms of how the interaction between humans and machines will take place. In particular, the mention of the ability of humans to exit the interaction with an AI system at any time, discontinuing the AI system, and accepting to interact with the AI system in the first place will have severe consequences for the large number of AI-infused products and services that are used daily across the most popular apps in China. How this comes into effect and how strict the enforcement will be will determine to what extent the guidelines achieve their intended goals.

**Between the lines**: Given the mandate at the Montreal AI Ethics Institute, it is very interesting to see "raising ethical literacy" be included as a core consideration in the AI ethics guidelines. We believe that achieving AI ethics in practice will require education and empowerment of all stakeholders, not just having guidelines and enforcing regulations for those who develop and deploy AI systems. Perhaps this is a harbinger of other countries adopting this as a core consideration as well.

# 3. Privacy

*We have a special contribution in Spanish for the first time in the State of AI Ethics Report as we make in-roads towards making our report more multilingual. We hope you enjoy this contribution! (An English version follows the Spanish text.)*

**Introduction** by Idoia Salazar, Cofounder, Observatory of the Social and Ethical Impact of Artificial Intelligence

**La importancia de la privacidad en un mundo dominado por los datos y la IA**

El impacto masivo de la Inteligencia Artificial y el Big Data que circula a nuestro alrededor está provocando la necesidad de un cambio importante en nuestra sociedad. Un cambio de mentalidad para adaptarnos a nuestra nueva realidad. Vivimos rodeados de datos. A cada paso que damos generamos datos. Y cada vez más procesos serán realizados y ejecutados en base a estos datos, con ayuda de la inteligencia artificial. Pero nosotros, como humanos, no estamos acostumbrados esta realidad digital. A esta vida híbrida en la que no solo cuentan los pasos físicos que damos en nuestra rutina diaria. Sino también la huella digital que vamos dejando, y que es, igualmente, parte de nosotros mismos. Nuestro gemelo digital. Poco a poco, a medida que vamos compartiendo nuestros datos, nos vamos volviendo cada vez más transparentes a nuestro entorno. ¿Esto es bueno o malo? ¿seremos capaces de aprender a vivir con ello como humanidad?

La privacidad es importante para las personas, pero también depende de lo que significa este concepto para cada una de ellas, y no siempre es lo mismo. Suele variar en función de la cultura o de tus propias experiencias personales. No todos le damos la misma importancia, en cada caso concreto. Pero lo que es un hecho es que, actualmente, todos sufrimos o disfrutamos de las consecuencias del "mundo de datos" en el que vivimos. Un ejemplo, muy ilustrativo, ocurrió ya en 2012, cuando un padre puso una queja a la cadena de grandes almacenes estadounidenses Target. Alegaba que había recibido publicidad, a nombre de su hija, sobre productos de bebés[5]. Le parecía ridículo que se lo enviaran porque la chica aún estaba terminando Bachillerato. Pero resultó que esta cadena de supermercados, a través de los datos, había detectado que su hija cumplía con el perfil de "embarazada" y, siguiendo su política comercial, enviaba esta publicidad. En efecto, la hija estaba embarazada y Target lo sabía antes que su padre.

---

[5] K. Hill. "How Target Figured Out A Teen Girl Was Pregnant Before Her Father Did". Forbes, 2012. https://www.forbes.com/sites/kashmirhill/2012/02/16/how-target-figured-out-a-teen-girl-was-pregnant-before-her-father-did/.

En el mundo laboral este seguimiento es habitual. Cada vez hay más empresas que investigan perfiles de sus candidatos para contratar en redes sociales, antes de tomar una decisión, y no es raro que, si ven algo que no les agrada ahí, toman una decisión negativa[6]. Tampoco es raro que nos llegue publicidad *online* sobre viajes a un país, sin haberlo solicitado, simplemente porque hemos hecho alguna búsqueda en el ordenador el día anterior. Se podría decir que es un mal necesario, ya que poca gente está disponible a pagar para leer las noticias, usar *apps,* como Google Maps, Facebook, Instagram, etc. Aunque no nos guste, hasta cierto punto, podríamos llegar a entender por qué explotan nuestros datos.

Sin embargo, hay algo que debe preocupar aún más: los casos de prácticas ilegales con los datos personales. Aunque haya muchísimos casos documentados, el más emblemático, hasta la fecha, ha sido el escándalo de Cambridge Analytica y los datos de Facebook de decenas de millones de usuarios de todo el mundo, ocurrido en marzo de 2018[7]. En este caso, un investigador de esta empresa transfirió, de manera ilegal (es decir en contra de la política de datos de Facebook), estos datos a la empresa Cambridge Analytica, especializada en el marketing político *online*, sobre todo en campañas electorales, usando tecnologías de *Big Data*.

**La importancia de anonimizar los datos**

En cualquier caso, y aunque sean prácticas habituales estos intercambios de datos, es muy importante en la actualidad mantener y cuidar la privacidad de las personas. Para ello, cada vez es más común que empresas y organizaciones, públicas y privadas, que utilizan o desarrollan IA, usen datos "anonimizados". Así, los datos son sometidos a un proceso en el que se desvincula los datos personales (nombre, IP, número de la Seguridad Social, número de cuenta bancaria…) del resto de valores. De esta manera, en principio, no se podrían volver a asociar de forma directa a campos considerados personales para identificar directamente a la persona individuo. Esto permite garantizar la privacidad del individuo en cuestión, a la vez que utilizas sus datos para la muestra. Este proceso de "anonimización" es importante porque los datos personales están protegidos, en toda Europa, por la Regulación General de Protección de Datos (la GDPR, por sus siglas en inglés). Una vez que "despersonalizas" ya no habría problema[8].

**Una responsabilidad individual**

En cualquier caso, y a pesar de tener estas protecciones legales, es cada vez más importante que cada persona tome cada vez más consciencia de su responsabilidad individual. Por supuesto, la empresa que produce o comercializa un producto o servicio con IA debe cumplir unos requisitos de ética y legalidad, pero también el que lo consume, en función de sus principios y de su propio criterio, debe asumir esa responsabilidad. Para ello, es importante que nos formemos convenientemente para tener criterios solidos ante las distintas situaciones, ya sean fáciles o difíciles. Si un algoritmo de IA nos hace una recomendación, es solo eso, una opción para nosotros. Somos nosotros, como personas y/o profesionales con criterio, los que debemos decidir seguir lo marcado por el sistema de IA o no. Por tanto, la responsabilidad recae sobre el consumidor, no sobre el algoritmo de IA[9]. Igualmente, debemos de ser toda la sociedad los que podamos decidir, de una forma sencilla, borrar nuestros datos cuando así lo consideremos. El consumidor, la persona individual, debe ser el responsable y dueño absoluto de los datos que genera. Y para que así lo pueda entender la educación en estas materias es la base fundamental para cimentar un futuro próspero que cada vez es más presente.

---

[6] M. Wood. "Not Getting Any Job Offers? Your Social Media Activity Could Be The Reason". Forbes, 2017.
https://www.forbes.com/sites/allbusiness/2017/06/22/not-getting-any-job-offers-your-social-media-activity-could-be-the-reason/.

[7] C. Cadwalladr y E. Graham-Harrison. "Revealed: 50 million Facebook profiles harvested for Cambridge Analytica in major data breach". The Guardian, 2018.
https://www.theguardian.com/news/2018/mar/17/cambridge-analytica-facebook-influence-us-election.

[8] Benjamins, R; Salazar, I (2020): El Mito del Algoritmo: cuentos y cuentas de la inteligencia Artificial. Anaya.

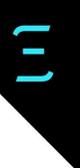

**English text:**

**The importance of privacy in a world dominated by data and AI**

The massive impact of Artificial Intelligence and Big Data circulating around us is provoking the need for a major change in our society. A change of mentality to adapt to our new reality. We live surrounded by data. Every step we take we generate data. And more and more processes will be performed and executed based on this data, with the help of artificial intelligence. But we, as humans, are not used to this digital reality. To this hybrid life in which not only the physical steps we take in our daily routine count. But also the digital footprint we leave behind, which is also part of ourselves. Our digital twin. Little by little, as we share our data, we become more and more transparent to our surroundings. Is this good or evil? Will we be able to learn to live with it as humanity?

Privacy is important to people, but it also depends on what this concept means to each person, and it is not always the same. It often varies according to culture or your own personal experiences. We do not all attach the same importance to it, in each individual case. But what is a fact is that, nowadays, we all suffer or enjoy the consequences of the "data world" in which we live. A very illustrative example occurred already in 2012, when a father complained to the US department store chain Target. He alleged that he had received advertising in his daughter's name for baby products[10]. He thought it was ridiculous that they sent it to him because the girl was still finishing high school. But it turned out that this supermarket chain, through the data, had detected that her daughter met the profile of "pregnant" and, following its commercial policy, sent this advertising. In fact, the daughter was pregnant and Target knew it before her father did.

---

[9] Salazar, I; Benjamins, R (2021): 'El algoritmo y yo: GuÍA de convivencia entre seres humanos y artificiales'. Anaya.
[10] K. Hill. "How Target Figured Out A Teen Girl Was Pregnant Before Her Father Did". Forbes, 2012. https://www.forbes.com/sites/kashmirhill/2012/02/16/how-target-figured-out-a-teen-girl-was-pregnant-before-her-father-did/.

In the professional world, such monitoring is commonplace[11]. More and more companies are researching profiles of their candidates for recruitment on social networks before making a decision, and it is not uncommon for them to make a negative decision if they see something they don't like there. It is also not uncommon to receive unsolicited online advertisements about travel to a country, simply because we have done some research on our computer the day before. You could say it's a necessary evil, as few people are willing to pay to read the news, use apps, like Google Maps, Facebook, Instagram, etc. Even if we don't like it, to a certain extent, we might come to understand why they exploit our data.

However, there is something that should be even more worrying: cases of illegal practices with personal data. Although there are many, many documented cases, the most emblematic, to date, has been the Cambridge Analytica scandal and the Facebook data of millions of users around the world, which occurred in March 2018[12]. In this case, a researcher from this company illegally (i.e. against Facebook's data policy) transferred this data to the company Cambridge Analytica, which specialises in online political marketing, especially in election campaigns, using Big Data technologies.

**The importance of anonymising data**

In any case, although such data exchanges are common practice, it is very important nowadays to maintain and protect people's privacy. To this end, it is increasingly common for companies and organisations, both public and private, that use or develop AI to use "anonymised" data. Thus, data are subjected to a process in which personal data (name, IP, Social Security number, bank account number...) are separated from the rest of the values. In this way, in principle, they can no longer be directly associated with fields considered personal in order to directly identify the individual person. This makes it possible to guarantee the privacy of the individual in question, while using his or her data for the sample[13].

This "anonymisation" process is important because personal data is protected throughout Europe by the General Data Protection Regulation (GDPR). Once you "depersonalise" it is no longer a problem.

---

[11] M. Wood. "Not Getting Any Job Offers? Your Social Media Activity Could Be The Reason". Forbes, 2017. https://www.forbes.com/sites/allbusiness/2017/06/22/not-getting-any-job-offers-your-social-media-activity-could-be-the-reason
[12] C. Cadwalladr y E. Graham-Harrison. "Revealed: 50 million Facebook profiles harvested for Cambridge Analytica in major data breach". The Guardian, 2018.
[13] Benjamins, R; Salazar, I (2020): El Mito del Algoritmo: cuentos y cuentas de la inteligencia Artificial. Anaya.

**An individual responsibility**

In any case, and despite having these legal protections, it is increasingly important that each person becomes more and more aware of his or her individual responsibility. Of course, the company that produces or markets a product or service with AI must comply with ethical and legal requirements, but also the consumer, according to his or her own principles and criteria, must assume this responsibility. To this end, it is important that we are properly trained to have solid criteria in different situations, whether they are easy or difficult. If an AI algorithm makes us a recommendation, it is just that, an option for us. It is up to us, as people and/or professionals with criteria, to decide whether to follow the AI system's recommendations or not. Therefore, the responsibility lies with the consumer, not with the AI algorithm[14]. Equally, it should be society as a whole that can decide, in a simple way, to delete our data when we consider it appropriate. The consumer, the individual person, must be the absolute owner and responsible for the data he or she generates. And in order for them to understand this, education in these matters is the fundamental basis for cementing a prosperous future that is increasingly our NOW.

---

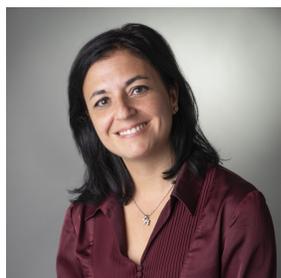

**Idoia Salazar**
Co-founder and President
Observatory of the Social and Ethical Impact of Artificial Intelligence (OdiseIA)

Salazar is the co-founder and president of the Observatory of the Social and Ethical Impact of Artificial Intelligence (OdiseIA). She is Principal Investigator of the SIMPAIR Research Group (Social Impact of Artificial Intelligence and Robotics)I. She is a specialist in Ethics in Artificial Intelligence and a professor in international degrees at CEU San Pablo University. She has authored the following books: 'The Algorithm and I: Guide to coexistence between human and artificial beings', 'The Myth of the Algorithm: Tales and truths of Artificial Intelligence (co-author with Richard Benjamins),' The Revolution of robots: How Artificial Intelligence and robotics affect our future 'and' The depths of the Internet: Access information that search engines cannot find and discover the intelligent future of the Internet ' (written in spanish), as well as scientific and informative articles oriented to investigate and raise awareness about the impact of Artificial Intelligence. She is in the list of experts to assist the European Parliament´s Artificial Intelligence Observatory (EPAIO); member of the Board of Directors of the Association 'Arco Atlántico para la Seguridad y el entorno Digital'; founding member of Springer AI and Ethics journal and member of the Global AI Ethics Consortium.

---

[14] Salazar, I; Benjamins, R (2021): 'El algoritmo y yo: GuÍA de convivencia entre seres humanos y artificiales'. Anaya.

# Go Wide: Article Summaries (summarized by Abhishek Gupta)

## The Limits of Differential Privacy (and Its Misuse in Data Release and Machine Learning)

[Original article by ACM Magazine]

**What happened**: We pitch the concept of differential privacy as a silver bullet to solve our struggle with wanting to share data (in the interest of building publicly beneficial technologies) with the desire to have strong privacy protections. Yet, there is no free lunch. Differential privacy has shortcomings: the more substantial the privacy protections, the less utility we get from the data as tuned by the epsilon parameter in the differentially private analysis. As per the original paper, we bring meaningful privacy protections when epsilon values stay less than 1 (lower values are better for privacy). Still, a lot of current uses of differential privacy use values as high as 30. In addition, the fundamental formulation of differential privacy works to protect individual records in a pool of records of many individuals. We violate this basic notion when we apply differential privacy to healthcare data from devices in conjunction with federated learning because all the records coming from a single device belong to the same person.

**Why it matters**: This sort of in-depth technical analysis and challenging dominant assumptions in the field is crucial if we want to achieve responsible AI in practice rather than just pay lip-service to it by articulating a set of principles.

**Between the lines**: Even though there is a somewhat valid diatribe against technical practitioners proposing solutions to address ethical challenges in AI, we cannot work without their expertise and help. We risk creating requirements and legislations with a limited understanding of the limits of proposed solutions, leading to more harm in the long-run.

## Should Families' Surveillance Cameras Be Allowed in Nursing Homes?

[Original article by The Markup]

**What happened**: A surveillance camera installed in a nursing home captured a death showing gruesome details with the victim crying out for help and the nursing staff idling and, in some cases laughing at the patient. It sparked a massive debate on whether it was legal to mount such cameras in nursing homes. Arguably there are privacy concerns, and it showcases a distrust in the staff at the nursing homes, while others have argued that it is a way to hold them accountable. As is usually the case with technological solutions to sociological problems, such a solution fails to address the underlying issues of under-compensated and overworked nursing staff, amongst other problems with the healthcare system.

**Why it matters**: Particularly as we start to automate the video processing captured from these cameras, the issues raised in this case will become even more pertinent to the overall discussion of surveillance. As the case made its way through the legislative process, the courts settled on deeming it ok to have visible cameras so that the people being surveilled are aware of the presence of a camera. In contrast, the use of hidden cameras was prohibited.

**Between the lines**: We see a collision of technology and society in ways that we couldn't have anticipated. Issues such as the use of cameras (which have become cheaper to deploy) coupled with the use of AI to automatically process the video feeds can help in unburdening the staff from having to constantly monitor patients, such as automatically detecting if a patient has fallen. On the other hand, it begins to normalize automated surveillance as an accepted part of our society which will have much more profound effects in the long term.

**Huge data leak shatters the lie that the innocent need not fear surveillance**

[Original article by The Guardian]

**What happened**: The firm NSO that is known to have sold surveillance software to organizations, including governments around the world, has come under fire for its software Pegasus which is under investigation for its implication in the surveillance on not just typical targets of spy-craft around the world but also everyday citizens Over the coming weeks, The Guardian's investigative team in partnership with other news organizations around the world will be releasing the names of the people who have been a target of the software and the compromises facilitated by it. The case that they seek to make is that anybody is susceptible to these intrusions given our over-reliance on our phones and why privacy as a core tenet of functioning in our digital society needs to be something that we pay a lot more attention to.

**Why it matters**: While there are some regulations perhaps in the use of surveillance technology when government agencies deploy them in their intelligence operations (though a lot of that was debunked with the Snowden leaks in 2013), the operations of a player like the NSO and their ability to sell their tools and services to anyone on the market change the equation significantly in terms of what privacy guarantees we can hope to have as individuals spending a chunk of our lives in the digital realm.

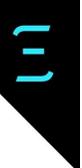

**Between the lines**: The seriousness of the matter is underscored by some phrases in the article that highlight the degree of protections that the staff behind this story had taken including not having any of their phones around during their meetings, sources, etc. They position this as something that will be as monumental as their investigation and publication of results at the time of the Snowden leaks that moved the needle of understanding for the public on what spy-craft capabilities exist and how they are being used. Now, that conversation will expand to include anyone in the world, innocent or not.

## The Inevitable Weaponization of App Data Is Here

**[Original article by [Vice](#)]**

**What happened**: A Substack publication called The Pillar bought location data from a data broker and combining it with data from the Grindr app outed a priest as potentially gay which led to his resignation. Even anonymized data without names attached to a specific person can be used to obtain information about a specific person. A very small number of location points are required to uniquely identify a person because of the patterns that we all follow in the places we visit and where we spend most of our time: our homes and offices.

**Why it matters**: Apps like Grindr defend themselves by saying that what was mentioned in the article that let to the ouster of the priest is "technically infeasible", the problem is that there are plenty of companies that offer data consulting services towards "identity resolution" as a way of unearthing data about specific individuals from troves of data that are sold by data brokers.

**Between the lines**: What was previously the domain of highly-resourced organizations is something that anyone with a little bit of money and motivation can execute with ease. Data brokers collect large amounts of data from all the apps that have any sort of in-app advertising and then package and sell that data over to anyone willing to fork over a few dollars. This is still a largely unregulated industry and calls from Senators like Wyden in the US to bring the force of the FTC to regulate this domain are essential if we want to get rid of the scourge of sensitive data exposing intimate details of our lives.

## Amazon will pay you $10 in credit for your palm print biometrics

**[Original article by [TechCrunch](#)]**

**What happened**: Amazon is rolling out payments in their physical stores using a contactless palm scanner and are offering $10 in store credit to those who enrol in the service. Contactless

payments do seem attractive in a pandemic but there are the obvious cybersecurity concerns, especially those of data privacy of immutable personally identifiable information, that is, your biometrics which cannot be changed unlike your phone number or home address.

**Why it matters**: Amazon doesn't have a great track record when it comes to the use of biometrics, we all remember their Rekognition program and the ensuing problems, including inaccuracies in the inferences generated by that system. In addition, they are a private company which can turn around and potentially sell this data to data brokers who can collate that with other information about you that is floating around on the internet to disastrous consequences. My paper from 2018 "The Evolution of Fraud: Ethical Implications in the Age of Large-Scale Data Breaches and Widespread Artificial Intelligence Solutions Deployment" talked about some horrifying consequences that might arise from leaked biometric data.

**Between the lines**: While the headline is sensationalist in its presentation of the scenario, namely that it brings focus to the monetary aspect of how much customers will be compensated for the use of biometrics, it detracts from the more important issue of how and when biometrics should be used and what regulations we need to develop to ensure their safe usage.

## Apple Walks a Privacy Tightrope to Spot Child Abuse in iCloud

[Original article by [Wired](#)]

**What happened**: Apple has introduced a new feature for the devices that use iCloud which will scan images to determine if there is any child sexual abuse material (CSAM) in them. This is being heralded as a win in the fight against child abuse online while some privacy activists believe that this weakens the privacy protections offered by the Apple ecosystem to its users. The determination process is split between the device and the cloud where hashes are computed on the images and these are compared against a known database of CSAM that is downloaded through a blinding process to the user's device. This prevents a user from reverse engineering all the hashes to prevent abuse and evasion of the detection system. It also uses something called NeuralHash that is robust to alterations in the images that abusers can use to evade detection. It also uses the notion of privacy set intersection to only alert the system when hashes are matched otherwise resting silent and preventing Apple from gaining access to hashes of all your images.

**Why it matters**: Online services have certainly made it easier to spread CSAM and this move by Apple is a huge win in combating this scourge. But, the concerns raised by privacy activists also have some merit in terms of what other demands might the company accede to in the future in the interest of law enforcement. The setting of the precedent is what scares the privacy scholars and activists more so than this particular instance which is quite clearly beneficial for the health of our information ecosystem.

**Between the lines**: The engineering solutions proposed to tackle this problem of CSAM detection in a privacy-preserving fashion will have lots of other positive downstream usage and it is a net win overall in designing technology that can keep harm at bay while still maintaining fundamental rights and expectations of users like privacy. Public discussion of such technology through open-source examination might be another way to boost the confidence that people have in the solutions being rolled out while also potentially pointing out holes in the technology leading to an overall more robust solution.

## Privacy in the Brain: The Ethics of Neurotechnology

[Original article by [Technology Networks](Technology Networks)]

**What happened**: The article points out the emerging regulatory challenges faced in the domain of neurotechnology as the quantified wellness industry has taken off with many mass-market devices like the Muse headband claiming that they are able to tap into brain signals to deliver better neurotech-enabled experiences to enhance productivity and improve meditation practices. What was previously the domain of the DIY community is now going mainstream and one of the interviewees in the article explicates that without undergoing the same stringent standards governed by the FDA as we have for other medical devices, we risk causing irreversible harm for those who use these novel devices in an experimental fashion.

**Why it matters**: The demonstration from Musk's Neuralink certainly brought neurotechnology to the attention of many more people than before. Pitched as being able to augment the brain and our ways of communicating with each other mediated by machines, in addition to the more immediate (and perhaps realistic) benefits of helping those vision and speech problems, the ethical concerns of commercial technologies without medical-grade approvals is very unnerving. The privacy implications, especially in the era of continual cyberattacks is another exacerbating factor.

**Between the lines**: While physical damage like skin burns from wrong usage of transcranial direct current stimulation (tDCS) devices can be measured to a certain extent, there is potential for damage that is hidden or alters the subjective experience of a person that can't be quantified. In such a case, understanding the burden of liability is tricky to resolve, especially with incomplete information and this has some bearing on a lot of scenarios we encounter in the domain of AI ethics as well.

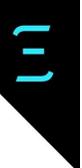

## This is the real story of the Afghan biometric databases abandoned to the Taliban

**[Original article by MIT Technology Review]**

**What happened**: As the US forces exited Afghanistan, the technical infrastructure and data trails left behind are causing problems as they fall into the hands of the Taliban. In particular, the article dives into the details of the Afghan Personnel and Pay System (APPS) used by the Ministry of the Interior and the Ministry of Defense that has access to highly sensitive information including biometrics of security personnel and their networks. The implications of that data extend beyond just breaches of privacy: there are real-world security concerns with how that data might be used against those who supported the previous regime.

**Why it matters**: Aside from the sensitive and immutable nature of the data that resides in these databases that have now been taken over by the Taliban, this is an unfortunate example of what happens when extensive data gathering operations are conducted without regard to what may happen when the information falls into the hands of malicious actors. In this case, there are incoming reports that mention how the data might have potentially been used to target individuals still within the country that had supported the previous regime. Given the immutability of the biometrics, the people who are captured in that data have no chance of erasing or escaping.

**Between the lines**: While the data is usually gathered under the guise of providing administrative services like access to government and social security benefits, without appropriate cybersecurity protections, and sometimes even with them, when data falls into the hands of those who can misuse that data or target individuals based on their identity or activity, there is little that people can do to escape the consequences of their presence in those datasets. This is one of the strongest reasons in support of data minimization and purpose limitation. As mentioned in the article, creating national ID schemes based on biometric data is not the best way to go about it, and this is one example where we see how this can go horribly wrong.

## The Downside to Surveilling Your Neighbors

**[Original article by The Markup]**

**What happened**: Apps like Citizen and Nextdoor that claim to make neighborhoods safer by empowering residents with more information through the deployment of surveillance infrastructure are under intense scrutiny as they have become platforms rife with racism and vigilantism. What is different compared to previous iterations of such neighborhood monitoring solutions is that these are now often tied in with official police departments, giving them a live feed into the happenings of a neighborhood. Exacerbating the problems is the fact that reported incidents can get blown out of proportion as residents might engage in racist behavior that disproportionately targets BIPOC members.

**Why it matters**: Some apps like Citizen are used in over 60 cities in the US. While a lot of apps have now severed connections with official police departments being able to read feeds directly, it doesn't prohibit police forces from having non-official accounts to monitor the feeds. More so, the problems of content moderation and health of the information ecosystem that they maintain suffers from the same challenges that other social media platforms suffer from.

**Between the lines**: This is a piece of technological solution that doesn't solve the real problem that it claims to provide a solution for. It just normalizes extreme views that some residents might hold on their neighbors by bringing those views out into the open and garnering support from other isolated pockets (hopefully) of the same viewpoints. By being loud and vocal on those platforms, they can spur hate and mistrust amongst the communities countering the goal of having safe and harmonious living. Ultimately, safer communities might just require steering away from technological fixes and more so focusing on community building IRL.

## Leaked Documents Show How Amazon's Astro Robot Tracks Everything You Do

**[Original article by Vice]**

**What happened**: Amazon has unveiled a new robot dubbed Astro that integrates with Alexa Guard and Ring (other products from Amazon) to provide automated home security solutions. It is a $999 robot that will patrol the home of the user and constantly surveil it for incidents that warrant the attention of the owner in the case of unusual activity and also to monitor strangers inside the house. If this doesn't already spook you, the article mentions a leaked memo that details the numerous flaws in the system that go counter to the marketing effort from Amazon.

**Why it matters**: Constant surveillance, data privacy, targeted advertising, and the list goes on in terms of concerns that arise from the use of a persistent sentry moving around the most private spaces of our lives: our homes. What is even more problematic is that people who have worked on Astro point out that the system is flawed in its person recognition capabilities, struggles to navigate spaces, and is sold as an accessibility enhancer within the home though it has notable failure modes where it is known to get into jams. What's even more interesting is that Amazon doesn't have a policy to allow for returning broken Astros as mentioned in the article referencing the leaked memo.

**Between the lines**: From a practical and technical standpoint, there are many challenges in getting social robots right: above all, getting it to operate in line with human expectations most of the time to be welcome in their most private spaces. The demonstrated failures of Astro along with the almost insurmountable combination of challenges of being able to respond to myriad voice commands from the owner, navigating a complex, dynamic, and uncertain environment, and interacting with dynamic live and static elements in that environment makes it highly unlikely that Astro succeeds in winning a place in people's home. As consumers become savvier about privacy and other ethical concerns regarding some of the tech that is required to power the Astro, Amazon will have to provide very robust guarantees before people are going to bring one home.

# "The power to surveil, control, and punish": The dystopian danger of a mandatory biometric database in Mexico

[Original article by Rest of World]

**What happened**: Mexico, empowered by a loan from the World Bank, is pushing hard to implement a unified national identity scheme with a view to make access to government services and public benefits linked through a single system. Similar initiatives have been funded by the World Bank in countries around the world, and in many places the implementation of such national identity schemes has led to less than desired outcomes. In particular, biometrics associated with the identities tend to result in failures at the point of receiving the services due to problems with the technology that is deployed to ascertain identity, such facial recognition and fingerprint scanning.

**Why it matters**: Once such a system is put in place, it is incredibly difficult to extricate the provision of the services from such a system. In a country where crime infiltrates various levels of government and where there is a risk for cybersecurity breaches, potentially compromising all the identities and biometrics of the people who have enrolled in the scheme, such a system might create more problems than it solves. Given all these problems, and the potential that all identities linked into a single scheme can cause a single point of failure and give too much authoritarian power to governments to surveil, we need to be careful before proceeding.

**Between the lines**: One of the things that stands out in the article is that such national identity schemes are pushed heavily in the Global South under the guise of improving access to public benefits and government services, but this might not be an ideal approach if there isn't adequate supporting infrastructure that can ensure that such a scheme is implemented with privacy safeguards and security measures in place. Also, we need to be cognizant of the fact that there continue to exist alternative ways for people to access services who aren't able to enroll in the scheme or experience failures in the verification process at the point of service provision due to hardware failures, denying them access to essential services like free rations and healthcare in certain places.

## The Popular Family Safety App Life360 Is Selling Precise Location Data on Its Tens of Millions of Users

[Original article by The Markup]

**What happened**: Data brokers are organizations that operate in the shadows, away from much scrutiny and regulations, furnishing a market worth billions of dollars with aggregated data from across various sources to fuel targeted advertising amongst other services that are meant to strip away at the consumer's agency, pocket, and autonomy. The article dives into details of Life360, a company that allows families to track the locations of their kids, pitched as a safety product. What is buried in the fine print, that parents sometimes gloss over, is that such data is sold downstream to 3rd parties, including government agencies. While they recently put in place a policy to not sell the data to law enforcement, they have been doing so for many years already, meaning data has potentially traveled far.

**Why it matters**: While such apps do provide a degree of comfort and utility to parents to monitor their kids, the cost is potentially too high, location data used to create a rich profile of their children that will follow them for the rest of their lives as these data brokers enrich such datasets with more information and sell that downstream for targeted advertising all the way up to changing insurance premiums and other higher stake situations. Given that such data is collected directly by the app and then sold later once it is centralized, typical approaches used by privacy researchers hunting for code that shows signs of linking to common data brokers doesn't really work here. Clients of Life360 have been flagged for problematic behaviour by many concerned entities, including Senator Wyden's office.

**Between the lines**: What is appalling about the whole situation is that such companies use instances like the pandemic and the veneer of providing a public service by gathering this data and selling it to organizations doing good work like the CDC to draw attention away from the fact that they are also selling this data to other unscrupulous parties. More so, they take advantage of parents' fears of child safety as a Trojan horse to more invasive data gathering which fuels their bottom lines, completely out of sync with their stated values. The business goals and stated values are often in conflict with the current trend indicating that business goals are winning by a mile.

## Singapore's tech-utopia dream is turning into a surveillance state nightmare

[Original article by Rest of World]

**What happened**: Technology is seen as an instrument to enable a fulfilling and meaningful life in Singapore, with large leeway provided to the government to impose technological solutions in service of building a utopia. But, as the pandemic rolled on, citizens in Singapore are chafing against the intrusions that constant surveillance and digital intrusions are now imposing on their lives. The inclination for technology remains so high in Singapore that the government offers regulatory sandboxes so that companies can experiment with novel technologies that can then be brought into mainstream society. Things like smart lamp posts to monitor traffic, environmental conditions, and people's movements, robots for elder care, biometric databases to process people at borders and improving security at banks and public services. With apps like TraceTogether and SafeEntry becoming mandatory and combined into a single experience, movement tracking in the interest of curbing the pandemic became ubiquitous.

**Why it matters**: These apps are quite detailed in the amount of data that they collect, especially when we consider that they are linked to the national identity system in Singapore. Even though the government provided assurances that such technology would only be used for contact tracing, it was later revealed that the data had been shared with law enforcement and cases have used this information as evidence. People have been slower to question and raise concerns because of the strict information ecosystem in Singapore with laws like POFMA and FICA, which are meant to protect against misinformation, being used to tightly control what is said about the government's technology use.

**Between the lines**: Migrant workers in Singapore, who form the backbone of physical labor that has been used to build the infrastructure powering and enabling everything, are the targets of technological experimentation and are often forced to live in situations with very limited rights. Once the technology is refined, it is deployed en masse to the rest of the Singapore population. Bundling solutions together and tying them with national identity solutions might have helped Singapore return to normalcy faster from the pandemic, but it has come with an astronomical cost of introducing extremely pervasive and intrusive surveillance technology that doesn't show any signs of going away.

# 4. Bias

**Introduction** by Abhishek Gupta, Founder and Principal Researcher, Montreal AI Ethics Institute

Bias mitigation continues to pose challenges to those who are designing and developing AI systems but also to those who are procuring these systems and integrating them into their own products and services. The opening article details a framework from the National Institute of Standards and Technology (NIST) which has a 3-step approach for bias mitigation focussed on the phases of pre-design, design and development, and deployment. Tying bias mitigation approaches explicitly to various stages of the AI lifecycle makes it so that we have actions that more closely map to the activities carried out by practitioners everyday in their work.

Carrying on with practical approaches to bias mitigation, the next piece in this chapter walks through co-designed checklists, a piece from Microsoft Research that highlights research work that draws from an understanding of how practitioners approach fairness concerns today, what are their desiderata for fairness checklists, and how they want them to be implemented. The checklist follows a similar approach to the NIST piece in the sense that there is a corresponding mapping to the AI lifecycle, as articulated in six steps: envision, define, prototype, build, launch, and evolve. Though they conclude by recognizing that a procedure alone cannot overcome the value tensions and incompatibilities in ethical practice.

Social media platforms have become the place where people self-organize to raise issues that are important to them. In this chapter, we see how selective filtering and content moderation policies can disproportionately negatively impact minorities. We see an example where making small changes to one's profile on TikTok to include phrases like "Black Lives Matter" can lead to flags but it isn't the same with white supremacy. Such biases are not just limited to profile information, for example, on Facebook, videos with Black men were wrongly tagged with "gorilla" making recommendations to watch videos with animals. This is a stark demonstration of how AI systems, even ones built by organizations on data repositories with huge amounts of user-uploaded content can fail spectacularly when there aren't appropriate guardrails in place.

Datasets like C4 that are used to train really large machine learning models like T5 and Switch Transformer are known now to have undergone severe filtering that disproportionately removed content from LGBTQ+ communities. This has direct impacts in terms of how strictly content with that material might be policed since it won't have good representation when these pretrained models are used in content moderation.

Bias remains an extremely important area of research and is deeply contextual requiring lots of interdisciplinary research before we get to a place where we can effectively put these ideas into practice. This chapter has some fascinating examples from the work of Cynthia Dwork on "fairness through awareness" and work that's been done at Vimeo to uncover biases in search and recommendation systems. I hope you find this chapter insightful and wide-ranging beyond the most commonly covered areas in the discussions about bias in AI systems.

---

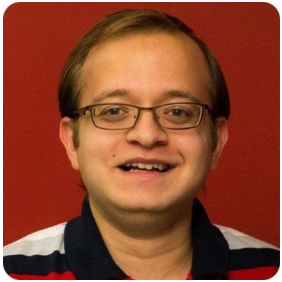

**Abhishek Gupta (@atg_abhishek)**
Founder, Director, & Principal Researcher
Montreal AI Ethics Institute

Abhishek Gupta is the Founder, Director, and Principal Researcher at the Montreal AI Ethics Institute. He is a Machine Learning Engineer at Microsoft, where he serves on the CSE Responsible AI Board. He also serves as the Chair of the Standards Working Group at the Green Software Foundation.

# Go Deep: Research Summaries

## "A Proposal for Identifying and Managing Bias in Artificial Intelligence". A draft from the NIST

[Original paper by Reva Schwartz, Leann Down, Adam Jonas, Elham Tabassi]
[Research summary by Connor Wright]

**Overview**: What does bias in an AI system look like? Is it obvious? How can we mitigate such threats? The NIST provides a 3-stage framework for mitigating bias in AI, with it being seen as key to building public confidence in the technology. Not only can such mitigation help us better reduce the effects of AI, but it can also help us better understand it, and the NIST wants to do just that.

**Introduction**

What does bias in an AI system look like? If we saw it, would we be able to mitigate it? The National Institute of Standards and Technology (NIST) tries to answer both of those questions as part of their pursuit for a framework for responsible and trustworthy AI. Mitigation, transparency, and public engagement are widely accepted as popular notions for building public trust in AI. For me, the most exciting points in the NIST's draft are their interaction with bias as a concept and their 3-stage framework. With bias proving one of AI's biggest problems, such frameworks can better expose this problem and better understand it.

**The problem of bias**

It's important to note how automated biases can spread more quickly and affect a wider audience than human biases on their own. Rather than being confined to those you interact with, the presence of AI systems that stretch across the globe means that those affected by its negative consequences are more numerous. Its effects are then heightened through AI's presence (and further potential presence) in our lives. For example, the proliferation of facial recognition technology and AI being used in job screening. As a result, the NIST finds it necessary to investigate how this can come about, and I wholeheartedly agree.

**Why is this the case?**

Bias can be seen to creep in when the object of study can only be partially captured by the data, such as a job application. Here, aspects such as the value gained from work experience and how it translates into the new role cannot be accounted for by just a simple keyword search.

At times, bias also enters into the fray through AI decisions being made using accessible rather than suitable data. Here, researchers are said to "go where the data is" and formulate their questions once they get there, rather than taking complete account of the necessary data for an informed and representative AI system. For example, it would be as if you were to look at a college application and solely focus on the academic data (grades) available, rather than also looking at the extra-curricular activities the candidate has undertaken.

To try and tackle this, the NIST proposes a 3-stage lifecycle to better locate how AI can enter the picture.

**Stage 1: Pre-design**
Here, the technology is "devised, defined and elaborated, " which includesto involve then framing the problem, the research, and the data procurement. Essential notions to consider can then be seen in identifying who's responsible for making the decisions and how much control they have over the decision-making process. This allows for a more evident tracking of responsibility in the AI's development and exposes the presence of any "fire, ready, aim" strategies. What is meant by this play on words is how, at times, AI systems are often deployed before they've been adequately tested and scrutinised. The second stage then becomes even more relevant.

**Stage 2: Design and development**
Usually involving data scientists, engineers and the like, this stage consists in the engineering, modelling and evaluation of the AI system. Here, the context in which the AI will be deployed must be taken into account. Simply deploying an accurate model does not automatically mitigate any problem of bias without this essential component. This is to say, a facial recognition system could be 95% accurate in identifying the faces of children between 5-11 years old, but being deployed in an adult context will render it useless.

In this sense, techniques such as "cultural effective challenge" can be pursued. This is a technique for creating an environment where technology developers can actively participate in questioning the AI process. This better translates the social context into the design process by involving more people and can prevent issues associated with "target leakage". To explain, "target leakage" is where the AI trains on data that prepares it for an alternative job than the one it initially intended to complete. To illustrate, training on past judicial data and learning the decision-making pattern of the judges and not the reasons for conviction. If such problems can then be avoided, the deployment stage will be less likely to run into any issues. However, this is not always the case.

**Stage 3: Deployment**
The deployment stage is probably the most likely stage for any harmful bias to emerge, especially given how the public now starts to interact with the technology. Given AI's accessibility, such interaction can also include malicious use on behalf of an unintended audience, such as using chatbot technology to spread fake news online. Even if this wasn't intentional, the general interaction by the public could also expose any problems to do with the technology.

This shouldn't be the case, however. Any such problems should instead be dealt with in the 2 previous stages, but the current AI ecosystem is geared towards treating the deployment phase as the testing phase. While this continues to be the case, the response to AI bias will not be mitigation but rather a delayed reaction.

**Between the lines**
For me, generating this kind of framework is definitely the right way to go. Having defined stages of the AI lifecycle can make the identification of responsible parties easier to manage and better expose how bias enters into the process. In my view, any approach to mitigating bias has to then involve the members of the social context in which it will be deployed. Such involvement can then lead to a more elaborate and deeper understanding of the societal implications of AI, rather than leaving that up to a select few in the design process. This technology is at its best when it's representative of all, rather than simply trying to represent all through the eyes of the few.

## Co-Designing Checklists to Understand Organizational Challenges and Opportunities around Fairness in AI

[Original paper by Michael A. Madaio, Jennifer Wortman Vaughan, Luke Stark and Hanna Wallach]
[Research Summary by Anne Boily]

**Overview**: Among the burgeoning literature on AI ethics and the values that would be important to respect in the development and use of artificial intelligence systems (AIS), fairness comes up a few times, perhaps as an echo of the very current notion of social justice. Authors Madaio, Vaughan, Stark and Wallach (Microsoft Research) have co-developed a checklist that seeks to ensure fairness, while recognizing that a procedure alone cannot overcome the value tensions and incompatibilities in ethical practice.

**Introduction**
Anyone interested in the ethics of artificial intelligence is aware of this: a plethora of position papers on AI ethics have emerged in the last five years. They come from private companies, civil society, universities, as well as governmental and international organizations.

Authors Michael A. Madaio, Jennifer Wortman Vaughan, Luke Stark, and Hanna Wallach (Microsoft Research) noted this sort of buzz around AI ethics, while remarking that the level of abstraction of many of these statements posed problems for their practical application (p.1).

To avoid this pitfall, these researchers participated in the development of an equity checklist with 48 AI practitioners from a dozen companies working in a variety of AI applications (pp.1, 5). Through semi-structured interviews as well as "[…] an iterative co-design process […]" (p.1), the authors were guided by three research questions:

*"RQ1: What are practitioners' current processes for identifying and mitigating AI fairness issues?*

*RQ2: What are practitioners' desiderata and concerns regarding AI fairness checklists?*

*RQ3: How do practitioners envision AI fairness checklists might be implemented within their organizations?" (emphasis in the text, p.4)*

**Key Insights**
The disconnect between principles and practice is a criticism that has been repeatedly leveled at AI ethical guidelines. Madaio et al. obviously intended to avoid this pitfall.

But how do we avoid this gap between ethics and technical practice (p.1)? Even with all the goodwill in the world, checklists for ethical AI development and deployment may be poorly followed by practitioners or ignored altogether. Even more, the items on the checklist may prove incompatible in practice (p.2), conflicting in potentially irreparable ways.

The authors are well aware of this, so much so that they admit that "[…] AI ethics principles can place practitioners in a challenging moral bind by establishing ethical responsibilities to different stakeholders without offering any guidance on how to navigate tradeoffs when these stakeholders' needs or expectations conflict." (p.2)

Would the solution to this challenge of compromise lie in a "technologization" of ethics? Madaio et al. do not think so (p.2.) One should not imagine that a simple answer to a binary question (p.3) captures the complexity of the ethical dilemma in which the developer may find himself.

Several contextual elements must be considered, for example the sector of activity or research (public or private), or the size of the company (p.10). There is often disagreement about the definition of the concepts themselves (p.3). In the context of developing their equity checklist, Madaio et al. explicitly acknowledge that the very concept of equity can be understood differently in different contexts: "Fairness is a complex concept and deeply contextual [and] […] There is no single definition of fairness that will apply equally well to different applications of AI" (p.15). For example, equity can be understood in personal or organizational terms (p.5).

The checklist proposed by Madaio et al., based on a previously developed checklist model (p.4) and modified according to the co-design workshops with the participants, consists of six main steps, which roughly correspond to the development of an artificial intelligence system (pp.16-20):

1. "Envision"

2. "Define"

3. "Prototype"

4. "Build"

5. "Launch"

6. "Evolve"

At all stages of the checklist, it is necessary to ensure that the criterion of fairness can be respected or, if compromises are necessary, to document them and to consider dropping the project if this would be preferable (pp.16-20). This proposal guarantees a great honesty in the development of AIS, orienting the approach not only towards the maximization of efficiency, but also towards the common good, encapsulated for these authors in the value of equity.

For Madaio et al, dialogue is central to the use of this list (p.16). The discussion must involve stakeholders as diverse as the people who will use the technology, who will be affected by it, the practitioners who develop it, their teams, and experts to consult at different stages of the process. Hence, the authors suggest that "[…] the most beneficial outcome of implementing an AI ethics checklist may be to prompt discussion and reflection that might otherwise not take place." (p.3)

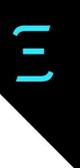

Another advantage of the checklist, according to the authors, is that it would make it possible to establish a preventive rather than a reactive ethic, which would have the mission of anticipating ethical glitches, while being adapted to the operating modes of the practitioners. With such a tool, we could possibly see a reduction in anxiety among developers (p. 6-7).

That said, the use of the checklist does not guarantee the eradication of equity problems, but their prevention and mitigation, as much as possible (p.15). In other words, one cannot eradicate tensions or value clashes in practice, but one can seek to minimize the negative effects of the trade-off one has found (cf. Blattberg 2018, 151). This view is not unlike the philosophical school of "value pluralists" such as Isaiah Berlin, Bernard Williams, or Stuart Hampshire.

One caveat, noted by the participants in this study, is that the checklist may be used merely as a formal, "minimal" process (p. 8), rather than as a means of generating deep conversations about the ethical implications of the technology being developed (p.8). It is important to clarify the role of the checklist. While it serves as a tool for discussion in the implementation of ethics, it is not understood in a fully procedural way. Indeed, this procedural understanding could be problematic, as one study participant noted: "[…] 'I'm a little bit suspicious of the checklist approach. I actually tend to think that when we have highly procedural processes we wind up with really procedural understandings of fairness'" (p.8).

The danger is there and, basically, it is difficult to reduce the richness of a concept such as equity to a definition and a procedure. The authors heard these concerns and adapted their model accordingly: "[…] our checklist items are intended to prompt critical conversations, using words like 'scrutinize' and asking teams to 'define fairness criteria' rather than including specific fairness criteria or thresholds to meet" (p.8).

**Between the lines**
Some will be skeptical of a proposal such as the Madaio et al.'s checklist, since it does not appear to "fix the problem" of AI ethics once and for all. On the contrary, this checklist would rather refer to "[…] a way to spur 'good tension,' prompting critical conversations and prying open discussion about AI fairness […]" (p.10). These conversations will be made possible by the organizational culture – an element that should not be overlooked (p.10).

To try to solve all ethical problems in advance by means of a procedure is probably idealistic. As the authors rightly suggest, "[t]here are seldom clear-cut answers. It is therefore important to document your processes and considerations (including priorities and tradeoffs), and to seek help when needed" (p.15).

Does this mean that AI ethics is, in a certain way, a moving target that no practitioner, theory, or procedure can immobilize once and for all? If that is the case, then dialogue does seem to be a good solution to practice ethics in the specificity of each context. I would add to this that the virtue of prudence would be a good guide for this type of discussion.

**Note**: To facilitate the location of the original information, the page numbers for the checklist are those of the PDF document (the continuation of the article).

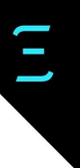

# Go Wide: Article Summaries (summarized by Abhishek Gupta)

**How TikTok's hate speech detection tool set off a debate about racial bias on the app**

[Original article by Vox]

**What happened**: TikTok is under fire again (as covered in a previous newsletter where it involuntarily changed people's faces) for flagging content that disproportionately impacts Black creators. The creator featured in the article mentions how while editing his bio on the platform, his content was flagged as inappropriate for including phrases like "Black Lives Matter." Phrases like "white supremacy" didn't have a similar effect. The creator called for strikes and his videos have had more than a million views with fellow Black creators understandably agitated about disproportionate harm. The company explained that such content wasn't against policy but their content moderation systems needed improvement to address these challenges.

**Why it matters**: In a COVID-19 world where a lot of activism is taking place online, such incidents impact historically marginalized people even more by stripping away their ability to organize and express their views. More so, it showcases how current automated systems for content moderation are quite brittle and unable to handle variances in the text where seemingly inappropriate content might actually be used to highlight and respond to key issues.

**Between the lines**: With the rise of people using social media platforms, human content moderation is only going to decrease over time since it is infeasible to check all the content that goes up on these platforms every minute. We need to have research into more robust automated methodologies and rely on community-driven moderation as an intermediate to still have some human-in-the-loop elements.

**We tested AI interview tools. Here's what we found.**

[Original article by MIT Technology Review]

**What happened**: With a lot of upheaval in the job market since the start of the pandemic, and limited staff capacities on the recruitment side of things for companies, many have resorted to the use of automated hiring tools. The authors of the article put two such systems to the test, CuriousThing and MyInterview to gain an understanding on how good they are in meeting their claims. To their disappointment and no surprise, they found these two tools to be opaque in what they evaluated. As an example, even when reading out paragraphs in German, the tool determined that the interviewee, one of the authors, was quite fluent in English. It also evaluated them on the Big Five personality traits and other attributes with varying results.

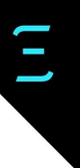

**Why it matters**: These tools are touted as a way to reduce bias in the hiring process and often the results from such tools are not the only data points in hiring decisions. But, they have the potential to skew the process, especially when they have glaring flaws and high sensitivity in how they evaluate some of these psychological traits using things like the intonation of someone's voice rather than the content of what they are saying.

**Between the lines**: While the founder of the company defended that the system is not meant to be used with German and hence the flawed results, it still raises an interesting question on what the degree of robustness of these systems is, particularly when they are used in the wild, as opposed to a controlled experiment in this article, where there might not be an opportunity to review how a certain person responded. This can lead to pre-emptive dismissal in a large pool of applicants in the interest of expediency, particularly affecting those who don't fit the mold that is determined to be ideal by the automated system.

## How Humans Can Force the Machines to Play Fair

[Original article by [Quanta Magazine](#)]

**What happened**: In this insightful interview with the inventor of the notion of differential privacy, we learn about the new challenges that Dwork is embarking on in her recent work. Tackling fairness in AI-infused systems, Dwork talks about her work titled "Fairness through Awareness" which takes into account both individual and group fairness and how to achieve both. She also talks about how this is a much more difficult challenge compared to her work in privacy but advocates taking a "sunshine" approach to the research work in this space. The article also has several great examples of how applying individual fairness isn't enough and how her experience with piano practice reinforced the importance of considerations for fair affirmative action to achieve group fairness.

**Why it matters**: The field of AI ethics is inundated with work on how to best achieve fairness in machine learning. Yet, a lot of it struggles to articulate how to account for tradeoffs that are bound to occur when offering preferential treatment to some over others in the interest of achieving fairness. Dwork's work and her history in providing clear metrics and methodologies for addressing challenges in achieving more responsible statistical systems is a good precedent and beacon for us to make meaningful progress in this space.

**Between the lines**: Bias and fairness are extremely challenging concepts when it comes to machine learning because they don't have clear metrics as is the case with privacy where we are fairly certain of what the outcomes need to be. In this case, there is widespread disagreement even about what fair outcomes look like presenting us with graver challenges. Work that claims to provide easy solutions is certainly something to guard against, especially in the face of this space becoming commercialized with startups and tools being offered that address, or worse "fix", bias in machine learning.

## The Secret Bias Hidden in Mortgage-Approval Algorithms

[Original article by [The Markup](#)]

**What happened**: There are strong biases against people of color in lending decisions made by financial institutions in the US as found out by a recent study conducted by The Markup on data from 2019. They found that even after controlling for new factors that are supposed to tackle racial disparities, the differences persisted. Even those with very high income levels ($100,000+) with low debt were rejected over White applicants with similar income levels but higher debt. This analysis was sent to the American Bankers Association and the Mortgage Bankers Association both of whom denied the results from the study citing that there were missing slices of information in the public data used by The Markup thus making the results incorrect. But, they didn't point out specific flaws in the analysis. Some of that data is not possible to include in the analysis because the Consumer Financial Protection Bureau strips it to protect borrower privacy.

**Why it matters**: While there are laws like the Equal Credit Opportunity Act and the Fair Housing Act that are supposed to protect against racial discrimination, with organizations like Freddie Mac and Fannie Mae driving how loans are approved through their opaque rating systems, it is very difficult to override decisions made by automated systems as mentioned in the case of the person discussed in the article who was denied a loan at the last moment, unresolvable by 15 or so loan officers who also looked at the loan application.

**Between the lines**: The algorithms being used by these organizations date back to more than 15 years and reward more heavily traditional credit which White people have more access to. They also unfairly penalize structural elements like missing payment reports filed by payday lenders who are disproportionately present in neighborhoods with people of color thus skewing the data on bad financial behavior while ignoring good financial behavior such as payment on time of utility payments. The lack of transparency on the part of organizations like Freddie Mac and

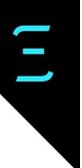

Fannie Mae, and the protections that they have in not disclosing outcomes from their systems in public data and the opaqueness around their evaluation methodology will continue to exacerbate the problem.

## Facebook Apologizes After A.I. Puts 'Primates' Label on Video of Black Men

[Original article by NYTimes]

**What happened**: Facebook provides automated recommendations for videos and other content on its platforms as users consume content. In a particularly egregious error, the platform showed a message prompting the user if they wanted to see more "keep seeing videos about Primates" when the video that they were watching was in fact a video of a few Black men having an altercation with some White people and police officers. The video had nothing to do with primates whatsoever. The spokespersons for the company said that they are doing a root cause analysis to see what might have gone wrong.

**Why it matters**: The use of an AI system trained without guardrails, especially when there are known issues of bias and racism due to the outputs from the system shows that there isn't yet enough being done to mitigate these issues that can instantly impact millions of people due to their scale and pace. Facebook as a platform has the largest repository of user-uploaded content and it uses that to train its AI systems. But, this recent incident demonstrates that there is much more to be done before this becomes something that we can safely deploy, if we ever get there.

**Between the lines**: Given the large number of problems that automated recommendation systems have today, it is a bit surprising that these are still used in deployment. What would be interesting to analyze is the extent to which these companies are willing to overlook such incidents in the interest of the gains that they get from keeping people engaged on the platform when the recommendations do work. Because the external research community and the public have no visibility on this tradeoff, it is increasingly difficult to hold such organizations accountable for such errors and to recommend corrective actions when it's not entirely clear how the system is being built and operated and what the internal costs and benefits analysis looks like.

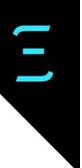

## Minority Voices 'Filtered' Out of Google Natural Language Processing Models

[Original article by Unite]

**What happened**: The article spotlights some findings from a recently published report that analyzed the filters that went into creating the C4 (Colossal Clean Crawled Corpus) dataset, a subset of the much larger Common Crawl (CC) dataset. The C4 was used to train Google's T5 and Switch Transformer, two massive language models that are used in downstream products and services. The essence of the findings were that in creating a non-toxic dataset, the aggressive filtering excluded material related to LGBTQ+ communities in non-sexual and non-offensive contexts along with a heavy filtering of colloquial and ethnicity-specific dialects like African-American and Hispanic-aligned English.

**Why it matters**: One of the reasons for poor performance of large language models on non-political, non-offensive, non-sexual material that discusses LGBTQ+ communities is that there is no representation of them in these curated datasets, or when it is there, it is heavily filtered. This has the impact of much stronger automated content moderation applied to that content compared to others on social media platforms. Other products and services that also consume such pretrained models for operations then suffer from biases because data related to these areas and dialects is excluded, rather than coming up with better approaches to moderation that don't just rely on a banned list of words.

**Between the lines**: One of the striking things about the research efforts that led to the report is that they've made the raw data available for C4 and provided different versions of it with various levels of filtering applied for people to further analyze the data. Even though the original authors of C4 (from Google) made the scripts available for that dataset, the computational costs are so high, that recreating the C4 from CC would be out of reach for many researchers. Not only do these minority communities suffer from the biases against them in content moderation, because of such misdirected filtering, they stand to miss out on legitimate benefits from ML like machine translation and search. As the authors of the study rightly pointed out, we need to do better in terms of how we process data because it has significant downstream effects.

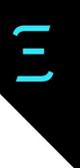

## Facebook, Citing Societal Concerns, Plans to Shut Down Facial Recognition System

**[Original article by NYTimes]**

**What happened**: In a move to "find the right balance," Facebook is going to be deactivating facial recognition technology within its ecosystem citing concerns with how this technology is used and what it powers. With the recent rebrand to Meta, Facebook is on a warpath to set its public image right. The feature was used to power automatic tagging of people in uploaded pictures and videos, something that would help the network deepen connections and make it more frictionless for users to associate their account with visual assets on the site. The technology was also used to power capabilities to detect if someone might be impersonating you on the site and to provide accessibility features like reading out descriptions of photos for blind users.

**Why it matters**: Given the fines that the company faced from the FTC and the state of Illinois citing violations of privacy, this is a win for privacy advocates to get Meta to shut down this feature. Approximately 1 billion facial recognition templates will also be deleted from the site and there are talks about controlling pictures' visibility as well to limit how external companies like PimEyes and ClearviewAI can use these assets to train facial recognition technology.

**Between the lines**: Despite this announcement, something of note in the article is that Meta has not ruled out completely the use of facial recognition technology in future products. Though the recently released glasses in partnership with Ray-Ban don't have it, this doesn't mean that future products will never again have facial recognition technology. We also need to continue to pay attention to how this data that has been collected will be removed and how other policies change on Meta and related sites like Instagram which continue to be the largest repositories of facial images in the world.

## Uncovering bias in search and recommendations

**[Original article by Vimeo Engineering]**

**What happened**: The team at Vimeo, the video streaming platform, talks about their work in assessing bias in the search results and the recommendations that they provide to the users of their platform. They do so for gender bias as an entry point to this assessment and utilize LTR (Learning to Rank) and BM25 approaches as the underlying search results ranking comparing results from gender-neutral search terms and checking for the presence of gendered terms in the returned result list and the ordering of those results. Chock-full of technical details, one of the things that stand out in the article is an interesting challenge on ground truth labels, which are hard to get because the relevance of search results, especially for items saved in a library are highly specific to the user themselves and hence it is difficult to generalize to the broader user base from that. So, they used clicks to form the signal for the ground truth in the supervised learning task.

**Why it matters**: The way the experiments are run (e.g. presenting two variations of blue text for buttons and judging which users prefer, only provides information about users' preferences for blue buttons and nothing about green buttons) and how data related to interactions is collected can have a tremendous impact on downstream tasks that might use this data. In addition, the small sample sizes of self-declared gender pronouns on the Vimeo user base and drawing conclusions from that to apply to the broader user base also poses challenges. For example, some might not choose to identify, the limited options of gender pronoun identification offered by Vimeo are also acknowledged by the team that did this analysis. But, this does offer a great starting point for diving into how bias may manifest in search results and recommendations.

**Between the lines**: For platforms that are even larger than Vimeo, the impact of bias in what kind of results pop up when a user types in a search result, and particularly how those search results are ordered (think back to how many times you navigate beyond the first page of search results on Google) have the potential to amplify gender and other biases significantly if clicks and other user-driven metrics are used to drive the modeling of relevance for any downstream tasks. Having more studies conducted by platforms themselves instead of by external organizations has the upside that the platforms have the deepest access to all the metrics and interactions; of course, this comes with the caveat that negative outcomes from such an investigation may be suppressed for business interests.

# 5. Social Media and Problematic Information

**Introduction** by Abhishek Gupta, Founder and Principal Researcher, Montreal AI Ethics Institute

The scourge of problematic information continued throughout the second half of 2022, not much has changed unfortunately since the publication of our **last report**. One of the biggest players in the space, Facebook rebranded as Meta and dominated quite a few news cycles towards the end of 2021 with its pivot towards the metaverse (I am personally still unclear on what it means exactly so if you have a clear explanation for it, feel free to drop me a line at abhishek@montrealethics.ai)

The opening piece in this chapter talks about deepfakes which continue to form a chunk of the problematic information spreading around on social media and it makes the case for finding those with capabilities to better analyze these deepfakes and not only find countermeasures but also ways to curb their spread in the first place.

The piece on how targeted ads can divide us even when they're not political was eye-opening; most of the time our focus is on political ads but harms can emerge in more subtle and equally pernicious ways elsewhere too. That said, it goes without saying that political ads still caused a ton of problems as a couple of pieces point out in this chapter. One of the pieces demonstrates the sheer scale of problematic information where more than 140 million people's attention was grabbed by troll farms. For context, that is almost half the population of the US. Documenting and analyzing the kind of content spreading on ~~Facebook~~ Meta isn't easy, especially as they revoked access for groups studying problematic information on the platform. The chapter shares some details on independent efforts like the one being run by The Markup called the Citizen Browser that gives them access to untouched Facebook feeds as participating users view them. The analyses coming out of that project are definitely ones to keep your eyes on in case you want to go beyond what is just published by the Meta team about the state of problematic information on their platform.

Finally, three other pieces really caught my attention in this chapter. The first one on how the underlying business models of both Google and Facebook that rewards content creators steers malicious actors into areas where they not only have political motivations but also financial ones (enough to sustain their operations!) that exacerbates the problem of problematic information on the platform. An initiative by the US Government that hired influencers to combat the spread of vaccine-related misinformation was particularly exciting since it taps into the native dynamics of the platform to bring about some positive change rather than solely relying on policy mechanisms. A heartbreaking piece concludes the chapter whereby the queer internet is gradually being erased completely from China and it documents the efforts of volunteers who dedicate their time and face potential persecution in attempts to try and preserve that and continue to maintain their social connections.

---

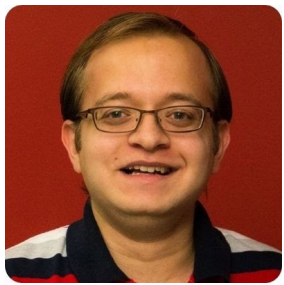

**Abhishek Gupta (@atg_abhishek)**
Founder, Director, & Principal Researcher
Montreal AI Ethics Institute

Abhishek Gupta is the Founder, Director, and Principal Researcher at the Montreal AI Ethics Institute. He is a Machine Learning Engineer at Microsoft, where he serves on the CSE Responsible AI Board. He also serves as the Chair of the Standards Working Group at the Green Software Foundation.

# Go Wide: Article Summaries (summarized by Abhishek Gupta)

## The World Needs Deepfake Experts to Stem This Chaos

[Original article by Wired]

**What happened**: In Myanmar, a video made claims that amplified corruption charges against Aung San Suu Kyi, but because of its grainy quality and the general distrust in government, people decried it as being a deepfake. People used online deepfake detectors to figure out the video's authenticity, and social media quickly made this opinion popular. The author of this article points out concerns in how malicious agents can easily manipulate untrained everyday citizens into believing whatever they want as the quality of deepfakes increase.

**Why it matters**: While the risk from deepfakes remains highest for unwanted, nonconsensual sexual images, their use for political manipulation is on the rise. Everyday citizens unaware of the limitation of deepfake detection run the risk of counter-forensic techniques that inject artifacts into videos to confound these free, online tools. Encouraging amateur forensics online can lead people down conspiracy rabbit holes exacerbating the problem of misinformation online.

**Between the lines**: Sam rightly points out that more advanced capabilities are limited to elite circles of academia, government, and industry in Europe and North America. We need more funding and sharing of knowledge and tools with other parts of the world, especially those vulnerable to such attacks. Inequity in the distribution of these capabilities will deepen the digital divide across regions.

## After Repeatedly Promising Not to, Facebook Keeps Recommending Political Groups to Its Users

[Original article by The Markup]

**What happened**: In The Markup's Citizen Browser project, which tracks the Facebook feeds of users paid by The Markup to send them data, researchers discovered that despite promises made by Facebook that they will stop recommending political groups to users, they haven't done so yet. In several responses to government agencies and in public, Facebook has claimed that they have applied measures to eliminate such recommendations but have let slip on occasion that they cannot do so entirely.

**Why it matters**: As documented in the article, about two-thirds of the people landing on politically-motivated Facebook groups arrive there through the recommendations made by the platform to its users. If, even after making public commitments to remedy that, we don't see changes, then that is a cause for severe concern.

**Between the lines**: As we've mentioned in this newsletter before, work from organizations like The Markup can help to hold companies like Facebook accountable. But, this requires funding and innovative research methods, especially when there aren't broad-access APIs available to researchers to scrutinize the activity on the platform.

## Targeted ads isolate and divide us even when they're not political – new research

[Original article by The Conversation]

**What happened**: When we think about divisive ads, political ads always come first to our minds. This article argues that commercial ads post an equally pernicious threat to the epistemic integrity of our information ecosystem online. Drawing on an interesting example regarding body positivity in the London underground, passengers complained to the regulator that the ads promoted unhealthy stereotypes and prompted action from the regulator in taking down the ad. Yet, out of the hundreds of thousands of passengers, only 387 filed such a complaint, presumably some were stirred by the graffiti on those ads prompting them to take action as well.

**Why it matters**: In the online world, we are neatly segmented into various categories (whether accurate and reflective of us or not) that make it difficult to understand and gain collective knowledge about whether some ads might be causing us harm without us even realizing, as was the case with the London underground example which manifested in the physical world. Commercial ads can cause harm both through their targeted messaging towards vulnerable populations like showing gambling addicts ads for casinos and omission of ads, say job postings to only a certain gender.

**Between the lines**: Given that most of our focus remains on tackling the problem of political ads on platforms, this article presents a compelling case for thinking more deeply about the impact that commercial ads have on us. There are certain policies and regulations, in the US for example there are regulations around disability, housing, and employment that might make some ads or omission of ads illegal, but for the most part it remains an understudied area. This problem is exacerbated by the fact that it is incredibly difficult to obtain the necessary information across a broad swathe of users without enrolling them in a study which can cost a lot of money, as is the case with the Citizen Browser project from The Markup.

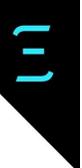

## Facebook Tells Biden: 'Facebook Is Not the Reason' Vaccination Goal Was Missed

**[Original article by NYTimes]**

**What happened**: The US had planned to have 70% of their population vaccinated by July 4, but it fell short of the target and the Biden administration laid some of that blame on the misinformation spread on Facebook as a cause for the vaccine hesitancy in the US. The platform responded by saying that they have undertaken many measures that have helped to inform the users of Facebook about vaccination such as notices and eradication of anti-vaccination ads on their platform. An adversarial dynamic is emerging between the administration and the social media platform as they are frustrated with each other's understanding of the efforts being made.

**Why it matters**: While it is not uncommon for such a divergence to emerge, the lack of transparency in the impacts of the efforts, especially in response to the continued concerns that misinformation is still spreading rapidly on the platform through groups. This is a continual platform where people who are already believers in conspiracy theories and other false content are suggested anti-vaccination groups given the meta-alignment. But, this only exacerbates the problem. A lot of engagement happens in these groups and until the platform is able to dramatically reduce these recommendations in addition to its other efforts, we will continue to see the problem prevail.

**Between the lines**: We need to find better ways of engaging the technology and government stakeholders in our information ecosystem. The stronger the adversarial dynamic, the more the risk of irreconcilable differences and non-resolution of the core issues. More structured experiments and transparency around the results from the efforts undertaken by the platform will help us build a better understanding of what actions are going to be effective in our fight against the infodemic which will ultimately help us fight the pandemic.

## To Fight Vaccine Lies, Authorities Recruit an 'Influencer Army'

**[Original article by NYTimes]**

**What happened**: While most of the time in this publication we talk about the negative effects of social media and the spread of disinformation and misinformation that it facilitates, this article highlights a great example of government effort to recruit the power of influencers in spreading "positive information" to get people vaccinated. The White House is working with influencers on TikTok, YouTube, and other platforms to get them to share the message of vaccination with their large follower bases.

**Why it matters**: The vaccination rates in the US have been higher in the older demographic than the younger ones. This is the exact audience that can be reached through influencers on social media. As pointed out by a survey cited in the article, people tend to be better persuaded by content creators that they watch / listen to on social media than other publication outlets. This is then the perfect channel to quash rumors and answer questions about vaccination, urging people to go out there and get the jab.

**Between the lines**: Borrowing on tactics that were used for political mobilization during the Biden campaign, the White House is now using the same insights and approach to get an important message out to the people to get vaccinated. As DiResta points out in the article that those looking to spread disinformation are more motivated and organic reach can exceed targeted measures like these, it is still a good first step in countering "negative information" with some action rather than just trying to suppress misinformation and disinformation on these platforms. A multi-pronged approach will always be more effective in countering pervasive problems like this one.

## Let's Keep the Vaccine Misinformation Problem in Perspective

**[Original article by Wired]**

**What happened**: An insightful article that talks about the complexities in trying to disentangle the effects of misinformation, which are myriad, from that of other confounding factors when it comes to the low rates of vaccination amongst certain demographics in the United States. In a study cited in the article, they find that there is a strong correlation between vaccine hesitancy and a general mistrust in mainstream institutions which pushes these people towards getting their news from social media rather than more trusted and reputable sources. The recent clash between the White House and Facebook on the role that Facebook has played in enhancing vaccine hesitancy and the US falling behind in meeting its vaccination goals is a demonstration of how collapsing the issue into easy to reason with binaries makes it a tough problem to solve. In particular, the article points out how some are advocating for dispensing this political capital in a better fashion to perhaps spark vaccine mandates by offices and schools and to treat misinformation's role in this as a broader issue that needs more long-term solutions.

**Why it matters**: As is the case with any socio-technical issue, there are a range of variables that impact the inputs and outputs of a problem. What we see here is that a lack of transparency on the parts of Facebook and YouTube for example limiting the ability of researchers to gather adequate data about the degrees of correlation between vaccine hesitancy and the content that they see online along with the efficacy of the measures undertaken by these platforms to limit the spread of misinformation, specifically here, related to vaccines.

**Between the lines**: The recent denial of access to researchers studying the Facebook platform is yet another blow that will only deepen the chasm between positive public health outcomes and the potential role that a company like Facebook can and is playing in that. The Klobuchar bill mentioned in the article serving as a messaging bill is a first step in establishing some baselines on how to tackle the issue but it raises even more questions and issues than it tries to solve: namely, transferring over the arbitration of what is and is not misinformation from the platform and their community of moderators to the government, which would certainly raise concerns around the violation of the First Amendment.

### Troll farms reached 140 million Americans a month on Facebook before 2020 election, internal report shows

[Original article by MIT Technology Review]

**What happened**: In a perhaps not so shocking report, a former senior-level data scientist revealed that troll farms continue to have significant audiences who are deeply engaged on Facebook. The report highlighted 3 key shortcomings in the existing platform design that allowed pages run by these troll farms, that have never engaged with, nor have knowledge of the communities that they influence, to shape their thoughts. Facebook doesn't penalize pages that post unoriginal content, allowing previously viral content to merely be copied and go viral again, perpetuating disinformation. Engaging content from pages that users don't even follow can still show in their feeds when a friend interacts with that piece of content. And finally, more engaging content is pushed up higher in the newsfeed no matter what the type of content or source. This incentivizes politically divisive and clickbait content to rise to the top.

**Why it matters**: In an ecosystem where a large number of people get their news updates from social media platforms rather than traditional media outlets, the combination of the above three forces entails a significant problem in our ability to maintain a healthy information ecosystem. More so, with a blatant disregard for the type of content and merely utilizing its engagement rates to disburse it, the platform specifically encourages the worst kind of behavior that troll farms in places like Kosovo and Macedonia are able to leverage for financial rather than political gains.

**Between the lines**: The report also provides some suggestions on how we can combat this scourge: using something called Graph Authority, one can get an understanding of how authentic and relevant a piece of content is based on the number of reputable in and out links, something that Google has done for several years already. Yet, as per the report, such efforts within Facebook have largely been ignored and it continues to prioritize content that has the highest likelihood of engagement driving usage on the platform rather than the quality of the content itself.

**The Facebook whistleblower says its algorithms are dangerous. Here's why.**

[Original article by MIT Technology Review]

**What happened**: Frances Haugen, the primary source for The Facebook Files included in the WSJ investigative series on the company, testified in a Senate hearing confirming a lot of things that people assumed about how Facebook operates and where it falls short in terms of practically addressing problems on its platform. One of the main arguments put forward by Haugen in the Senate hearing was that the company knew about the problems, and didn't act on them. More so, the emphasis on content moderation as a tool for creating a healthier information ecosystem is inherently flawed and instead we should be focusing on the design of the algorithms powering the platform to address the root causes of the problems plaguing the platform.

**Why it matters**: Scathing in its criticism of the platform and what it is doing to address the challenges including misinformation, polarization, addictive engagement, data misuse for targeted advertising, and others, the fact that existing mechanisms like content moderation because of limitations in their language and context capabilities are just proverbial band-aids on a broken dam are the call-to-action that should spur Facebook to make more investments in reshaping the platform to mitigate the emergence of such problems in the first place. What this also does is shows that presented evidence of investments into content moderation, we should be more cognizant of the actual impact that such measures will have in solving the root problems at the heart of the platform.

**Between the lines**: At the center of all the proposed mechanisms, the problems, proposed regulations, and everything else to create a more healthy ecosystem is a fundamental tension: the business incentives of the platform in realizing profits are stacked against the interests of the users of the platform. Yes, there might be ways of giving one side more of an edge but the tension remains because of the business model which ultimately drives a lot of the activity on the platform's design, development, and deployment. Without acknowledging that more fully, and working towards resolving that tension, the solutions will only address the root problems in a piece-meal fashion.

### How We Investigated Facebook's Most Popular Content

**[Original article by [The Markup](#)]**

**What happened**: The team at The Markup reviewed the recently published "Widely Viewed Content Report" from Facebook, a document that is published by the company in the interest of transparency, to see how content disperses online, what the frequency of that content is on the platform, and how the rankings of various websites change based on what kind of methodology is used. They utilized their Citizen Browser project to simulate the calculations done by the team at Facebook, and applying statistical methods, determined that the sample size and approach that they are using lines up quite well to draw statistically significant conclusions about the performance of top performing content on the platform.

**Why it matters**: The article dives into methodological details that are well worth reviewing, but more importantly, they highlight the lack of transparency, ironic given the purpose of the report, in the methodology published accompanying the report from Facebook. The focus of that report was solely on the views for the content and much less so on the frequency with which that content might have appeared in the newsfeed of a user. This is an important distinction to make, since the frequency with which someone sees a piece of content, the chances that they assimilate its message increases, and it also correlates with the probability of them sharing that piece of content, thus amplifying its reach. Without that level of granularity, we get a poor facsimile of the actual distribution and influence of content on the platform.

**Between the lines**: In an effort to provide transparency, Facebook's report is a great first step, but as the investigation done by The Markup points out, the report obscures quite a bit, especially given the limited information about the underlying methodology that was used to arrive at the final numbers. More so, something that is now backed by empirical evidence as per the work done by The Markup, is that sensationalist content and opinion sites do outperform mainstream news content on the platform and this isn't apparent in the report from Facebook because of the way the calculations are carried out.

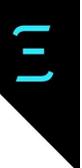

## How social media companies help African governments abuse "disinformation laws" to target critics

[Original article by [Rest of World](#)]

**What happened**: The article describes how the combination of vaguely defined disinformation laws in countries like Kenya, Uganda, Malawi, and Nigeria, instead of clamping down on disinformation actually restricts legitimate speech more. This is exacerbated by the fact that social media platforms are also limited in their approach of addressing disinformation problems on their platforms, such as narrow approaches like simply taking down content. Sometimes, these vague laws have also led to internet shutdowns in African nations. In the midst of all this, the regulations mostly serve the interests of the government while the policies of social media platforms mostly serve the companies themselves. The fundamental rights of end users are mostly ignored.

**Why it matters**: The article does point to some fundamental texts in the space like the Santa Clara Principles (MAIEI provided comments to it) that can serve as guides on effective regulation of the problem of disinformation such that fundamental rights of end users are still protected. Concretely, creating something that is soft law in the beginning based on these guidelines and then determining which parts of it work and which don't, can be moved into the hard law territory.

**Between the lines**: A shared responsibility model where we have many actors who are jointly responsible for governing how the disinformation challenge is addressed on social media platforms is going to be essential. More so, I believe that elevating media and digital literacy will offer yet another effective avenue to combat this problem, further bolstering the efforts that emerge on the technical and policy fronts to address these challenges.

## The Metaverse Is Mark Zuckerberg's Mobile Do-Over

[Original article by Wired]

**What happened**: The metaverse has taken the world of tech-related discussions by storm ever since the rebranding announcement from Facebook becoming Meta. The article dives into the details of previous attempts by Meta in trying to establish itself in the leagues of other companies that have a stronger grip on the underlying infrastructure and plumbing that enables us to enjoy all the apps and other services built on top of them, so things like OSes, devices, and platforms (which is its domain for now). The article examines past endeavors from the companies in trying to introduce mobile OS, a Facebook phone, the Facebook Home that was supposed to become the central thing on your phone, and finally what they're trying to achieve with their vision for the metaverse and how they are approaching it.

**Why it matters**: If the metaverse is something that takes hold (though some argue that we are already in a metaverse with all the other online activities that we are engaged in and how we define the metaverse in the first place!), Meta argues that it will only become successful if it involves open standards and other companies providing services and solutions that can all plug into a single ecosystem. Of course, there are undercurrents to this approach in that Meta would be delighted if it is based on their vision and platform + infrastructure that would make them a central player in this future if it comes to pass.

**Between the lines**: With all the scrutiny that the company has faced in the US, the rebranding and moving away from the social media platform to something more nebulous like the metaverse might seem like a mechanism for drawing away attention. But, as technology becomes more ubiquitous, and the possibility of realizing the metaverse, at least in the form that Meta imagines it, becomes more likely, this is a good call for our community to start thinking about what the ethical consequences of this might be so that we are prepared and respond proactively.

## How Facebook and Google fund global misinformation

[Original article by MIT Technology Review]

**What happened**: Algorithmic amplification of problematic information online is nothing new to the readers of this newsletter. We've covered it time and again. But, this article sheds a new light on the machinery that feeds this information ecosystem, in particular, it highlights some of the funding mechanisms that power malicious actors to continue their activities. In particular, the introduction of Instant Articles by Facebook brought into the fray additional incentives, financial ones, that directed the energies of otherwise undirected malicious actors into the political arena, given the high engagement rates of political content on the website. What this meant is that not only were there politically motivated malicious actors, but also those who aren't really connected with any political objectives and are seeking to eke out a profit by milking the content dissemination machinery that Facebook and Google proffer.

**Why it matters**: While addressing the algorithmic basis of how information spreads online is one way of going about tackling the proliferation of problematic information, we need to also focus on the underlying business mechanisms. Especially as highlighted by this article when new tools like Instant Articles propel clickbait and non-mainstream media outfits to outcompete and overcome the platform when it comes to the content that is viewed and engaged with by the users. The fact that the information ecosystem is dominated by a few giants and that what happens on one platform (say YouTube) has a dramatic impact on content that shows up on and dominates another platform (videos trending on Facebook), tells us that we also need to examine what such a monopolization means for the health of the information ecosystem.

**Between the lines**: Adding financial incentives to an already charged ecosystem where there are many motivations for adversaries and malicious actors to pollute the information ecosystem demonstrates a worsening state of affairs. Having higher transparency on who is paid out and how much from monetization mechanisms, along with access to external auditors and researchers (who have had their access taken away from conducting independent research on Facebook) and demonstration of action on the recommendations that are provided by civil society organizations and individual watchdogs is going to be essential to curb the spread of misinformation online, and reduce the very real harms inflicted on people as a result of this proliferation as seen in Myanmar amongst other places.

### China's queer internet is being erased

[Original article by Rest of World]

**What happened**: In July 2021, some of the most prominent and well-connected social media accounts for members from the LGBTQI communities were banned, disconnecting folks from across the nation who relied on these to coordinate their online activities and exchange on issues they face. Some of the people interviewed for the article mentioned that they saw such a ban coming given the slow erosion of safe spaces for them both offline and online to organize. In the early days, there was support from organizations like universities who supported these online accounts as a way to showcase that they were open and progressive. But, the communities started to get called out for anti-national sentiments and these were used as reasons to start censoring them. Even apps like Blued that allegedly served the needs of the communities have started acting in alignment with national government interests; embarking on things like assigning credits for good behaviour on the platform and stripping those credits away as the accounts posted anything that had the potential to draw ire from the government censors.

**Why it matters**: Given the continued taboo around LGBTQI identity in China, online spaces, through their anonymity, offered a safe space to explore and discuss issues while moving towards securing greater rights to be more open about this subject. But, as mentioned by the interviewees in the article, anything that relates to rights is something that is quicker to attract the censors and has a higher likelihood of getting their accounts shut down. Compromise in the form of using their accounts (at least the ones that are still active) as a medium to share resources on sexual health and other topics that aren't rights related is still a way to convene these communities without receiving outright bans.

**Between the lines**: As more and more of everyone's activities leave digital traces, the situation in China doesn't bode well for how people organize around interests and identities that aren't acceptable to the government. This is particularly problematic in areas where there are few people that identify as one does in the LGBTQI community, making it incredibly difficult to find others to share their struggles with. This has the potential to make things worse even from a mental health perspective for those who are not able to find a like-minded community. Moving towards offline, in-person community gatherings is a way to counter this force, but it comes with the cost of loss of anonymity, and with the ongoing pandemic, elevated health risks.

# 6. AI Design and Governance

**Introduction** by Michael Klenk, Assistant Professor, Philosophy, Delft University of Technology

We lived in a beautiful, century-old house at a Dutch canal. The problem was, we had mice in the building. Their squeaking and shuffling woke us up at night in sometimes eerie and always annoying episodes. I had designed things before, like a kitchen table, but now I needed an elaborate mechanism to catch the critters. I had my design problem cut out for me.

The design of mouse traps shares its two fundamental questions with any design activity, including the design of AI. *What* problem should be solved? And *how* should the problem be solved? As much as I wanted to get rid of the mice, I did not want a lethal trap, nor did I wish to turn the whole house into a warzone. Looking back, I recognise how I implicitly combined the obvious functional requirements for a trap with social and ethical requirements. I balanced the condition to '*catch mice a.s.a.p.*' with '*unnoticeable in daily life*' and '*not lethal*.'

Design is an inherently normative activity. Designing something means deciding how the world should be like. Your individual decision may be small and inconsequential. For example, in the case of my humane mouse trap, little was at stake except for the mice and my sleep quality. But take together all our design choices, however small, and you see them giving shape to our world, including our social world. For example, imagine how the AI co-worker discussed in the report below may influence and change how you experience your work.

As a designer, you orient yourself in a design space that contains everything you could do. No design problem has a single solution. There are always multiple options for thinking about a problem and different resolutions (otherwise, I would not deem it a design activity). Obviously, that you could design something in some way does not imply that you should do it that way. For example, I could have burned down the house to get rid of the mice, but that option is patently absurd. Possible, but not what ought to be done. There are so many ways our world could be, and naturally, some are better than others. At this point, design meets ethics.

The normativity of design means that we must be critical about the problems we want to solve with AI. When a design problem is given to you (say, at work), it may seem like the only question is *how* to solve it. But in that case, the normative question – *what* problem should be solved? - has already been answered for you. The question may often escape our view, but it is there nonetheless. Design thinking capitalises on this simple insight into the normative nature of design. It is never the case that even the problem definition is certain. There is always an open question: Is that the problem we *should* solve? Good designers ask that question. When you read the summaries below, you can ask yourself whether AI is being used to solve a problem that we ought to solve. For example, what problem does the AI co-worker solve, and is it a good problem that needs a solution?

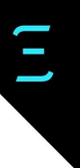

Moreover, the criteria we should use in assessing our design decisions are not set in stone and certainly not limited to technical and economic considerations. The surging attention to AI Ethics is based on a fundamentally similar realisation. Rather than myopically focusing on technical and financial factors, people now realise that ethical, social, and ecological considerations determine what problems we ought to design AI for and how we should solve them. The summaries of this report illustrate this in several ways. For example, the call for sustainable AI is a specific design requirement that constrains the design space for AI solutions by demanding that AI training and execution be done sustainably. Another example is the AI certification summarised below. Notwithstanding the value of such certifications, the requirement that AI meet specific ethical requirements shapes which options in design space are feasible.

The deep and hard work is to find out the proper criteria that we can use to evaluate our design choices. This is most explicitly discussed in the literature on Value Sensitive Design and Value Alignment in AI. In both fields of literature, you will re-discover variants of our two fundamental design questions: What should we design AI for? And how can we ensure it aligns with the targets we chose? I am glad to see the summary of DeepMinds work on the topic. It illustrates that serious attention is given to this topic in academia and industry. The summary of mapping AI4Good principles for a Value Sensitive Design approach is also helpful. It illustrates how design principles are at their heart attempts to suss out which options in design space are feasible for us. Going forward, we need more integration of philosophical perspectives about what values are to make progress on the value alignment problem.

If we bracket for the moment the technical question of how we can align AI with given values, we first need to find out what these values are. And that is a question that philosophy has done much to clarify. So we should use this resource. I am often amazed how my students, most of whom have a STEM background and come to my courses at best sceptical about the value of philosophy, turn into epistemologists and metaphysicians within a few sessions. They ask difficult and excellent questions about the nature of the values we are supposed to be aligning AI with and the possibilities for finding out about them. For example, are values discoverable by science, and how should we deal with value disagreements? I suspect that the scepticism expressed by AI practitioners about the widespread implementation of AI Ethics summarised below is partly a reflection of the difficulty of these questions.

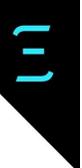

Some progress can be made even when a final answer is not yet settled from a philosophical perspective. For example, the idea that ethics applies to design decisions in the first place is not trivial and still supported by several lines of philosophical argument. And one thing I know for sure. Our shared aim of using AI to shape the world positively is not helped by simplifying things. On the one hand, it appears that there is a strong tendency in AI Ethics to oversimplify in a subjectivist direction. AI Ethicists often seem overly impressed by apparently deep, unassailable differences in values. But, of course, there are legitimate questions about how deep these disagreements really are. We should not be led to deny that there are better or worse aims to strive for in the design of AI. On the other hand, we must not oversimplify in the direction of objectivism and uncritically suggest that a certain set of principles is evidently clear and set in stone. In any case, we need a good explanation for why specific values hold and why they should shape our design choices.

When we see the design of AI for what it is, an inherently normative activity, we can ask the right questions. What problems should we solve? And how should we do it? With these questions in clear sight, we can start seeking answers. I urge that we take the philosophical fundamentals of these questions seriously. Though there are no neat and straightforward answers to be expected, we will avoid repeating mistakes that years of philosophical theorising have laid bare. Whatever happened to the mice in our house, however, you'll have to ask me in person.

---

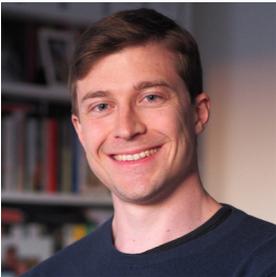

**Michael Klenk**
Assistant Professor, Philosophy
Delft University of Technology

Klenk is Assistant Professor of Philosophy at Delft University of Technology. His research is part of the ERC-funded Value Change project and the Ethics of Socially Disruptive Technologies research programme. He is a member of the Delft Design for Values Institute and as an advisor for Ethical Intelligence. He also works as a columnist at 3Quarks Daily.

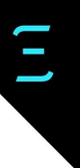

# Go Deep: Research Summaries

**Experts Doubt Ethical AI Design Will Be Broadly Adopted as the Norm Within the Next Decade**

[Original paper by Lee Rainie, Janna Anderson and Emily A. Vogels]
[Research Summary by Connor Wright]

**Overview**: How would you answer the following question: "By 2030, will most of the AI systems being used by organizations of all sorts employ ethical principles focused primarily on the public good?" An overwhelming majority (68%) say no, and there are more than just ethical reasons why this is the case.

**Introduction**
How would you answer the following question: "By 2030, will most of the AI systems being used by organizations of all sorts employ ethical principles focused primarily on the public good?" A resounding 68% of the experts involved in this research paper answered no. Positives are few and far between within the research presented, despite some clear examples. So, let's look into why that is the case.

**Ethics is both vague and subjective**
One prevalent theme throughout this piece is the frustratingly vague and subjective nature of ethics. There is no consensus over what ethical AI looks like, nor is there any agreement over what is a moral outcome. In this sense, it could be rightly said how our ethical frameworks are only 'half-written books, missing some crucial pages and chapters to guide us. As a result, ethics turns out to be an iterative rather than dogmatic process, requiring us to be okay with not knowing the potential outcomes and answers of a situation. Unfortunately, this does not bode well with trying to encode ethical systems into AI.

What I mean by this is how real-life situations can be seen as being too situational to programme into an ethical AI framework, whereby actual ethical dilemmas do not possess any correct answers. For example, views of what is ethical differ worldwide, where countries such as China values social stability more than, say, Western countries. Thus, when AI is applied (such as in warfare), it is unlikely that both sides of the conflict would employ the same ethical framework. Hence, finding a common ethical thread can better help fuse a potentially fractured AI regulation approach, which I believe lies in identifying the human in the AI process.

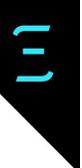

**Identifying the human in the process**
Here, the paper rightly points out the false claim that technological solutions are better than human solutions as they're based on 'cold computing' and not 'emotive human responses'. Instead, it should be noted how, perhaps, when we talk about AI ethics, we should referring to human ethics mediated through AI. By this, I mean how there are no inherently good or evil mathematical functions, whereby it is rather the human presence that determines the ethical propensity of the AI application. The obligation to be moral lies in the hands of corporations and system designers rather than in what the AI does.

As a result, the role the human plays in 'feeding and nurturing' their AI is to be acknowledged. Supplying the system with adequate data for it to train on and proper privacy protections are two ways in which this role can be carried out meaningfully. Without such measures in place, AI then has the potential to become the medium through which our lack of understanding of human bias and bias in itself is expressed. One environment in which this has become too apparent is in AI innovation.

**Ethics doesn't drive AI innovation**
Effective AI has been seen to be prioritised over ethical AI. Looking at facial recognition systems such as Amazon's Rekognition and IBM, it becomes clear that companies are prioritising the 'E' word, but not the one that should be emphasised. Thus, Techno-power has become the main driver behind the pursuit of AI instead of ethical considerations. As a consequence, those few at the helm of AI innovation have proliferated the techno-solutionist mindset throughout the practice, allowing AI to be used as the new manifestation to masquerade and hide the business interests and biases of the institutions and people involved. In this sense, AI has become the digital representation of the collective corporate mindset, meaning that, as some experts in the paper observed, so long as AI is owned those who have access to it will benefit and those who do not will suffer the consequences.

In this sense, perhaps taking the view of seeing the wood for the trees and observing what AI is at its core is now worth exploring.

**Taking AI as it really is**
One of the lures of AI is how it almost creates its own separate reality, filled with the promise of what can be in a different world separate from the current reality. However, this distracts from what AI is in essence. For example, AI applications in different sectors such as law enforcement do what they're told to do. It does not possess a moral compass nor social awareness. In this sense, AI can be seen to lack contextual understanding as it sets out to achieve its goal. To illustrate, the paper included how an AI tasked to keep you dry would not be fussed about stealing an umbrella from an old lady in the street when it starts to rain. In this sense,

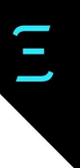
recognising AI as a tool, or even yet, potentially going as far as saying that it's an elongation of previous statistical techniques and innovations, could serve to help cut away the confusing mist surrounding such technology. Perhaps viewing it as a tool can then help to influence the future applications of such a tool, including in the incentives to action it brings with it.

**The problem of incentives**

One potential way to correct the mentioned corporate prioritisation of efficiency could then be to look into what incentivises businesses to act this way. In this sense, the experts involved in the paper observe how, in its current state, the corporate world is not offered any benefits from ethically coordinating AI, with businesses tending to prioritise efficiency, scale and automation, rather than augmentation, inclusion and local context. If this can be achieved, there certainly is a bright side to AI.

**The positives**

AI has been showing promise in its use in education and health, allowing the prioritisation of accessible and necessary digital skills in education programmes, as well as improving the accuracy of certain diagnoses. In this way, it has been observed in the paper how the more we develop AI, the more we appreciate the unique traits and special qualities of humans that are so hard to code. Such qualities such as compassion, contextual understanding and decision-making are common throughout the human world, meaning that AI could also prove the median through which we are able to bridge the conversation between countries. While these positives are few in the paper, they are worth keeping in mind nonetheless.

**Between the lines**

From my perspective, what kind of humans we want to be should be reflected in how we go about designing our AI systems. In this sense, there should be a lack of cheap and subversive techniques to avoid complicated issues like justice, with the social good and social infrastructure over innovation and the good of the governments. For me, this comes through acknowledging the human in the process, both in its role as the protagonist in the AI process, as well as the eventual recipients of both its positives and its negatives.

# The Logic of Strategic Assets: From Oil to AI

[Original paper by Jeffrey Ding and Allan Dafoe]
[Research Summary by Connor Wright]

**Overview**: Does AI qualify as a strategic good? What does a strategic good even look like? The paper aims to provide a framework for answering both of these questions. One thing's for sure; AI is not as strategic as you may think.

**Introduction**

Is AI a strategic good for countries? What is strategic nowadays? The theory proposed serves to aid policymakers and those on the highest level to identify strategic goods and accurately interpret the situation. What a strategic good involves will now be discussed, both in terms of the importance of externalities and whether AI qualifies.

**What is a strategic good?**

The crux of the paper centres on the problem of accurately identifying a strategic good. The paper suggests that such goods "require attention from the highest levels of the state to secure national welfare against interstate competition". While this may be wide-reaching, the authors offer the following formula:

*"Strategic level of asset = Importance x Externality x Nationalization"*

The importance of the asset is based on both military and economic terms. For example, oil that fuels a country's naval fleet vs cotton being used to manufacture high-end fashion.

The externality part is about positive externalities. Here, the more positive externalities produced, the more strategic the product. Private actors are discouraged from investing in the good as they cannot receive all the positive externalities exclusively. For example, wind turbines offer positive externalities in clean energy, but private actors can't exclusively own this.

Nationalisation then focuses on how localised the externalities are. The good becomes less strategic if the externalities derived from it can spread to other countries.

**Strategic goods in terms of externalities**

The externalities brought by strategic goods can be classed in three ways: cumulative-strategic logics, infrastructure-strategic logics and dependency-strategic logics:

**Cumulative-strategic logics** term how strategic goods are to possess high barriers to entry. This leads to low market investment and the need for government consent for the product to be purchased (such as aircraft engines). On the other hand, Uranium isn't a cumulative-strategic logic as a country's purchasing of uranium doesn't put up barriers to entry for others.

**Infrastructure-strategic logics** note how strategic goods in the form of fundamental technologies tend to upgrade society. The diffuse positive externalities produced echo throughout the community and the military, such as the steam train in the Industrial Revolution.

**Dependency-strategic logics** focus on whether extra market forces and few substitutes determine the supply of a good or not. For example, the good becomes more strategic if a nation can cut supplies of a specific good to other countries (such as lithium).

As a result, a strategic good is based on the good itself and the country's strategy with it. For example, the US's use of oil in 1941 allowed them to be the supplier of 80% of Japan's oil. Hence, when the US decided to cut the oil supply to Japan as part of the war effort, it had devastating effects on the Japanese military.

It's important to note how the good's positive externalities must be both important and strategic, as seen in this case. For example, oil was able to produce positive externalities in the form of modernising travel. However, standard-issue military rifles can be necessary for a country's military, but not strategic. They are easy to manufacture (cannot produce a dependency-strategic logic), do not have high barriers to entry, and do not change society too much. Hence, the more logics employed at the same time, the more strategic the good is.

**What this theory means for strategic goods**
A strategic asset is then where "there is an externality that is both important and rivalrous [(strategic)].". Strategic goods are no longer based on military significance, where a good would be strategic if it could be used in the war effort. Under this framework, such goods would not require a high level of attention, so they would not be classed as strategic. Instead, the important and rivalrous externalities derived from technology that can reduce $CO_2$ emissions solely in the country that uses it can be tagged as strategic.

**The strategic aspect of the development of AI**
AI then becomes an interesting case in determining whether it is a strategic asset or not. Here, there is a low rate of cumulative-strategic logics. There are no high barrier entries to AI while also possessing high infrastructural logics through its potential to modernise society. From there, a potential emerging dependency-logic between the US and China could begin to surface, with time only telling whether the US's computing power can be restricted to China. If so, a dependency-logic can be taken advantage of, and if not, China can continue to surge in the AI power rankings.

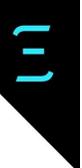

**Between the lines**
AI can certainly be classed as a strategic good in my book, but I thought it would be classified more strongly according to the formula at hand. At times, the lower barrier to entry to gain a foothold in the AI arena is often overlooked. This sobering realization can contribute to what I believe in strongly: seeing AI for what it is.

## Corporate Governance of Artificial Intelligence in the Public Interest

[Original paper by Peter Cihon, Jonas Schuett, Seth D. Baum]
[Research Summary by Jonas Schuett]

**Overview**: How can different actors improve the corporate governance of AI in the public interest? This paper offers a broad introduction to the topic. It surveys opportunities of nine types of actors inside and outside the corporation. In many cases, the best results will accrue when multiple types of actors work together.

**Introduction**
Private industry is at the forefront of AI research and development. AI is a major focus of the technology industry, which includes some of the largest corporations in the world. As AI research and development has an increasingly outsized impact on the world, it is essential to ensure that the governance of the field's leading companies supports the public interest.

**Opportunities to improve the corporate governance of AI**
The opportunities to improve AI corporate governance are diverse. The paper surveys opportunities for nine different types of actors:

Management can establish policies, translate policies into practice, and create structures such as oversight committees.

Workers can directly affect the design and use of AI systems, and can have indirect effects by influencing management.

Investors can voice concerns to management, vote in shareholder resolutions, replace a corporation's board of directors, sell off their investments to signal disapproval, and file lawsuits against the corporation.

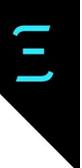

Corporate partners can use their business-to-business market power and relations to influence companies, while corporate competitors can push each other in pursuit of market share and reputation.

Industry consortia can identify and promote best practices, formalize best practices as standards, and pool resources to advance industry interests, such as by lobbying governments.

Nonprofit organizations can conduct research, advocate for change, organize coalitions, and raise awareness.

The public can select which corporate AI products and services to use, and also support specific AI public policies.

The media can research, document, analyze, and generate attention to corporate governance activities and related matters.

**Coordination and collaboration**
In many cases, the best results will accrue when multiple types of actors work together. The paper shows this via extended discussion of three running examples:

First, workers and the media collaborated to influence managers at Google to leave Project Maven, a drone video classification project of the US Department of Defense. Workers initially leaked information about Maven to the media, and then signed an open letter against Maven following media reports.

Second, nonprofit research and advocacy on law enforcement use of facial recognition technology fueled worker and investor activism and public pressure (especially the 2020 protests against racism and police brutality) that ultimately pushed multiple competing AI corporations to change their practices.

Third, workers, management, and industry consortia have interacted to develop and promote best practices concerning the publication of potentially harmful research.

**Between the lines**
The paper will be of use to researchers looking for an overview of corporate governance at leading AI companies, levers of influence in corporate AI development, and opportunities to improve corporate governance with an eye towards long-term AI development.

# AI Certification: Advancing Ethical Practice by Reducing Information Asymmetries

[Original paper by Peter Cihon, Moritz J. Kleinaltenkamp, Jonas Schuett, Seth D. Baum]
[Research Summary by Jonas Schuett]

**Overview**: How can we incentivize the adoption of AI ethics principles? This paper explores the role of certification. Based on a review of the management literature on certification, it shows how AI certification can reduce information asymmetries and incentivize change. It also surveys the current landscape of AI certification schemes and briefly discusses implications for the future of AI research and development.

**Introduction**
Certification is widely used to convey that an entity has met some sort of performance standard. It includes everything from the certificate that people receive for completing a university degree to certificates for energy efficiency in consumer appliances and quality management in organizations. As AI technology becomes increasingly impactful across society, there can be a role for certification to improve AI governance. This paper presents an overview of AI certification, applying insights from prior research and experience with certification in other domains to the relatively new domain of AI certification.

**Certification can reduce information asymmetries**
A primary role of certification is to reduce information asymmetries. Information asymmetries are acute in AI systems because the systems are often complex and opaque and users typically lack the data and expertise necessary to understand them. For example, it is difficult or impossible to evaluate from the outside how biased or explainable a model is, or whether it was developed according to certain ethics principles.

**Certification can incentivize change**
In reducing the asymmetry of information between insiders and outsiders, certification can further serve to incentivize good behavior by the insiders. For example, corporations may be more motivated to achieve ethics standards if they can use certification to demonstrate their achievements to customers who value these achievements.

**The current landscape of AI certification**
The paper surveys the landscape of AI certification from 2020, identifying seven active and proposed programs:

- the European Commission White Paper on Artificial Intelligence (this is outdated, see the proposed Artificial Intelligence Act),
- the IEEE Ethics Certification Program for Autonomous and Intelligence Systems,
- the Malta AI Innovative Technology Arrangement,
- the Turing Certification proposed by Australia's Chief Scientist,
- the Queen's University executive education program Principles of AI Implementation,
- the Finland civics course Elements of AI, and
- a Danish program in development for labeling IT-security and responsible use of data.

These programs demonstrate the variety of forms AI certification can take, including both public and private, certifying both individuals and groups, and covering a range of AI-related activities.

**The value of certification for future AI research and development**
Finally, the paper addresses the potential value of certification for future AI technology. Some aspects of certification will likely remain relevant even as the technology changes. For example, the various roles of corporations, their employees and management, governments, and other actors tend to stay the same. Likewise, certification programs can remain relevant over time by emphasizing human and institutional factors. Programs can also build in mechanisms to update their certification criteria as AI technology changes. Looking further into the future, certification may play a constructive role in governance of the processes that lead to the development of advanced systems. Certification could be especially valuable for building trust among rival AI development groups and ensuring that advanced AI systems are built to high standards of safety and ethics.

**Between the lines**
In summary, certification can be a valuable tool for AI governance. It is not a panacea for ensuring ethical AI, but it can help especially for reducing information asymmetries and incentivizing ethical AI development and use. The paper presents the first-ever research study of AI certification and therefore serves to establish essential fundamentals of the topic, including key terms and concepts.

## Collective Action on Artificial Intelligence: A Primer and Review

[Original paper by Robert de Neufville and Seth D. Baum]
[Research Summary by Robert de Neufville]

**Overview**: The development of safe and socially beneficial AI will require collective action, in the sense that outcomes will depend on the efforts of many different actors. This paper is a primer on the fundamental concepts of collective action in social science and a review of the collective action literature as it pertains to AI. The paper considers different types of AI collective action situations, different types of AI race scenarios, and different types proposed solutions to AI collective action problems.

**Introduction**

The development of safe and socially beneficial AI will require many different people working together. Social scientists have extensively studied different types of "collective action" situations that require actors to cooperate in some way to achieve the best outcomes for the group as a whole. How difficult it will be to achieve the best outcomes may depend on structural factors, like the extent to which the interests of individuals diverge from the interests of the group as a whole, the nature of the goods involved, and the degree to which they hinge on the efforts of a single actor or on some combination of different actors.

In this paper, we first present a primer on the theory of collection action and relate it to the different types of AI collective action situations. The paper looks in particular at AI race scenarios, which have been a major focus of the literature on AI collective action literature. AI races could hasten the arrival of beneficial forms of AI, but could be dangerous if individual actors rush development in order to be the first to develop a particular AI technology. Second, we review the three primary types of potential solutions to AI collective action problems: government regulation, private markets, and community self-organization.

**Collective Action and AI issues**

The impact of AI on society will ultimately depend on the actions of many different people and groups. In some cases, the interests of individual actors will align with the interests of society as a whole, so that good outcomes will result from individual actors pursuing their own interest. In other cases, some actors will be able to benefit individually from acting against the interest of society.  In these cases, AI outcomes may depend on the extent to which the interests of individuals and society as a whole can be reconciled.

In public choice theory, collective action is required where outcomes depend on the actions of different people with different interests. Many aspects and applications of AI will require collective action. In particular, collective action will be needed (1) to reach agreement on rules and standards, (2) to develop AI that is broadly beneficial rather than merely profitable or otherwise advantageous for particular developers, and (3) to avoid competition or conflict that could lead to AI be developed or used in a way that is unsafe.

In recent years, a large but disparate literature has looked at the challenges of collective action with respect to AI. One important distinction is between coordination problems like the development of common AI platforms, in which individual and collective interests mostly align, and competitive situations like competitive AI races, in which individual and collective interests diverge. In general, collective action is easier to achieve when the interests of individuals align with the interests of the group. The type of collective action problem can in turn depend on whether the goods involved are "excludable" (that is, can be restricted to particular consumers) or "rivalrous" (that is, is used up when its benefits are enjoyed). Typically, the interests of individuals and the group are easy to align when goods are excludable—because their use can be limited to those who have paid for them in some sense—and non-rivalrous—because their supply is not limited. Another important issue is the degree to which addressing a collective action situation depends primarily on the effort of a single actor or requires many actors to contribute something.

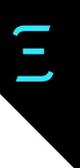

One type of collective action situation that has received a lot of attention in the literature is AI race scenarios. AI races could be dangerous if individual actors' interest in winning the race is at odds with the general interest in developing AI that is safe and socially beneficial. The paper looks at both near-term and long-term AI races. The literature identified in this paper focuses in particular on near-term races to develop military AI applications and long-term AI races to develop advanced forms of AI like artificial general intelligence and artificial superintelligence. The two types of races are potentially related since near-term races could affect the long-term development of AI.

Finally, the paper evaluates three different types of potential solutions to collective action problems: government regulation, private markets, and community self-organization. All three types of solution can address collective action problems, but no single approach is a silver-bullet solution to the entire range of collective action problems. It may be better to pursue a mix of different types of solutions to address AI collective action in different ways and at different scales. Governance regimes will also need to account for other factors, like the extent to which AI developers are transparent about their technology.

**Between the lines**
The collective action issues raised by AI are increasingly pressing. Collective action will be necessary to ensure that AI serves the public interest rather than simply serving the narrow interests of those who develop it. Collective action will also be necessary to ensure that AI is developed with appropriate risk management protocols and adequate safety measures. The institutions we develop now to help resolve the AI collective action problems that arise today could have long-lasting and far-reaching consequences. The literature on AI collective action situations is still young; a great deal more work on designing systems to govern specific AI collective action problems still remains to be done.

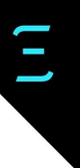

# AI Ethics Maturity Model

[Original paper by Kathy Baxter]
[Research Summary by Connor Wright]

**Overview**: With ethical AI certainly being a hot topic in the business world, how can this be achieved? The Ethical AI Practice Maturity Model sets out 4-steps towards achieving the end goal of an "end-to-end-ethics-by-design" model. With that in sight, the need for company-wide participation and the passion for building ethical AI are a must.

**Introduction**

Does your company engage in AI products? Does it have an ethical AI team? If not, how would such a team be established? The Ethical AI Practice Maturity Model aims to answer the latter. Stretching from the inception of an ad hoc review from a group of employees, to having ethical AI awareness coursing through a company's veins, it offers us a roadmap. Calling on company-wide engagement alongside a passion for ethical AI, the end goal includes having ethical thresholds necessary for the AI product to pass in order to be launched. The best way to illustrate how this can be achieved is to go through the model itself.

**Ad Hoc**

Questioning of the AI process at hand begins to take hold. Certain issues arise and are then brought into question on an ad hoc basis. The question no longer becomes "can we do this?", but instead "should we do this?". The resultant conversations can prove good fuel for informal talks about the technology, helping to clarify the importance of these problems. Once these issues are known and employees can see them being dealt with, trust can start to be developed between those designing the AI and the wider company.

However, the desired confidence takes time to develop. So, building an ethical AI team that accumulates "small wins" can help consolidate their position in the AI process. Churning out results, big or small, will help create more advocates throughout the business and cultivate pivotal involvement from those at the top.

**Organized and repeatable**

Arriving at this stage means executives are now on board, and responsible AI practices are now being rewarded. As a result, the next step lies in convincing internal stakeholders to join the process as well. Demonstrating why getting involved is crucial by explaining the risks involved with AI is a sure way to get more employees to sign up. Moreover, contextualizing AI in the company context and within its ethical principles to explain its importance could prove even more gripping.

What executives must not do is simply "ethics washing" company employees. This entails placing broad ethical principles, such as 'AI must always do the right thing', and sticking it on the company's website. Instead, how these company principles that apply to AI will be achieved is paramount for forming a successful ethical AI team.

Hence, the stage also includes the formation of the team itself. Given the different situations the team will face, different expertise will be required. Accordingly, the team should be composed of diverse skill sets, backgrounds, and understandings. Furthermore, the metrics for evaluation should not be classic "revenue generation" and the like, but rather making sure the AI systems are safe and not being penalized when they identify ethical risks.

Given the need to identify these risks, considerations on questions of scale can be helpful. To join the team, what is the base knowledge all employees should have of AI? How would you design formal training to convey this knowledge? Would teams working on AI be able to loop in the ethical AI team? Whatever the answers to these questions are, it needs to be sustained and managed in the long run.

**Managed and Sustainable**
The training required for the desired base level of knowledge must only include mandatory elements for all employees if it's necessary. The company's ethical principles ought to be common knowledge, but knowing how to mitigate AI system bias is only relevant for data scientists. Managing what the training allows employees to find is the next important step.

Coming across an AI ethical risk is not to be frowned upon completely. No AI system can be 100% bias-free, so saying what bias there is, how it's being mitigated, and the potential harms it could cause is the best way to deal with the problem. Any damages that are then caused (which can vary depending on the person) need to have appropriate channels to be brought up. Should your business stretch across different countries, the ethical review must ensure the AI system includes other languages and cultures. Dealing with bias for your American clients will not be the same when approaching your Taiwanese partners.

**Optimized and innovative**
The final stage is the one to be most desired to achieve. The ethical AI team is no longer a central hub but rather dispersed throughout the whole company. Products and resources require that ethical debt is resolved to be realized, ensuring an "end-to-end-ethics-by-design" model. However, this does not mean that striving for perfection is halted. With "practice" being the keyword, ethical AI practice never reaches its conclusion. New innovations bring new techno-ethical issues, requiring even more elaboration from the diverse backgrounds of the ethical AI team. This stage may be the end goal, but the end goal is a refined process, not a product.

**Between the lines**
In my view, ethical AI practice is both necessary and sufficient to operationalise principles like transparency, fairness and equality. Subsequently, any ethical review needs to happen early in the design process; otherwise, there's no time to make significant changes. Should this not be the case, the rise of "ethical debt" from unethical AI models, though almost invisible during the AI design, will become very tangible in the form of harm to the public. The Ethical AI Practice Maturity Model gives a company a roadmap to follow and harbors the vital point that change must come from all. Bravery is required, and it all starts with that first small win.

## Mapping value sensitive design onto AI for social good principles

[Original paper by Steven Umbrello, Ibo van de Poel]
[Research Summary by Marianna Ganapini]

**Overview**: Value sensitive design (VSD) is a method for shaping technology in accordance with our values. In this paper, the authors argue that, when applied to AI, VSD faces some specific challenges (connected to machine learning, in particular). To address these challenges, they propose modifying VSD, integrating it with a set of AI-specific principles, and ensuring that the unintended uses and consequences of AI technologies are monitored and addressed.

**Introduction**
How do we bridge theory and practice when it comes to following ethical principles in AI? This paper aims at answering that very question by adopting Value sensitive design: a set of steps to implement values in technological innovation. Value sensitive design potentially applies to a vast range of technologies, but when used in AI and machine learning, it inevitably faces some specific challenges. The authors propose a way to fix these problems by integrating Value sensitive design with other actionable frameworks.

**Value sensitive design (VSD)**
Value sensitive design (VSD) is a method originally developed by researchers at the University of Washington and it lays out actional steps for designing technology in accordance with our values. These steps are grouped in three main categories: conceptual, empirical and technical investigations. Conceptual analysis determines the appropriate set of values (coming from the philosophical literature and/or from the stakeholders' expectations), whereas empirical investigations may survey direct and indirect stakeholders to understand their values and needs. The third set of steps looks into potential technical limitations and resources to design a technology following the appropriate set of values.

Unfortunately, the self-learning capabilities of AI pose some specific challenges for VSD. Notoriously, models developed through machine learning can have features that were not initially designed or foreseen, and some of these features may be opaque and thus not easily detectable. This could mean that AI systems, originally designed following VSD, "may have unintended value consequences, […] or unintentionally 'disembody' values embedded in their original design." As the authors explain, this means that we need design principles specific for this kind of technology and expand VSD to address those challenges. The question is how to do that.

**Solutions**
The authors propose to modify VSD in the following three ways: (1) VSD should include a set of AI-specific principles (AI4SG); (2) for VSD, the goal should be not only to promote outcomes that avoid harming but also to contribute to social good overall; (3) VSD should look at the downstream consequences of adopting a certain AI system to make sure the designed values are in fact respected.

**2.1 VSD & AI4SG**
Let's start witht the first point. The authors propose to adopt AI-specific principles in VSD. In particular, they look at AI4SG (AI for social good) principles, which are actionable guidelines, inspired by the more high level values of "respect for human autonomy, prevention of harm, fairness, and explicability". These are the principles:

"(i) falsifiability and incremental deployment; (ii) safeguards against the manipulation of predictors; (iii) receiver-contextualized intervention; (iv) receiver-contextualized explanation and transparent purposes; (v) privacy protection and data subject consent; (vi) situational fairness; and (vii) human-friendly semanticization."

The authors of the paper point out that applying these specific principles in the design of AI systems would address some of the concerns mentioned above. This is because these steps are not only more practical than the high-level values but they are also specific to AI and so are the right tools to avoid the challenges raised by this kind of technology. These principles are, in other words, a more concrete application of the key values (e.g. beneficence) we want to see as part of the design of AI going forward.

**2.2 VSD & the social good**

Here's the second issue: VSD should be not only to promote outcomes that avoid doing harm but also to contribute to social good and so "there must be an explicit orientation toward socially desirable ends." To promote this, the authors recommend that VSD adopts "the Sustainable Development Goals (SDGs), proposed by the United Nations, as the best approximation of what we collectively believe to be valuable societal ends". Again, this is a matter of complementing and enriching VSD with a set of principles that actively try to promote social good, and as such, they should be part of the design of AI systems.

**2.3 VSD and downstream consequences**

Finally, ongoing monitoring is needed to address possible unintended consequences of adopting AI systems. Indeed, when employed, AI systems may not respect the original design values (see here for more). This is why there is the need to apply VSD to the entire "life cycle of an AI technology", monitoring systems, and adopt the necessary design changes when needed. The authors point out that prototyping and small scale testing could really help address unforeseen consequences.

By combining these principles and ideas, the authors embrace a framework that encompasses the following recursive loop:

*Context Analysis (e.g. societal challenges, values for stakeholders)* → *Value Identification (e.g. beneficence, autonomy, SDGs, case specific values)* → *Design Requirements (e.g. AI4SG),* → *Prototyping (e.g. small-scale testing)*

This proposed framework is meant to be taking into account the various aspects of VSD while also addressing some of its shortcomings.

**Between the lines**

It is important to find a way to bridge theory and practice when it comes to building ethical AI systems. This paper is charting a way forward to address this need. It brings together different methods and approaches by explaining how to integrate action steps within the VSD framework while also making sure social good is taken into account. Now that we have a fairly comprehensive set of high-level values, future research will need to establish more precise, actionable and concrete steps to embody those values within AI systems, and it will need to find new ways to determine the ethically relevant, downstream consequences of the use of those systems.

# Embedding Values in Artificial Intelligence (AI) Systems

[Original paper by Ibo van de Poel]
[Research Summary by Andrea Pedeferri]

**Overview**: Though there are numerous high-level normative frameworks, it is still quite unclear how or whether values can be implemented in AI systems. Van de Poel and Kroes's (2014) have recently provided an account of how to embed values in technology. The current article proposes to expand that view to complex AI systems and explain how values can be embedded in technological systems that are "autonomous, interactive, and adaptive".

**Introduction**
Though there are numerous high-level normative frameworks, it is still quite unclear how or whether those frameworks can be implemented in AI systems. Van de Poel and Kroes's (2014) have recently provided an account of how to embed values in technology in general. The current article proposes to expand that view to AI systems which, according to the author, have five building blocks: "technical artifacts, institutions, human agents, artificial agents, and technical norms". This paper is a very useful guide to understanding how values can be embedded in a complex system composed of multiple parts that interact in different ways.

**Embedding Values**
Organizations such as the EU High-Level Expert Group on AI and the IEEE have provided a list of high-level ethical values and principles to implement in AI systems. Whatever your views on values might be, the paper points out that we need an account of what it means for those values to be embedded. To start, a set of values is said to be 'embedded' only if it is integrated into the system by design. That is, those who design the system should intentionally build that system with a specific set of values in mind. More is needed, though, because even if a system is designed to comply with certain values, that does not mean it will really realize those values.

So the paper proposes the following definition of "embodied values": "The embodied value is the value that is both intended (by the designers) and realized if the artifact or system is properly used."

Drawing both from the current paper and Van de Poel and Kroes's (2014), we have the following set of useful definitions:

**Designed value**: any value that is intentionally part of the design of a technological system

**Realized value**: any value that the (appropriate) use of the system is prone to bring about

**Embedded value**: any value that is both designed and realized. Thus, a value-embedded system is a system that, because of the way it was designed, will bring about certain values (when it is properly used).

As the paper explains, this opens the door to the idea of a feedback loop: when an intended value is not realized, there has to be some change in the way it is used and/or designed. Similarly, if a system is used in a way that is contrary to intended values, a re-design might be in order. As the author points out, the practice of re-design systems to avoid unintended consequences "is particularly important in the case of AI systems, which due to the adaptive abilities of AI, may acquire system properties that were never intended or foreseen by the original designers."

**Embedding Values in AI systems**
This account provides a way to understand how values can be embedded in AI by looking both at the components and the system level. More specifically, the paper understands AI systems as socio-technical systems composed not only of "technical artifacts, human agents, and institutions" but also "artificial agents and certain technical norms that regulate interactions between artificial agents and other elements of the system." To clarify, a socio-technical system is a system that depends "on not only technical hardware but also human behavior and social institutions for their proper functioning (cf. Kroes et al. 2006)."

To start, the paper clarifies that an AI system will be the result of both social institutions and human agents interacting to design technological artifacts in accordance with certain values. Importantly, the paper points out that those social institutions will also be embedded with values. As such, the role of humans is key: they need to monitor and evaluate the outcomes and use of both the technological artifacts and the social institutions that influence the production and design of those technological artifacts. In addition, because of how AI systems work, there will also be technical norms that regulate how artificial agents interact with humans and social institutions. As such, these norms will embed and promote certain values.

Therefore, in conclusion, an AI system promotes a set of values if and only if all five of its main components (i.e. technical artifacts, institutions, human agents, artificial agents, and technical norms) will either embody or intentionally promote V. As the author rightly points out then, "AI systems offer unique value-embedding opportunities and constraints because they contain additional building blocks compared to traditional sociotechnical systems. While these allow new possibilities for value embedding, they also impose constraints and risks, e.g., the risk that an AI system disembodies certain values due to how it evolves. This means that for AI systems, it is crucial to monitor their realized values and to undertake continuous redesign activities."

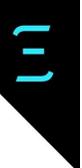

**Between the lines**
The paper is a very useful guide to understanding how values can be embedded in a complex system composed of multiple parts that interact in different ways. The next step is to figure out how this analysis connects to the debate on trust and trustworthy AI: given the current way we understand value-embedded AI, is it possible to build an AI we can actually trust?

## Moral consideration of nonhumans in the ethics of artificial intelligence

[Original paper by Andrea Owe and Seth D. Baum]
[Research Summary by Andrea Owe]

**Overview**: As AI becomes increasingly impactful to the world, the extent to which AI ethics includes the nonhuman world will be important. This paper calls for the field of AI ethics to give more attention to the values and interests of nonhumans. The paper examines the extent to which non-humans are given moral consideration across AI ethics, finds that attention to nonhumans is limited and inconsistent, argues that nonhumans merit moral consideration, and outlines five suggestions for how this can better be incorporated across AI ethics.

**Introduction**
Is the field of AI ethics adequately accounting for nonhumans? Recent work on AI ethics has often been human-centered, such as on "AI for people", "AI for humanity", "human-compatible AI", and "human-centered AI". This work has value by shifting emphasis away from the narrow interests of developers, but it does not include explicit consideration of nonhumans. How do AI systems' resources and energy use impact nonhumans? What is the potential of AI for environmental protection or animal welfare? Social algorithmic bias is currently a major topic but are there important nonhuman algorithmic biases? How may we incorporate nonhuman interests and values into AI system design? What might be the risks of not doing so?

This paper documents the state of attention to nonhumans in AI ethics and argues that the field can and should do more. The paper finds that the field generally fails to give moral consideration to nonhumans, such as nonhuman animals and the natural environment, aside from some consideration of the AI itself. The paper calls on the field to give more attention to nonhumans, suggesting five specific ways AI researchers and developers can accomplish this.

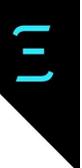

**What it means to give moral consideration to nonhumans**

Moral consideration of nonhumans means actively valuing nonhumans for their own sake. In moral philosophy terminology, to "intrinsically value" nonhumans. One can fail to give moral consideration to nonhumans by actively denying their intrinsic value or by neglecting to actively recognize their intrinsic value. There are many conceptions of which nonhumans merit moral consideration, such as the welfare of nonhuman animals or sentient AI systems, or the flourishing of ecosystems. Moral consideration of nonhumans does not require any one specific conception of which nonhumans merit moral consideration. It also does not require a specific type of moral framework, such as consequentialism, deontology, or virtue ethics.

**Why it matters that AI ethics morally consider nonhumans**

Moral consideration of nonhumans is a practical issue for real-world AI systems, with several matters at stake. For example, AI can be applied for the advancement of nonhuman entities, such as for environmental protection. On the other hand, AI can inadvertently harm the nonhuman world, such as via its considerable energy consumption. Certain algorithmic biases could additionally affect nonhumans in a variety of ways. Further, the long-term prospect of strong AI or artificial general intelligence may radically transform the world for humans and nonhumans alike. The extent to which non-humans are morally considered can play an important role in assessing how AI systems should be designed, built, and used.

**Empirical findings: Limited attention to nonhumans**

The paper surveys a variety of prior work in AI ethics in terms of the extent to which it gives moral consideration to nonhumans. Overall, the paper finds that the field generally fails to give moral consideration to nonhumans. The primary exception is the line of research on the moral status of AI. The paper finds no attention to nonhumans in 76 of 84 sets of AI ethics principles surveyed by Jobin et al., 40 of 45 artificial general intelligence R&D projects surveyed by Baum, 38 of 44 chapters in the Oxford Handbook of Ethics of AI, and 13 of 17 chapters in the anthology Ethics of Artificial Intelligence. In the two latter examples, any dedicated attention is on the moral status of AI itself. No other types of non-humans are given dedicated attention.

**The case for moral consideration of nonhumans**

Modern science is unambiguous in documenting that humans are members of the animal kingdom and part of nature. Attributes of humans that are commonly intrinsically valued, such as human life or human welfare, are also found in many nonhuman entities. It would very arguably be an unfair bias to intrinsically value something in humans but not intrinsically value the same thing in nonhumans. Additionally, compelling arguments can be made for intrinsically valuing things that inherently transcend the human realm, such as biodiversity. To insist on only giving moral consideration to humans requires rejecting all of these arguments. The paper posits that this is untenable, meaning that nonhumans merit moral consideration.

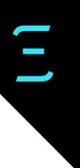

**What can be done? Five suggestions for future work**
AI ethics research needs a robust study of the moral consideration of nonhumans, focusing on issues such as how to balance between humans and nonhumans, the handling of the natural nonhuman world, and the role of nonhumans in major AI issues. For example, research in ecolinguistics shows that English—the primary language for AI system design—contains biases in favor of humans over nonhumans. This insight could be applied to the study of nonhuman algorithmic bias in, for example, natural language processing.

Statements of AI ethics principles should give explicit attention to the intrinsic value of nonhumans. The Montréal Declaration for the Responsible Development of Artificial Intelligence is one example, with the principle stating: "The development and use of artificial intelligence systems (AIS) must permit the growth of the well-being of all sentient beings." For illustration, an even stronger statement would be: "The main objective of development and use of AIS must be to enhance the wellbeing and flourishing of all sentient life and the natural environment, now and in the future."

AI projects that advance the interests and values of nonhumans should be among the projects considered when selecting which AI projects to pursue. The Microsoft AI for Earth program is a good example of AI used in ways that benefit nonhumans, and further serves as an example of how to operationalize moral consideration for nonhumans in AI project selection. The program supports several projects for environmental protection and biodiversity conservation that give explicit moral consideration to nonhumans, including Wild Me, eMammal, NatureServe, and Zamba Cloud.

The inadvertent implications for nonhumans should be accounted for in decisions about which AI systems to develop and use, such as the material resource consumption and energy use of AI systems. AI groups should acknowledge that if an AI system will/could cause sufficient harm to nonhumans, it would be better to not use it in the first place.

AI research should investigate how to incorporate nonhuman interests and values into AI system designs. How to incorporate human values is currently a major subject of study in AI, but some of the proposed techniques do not apply to nonhumans. AI ethics design is of particular importance for certain long-term AI scenarios in which an AGI takes a major or dominant position within human society, the world at large, and even broader portions of outer space.

Even the most well-designed AGI could be catastrophic for some nonhumans if it is designed to advance the interests of humans or other nonhumans.

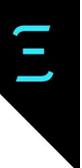

**Between the lines**

In summary, accounting for nonhumans in AI R&D is critical to ensure that AI benefits more than just humans. This can prevent further harm to nonhuman entities already under immense pressure from human activities. Furthermore, this will enable the field to better handle future moral issues, such as the potential of artificial entities like AI to merit moral consideration themselves. In addition, there are plenty of opportunities for AI to mitigate existing harm to nonhumans and enable benefits to also nonhumans. As documented by this paper, the AI ethics field has given little attention to nonhumans thus far. Therefore, there exist manifold opportunities for work addressing the implications of nonhumans across AI design and use.

## Governance of artificial intelligence

[Original paper by Araz Taeihagh]
[Research Summary by Angshuman Kaushik]

**Overview**: The various applications of AI not only offer opportunities for increasing economic efficiency and cutting costs, but they also present new forms of risks. Therefore, in order to maximize the benefits derived from AI while minimizing its threats, governments' worldwide need to understand the scope and the depth of the hazards posed by it, and develop regulatory processes to address these challenges. This paper describes why the governance of AI should receive more attention, considering the myriad challenges it presents.

**Introduction**

The internet is full of a plethora of websites catering to the diverse needs of its users. Most of these websites use complex machine learning algorithms to make the browsing experience of a surfer 'seamless' (as the marketers would love to call it). For example, there are content recommendation algorithms powering certain websites, which play a considerable role in shaping the 'thought processes' of its users. These algorithms apart from being used for predicting and evaluating human behavior are also used for profiling and ranking people. However, there have been instances, when these content recommendation algorithms have been criticized for leading and exposing users to extreme content. Since the modus operandi of these algorithms is built to engage users and keep them on the platform ('dollars for eyeballs' mentality) it creates a 'feedback loop', by suggesting content that users have expressed interest in. The consequence is that the users migrate from milder to more extreme content. The situation becomes grave, when say, for example, it becomes a fertile ground for any insurrectionist group to broadcast propaganda upon young and impressionable minds, thereby, attracting devastating consequences. Hence, in such scenarios, the governments need to step in and keep the system within bounds, by formulating effective policies and regulations. This paper starts off with an introduction to the all-pervading and omnipresent AI, replete with its various value-laden decisions for the society, be it, in clinical decision support systems, policing systems, provision of personalized content etc. It then enters into the awfully difficult territory of unexpected consequences and risks (in the form of bias, discrimination etc.), associated with the use of AI systems and then, proceeds to address the challenges encountered during its governance, and steps forward.

**AI - General**

Conceptions of AI date back to the earlier efforts in developing artificial neural networks to replicate human intelligence, which can be referred to as the ability to interpret and learn from the information. The present AI capabilities have expanded to include computer programs that can learn from massive amounts of data and make decisions without human guidance, commonly referred to as Machine Learning algorithms (ML). Although these algorithms are quite fast and efficient, there is a broad consensus that it still falls short of human cognitive abilities, and most of the AI systems that have been successful till now, belong to the category of 'narrow or weak AI.' As per the researcher, some of the incentives for deploying AI include increasing economic efficiency and quality of life, meeting labor shortages, tackling aging populations etc.

**Understanding the risks of AI**

One of the biggest challenges faced by most of the AI systems is what is widely referred to as 'corner cases' i.e., unexpected situations, that the system had not been trained to handle. Further, the decision-making autonomy of AI significantly reduces human control over their decisions, creating new challenges for ascribing liability for the harms imposed by it. Moreover, given the value-laden nature of the outcomes reached by the algorithms, AI systems can potentially exhibit behaviours that conflict with societal norms and values, prompting concerns regarding the ethical issues that can crop up from its adoption. The paper also highlights the hazards of data privacy, surveillance, unemployment and social instability arising from the deployment of AI applications.

**Challenges to AI Governance**

According to the paper, the reason why the governments face innumerable difficulties in designing and implementing effective policies to govern AI, is due to its high degree of inherent opacity, uncertainty and complexity, which makes it challenging to ensure its accountability, interpretability, transparency and explainability. Another key issue surrounding the debate on AI governance is data governance, as multiple organizational and technical challenges exist that impede effective control over data and attribution of responsibility to data-driven decisions made by AI systems. To add to the above, the existing regulatory and governance frameworks are ill-equipped to manage the unique and novel societal problems introduced by the AI systems. The Regulators being generalists, struggle enormously when it comes to comprehending the subtle nuances of the ever evolving AI landscape. Hence, an information asymmetry and a chasm is created between tech companies and regulators which prove to be a major hindrance for the latter in formulating policies and regulations that are specific to the issue in hand. Further, considering the issues associated with 'hard' regulatory frameworks, the discussion in the paper veers towards the adoption of self-regulatory or 'soft law' approaches, espoused by the various industry bodies and governments to govern AI. 'Soft law' approaches refer to non-binding norms that create substantive expectations that are not directly enforceable. For example, industry bodies like IEEE and the High-Level Expert Group on AI formed by the European Commission have released their own Ethics Guidelines for Trustworthy AI. The paper also raises question marks at the efficacy of such self-regulatory initiatives and standards, considering their voluntary nature. Another challenge faced by the governments is the significant influence exerted by the big technology companies in the formulation and implementation of efficacious AI Policies, through their lobbying efforts, and their inclusion in the AI expert groups formed by the governments. Studies have highlighted the risks of regulatory capture by AI developers due to their substantial informational advantages, which makes their technological expertise particularly valuable to the regulators. The paper also calls for more research in the field to ensure greater inclusivity and diversity in AI governance.

**Steps forward for AI Governance**
According to the author, as AI is still developing with the potential to grow more salient and diverse, the complexity of its challenges suggests that its decision-making needs to be carefully conceptualized according to their context of application, and these framing processes should be subject to public debate. In fact, there are increasing calls for the adoption of innovative governance approaches, such as, adaptive governance and hybrid or 'de-centered' governance to address the governance challenges posed by the complexity and the uncertainty of the AI systems. The characteristic of adaptive and hybrid governance is the diminished role of the government in controlling the distribution of resources in the society. Another area of emphasis pointed out is the presence of flexibility, which is imperative to enable diverse groups of stakeholders to build consensus around the norms and trade-offs in designing AI systems, as well as for global AI governance to be applicable across different geographical, cultural and legal contexts, and aligned with existing standards of democracy and human rights. Further, the paper calls for learning from the experiences of governing previous emerging technologies, such as, the internet, nanotechnology, aviation safety and space law. Reference is also made towards an emerging body of literature that has proposed governing AI systems through their design, where social, legal and ethical rules can be enforced through code to regulate the behaviour of AI systems. According to the author, the trend common to recent studies in their proposed frameworks for AI governance is the emphasis on building broad societal consensus around AI ethical principles and ensuring accountability, but there is a need for studies examining how these frameworks can be implemented in practice. He goes on to refer to different frameworks, such as, the society-in-the-loop framework, where society is first responsible for finding consensus on the values that should shape AI and the distribution of benefits and costs among different stakeholders. Another approach includes the centralization and cross-cultural cooperation to improve coordination among national approaches. However, the various AI governance frameworks call for producing more concrete specifications on implementing these governance frameworks in practice, and identifying the parties in government that are responsible for leading different aspects of AI governance.

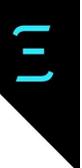

**Between the lines**
The paper quite methodically, decrypts the trials and tribulations faced in governing AI. Even the solutions envisioned in order to tackle the perils associated with AI systems by the government, seem workable, to a large extent. In fact, the paper lays out a very feasible and pragmatic path for the governments to follow, while formulating their various policies and regulations, concerning AI. More importantly, the findings are extremely crucial, considering the situation created by certain unbridled AI systems.

## Avoiding an Oppressive Future of Machine Learning: A Design Theory for Emancipatory Assistants

[Original paper by Gerald C. Kane, Amber Young, Ann Majchrzak, and Sam Ransbotham]
[Research Summary by Sarah P. Grant]

**Overview**: Broad adoption of machine learning systems could usher in an era of ubiquitous data collection and behavior control. However, this is only one potential path for the technology, argue Gerald C. Kane et al. Drawing on emancipatory pedagogy, this paper presents design principles for a new type of machine learning system that acts on behalf of individuals within an oppressive environment.

**Introduction**

*"It is capitalism that assigns the price tag of subjugation and helplessness, not the technology,"* asserts Shoshana Zuboff in her bestselling book, The Age of Surveillance Capitalism: The Fight for a Human Future at the New Frontier of Power.

In contrast, some academics argue that technology itself can be inherently oppressive. In their paper about emancipatory assistants, Kane et al. demonstrate that machine learning systems have several oppressive features, including the tendency to optimize "on outcomes for large samples at the expense of [individual users]."

These oppressive characteristics mean that as systems fuelled by machine learning seep into every aspect of life–from job searching to reading the news–individuals face more limits on their freedoms. Addressing this problem, argue Kane et al., requires innovative approaches to machine learning system design.

Using the emancipatory pedagogy of Brazilian educator and philosopher Paulo Freire as a foundation, the paper presents design principles for a new type of machine learning system called the "emancipatory assistant"–an agent that "would help individuals express and enact their preferences" in a world of pervasive data extraction and behavior manipulation.

**The rise of informania**
To illustrate how an emancipatory machine learning system would work, the authors paint a picture of a dystopian future called Informania. In Informania, machine learning systems "optimize on outcomes for millions (or billions) of users, with little regard for individual rights within the collective."

Such a future is becoming more likely, the authors state, pointing to China's social credit system and the US Justice Department's COMPAS algorithm. The authors also describe how, in a system of unchecked free-market capitalism, "multiple organizations could develop [machine learning] infrastructures….resulting in a massive [behavior] control infrastructure."

While the authors note that such an outcome represents "the logical conclusion of our current trajectory," they also emphasize that Informania's oppression "need not necessarily arise from malicious intent." Acting on behalf of the individual, an emancipatory assistant would help redress power imbalances within Informania.

**Machine Learning Systems: Oppressive Features**
Before describing in detail what an emancipatory system would look like, the authors demonstrate how machine learning systems are "inherently oppressive" by applying theoretical constructs of emancipation and oppression. For example, many algorithms use past behaviors to filter the information that appears on newsfeeds and product recommendations. This impacts a person's "freedom to think" by controlling the amount and type of information available for making decisions.

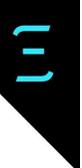

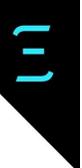

A significant share of the paper is devoted to specifying how the basic machine learning model is oppressive by nature as compared to code-based systems. The authors state that machine learning systems are oppressive because they:

- optimize on outcome variables, which typically benefit the platform above individual users;
- are based on training data that may reflect historical biases;
- are opaque and difficult to understand;
- typically don't incorporate user feedback.

Later on in the paper, the authors describe four modifications to the basic machine learning model that will "yield distinctive design features" of an emancipatory assistant.

**The role of emancipatory assistants**
Referring to past research, the paper describes emancipation as "a theoretical state in which power dynamics between agents are neutral or equal." Within an oppressive machine learning environment, an emancipatory assistant could act as an intermediary that helps individual users achieve more power.

The authors argue that critical social theory is well-suited for the development of new machine learning design principles. Freire's emancipatory pedagogy in particular "provides ready-made pedagogical steps to foster concrete gains of emancipation."
For example, Freire did not push for the oppressed to overthrow the oppressors, but rather that they work together in a new type of co-education. In a similar way, the emancipatory assistant could facilitate a process of mutual inquiry, "first by helping an individual uncover his or her authentic preferences and desires and then by providing Informania with a mechanism to factor those desires into its optimization function."

**Key Design Principles**
The authors identify key design principles for emancipatory assistants, which optimize for:

**#1. Richness of Preferences**
Emancipatory assistants can help users provide Informania with more details about the individual's interest. For example, the assistant could help an individual who wants to change careers overcome Informania's assumption that job history indicates future job preferences.

**#2. Recognizing Conflict**

An emancipatory assistant can help users recognize when their goals conflict with the goals of Informania. For example, the assistant could help users present attributes to Informania that lead to better pricing when finding the best loan options for purchasing a home.

#3. Personalized Storytelling
Emancipatory assistants can help users manage information sharing based on different contexts. According to the authors, "users might be more comfortable with complete information to a spouse but might restrict slightly to children and restrict even further to potential employers."

#4. Alternative Perspectives
Rather than encouraging people to click on the same types of news articles they have read in the past, the assistant "can provide a richer article landscape and indirectly encourage critical consciousness." Emancipatory assistants can help individuals "develop the robust rationality needed to think critically about the world around them."

While the authors predict that Informania will dominate in the shorter term, they envision a longer term future where there is a more balanced power dynamic between emancipatory assistants and Informania. This would necessitate the establishment of a certification body as well as audit committees to promote compliance to standards for the newer types of machine learning systems.

**Between the lines**
This is an important paper because it encourages more expansive thinking within the field of machine learning. By drawing on established theories from multiple domains, it could also foster more interdisciplinary collaboration.

While the authors do touch on the subject of algorithmic literacy in this paper, further research could investigate the implications of divisions in algorithmic awareness. For example, one survey of internet users in Norway (where 98% of the population has internet access), found that education is strongly linked to algorithm awareness, with low awareness highest among the least educated group. Groups with low algorithm awareness were more likely to hold neutral attitudes towards algorithms.

It could be argued, then, that many individuals who would benefit from emancipatory assistants may not be motivated or may not have the resources to use such systems. Future research could address how new types of machine learning systems could yield emancipatory outcomes for all users of Internet-based platforms–and not just a privileged few.

## Against Interpretability: a Critical Examination

[Original paper by Maya Krishnan]
[Research Summary by Andrea Pederferri]

**Overview**: "Explainability", "transparency, "interpretability" … these are all terms that are employed in different ways in the AI ecosystem. Indeed, we often hear that we should make AI more "interpretable" and/or more "explainable". In contrast, the author of this paper challenges the idea that "interpretability" and the like should be values or requirements for AI. First, it seems that these concepts are not really intuitively clear and technically implementable. Second, they are mostly proposed not as values in themselves but as means to reach some other valuable goals (e.g. fairness, respect for users' privacy). And so the author argues that rather than directing our attention to "interpretability" or "explainability" per se, we should focus on the ethical and epistemic goals we set for AI while also making sure we can adopt a variety of solutions and tools to reach those goals.

**Introduction**

As we mentioned in a previous summary, back box's opaqueness poses both epistemic (are the algorithms in fact reliable?) and ethical (are the algorithms ethical?) challenges. Relatedly, they also seem to violate people's claim-right to know why a certain algorithm has produced some predictions or automated decisions that concern them. For many the epistemic and ethical risks back box's opaqueness poses could be mitigated by making sure that AI is somehow interpretable, explainable and/or (as some say) transparent. Contrary to the received view on this issue, the author of this paper challenges the idea that there is a black box problem and denies that "interpretability", "explainability" or "transparency" should be values or requirements for AI (see also here).

**Key Insights**

The author of this paper challenges the idea that there is a black box problem and that "interpretability", "explainability" or "transparency" should be criteria for evaluating an AI system.

The first problem the author points out is that the terms above ("explainability", "transparency, "interpretability") are often unclear and poorly defined. Let's take "interpretability": AI is interpretable when, roughly, it is understandable to consumers. The author notices that this definition is not really helpful: it does not clarify what "understandable" means and does not offer any insight on what the term could mean when applied to algorithms. Also, there is confusion on what should be understandable. Here are a few candidates: the prediction of the algorithm itself, the inner workings of the AI that produced its prediction, or the reasons why / the justification for the algorithm making that prediction. Which one is key for human understanding?

Many seem to believe that explaining how an algorithm reached a certain outcome is tantamount to making AI understandable. However, it should be noted that causal explanations are not the same as justifications. That is, the reason or justification for a given outcome might not clearly map into the causal path that brought the algorithm to that conclusion. As the author puts it, "[t]his point is particularly apparent in the case of neural networks. The causal process by which some input triggers a certain pathway within the network does not straightforwardly map on to justificatory considerations." Indeed it would be like asking "a person why they have given an answer to a particular question and they respond with an account of how their neurons are firing". The causal story is not a rational explanation per se. Thus, if the explanation we look for tells us only about the causal path that gets the algorithm to a certain conclusion, this story would not provide the right level of explanation needed to rationally understand that very outcome.

Finally, the author notices that interpretability, explainability and the like are a means to an end, i.e. ensuring that AI is ethical and trustworthy. The author recommends that we focus on those goals instead of treating interpretability and the like as they were ends in themselves. Since there might be other ways to reach those goals, it seems unhelpful to focus just on one set of solutions.

**Between the lines**
The paper rightly points out that we need a more coherent and precise analysis of concepts such as interpretability and explainability. And the author also clarifies that "[w]hile this paper questions both the importance and the coherence of interpretability and cognates, it does not make a decisive case for the abandonment of the concepts." We agree with this too, as we appreciate the importance of ensuring explainability in AI as a way to both assess whether algorithms are ethical, robust and reliable, and to protect people's right to know and understand how assessments are made. To do so, however, we first need to make sure that we agree on what we mean by 'explanation' and on what kind of explanation is needed for real human understanding.

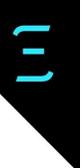

# Transparency as design publicity: explaining and justifying inscrutable algorithms

[Original paper by Michele Loi, Andrea Ferrario, and Eleonora Viganòngsma]
[Research Summary by Marianna Ganapini]

**Overview**: It is often said that trustworthy AI requires systems to be transparent and/or explainable. The goal is to make sure that these systems are epistemically and ethically reliable, while also giving people the chance to understand the outcomes of those systems and the decisions made based on those outcomes. In this paper, the solution proposed stems from the relationship between "design explanations" and transparency: if we have access to the goals, the values and the built-in priorities of an algorithm system, we will be in a better position to evaluate its outcomes.

## Introduction

How can we make AI more understandable? According to the authors of the paper, we care about making AI more intelligible mostly because we want to understand the normative reasons behind a certain AI prediction or outcome. In other words, we want to know: what justifies the outcome of a certain algorithmic assessment, why should I trust that outcome to act and form beliefs based on it? In the paper, the solution proposed stems from the relationship between "design explanations" and transparency: if we have access to the goals, the values and the built-in priorities of an algorithm-system, we will be in a better position to evaluate its outcomes.

## Key Insights

The starting point for talking about transparency and explainability in AI is Lipton's (2018) claim that interpretations of ML models are divided in two categories: model-transparency and post-hoc explanations. Post-hoc explanations look at the prediction of a model and include, most prominently, counterfactual explanations (Wachter et al. 2017). These are based on certain "model features" which, if altered, change the outcome of the model, other things being equal. By looking at the features that impacted a certain outcome, one can in theory determine the (counterfactual) causes that produced that outcome. Though these tools are often used in explainable-AI, the authors of the paper are skeptical: they believe counterfactual explanations do not provide the necessary insights to understand the normative aspects of the model.

Transparency should somehow tell us how the model works, at least in Lipton's definition. However, the authors of the paper have something slightly different in mind: they believe transparency is really the result of making "design explanations" explicit. That is, we need to

The State of AI Ethics Report, Volume 6 (January 2022)                        183

know what the system's function is and how the system was designed to achieve that function. As the authors put it, "explaining the purpose of an algorithm requires giving information on various elements: the goal that the algorithm pursues, the mathematical constructs into which the goal is translated in order to be implemented in the algorithm, and the tests and the data with which the performance of the algorithm was verified."

Parallely, they see "design transparency of an algorithmic system to be the adequate communication of the essential information necessary to provide a satisfactory design explanation of such a system." The most prominent type of transparency in this context is value transparency: we need an accessible account of what values were designed in the system, how they were implemented and to what extent (what tradeoffs were made). Embedded values are values that are designed as part of an algorithmic system and that the system is also able to show in its output. As the authors explain, "[o]nce the criteria to measure the degree of goal achievement are specified" the "a design explanation of an algorithm should provide information on the effective achievement of such objectives in the environment for which the system was built." That is called "performance transparency" in the paper.

This approach is meant to shed light on the goals algorithmic systems are designed to achieve, the values and tradeoffs built into the systems, the set of priorities the system is designed to have and the benchmarks for evaluating success and failure of this design. The goal of transparency is ultimately to provide "the public with the essential elements that are needed in order to assess the justification […] of the decisions" that are based on automated evaluations. If the decisions are based on a system not designed – either intentionally or at the level of how the values are translated – to foster some ethical values, then one might reasonably suspect the decisions made won't match some ethical requirements. More importantly, these decisions cannot be morally acceptable since they are not motivated by the right set of priorities. Understanding all this is a key requirement for evaluating AI and the decisions made based on its recommendations.

**Between The Lines**
In this very interesting paper, the authors offer some actionable recommendations for how to make AI more understandable which seem fully in line with the idea of achieving an "ethics by design" approach to AI. Yet, we also believe that counterfactual and post-hoc explanations could be part of this approach with the goal, for instance, of checking for things that might have gone wrong. Therefore, we would not exclude them from an account of explainability in AI and we recommend a comprehensive approach to make AI understandable to humans.

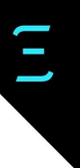

# Ethics-based auditing of automated decision-making systems: intervention points and policy implications

[Original paper by Jakob Mökander and Maria Axente]
[Research Summary by Angshuman Kaushik]

**Overview**: The government mechanisms currently used to oversee human decision-making often fail when applied to automated decision-making systems ("ADMS"). In this paper, the researchers propose the feasibility and effectiveness of ethics-based auditing ("EBA") as a 'soft' yet 'formal' governance mechanism to regulate ADMS and also discuss the policy implications of their findings.

**Introduction**
We are aware of the ethical hazards associated with ADMS, which are in fact, well-documented. In such a scenario, the capacity to address and mitigate these ethical risks posed by ADMS is essential for good governance. This paper, keeping aside the underlying technologies powering ADMS, focuses on its features, for e.g., autonomy, adaptability and scalability that underpin both its socially beneficial and ethically challenging uses. In fact, it narrows down its focus on how organizations can develop and implement effective EBA procedures in practice. While the analysis suggests that EBA is subject to a range of conceptual, technical, economic, legal and institutional constraints, the researchers nevertheless conclude that, EBA should be considered as an integral component of multi-faced approaches to managing the ethical risks posed by ADMS.

**EBA: What is it?**
The emphasis of this paper is entirely on EBA, which is functionally understood as a governance mechanism that helps organizations operationalize their ethical commitments. It concerns what ought and ought not to be done over and above existing regulation. Operationally, EBA is characterized by a structured process whereby an entity's present or past behavior is assessed for consistency with a predefined set of principles. Throughout this process, various tools and methods such as software programmes, stakeholder consultation etc. are employed to verify claims and create documentation. In fact, different EBA procedures employ different tools and contain different steps. However, an EBA differs from simply publishing a code of conduct since its main activity consists of demonstrating adherence to a predefined standard. The paper also emphasizes on how organizations can develop and implement effective EBA procedures in practice instead of concentrating only on what EBA is and why it is needed. The objective is twofold. First, the researchers seek to identify the intervention points, both in organizational governance as well as in the software development lifecycle, at which EBA can help inform ethical deliberation and thereby make a positive difference to the ways in which ADMS are designed and deployed. Second, they seek to contribute to an understanding of how policymakers and regulators can facilitate and support the implementation of EBA procedures in organizations that develop ADMS.

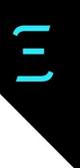

**EBA: Different approaches**

The paper distinguishes between different approaches for EBA, such as functionality audits, which for example, focuses on the rationale behind decisions. In contrast, code audits entail reviewing the source code of an algorithm. Finally, impact auditing investigates the types, severity, and prevalence of effects of an algorithm's outputs. These approaches are complementary and can be combined into holistic EBA procedures. According to the researchers, since autonomous and self-learning ADMS may evolve and adapt over time as they interact with their environments, EBA needs to include at least the elements of continuous, real-time monitoring i.e. impact auditing.

**Governing STS and identifying intervention points for EBA**

The paper then dwells upon Socio-Technical Systems (STS), which comprises both social entities, like people and organizations, and technical entities, like tools, infrastructures, and processes. ADMS, then, refers to technical systems that encompass decision-making models, algorithms that translate models into computable code, as well as methods to acquire and process input data. Further, ADMS interact with the entire political and economic environment surrounding their use. The paper then goes on to analyze how complex STS are governed today and discusses how EBA procedures can be designed to complement and enhance existing governance structures. Governance consists of both hard and soft aspects. Hard governance mechanisms are systems of rules elaborated and enforced through institutions to govern the behavior of agents. When considering ADMS, examples of hard governance mechanisms range from legal restrictions on system outputs to outright prohibition of the use of ADMS for specific applications. Soft governance, on the other hand, embodies mechanisms that abide by the prescriptions of hard governance while exhibiting some degree of contextual flexibility. A further distinction is also made between formal and informal governance mechanisms, where formal governance mechanisms refer to official communications. The researchers go on to advocate EBA as a soft yet formal governance mechanism to complement and strengthen the congruence of existing governance structures within organizations that develop and use ADMS. Further, the paper looks at some of the potential intervention points (points at which decisions, actions, or activities are likely to shape the design and behavior of ADMS) at which EBA can help shape the design and deployment of ethical ADMS by informing ethical deliberation. They are as follows:

- value and vision statement ;

- principles and codes of conduct;
- ethics boards and review committees;
- stakeholder consultation;
- employee education and training;
- performance criteria and incentives;
- reporting channels;
- product development;
- product deployment and redesign;
- periodic audits; and
- monitoring of outputs

**Recommendations to policymakers**

The paper not only identifies limitations and risks associated with EBA but also discusses how policymakers and regulators can facilitate the adoption of EBA by organizations that design and deploy ADMS. According to the researchers, the organizations that design and deploy ADMS have good reasons to subject themselves and the systems they operate to EBA. For example, ensuring the ethical alignment of ADMS would help organizations manage financial and legal risks, help them gain competitive advantage etc. In fact, the documentation and communication of the steps taken to ensure that ADMS are ethical can play a positive role in both marketing and public relations.

The paper also highlights eight policy recommendations for policymakers and regulators to follow:

- Help provide working definitions for ADMS – regulators shall define for organizations the material scope for EBA by providing working definitions or risk classifications of ADMS that enable proportionate and progressive governance;
- Provide guidance on how to resolve tensions – when designing and operating ADMS, conflicts may arise between different ethical principles such as fairness, privacy etc., for which there are no fixed solutions. In such a scenario, regulators shall provide guidance on how to resolve tensions between such conflicting values in different situations;
- Support the creation of standardized evaluation matrices and reporting formats – while organizations should be free to adopt different EBA procedures, regulators can also support the creation of standardized evaluation metrics and reporting formats;
- Facilitate knowledge sharing and communication of best practices – regulators can not only provide digital platforms where software code and data could be shared but also create forums where stakeholders could discuss and share best practices for EBA of ADMS;

- Create an independent body to oversee EBA of ADMS – create an independent body that authorizes organizations who, in turn, conduct EBA of, or issue ethics-based certifications for, ADMS;
- Create incentives for voluntary adoption of EBA – implementing EBA across organizations would involve costs. Therefore, to incentivize the voluntary adoption of EBA, regulators should encourage and reward demonstrable achievements;
- Promote trust through transparency and accountability – regulators can strengthen trust in emerging EBA procedures by ensuring accountability, e.g., by imposing sanctions where trust is breached; and

Provide governmental leadership – political leaders can help strengthen the feasibility and effectiveness of EBA as a governance mechanism by explaining and endorsing it.Therefore, in order to demonstrate their commitment to officially stated policies, governments can consider conducting EBA of ADMS employed in the public sector and include ethics-based criteria in the public procurement of ADMS.

**Between the lines**
This paper provides a very holistic and process-oriented approach to EBA. In fact, many of the intervention points listed in the paper already exist within organizations that design and deploy ADMS. Hence, implementing EBA would not entail imposition of any additional layers of governance upon them. It is pertinent to mention here that the key to developing feasible and effective EBA procedures is to combine existing conceptual frameworks into structured processes that monitor each phase of the ADMS lifecycle to identify and correct the points at which ethical failures may occur. To sum up, the recommendations delineated in this paper would definitely go a long way in mitigating some of the ethical hazards posed by ADMS.

## Trustworthiness of Artificial Intelligence

[Original paper by Sonali Jain, Shagun Sharma, Manan Luthra, Mehtab Fatima]
[Research Summary by Connor Wright]

**Overview**: If you are new to the space of AI Ethics, this is the paper for you. Offering a wide coverage of the issues that enter into the debate, AI governance and how we build trustworthy AI are explored by the authors.

**Introduction**
One of the strengths of this paper is how it proves a productive introduction for those who are new to the AI Ethics space. Touching upon governance (as we have done), how we create trustworthy AI is explored. What we mean by 'trustworthy' is open for review, but some aspects must enter the debate. Three of these are highlighted below.

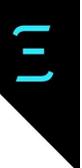

**Key Insights**
The authors appeal to how AI should be compliant in the following 3 ways:

1. Lawful: The AI system should be compliant with various rules and laws.

2. Ethical: It should contain morals and ethics and adhere to moral values and principles.

3. Robust: AI should be sturdy in both social and technical sense.

**How AI can be made lawful: A rights approach to AI**
The benefit of such an approach is its ability to put humanity at the center of AI considerations while maintaining respect for human dignity. One example of how this works is the right to freedom from coercion. Focused on preventing manipulation, laws such as the California Law try to make sure that "AI systems must not in any case dominate, force, deceive or manipulate human beings" (p.g. 908).

The approach becomes even more intriguing when applied to harm. Often, AI systems are said to be designed not to harm humans. While being an intuitive claim, such an approach does require the AI to be aware of humans alongside the context in which it finds itself.

Furthermore, the depth of awareness required depends on which AI system you're talking about. You can imagine that the AI used in CV screening does not need to have an acute sense of other humans compared to facial recognition (especially at Facebook).

However, a rights-based approach can't do it all on its own.

**Ethical principles in the AI space**
The importance of privacy, explainability and transparency were rightly explored here, staple products in building trustworthy AI. However, what jumped out at me was how the authors did not advocate for complete transparency. Instead, transparency is to be pursued in the name of fueling explainability, but some information should only be accessible to those in the appropriate positions.

Nevertheless, those in these positions should be both interdisciplinary and diverse.

**The importance of universal design**

Given AI's wide-reaching effects, the design should be accessible to all genders, ages and ethnicities. This comes from designing the AI with diversity already in the team, a token of its all-encompassing nature. Furthermore, the 'common AI fight' is shown in the paper's methods for trustworthy AI involving cross-business and cross-sector collaboration. With AI's impact being both mental and physical, the AI space needs all the collaboration it can get.

**Between the lines**
While a good introduction into the AI space, I would've liked a deeper exploration into the practical side of these approaches. For example, how human intervention in AI processes can be beneficial, rather than having it assumed to be so. Nevertheless, should any human intervention have a chance of success, the correct education would be required. Here, I liked how the paper mentioned AI's potential call for the educational system to be more job orientated and reflect the state of the world it will be creating. While this may not be the actuality, it will soon convert into a necessity.

## Getting from Commitment to Content in AI and Data Ethics: Justice and Explainability

[[Original paper](#) **by John Basl, Ronald Sandler and Steven Tiell]**
**[Research Summary by Angshuman Kaushik]**

**Overview**: AI or data ethics principles or frameworks meant to demonstrate a commitment to addressing the challenges posed by AI are ubiquitous and are an 'easy first step'. However, the harder task is to operationalize them. This report, inter alia, stipulates strategies for putting those principles into practice.

**Introduction**
Amidst the chaotic AI Ethics principles landscape, this report emerges as a much-needed guide in understanding the entire gamut of issues related with the application of those principles in governance scenarios. It emphasizes the complexities associated with moving from general commitments to substantive specifications in AI and data ethics. According to it, much of this complexity arises from three key factors:

- ethical concepts such as justice and transparency that often have many senses and meaning;
- which senses of ethical concepts are operative or appropriate is often contextual; and
- ethical concepts are multidimensional e.g., in terms of what needs to be transparent, to whom, and in what form.

Further, the objectives of the report are to:

- demonstrate the importance and complexity of moving from general ethical concepts and principles to action-guiding substantive content, i.e., normative content;
- provide detailed analysis of two widely discussed and interconnected ethical concepts, justice and transparency; and
- indicate strategies for moving from general ethical concepts and principles to more specific normative content and ultimately to operationalizing that content.

**AI Ethics – Understanding the challenges**
The report talks about considerable convergence among the many AI ethics frameworks that have been developed. They coalesce around core concepts, some of which are individual-oriented, others society-oriented and still others system-oriented. However, according to the researchers, enunciating ethical values and principles is only the first step in addressing AI and data ethics challenges and it is in many ways the easiest. The much harder work is the following:

- substantively specifying the content of the concepts, principles and commitments; and
- building professional, social and organizational capacity to realize these in practice.

**An example from the field of bioethics**
In order to better comprehend the obstacles encountered in moving from general ethical concepts to a functioning AI framework (normative content), the paper takes the case of informed consent in bioethics, which is widely recognized as a crucial component of ethical clinical practice. Informed consent operationalizes the principle of individual autonomy.

Practically, it requires the fulfillment of three conditions namely:

- **disclosure** – provision of clear, accurate and relevant information to the subjects;
- **comprehension** – information is provided to the subjects in a way that they can understand; and
- **voluntariness** – the subjects make the decision without undue influence or coercion.

The enforcement of these three conditions is the task of bioethicists, hospital ethics committees and institutional review boards. They prepare guidelines, best practices, procedures etc., for meeting the above informed consent conditions.

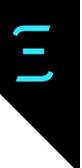

According to the researchers, while informed consent is meant to protect the value of autonomy and express respect for persons, a general commitment to the principle of informed consent is just the beginning. The principle must be explicated and operationalized before it is meaningful and useful in practice. The same is true for principles of AI and data ethics. The researchers then narrow down their focus on the complexities involved in moving from core concepts and principles to operationalization of the normative content for two prominently discussed and interconnected AI and data ethics concepts: justice and transparency.

**Meaning of justice in AI**
The report mentions that the concept of justice is a complex one, and can mean different things in different contexts. To determine what justice in AI and data use requires in a particular context, it is imperative to clarify the normative content and underlying values. Only then it is possible to specify what is required in specific cases, and in turn how or to what extent justice can be operationalized in technical systems. According to the report, the general principle of justice is that all people should be equally respected and valued in social, economic and political systems and processes. However, there are many ways this very general principle of justice intersects with social structures and systems. As a result, there is a diverse set of more specific justice-oriented principles such as procedural, distributive and recognition justice.

**What does committing to justice mean?**
The researchers consider context to be critically important in determining which justice-oriented principles take precedence. Therefore, the first step in specifying the normative content is to identify the justice-oriented principles that are crucial to the work that the AI system does. Only then can a commitment to justice be effectively put into practice. Articulating the relevant justice-oriented principles will also require considering organizational missions, the types of products and services involved, how those products and services could impact communities and individuals etc. In identifying these, it will be helpful to reflect on similar cases and carefully consider the sorts of concerns that people have raised about AI systems. The researchers have cited two hypothetical cases to illustrate this. Further, the report states that the diversity of the justice-oriented principles and the need to make context-specific determinations about which are relevant and which to prioritize expose the limits of a strictly algorithmic manner in incorporating justice in AI systems. The reason being, firstly, there is no singular, general justice-oriented constraint, optimization or utility function and secondly, there will not be a strictly algorithmic way to fully incorporate justice into decision-making, even once the relevant justice considerations have been identified. The report then goes on to ask the question as to how and to what extent can the salient aspects of justice be achieved algorithmically. According to the researchers, accomplishing justice in AI will require developing justice-informed, techno-social or human-algorithm systems. AI systems can support social workers in service determinations, admissions officers in college admissions determinations, or healthcare professionals in diagnostic determinations and they might even be able to help reduce biases in those processes. According to the researchers, a commitment to justice in AI involves remaining open to the possibility that sometimes an AI-oriented approach might not be a just one. They stress on the fact that, organizations that are committed to justice in AI will require significant organizational capacity and processes to operationalize and implement their commitment, in addition to technical capacity and expertise. Reliance upon techno solutionism or on standards developed in other contexts is not desirable.

**Transparency in AI**

In the view of the researchers, in spite of the role that transparency plays in helping to achieve justice, it can also play an important role in realizing other concepts and values. They also lay down the many ways in which a decision system could be made transparent. The forms that commitments to transparency may take are as follows:

- **Interpretability** – requiring AI systems to be interpretable;
- **Explainability** – a decision-making system is explainable when it is possible to offer stakeholders an explanation that can be understood as justifying a given decision;
- **Justified Opacity** – transparency about the reasons for adopting opaque systems can serve to justify other forms of opacity; and
- **Auditability** – a carefully constructed audit can provide assurance that decision-making systems broadly are trustworthy, reliable and compliant.

**Way forward**

The researchers point out that for organizations to be successful in realizing their ethical commitments and accomplishing responsible AI, they must think broadly about how to build ethical capacity within their organizations. Some of the initiatives cited are as follows:

- creating AI and data ethics committees that can aid in developing policies and other governance measures;
- meaningfully engaging with impacted communities to better comprehend ethical issues and other ways to broaden perspectives and collaborations;
- training and education;
- integrating ethics into practice; and
- building an AI and data ethics community.

**Between the lines**

The plethora of vaguely formulated AI Ethics principles, guidelines, standards etc., that have come to dominate the AI Ethics space in the last few years have hardly aided in operationalizing ethical AI in practice. With the passage of time such principles have begun to sound banal and an appendage to an organization's other 'significant documents'. In such a scenario, this report serves as a guidepost by laying down strategies for moving from general ethical concepts and principles to more specific normative content and ultimately to operationalizing that content. Further, the report simplifies comprehension of the complexities involved in such a transition, by the use of illustrations. The report can prove handy and a 'go-to guide' not only for those entities that are struggling to formulate ethical principles but also to those that are trying to get an AI Ethics framework up and running.

## Foundations for the future: institution building for the purpose of artificial intelligence governance

[Original paper by Charlotte Stix]
[Research Summary by Angshuman Kaushik]

**Overview**: To implement governance efforts for artificial intelligence (AI), new institutions require to be established, both at a national and an international level. This paper outlines a scheme of such institutions and conducts an in-depth investigation of three key components of any future AI governance institution, exploring benefits and associated drawbacks. Thereafter, the paper highlights significant aspects of various institutional roles specifically around questions of institutional purpose, and frames what these could look like in practice, by placing these debates in a European context and proposing different iterations of a European AI Agency. Finally, conclusions and future research directions are proposed.

**Introduction**
The paper begins by drawing the attention of the readers to the fact that the governments around the world have begun to approach the governance of AI through multiple controls. One example being the European Union's recent Proposal for a Regulation of the European Parliament and of the Council Laying Down Harmonized Rules on Artificial Intelligence and Amending Certain Union Legislative Acts ("Artificial Intelligence Act") which puts forward a regulatory framework for high-risk AI systems and the other being the Trade and Technology Council co-established by the US and the EU with the mandate to cooperate on the development of suitable standards for AI. Further, as the field of AI governance is relatively new, as such, there exist only a few specialist governmental institutions exclusively dedicated in the area. According to the author, to properly develop, support and implement new AI governance efforts, it is likely that a number of new institutions will need to be established in the future. There are broadly two types of institutions that one could investigate: those that exist and may be adapted and those that do not exist yet but will eventually come into existence to fill the void created by new governance initiatives. This paper puts emphasis on the latter type of institutions, with particular focus on institutions set up by governments. In order to proceed with its objective, the paper builds on recent academic calls for an international governance coordinating committee for AI, for an international regulatory agency for AI etc., and draws on existing scholarship in the area, and addresses itself to those individuals who will be involved in setting up new institutions and those who are interested in conducting further research on pragmatic institution building for AI governance.

**Building new AI governance institutions**
The paper states that AI-specific governance institutions working on soft governance mechanisms with non-binding rules have already come into existence. A few examples are OECD, G7, and the Global Partnership on AI etc. However, there has been mounting pressure to develop and implement stronger and more binding AI governance mechanisms than those covered by ethical principles. As countries move towards harder governance efforts, they are likely to require increasingly specialized institutions to oversee their implementation. Moreover, as AI governance efforts soar and more coordination, action and policy proposals become necessary within a nation as well as at an international level, it is likely that there will be a need for more specialized governmental agencies to handle an increasingly diverse set of tasks on top of the existing work. It might be overall quicker, cheaper and more effective to build a new institution from scratch that is 'fit for purpose' rather than exert time, effort and political goodwill to change the structure of an existing institution. The author then puts forward a selection of axes that need to be considered in building new AI governance institutions, namely, purpose, geography and capacity, with particular emphasis on purpose.

**Purpose**
The first question that needs to be answered is the purposes of the new institution i.e., what is it meant to do? Under the broad heading of purpose, the paper introduces the outline of four different roles an institution for AI governance could take. The roles are namely, coordinator, analyzer, developer and investigator.

**The coordinator institution**
The task of a coordinator institution could, for instance, include working with the rising number of ethical guidelines and attempting to operationalize them more clearly. It could also serve as an umbrella organization and coordinate activities amongst different groups. Some examples of coordinator institutions are the UN, the G20, and NATO etc. The paper goes on to highlight the fact that the actions of the coordinator institution shall be timely and appropriate and proposes that a future AI Agency in the EU might take up the role of a coordinator institution.

**The analyzer institution**
The duties of an analyzer institution could be varied, such as mapping existing efforts and identifying gaps across various governments (for instance the European Commission), compiling data sets and information on the technical landscape and sketching technological trajectories (for instance the AI Index) etc. The role of an analyzer institution is more active than that of a coordinator institution, in that it interferes more directly with the governance or policy making process by way of providing crucial information that can inform and shape those decision-making processes.

**The developer institution**
A developer institution shall provide either directly actionable measure or formulate new policy solutions to existing issues. It may take up the role of examining blind spots and proposing solutions for those by way of its own initiative, in addition to work it might be asked to undertake by various government agencies.

**The investigator institution**
It is envisioned as a 'watchdog' assigned with the task to investigate whether or not actors such as governments, companies or specific organizations are adhering to the relevant standards, procedures, laws or not. One example of an investigator institution is the Human Rights Council. The most important requirement of such an institution would be its independence and impartiality.

**Geography**
The effects of AI systems transcend geographies and are not confined within national borders. Therefore, many AI governance issues could be seen as multi-country concerns. The couple of broad considerations with respect to geography that the paper delves into are: what is the benefit or downside of a new multi-country institution and how does it fare in comparison to nationally 'restricted' institutions? A multi-country institution must consider questions of access, inclusion and participation. One model proposed is if several nations expect that their position towards AI governance is broadly more beneficial than that of other nations, it may be reasonable for them to cooperate and coordinate to establish a dedicated institution. Conversely, if nations choose not to form a new institution, a proliferation of similar but distinct institutions could affect fragmentation of global AI governance regimes.

**Capacity**
The third axis is capacity which relates to the previous two axes i.e., purpose and geography. It concerns what the institution needs in terms of capacities for it to thrive, both on the technical and non-technical side. The paper proposes that access to technical infrastructure could play an important role for future AI governance institutions. The said technical infrastructure may include access to compute, available datasets, testing and experimentation facilities etc. It can minimize bottlenecks in terms of information exchange and increase speed between what is to be governed and the associated governance actions and decisions, thereby contributing to more agility, specificity and foresight in policy making for AI. On the non-technical side, the paper underscores the need to build up human capacity which could broadly take two forms: (a) out of the house capacity with either (1) a network of individual experts to draw upon when needed, or (2) expert groups and external panels and (b) in-house capacity with a team having a diverse background with relevant experience in technical, legal and ethical areas.

**Between the lines**

The rapid and unbridled increase in the use of AI systems has necessitated its effective governance. In fact to govern efficiently, the need of the hour is institutions that can deliver. This paper, instead of making normative assessments of the various institutional setups, charts out a pragmatic approach in building up institutions for AI governance at a time when proposals for setting up such institutions are gathering steam. More importantly, it provides a framework to start with. Another highlight of the paper is that it shows the way for future research direction on the topic.

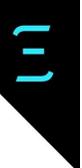

# Go Wide: Article Summaries (summarized by Abhishek Gupta)

## Five Recommendations For Creating More Ethical AI

**[Original article by Forbes]**

**What happened**: The article proposes some foundational steps that will be important in creating more ethical AI systems and make them a normalized practice. Hiding behind technicalities and shirking moral obligations as not part of the leadership role should be avoided. Instead, the author asks companies to make long-term research commitments and working with monied partners who understand this approach rather than asking them to take shortcuts. Empowering employees not only so that they can raise issues as they arise but also so that they can propose innovative solutions and see them implemented is another crucial step. Finally, being transparent about one's approach to AI ethics and holding oneself accountable for following that approach will also help build public trust in one's work.

**Why it matters**: Having more actionable approaches to AI ethics, especially guidance for leadership, will be essential for the actual implementation of these ideas in practice. The shortlist provided here serves as a reminder to practitioners and researchers in AI ethics that the organizational challenges are just as significant as the technical and socio-technical challenges in building more ethical, safe, and inclusive AI systems.

**Between the lines**: I've found the approach undertaken at Microsoft as outlined in this WEF Case Study to effectively marry the organizational, technical, and socio-technical methods to achieve Responsible AI objectives. We need more examples where Responsible AI methods and recommendations such as the ones highlighted in this article are trialed and analyze those results to learn what works and what doesn't.

## Police Are Telling ShotSpotter to Alter Evidence From Gunshot-Detecting AI

**[Original article by Vice]**

**What happened**: An AI-powered tool that is used to detect whether shots were fired in a neighborhood was used as evidence in a case in Chicago but the accused was acquitted when it was discovered through cross-examination and deeper investigation that the alerts from the system were modified to better align with the narrative that was being presented by the prosecution. As the article goes on to show, this wasn't the first time that this happened, and that trust in the system has been declining over time. In particular, there are many false alerts that are issued by the system, but more so that the "humans-in-the-loop" that work for the company receive requests from law enforcement to dive deeper and have modified the actual alerts to bolster the case being presented against the defendant.

**Why it matters**: In matters of someone's life, relying on flimsy evidence, especially one that is not subject to more rigorous tests and supported by studies that have been funded by unrestricted funding from the very company selling the tool to attest to its efficacy should be taken with more than just a grain of salt. Just as we wouldn't trust a thoroughly tested DNA test as evidence in a court, digital forensics tools should face similar scrutiny. As the article mentions, the city of Chicago is the second-largest client for the company and with their contract coming up for renewal, a more unbiased, and scientifically grounded analysis should be conducted before engaging their services again.

**Between the lines**: Something that really jumped out in the article was the mention of how unevenly such systems are deployed across different neighborhoods in the city with Latinx, Black, and Brown neighborhoods facing the brunt of this form of policing while being notably absent from more affluent and White neighborhoods. More so, residents of these policed neighborhoods raise a very pertinent point: if law enforcement just asked them if a shot was fired, as responsible neighbors, they would share that with them rather than having to rely on flimsy technology.

### Optimizing People You May Know (PYMK) for equity in network creation

[Original article by LinkedIn Engineering]

**What happened**: LinkedIn has applied two fairness measures of equality of opportunity and equalized odds to make the recommendations for potential connections more equitable across the users of the platform, especially for those who don't have as "influential" profiles as some of the more frequent members (FM) of the platform. In applying these fairness measures on top of their ranking algorithms, they've found that engagement on the platform didn"t go down, showing that fairness objectives don't necessarily have to stand against the business objectives.

**Why it matters**: All social media platforms are prone to having bias in terms of the benefits of the platform skewing towards those who have amassed influence both on and off the platform. Creating opportunities for newer entrants is a way of more fairly distributing the opportunities on the social media platform, as is shown in this article for LinkedIn where recommendations are now being made of those profiles that have fewer pending requests amongst some other balancing factors.

**Between the lines**: As social media platforms take on a more pervasive presence in our lives, those that seek to aid the less influential participants of that ecosystem have the potential to become more prominent. In a sense, we have seen this happen with TikTok where even new entrants on the platforms have a chance to quickly amass huge followings compared to other platforms where having a large following predisposes your presence and dominance in the newsfeed of that platform.

### AI datasets are prone to mismanagement, study finds

[Original article by [VentureBeat](VentureBeat)]

**What happened**: Researchers from Princeton found that popular datasets containing images that are used to train computer vision models contain data for which they might not have had consent. In addition, they found misuses of the datasets through modifications made to it where it wasn't clear if that was allowed under the licenses and even when they were not clearly allowed by the licenses on the original datasets. While two out of the three datasets have been taken down from their original sources, one continues to exist with a disclaimer that the data shouldn't be used for commercial purposes. The other two datasets though are still accessible through non-official means via torrents and other places that have archived it.

**Why it matters**: Ethically dubious applications are powered using the training data offered by these datasets, often going beyond the original intentions and purposes for which they were created. The authors of the paper recommend stewardship of datasets throughout their existence, and being more proactive about potential misuses. They also recommend being more clear in the licenses associated with these datasets. This will hopefully reduce consent violations and downstream misuses.

**Between the lines**: What is still missing from the conversation is how the solutions mentioned in the article are non-binding, voluntary, and won't actually lead to any change as long as the benefits to be derived from using the training datasets outweigh the (non-existent) costs associated with their misuses. If this problem is to be tackled effectively, the suggestions need to be a lot more rigorous and have elements of enforceability and legal might that will deter misuse and strongly mandate consent for any data used to compose the dataset.

### Apple says collision in child-abuse hashing system is not a concern

**[Original article by The Verge]**

**What happened**: Apple recently unveiled the NeuralHash perceptual hash algorithm that will be applied to iCloud backed content on Apple devices to detect child sexual abuse material (CSAM). This has been met with backlash from privacy-minded organizations and activists who have called it out for setting a precedent that might allow for more invasive monitoring of people's private content on their devices. As the details of the system have come to light, researchers have reverse engineered the algorithm and have demonstrated hash collisions (when two images that are different produce the same hash - a representation code for that image) that will befuddle the system into giving out false positives. Apple has mentioned that there are secondary checks in place that will minimize the impact from such false positives through the use of an additional server-side algorithm different from NeuralHash and more than 30 images need to be flagged before they are passed on as an alert for human intervention.

**Why it matters**: Robustness in systems that detect and automatically flag content is important, especially if there is analysis being performed on private content. Yet, it would appear that there are some flaws in the system as demonstrated by researchers. More importantly, a lack of complete transparency on the secondary systems and what the real-world probabilities of these collisions is going to be like further exacerbate the doubts that people have about the effectiveness of such a system.

**Between the lines**: While the intention behind the deployment of such a system stands to make the information ecosystem safer, especially as it relates to CSAM, without trust from the users who form that ecosystem, there is bound to be pushback and hesitation in full participation. Apple can of course railroad ahead since they own the software and hardware stack but that will be harakiri in a competitive marketplace where they have always prided themselves on keeping the privacy of their users above all else, even in the face of mounting pressure from law enforcement agencies in the past.

### Deleting unethical data sets isn't good enough

**[Original article by MIT Technology Review]**

**What happened**: The rapid progress of AI systems' capabilities has been fueled by the availability of large-scale, benchmark datasets. A lot of those were collected by scraping data from the internet, often without the consent of subjects whose data was scooped up in that scraping process. The paper highlighted in this article talks about DukeMTMC, MS-Celeb-1M, and other datasets which have been retracted after concerns were raised by the community. But, they continue to linger online in various forms, often morphed to serve new purposes: adding masks to faces in these datasets to improve the capability of facial recognition technology during the pandemic.

**Why it matters**: Some datasets do come with warnings and documentation on their limitations but these end up being ignored and derived datasets even lose those pieces of documentation. This means that we have problematic datasets, replete with biases and privacy violations, continuing to exist in the wild, with hundreds of papers being written and published at conferences based on the results from training AI systems on that data.

**Between the lines**: A potential solution mentioned in the article talks about data stewardship where a potentially independent organization can take on the role of stewarding the proper use of that data throughout the lifecycle of its existence. While noble, there are tremendous challenges in sourcing funding for such organizations and allocating sufficient recognition to such work where the emphasis in the academic domain continues to center on publishing state-of-the-art results and work such as stewardship would face an uphill battle. I'd be delighted if I'm proven wrong on this front and hope that we start to recognize the hard work that goes into preparing, maintaining and retiring datasets.

## Six Essential Elements Of A Responsible AI Model

**[Original article by [Forbes](#)]**

**What happened**: The article presents a simple model with 6 items to think about in Responsible AI: accountable, impartial, transparent, resilient, secure, and governed. This is a combination and perhaps rehash of many existing frameworks, guidelines, and sets of principles already out there. The author admits to as much. What is interesting in the article is the list of questions that are provided when thinking about whether or not to have an AI ethics board and those are the biggest takeaways from the article. In particular, splitting up who should be held accountable when something goes wrong and who is responsible for making changes to address undesirable outcomes is something that is important.

**Why it matters**: A lot of frameworks in Responsible AI can end up being overly complicated or overly simplistic. This model perhaps has the right level of granularity but more than that what is useful is that it provides a great starting point for someone who is early in their Responsible AI journey to get started with a core set of priorities.

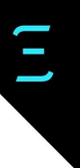

**Between the lines**: What we really need next is a comprehensive evaluation of the effectiveness of this framework and all the other frameworks that are out there in terms of meeting the stated goals of actually achieving responsible AI in practice at their organization. Unless that happens, we can't meaningfully pick one framework over another because all evaluations as of yet would be theoretical in nature.

## Why you should hire a chief AI ethics officer

[Original article by World Economic Forum]

**What happened**: As we have more organizations moving from principles to practice, this article gives a quick overview of the Chief AI Ethics Officer role including what it should include in its purview and how it can accelerate adoption of AI ethics within organizations. The role should drive the broad definition of ethics principles for the organization, define suitable properties for the AI systems, and drive tooling and processes within the organization for practitioners. A wide range of expertise is also required to take on this role including a multi-disciplinary background, the ability to effectively communicate with a diversity of stakeholders, driving company-wide engagement, and helping to make the business case for AI ethics as a core consideration.

**Why it matters**: In a recently published article, we highlighted what it would take for a Chief AI Ethics Officer to succeed within the organization and why it is an important role. In particular, the current problem with the move from principles to practice is that such efforts are often ad-hoc or don't have enough executive support to really drive meaningful change within the organization. The appointment of such a position helps to overcome some of these challenges.

**Between the lines**: But, the mere appointment of such a position to meet public appearance requirements will only cause more harm in the long run. What needs to be carefully considered is how much actual power is allocated to this person and whether they have the necessary background and skills to be able to drive change across the organization. One of the most problematic issues at the moment is a lack of sufficient technical and operational expertise in such roles that leads to a breakdown of strategy when it comes time to actually put these ideas into practice.

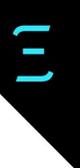

## Thinking Through the Ethics of New Tech…Before There's a Problem

[Original article by Harvard Business Review]

**What happened**: The article points out how society has typically stopped to address ethical issues with the use of new technology after it rapidly permeates our life; the author asks us to imagine what would happen if that is not the case? Taking the examples of automobile safety features like the seatbelt that appeared many years after which if implemented earlier would have saved many lives. We are on a similar cusp with AI rapidly permeating many parts of our lives. By bringing in specialists who are domain experts, eschewing haste in the deployment of a new piece of technology because it seems to offer immediate gains, but one that might have delayed, severe downstream consequences and assigning accountability to stakeholders in different parts of the lifecycle and having someone in a leadership position take this on as a core responsibility are some proposed ways that we might be able to mitigate these issues.

**Why it matters**: As organizations struggle to move from principles to practices, the advice offered in this article is a great starting point for those who want to realize some early wins in the ethical, safe, and inclusive deployment of AI systems. Reframing the challenges as opportunities to do better and guide others along the way might be yet another benefit emerging from adopting these practices.

**Between the lines**: I think another layer of nuance needs to be added to the advice of "pausing and thinking," which is to understand the incentives that guide employee and stakeholder behavior within the organization. In particular, if KPIs are such that a certain number of users need to be secured or a certain sales quota needs to be met to secure a bonus at the end of the year, then we need to make sure that these ideas are discussed keeping this in mind, otherwise implementations of tech ethics are doomed to fail.

# 7. Laws and Regulations

**Introduction** by Abhishek Gupta, Founder and Principal Researcher, Montreal AI Ethics Institute

The legal landscape is getting busy with many regulations coming forth to regulate the use of AI. This chapter opens with an analysis of the EU AI Act and what the introduction of the Act means for cooperation between different parts of the world, such as the US and EU. The next piece on "Algorithmic accountability for the public sector" resonated with me a lot given our focus on the rising importance of context and public participation as a pivotal factor in the success of new policy mechanisms. Going further with those ideas, this chapter also covers the Responsible AI principles from NATO that applies to the military use of AI. There remain several ambiguous terms in the strategy document (or at least the summary version of it which is the publicly available version) which water down the ability for people to implement those ideas in practice. Nonetheless, given the prominence of NATO, this is a great start to move countries towards coalescing around a shared set of principles for the responsible development and deployment of AI systems, especially in the military context.

As we enter the era of co-creation and co-invention with machines, it is important to examine, with a legal lens, how we grant patents and to whom for what kinds of invention. In fact, in my view, it even raises questions about what the usefulness of patents is, at least purely in the domain of algorithmic advances within the domain of AI. "Summoning a New Artificial Intelligence Patent Model: In the Age of Pandemic" proposes creative solutions to the hurdles of patenting AI technology by establishing a new patent track model for AI inventions.

So far, regulations have had a mixed track record on curbing the spread of harm from technological systems. Uneven enforcement and ambiguity seem to be the culprits for this problem. What we see from the rest of the chapter are a few examples such as missing geofence warrants in the California DoJ Transparency Database that showcases how vulnerable we are still to the actual following of practice even if there are mandates in place to ensure accountability and citizen welfare. Other examples include the changes that we need in Congressional practices today so that regulation for Big Tech happens in a more timely and meaningful fashion. With the launch of CoPilot from GitHub, code generation and verbatim replication of code samples has started posing legal challenges, especially since there are slightly varying interpretations of how licenses are placed in each code repository and the overall license and fair use policies that govern the entire GitHub platform.

It suffices to say that there is a lot of movement in the legal domain and it will be interesting to see what other developments take place in the rest of 2022 as we seek to concretize regulations and search for effective mechanisms to enforce them so that it doesn't just happen in letter but also in spirit.

MAIEI is also happy to endorse the upcoming Algorithmic Accountability Act (2022) from Senator Wyden's office in the US and is happy to engage with other policymakers from around the world. Please don't hesitate in reaching out to us to collaborate!

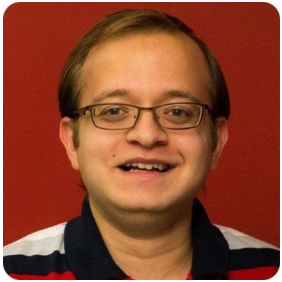

**Abhishek Gupta ([@atg_abhishek](https://twitter.com/atg_abhishek))**
Founder, Director, & Principal Researcher
Montreal AI Ethics Institute

Abhishek Gupta is the Founder, Director, and Principal Researcher at the Montreal AI Ethics Institute. He is a Machine Learning Engineer at Microsoft, where he serves on the CSE Responsible AI Board. He also serves as the Chair of the Standards Working Group at the Green Software Foundation.

# Go Deep: Research Summaries

## The European Commission's Artificial Intelligence Act (Stanford HAI Policy Brief)

[Original paper by Marietje Schaake]
[Research Summary by Abhishek Gupta]

**Overview**: With the recently released Artificial Intelligence Act in the EU, a lively debate has erupted around what this means for different AI applications, companies building these systems, and more broadly, the future of innovation and regulation. Schaake provides an excellent overview of the Act with an analysis of the implications and sentiments around this Act, including global cooperation between different regions of the world like the US and EU.

**Introduction**

Just as we had a scramble in the wake of the GDPR in 2018 as companies rushed to become compliant, the announcement of the AI Act has triggered a frenzy amongst organizations to find ways to become compliant while maintaining their ability to innovate. The current paradigm of AI applications incentivizes more invasive data collection to power these applications while providing recommendations, decisions, and influencing people's lives in more and more significant ways.

The policy brief provides a quick overview of the definition of AI used in the AI Act, which kinds of applications it applies to (high-risk), what high-risk means, some banned use cases, some exceptions to those banned use cases, what conformity assessments are, the implications of the AI Act on the rest of the world, and how civil society and other organizations have reacted to the Act. There are mixed reactions, but Schaake concludes on an optimistic note that the Act can become a rallying point to achieve more consistency in cybersecurity and other practices in addition to AI development across the world. We shouldn't treat the harms from AI systems as inevitable.

The definition of AI utilized in the Act follows an interesting path of using a broad, overarching definition with some specifically defined categories and use cases. This hybrid approach is supplemented by the power to amend these definitions as we go along to make them more compatible with technical and sociological developments in the future. This will be critical for the continued applicability of the Act, which is lacking in many other proposed regulations that are either too vague or too specific.

**Risk and unacceptable uses**
The central operating mechanism of the Act is to look at high-risk AI uses-cases which include biometric identification, critical infrastructure that can significantly impact human lives, determining access to education and employment, worker management, access to private and public services (e.g., finance), law enforcement, migration and immigration, and administration of justice and democratic processes. Article 7(2) gives more details on how to make these assessments. For such high-risk systems, they cannot be released to the public before undergoing a conformity assessment which determines whether all the needs of the AIA risk framework have been met.

Distorting human behavior, exploiting vulnerabilities of marginalized groups, social scoring, and real-time biometric identification in public spaces (except in certain circumstances like those mandated by national law, or for tracking terrorist activities, searching for missing persons, etc.) are prohibited use cases.

**Complying with the AIA**
Articles 9 through 15 of the AIA provide guidance on how to comply with the Act and include practices like maintaining a risk management system, data governance and management, transparency via constantly updated documentation of the high-risk AI system, logging and traceability through the AI system, appropriate human oversight, and balancing accuracy of the system with other desired properties like robustness and explainability of the system. Some of these requirements will sound familiar to those who had worked in compliance before and helped their organizations transition into the GDPR era. Others emerge from best practices in the MLOps domain as well. A combined policy and technical approach is the way forward to build AIA-compliant systems. This will help in meeting the post-market monitoring requirements as proposed in the AIA.

We can expect there to be some intense lobbying from different corporations and other organizations to tailor the AIA to align better with their needs. Standard-setting organizations will become more potent through economic, legal, and political levers, and we must account for the potential power imbalances that occur through this channel. Finally, through the Brussels effect, we will potentially see a more positive change in the attitude towards building more ethical, safe, and inclusive AI systems worldwide.

**Between the lines**
In line with the work done at the Montreal AI Ethics Institute in creating research summaries, such policy briefs provide a great avenue to catch up on pertinent issues without diving into all the details until needed. These are especially valuable for those impacted by policy and technical changes in the field but might lack the time and resources to parse through the

The State of AI Ethics Report, Volume 6 (January 2022)                                                         208

fast-moving field. The next step in making such pieces more actionable is to analyze case studies. In the case of the AI Act, it would be great to see how this impacts currently deployed high-risk AI systems, what that means for the process, and technical changes required to make these systems conform with the requirements to be allowed deployment in the field. Companies that are fast to act on these compliance requirements will surely gain a competitive market edge, essentially mimicking the changes during the transition to the GDPR era.

## Algorithmic accountability for the public sector

[Original paper by Ada Lovelace Institute, AI Now Institute and Open Government Partnership]

**Overview**: The developments in AI are never-ending, and so is the need for policy regulation. The report exposes what has been implemented, their successes and failings while also presenting the emergence of two pivotal factors in any policy context. These are the importance of context and public participation.

**Introduction**
With AI developing at warp speed, what is the current situation in the algorithmic space? Do we know what works in terms of regulation? Due to the lack of policy and data about algorithmic regulation in the Global South, the paper adopts a European and North American focus. Nevertheless, this report aims to understand the success of algorithmic accountability policies from different actors' perspectives. While exposing what has been attempted (alongside its successes and failures), two crucial factors emerged: public participation and context. The latter is where we are going to begin.

**The importance of context when implementing policy**
The literature review conducted in the report showed that people understand algorithmic accountability but not so much about implementing it. Nevertheless, one key element in realizing policy is the context in which it is deployed. The Canadian ADM directive requires any custom source code owned by the Government to be made public. Yet, the New Zealand Aotearoa NZ Algorithm Charter asks how the data was collected and stored to be made available.

With this in mind, the effectiveness of the same policy can be drastically different in two different contexts. Hence, what has been implemented and what are the general problems with these approaches?

**What has been attempted, and what are their faults?**
In this section, I will list a broad overview of the policy methods carried out by different actors in the report and their associated problems.

**High-level ethical policies**: provide a helpful frame of reference to approach algorithmic issues.

**Problem**: doesn't provide any form of obligation to specific actions.

**Prohibitions and moratoria**: prevent harmful technologies from being used entirely, or gives regulators time to catch up to their development. have also been attempted to be implemented.

**Problem**: they rest on the assumptions that either the technology should never be used and that the policy and regulation efforts will be adequate in a couple of years.

**Impact assessments**: aim to expose how the agents have subjectively defined what harms, and risks are.

**Problem**: they struggle to provide clear avenues for public participation.

**Audits**: standardize and scrutinize the efforts being made to generate an environment of algorithmic accountability.

**Problem**: the company must provide adequate data to be audited, and the performance during auditing is the same as afterwards.

**Oversight bodies**: possibility of influencing the behavior of prominent actors.

**Problem**: the influence may only be minute.

**Appeals to human intervention**: involving humans in the process to better ensure fairness and establish some form of responsibility.

**Problem**: assumes that having a human in the process does help to ensure fairness and doesn't acknowledge how algorithmic data can influence human decision making.

**The role of the public**

Given the last point of human intervention, the role of public intervention should not be underestimated. The intervention helps to match better governmental actions with the needs of the people.

What's still noteworthy is how different people have varying resources that allow them to get involved. Here, access to the media can help level this playing field.

**The role of the media**
Legal frameworks don't just rely on the law to be effective, but also on other factors such as "political will and cultural norms". Pressure from media outlets can help to reinforce the need to implement and maintain policies beyond just their legally binding status. Such intervention can make the policies 'societally binding', fixing the need for communication between policymakers and the public.

**Between the lines**
For me, the key findings are the importance of the public and the context within policymaking. No longer can a 'one size fits all' attitude be adopted in the algorithmic space, bringing in the need for an appropriate scope. Regulating individual actors too closely can ignore the systemic and social pressures present. Adopting too broad a viewpoint can then generalize important peculiarities that need attention in different contexts. What's for sure, in my eyes, is that while policy aims to serve the public, it must first learn from the public.

## Summoning a New Artificial Intelligence Patent Model: In the Age of Pandemic

[Original paper by Shlomit Yanisky-Ravid and Regina Jin]
[Research Summary by Avantika Bhandari]

**Overview**: The article analyzes the challenges posed by the current patent law regime when applied in the context of Artificial Intelligence (AI) in general and especially at the time of covid pandemic. The article also proposes creative solutions to the hurdles of patenting AI technology by establishing a new patent track model for AI inventions.

**Introduction**
Covid-19 has created a worldwide pandemic, causing millions of deaths within months. Lack of vaccines or the FDA approved drugs in the initial days have all aggravated the global health crisis. On the forefront against Covid-19, AI technology is proving to be an effective and powerful tool in developing new drugs, vaccines, and diagnostic methods. Also, AI mechanisms are built for the purposes of tracking and forecasting the outbreaks, processing health claims, managing drones to deliver supplies, and identifying high-risk individuals. For instance, the Korean biotech company Seegene utilized an AI system to create 'a novel coronavirus testing method—an unprecedented short period of time as it usually takes several months with a large group of scientists to develop such testing protocol.' Alibaba developed an AI-based platform to detect complications in CT scans of patients' chests with 96% accuracy. Furthermore, the AI system created by the Canadian startup Bluedot successfully predicted the virus outbreak even before the World Health Organization (WHO) officially declared the discovery of a novel coronavirus. These examples bring to light many crucial features of AI tools- they are efficient, creative, accurate, evolving, and rapid. Acknowledging its benefits the White House has urged researchers to employ AI to analyze tens and thousands of documents to decipher the origins of coronavirus.

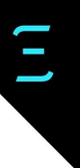

**Key Insights**
The researchers of the article argue that AI-made inventions must be patentable. However as per the patent law only 'human inventors' are eligible for patent ownership, therefore, a new model is needed. When talking about AI inventions there are generally two (2) types of innovative AI applications. First, when AI inventions are creative AI systems themselves, referred to as 'creativity machines' that are capable of generating new inventions themselves. Second, when AI inventions are AI-made inventions, in other words, the resulting inventions generated by the AI systems. The AI-made inventions have posed challenges for the current patent law regime, which was drafted in an era when AI technology was absent. Realizing the absence of AI-made inventions in patent laws, the United States Patent and Trademark Office (USPTO) published a Request for Comments on Patenting AI Inventions on the Federal Register in August 2019. However, USPTO has not issued any guidelines regarding patent rights in respect of AI inventions.

The researchers suggest a complete novel model to incorporate AI-made inventions. They argue that the current patent model is inapplicable and propose a new legal paradigm for examining AI inventions. They assert that a revolution is needed to establish a 'distinct AI patent track model separating from the current patent regime applied to human-made inventions.' A new track model is crucial as many factors of the current patent law regime are inapplicable in the AI context and amending would not address all the existing concerns. The researchers believe that the new AI patent track model would provide a distinct scope of protection for creative AI systems and AI-made inventions- all of which might not be patentable under the current patent regime.

AI's creativity can be found in the pharmaceutical industry, where AI tools are employed in the process of drug discovery to disease target identification. Many new drug targets based on RNA binding proteins were discovered by IBM Watson to cure a neurodegenerative disease. A drug design AI held by AstraZeneca, U.K. has formulated a large number of new drug structures catering to the chemical space that the human may not have thought of. These examples prove that AI systems not only facilitate creativity but also help in speeding the drug-discovery process in an accurate and efficient manner.

**Protection of Creative AI Systems and AI-Made Inventions**
The researchers argue that such AI-made inventions should be patentable to incentivize innovation and to reward the labor. Allowing patenting of algorithms, a part of AI creative systems, would incentivize the research on fundamental AI building blocks. This will help boost the advancement of AI technology and encourage technological development in the fields like medical, engineering, and science. The researchers also maintain that allowing the patenting of AI-trained models would incentivize data scientists and trainers to generate new resourceful AI models in an attempt to solve practical problems. AI-trained models are capable of finding answers by learning from the data and target attributes. DeepMind, for instance, is a trained model that learns how to solve 'problems and advances discovery in various fields such as science, medicine, and energy.' The patenting of AI-made inventions would boost efficiency in research and development, leading to more innovation in useful products and processes. The investors are motivated by economic returns through licensing and sales from the exclusive patent rights in AI-made inventions.

The researchers in this article propose a new patent track model 'to incorporate a wider scope of patent protection for AI inventions.' They recommend the following rules for the new track models:

- **Change of person of ordinary skill in the art (POSITA) standards**: The POSITA standard may not be applicable in the obviousness assessment for AI inventions under both the motivation test and the 'obvious to try' analysis. This is because the invention is intended to address the intricate problems in a seemingly unforeseen way. To settle the implication of the 'obviousness requirement' in respect to AI inventions, the researchers propose the POSITA standard of 'a skilled person using an ordinary AI tool in the art.' This would help a professional understand the complexity of the AI algorithm, the versatility of the AI system, and the complication of the problem in a patent application.
- **Expedited Patent Examination**: The time taken to acquire a patent is crucial in the COVID-19 urgency. Given the constraints of the patent system the long waiting period for examination may discourage organizations from investing in researching a cure for the virus. By the time a drug is granted a patent, a pharmaceutical company may have already missed the peak in demand for the drug and may not be able to reap the highest rewards. This is why expedited patent examination for AI inventions is needed.

- **Use of AI for Patent Examination**: The researchers advise the use of AI tools for patent examination to review the difficult algorithms and huge amounts of data which may be overwhelming for humans to handle, as the AI tools would boost efficiency and can accelerate the patent examination process. The USPTO uses the AI system Unity to increase the efficiency of patent examination. However, the application of Unity seems limited to searching patents, publications, and images, rather than examining patents.
- **Shortened Patent Lifetime**: "In the AI industry, the invention process as well as product life cycles can sometimes be extremely short." A patent is granted for 20 years however, researchers believe in shortening the patent lifetime for AI patents as this would allow the technology to come to the public domain faster for the benefit of knowledge dissemination.

**Between the lines**
The article rightly evaluates the need to establish a new AI patent track as the current patent law regime poses substantial hurdles and uncertainties for patenting AI-made inventions. The new track addresses many ambiguous elements of the patent law to be more in sync with the 3A era digital tools, such as the 'person skilled in the art' standard, the examination of timing and method, and the patent lifetime.

## NATO Artificial Intelligence Strategy

[[Original document](#) by NATO]
[Research Summary by Angshuman Kaushik]

**Overview**: On October 21-22, 2021 during the NATO Defence Ministers Meeting, held in Brussels, the ministers agreed upon to adopt the NATO Artificial Intelligence Strategy ("hereinafter the strategy"). The strategy is not publicly available and what is accessible is a document titled 'Summary of the NATO Artificial Intelligence Strategy'. This write-up provides an overview of the said summary.

**Introduction**
"We see authoritarian regimes racing to develop new technologies, from artificial intelligence to autonomous systems," NATO Secretary General Jens Stoltenberg said during a media conference at NATO headquarters in Brussels on October 20, 2021, a day prior to the aforesaid Defence Ministers Meeting. No prizes for guessing as to who he was referring to by the use of the phrase 'authoritarian regimes'. Although, putting out a strategy on AI is a step in the right direction, how far the same would be implemented in practice is the sixty-four thousand dollar question. Nonetheless, the fourfold aim of the strategy is as follows:

- to provide a foundation for NATO and Allies to lead by example and encourage the development and use of AI in a responsible manner for Allied defence and security purposes;
- to accelerate and mainstream AI adoption in capability development and delivery, enhancing interoperability within the Alliance, including through proposals for AI Use Cases, new structures, and new programmes;
- to protect and monitor our AI technologies and ability to innovate, addressing security policy considerations such as the operationalisation of our Principles of Responsible Use; and
- to identify and safeguard against the threats from malicious use of AI by state and non-state actors.

**The Strategy**
The strategy talks about AI changing the global defense and security environment and offering an unprecedented opportunity to strengthen NATO's technological edge and at the same time, escalate the speed of the threats it faces. It further mentions that AI will likely affect the full spectrum of activities undertaken by the Alliance in support of its three core tasks; collective defense, crisis management, and cooperative security. In the future, the NATO Alliance aims to integrate AI in an interoperable way to support its three core tasks. The strategy recognizes the leading role played by the private sector and the academia in the development of AI and envisages significant cooperation between NATO, the private sector and academia; a capable workforce of NATO technical and policy-based AI talent; a robust, relevant, secure data infrastructure; and appropriate cyber defenses. According to the footnote in the strategy, 'private sector' includes Big Tech, start-ups, entrepreneurs and SMEs as well as risk capital (such as venture and private equity funds). It is obvious that the AI revolution is being spearheaded by the private sector and the academia and NATO plans attracting the best talent to join its workforce. Under the forthcoming Defence Innovation Accelerator for the North Atlantic (DIANA), NATO aims to support its AI ambition through the national AI test centers and also intends to conduct regular high-level dialogues, engaging technology companies at a strategic political level. At the forefront of the strategy lie the NATO Principles of Responsible Use for AI in Defence, which are based on existing and widely accepted ethical, legal, and policy commitments.

**NATO Principles of Responsible Use of AI in Defense**

NATO and the Allies commit to ensuring that the AI applications they develop and consider for deployment will be at the various stages of their life cycles, in accordance with the following six principles:

A. **Lawfulness**: AI applications will be developed and used in accordance with national and international law, including international humanitarian law and human rights law, as applicable;

B. **Responsibility and Accountability**: AI applications will be developed and used with appropriate levels of judgment and care; clear human responsibility shall apply in order to ensure accountability;

C. **Explainability and Traceability**: AI applications will be appropriately understandable and transparent, including through the use of review methodologies, sources, and procedures. This includes verification, assessment and validation mechanisms at either a NATO and/or national level;

D. **Reliability**: AI applications will have explicit, well-defined use cases. The safety, security, and robustness of such capabilities will be subject to testing and assurance within those use cases across their entire life cycle, including through established NATO and/or national certification procedures;

E. **Governability**: AI applications will be developed and used according to their intended functions and will allow for: appropriate human-machine interaction; the ability to detect and avoid unintended consequences; and the ability to take steps, such as disengagement or deactivation of systems, when such systems demonstrate unintended behavior; and

F. **Bias Mitigation**: Proactive steps will be taken to minimize any unintended bias in the development and use of AI applications and in data sets.

The commitment to abide by the principles at the various stages of a lifestyle of AI systems is a substantial one, and only time will tell as to the operationalization of the same. Moreover, terms like 'appropriate levels', 'judgment and care', and 'appropriately understandable' etc. need exposition. Further, the strategy also talks about NATO operationalizing its Principles of Responsible Use to ensure the safe and responsible use of AI. It lays emphasis on consciously putting bias mitigation efforts into practice, which will seek to minimize biases such as gender, ethnicity or personal attributes. There is a further commitment to conduct appropriate risk and/or impact assessments prior to deploying AI capabilities. The strategy also takes note of the fact that some state and non-state actors will likely seek to exploit defects or limitations within NATO's AI technologies. Hence, it must strive to protect the AI systems from such interference, manipulation, or sabotage, in line with the 'Reliability Principle of Responsible Use'. Adequate security certification requirements, such as specific threat analysis frameworks and tailored security audits for purposes of 'stress-testing', also find mention in the strategy. The strategy also refers to AI's impact on critical infrastructure, capabilities and civil preparedness, including those covered by NATO's seven resilience Baseline Requirements, creating potential vulnerabilities that could be exploited by certain state and non-state actors. Issues such as disinformation and public distrust of military use of AI by state and non-state actors are also stressed. The strategy envisions further working with relevant international AI standards setting bodies to help foster military-civil standards coherence with regards to AI standards.

**Between the lines**
Some of the key areas that need elucidation with respect to the aim of the strategy include firstly, the position of NATO with respect to the use of Lethal Autonomous Weapon Systems (LAWS) in a 'responsible manner'. In fact, the strategy does not even mention anything about LAWS. Secondly, the aspect of 'interoperability' needs further clarity with regard to its scope. Thirdly, elaboration on how security policy considerations come under the ambit of 'operationalisation of Principles of Responsible Use'. Fourthly, whether a NATO member state will fall within the meaning of a 'state actor' if it is involved in the malicious use of AI needs to be clarified? For instance, what happens in a scenario like Turkey's use of AI-controlled drones (read LAWS) in the Libyan skies in the recent past?

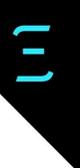

# Go Wide: Article Summaries (summarized by Abhishek Gupta)

## Judge Throws Out 2 Antitrust Cases Against Facebook

**[Original article by NYTimes]**

**What happened**: In a monumental setback to efforts to rein in BigTech, a judge dismissed antitrust cases against Facebook because more evidence was needed and that the regulators filed their lawsuits late, given that the acquisitions that they refer to happened 6 and 8 years ago (Whatsapp and Instagram respectively). While the regulators have 30 days to file again, they face a stiff challenge as the courts have narrowed their interpretation of antitrust law over the last few years. The courts also took the position that if a monopoly emerged from Facebook's acquisitions, then they should have acted years ago rather than now.

**Why it matters**: As principles for technology use promulgate, this is a reminder that what is enshrined in regulations and law is ultimately what holds a significant amount of sway on whether we can generate the socially friendly outcomes that we desire. The call from senators and lawmakers arguing for broadening the scope of antitrust regulations is a step in the right direction, especially as they are applied to Internet companies who might not have the same hallmarks of traditional monopolies, for example, pricing where a lot of these services are offered for free to the users.

**Between the lines**: The call for breaking out Instagram and Whatsapp from Facebook addresses only a tiny part of the more significant problem. Such monopolies are bound to arise again and again due to the network effects and structure of social media networks today. This will help to stem the tide with the current crop of companies but it does little to change what will inevitably happen again in the future. A more systematic overhaul of the regulatory ecosystem is perhaps what is needed.

## To regulate AI, try playing in a sandbox

**[Original article by Morning Brew]**

**What happened**: Sandboxes have been proposed as a part of EU regulations on getting AI systems to comply with requirements from the GDPR among others. The article details some of the engagements with companies that Norway has embarked on to figure out these challenges. In particular, complying with requirements like privacy-by-design and reporting on that compliance in an understandable manner requires cooperation between legal and technology stakeholders. The initiative in Norway is helping to facilitate that. Similar efforts are also underway in the US with Broussard from NYU and O'Neill from ORCAA attempting to create sandboxes that can help unearth concrete practices that will help address regulatory needs.

**Why it matters**: This movement towards trying to figure out tangible solutions through trial-and-error is a welcome change compared to the incessant merry-go-round that we have today talking about problems and regulations in the abstract with untested solutions being put forward without empirical evidence to back up how they might work and in which contexts.

**Between the lines**: It will be interesting to see the lessons learned both from the Norway experiment and some of the ones being run in the US. The regulatory ecosystem is vastly different between the two regions and I foresee that the approaches that emerge from sandbox efforts in both places will be quite different. But, there should be some common threads from both experiments that will help practitioners put these regulatory requirements into practice rather than just running in circles trying to make their systems compliant.

## What Is Congress's Plan to Crack Down on Big Tech?

[Original article by The Markup]

**What happened**: Six bills are being introduced in the US Congress that come from a 16-month House Judiciary Committee investigation into the antitrust behaviour of tech giants. Two of the proposed bills have very little controversy and are expected to pass without much furore: one that increases merger filing fees and another that limits the moving around of antitrust cases from one state to another, something that has been misused in the past by tech companies to obtain more favorable jurisdictions and judiciaries along with a delay in the process and increase in cost for the case. The other 4 bills are expected to raise quite a bit of fuss since they target antitrust and anti-competitive behaviour of tech giants including things like more stringent limits on the favoritism of companies to feature their own products and services on their platforms, limits on using insights from competitor behaviour on their platform to develop and promote their own offerings, limits on mergers and acquisitions that reduce market competition and a push for increasing interoperability and data portability between different services in the market, thus increasing consumer choice.

**Why it matters**: Each of these bills present solid cases for what can be achieved through the legislative process in reining in Big Tech and making sure that consumer welfare is kept top of mind in a world where monopolies abound and unethical behaviour is hard to control, especially when there are strong network effects and platform lock-in for almost all of the products and services that we use.

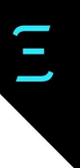

**Between the lines**: While there is bipartisan support for the bills, what will be interesting to see is how successful these bills end up being in creating an ecosystem where ethical and competitive practices become the norm rather than the exception. Adequate enforcement mechanisms also need to be envisioned if these are going to become a success.

## Analyzing the Legal Implications of GitHub Copilot

[Original article by FOSSA]

**What happened**: With the recent release of the AI-powered pair programming / code-completion tool from GitHub, many have raised concerns about whether the outputs from the system have copyright infringements. This article provides some nuance stating that despite the licensing on specific repositories, under the terms of service of hosting code on GitHub, there wouldn't be a strict violation per se. Furthermore, given the length of code outputs from the system, for smaller snippets, there might not be copyrightable material, as the interviewee describes them as Lego blocks that are common everywhere in the programming ecosystem. Finally, from a legal standpoint, there are arguments to be made similar to how Google Books used copyrighted book material as a part of its service allowing people to search books. This was acceptable because it was a "transformative" use and created new value that was different from the original text in the books themselves.

**Why it matters**: As tools like this become more common and more powerful, especially being able to produce longer segments of working and coherent code, the legal implications of such code generation will become more relevant. Precedents like Google's use of copyrighted book material serves as a very loose analogue and we'll need more scholarship and legal precedents before we are able to better understand the implications of tools like Copilot.

**Between the lines**: A lot of the incensed discussions on Twitter and elsewhere have focused on the surface-level argument that they feel a lot of the longer code snippets are just reproductions of code snippets from the training corpus. The paper published by OpenAI explains that the probability of that happening is approximately 0.1%. More so, a lot of the longer code snippets that have been generated are what is called boilerplate code (code that isn't a direct copy-paste but it requires little cognitive effort) meaning that there is a diminished risk of copyright infringement since such boilerplate code is made freely available on tutorial pages for packages and libraries. We need a lot more nuance in the discussion before we are able to say anything definitively about the legal and other implications of tools like Copilot.

## Should Doxing Be Illegal?

[Original article by The Markup]

**What happened**: Doc-dropping, shortened to doxxing, is the process of releasing private information about an individual such as their address, phone number, identifiers, and other information with the intent of targeting the internet mob to harass the individual through threats and unwanted contact. The article documents the case of an individual in Montana who suffered tremendously at the hands of such an attack and successfully sued the individual that instigated this attack, though she is yet to receive the court-awarded compensation.

**Why it matters**: As outlined in the article, the law implemented in several states in the US makes doxxing a civil offense in some, criminal in others. In some cases, the law is also geared towards protecting specific kinds of people from doxxing attacks like reproductive healthcare workers, police officers, etc. Each of the approaches come with their own pros and cons, in the case of civil offenses, the burden of proof remains lighter making it perhaps easier to obtain justice but criminal offenses carry a higher punitive burden offering a stronger deterrent.

**Between the lines**: In the case of the person mentioned in the article, she believes that such laws would have stemmed the hateful outpour against her by making it clear that perpetrators cannot hide behind a screen and keyboard. These virtual attacks have very real consequences for the victims and stronger legislation that offers protections against such behavior to all citizens has the potential to make our interactions in the virtual world much safer.

## We need concrete protections from artificial intelligence threatening human rights

[Original article by The Conversation]

**What happened**: The article makes a succinct case for how human rights based approaches might be better in getting more robust adoption of responsible AI rather than relying on ethics principles alone. First, it argues that since ethics are grounded in values, there is an indirect path to their enforcement. Second, since ethics depend on values and values can differ significantly, the enforcement becomes even harder as there is a lack of consensus and unified framework. Finally, given that human rights have precedents established in law already and there is some degree of universal agreement on them, designing more just AI systems might benefit more from following this path instead.

**Why it matters**: The friction between wanting responsible AI systems and actually having them in practice has been an ongoing theme. What we lack currently is the absence of concrete enough regulation that can enforce all these ideas of privacy-by-design, ethics-by-design, and many other X-by-design framings that are present in the domain of AI ethics.

**Between the lines**:  A human-rights based approach has been proposed many times before but where it runs into trouble as well is lacking the connection with more concrete practices which can translate those ideas into better designed technologies. Whatever approach we choose to take, we need to make sure that practitioners are consulted and made an integral part of the process, otherwise the solutions proposed will fall flat when it comes time to implement them.

## Americans Need a Bill of Rights for an AI-Powered World

[Original article by [Wired](#)]

**What happened**: From the Office of Science and Technology Policy in the US, this article makes a strong case for including in the Bill of Rights considerations for how technology, especially AI, impacts the ability of people to enjoy their freedoms and exercise their rights. They make the case that codifying that technology respects fundamental democratic values will help us adhere to the rights and freedoms that people are entitled to without leaving it up to market forces and private interests doing so out of goodwill. There are precedents when the Bill of Rights has been reinterpreted, reaffirmed, and expanded to keep up with the times as changes happened in society powered by technology and otherwise.

**Why it matters**: What is different with the current wave of technology, in particular AI, is the scale and pace of its impact. Hitherto it took a while before technology moved from labs to products, but that timeline has now been shortened down to a few months with integrated research labs within industry firms. The internet and smartphones with ample compute and storage become ready vectors for the dissemination of these technological advances; far more rapidly than ever before, not allowing us a chance to grasp their impact before they embed themselves into all facets of our lives.

**Between the lines**: It is great to see the leaders of government institutions at the highest levels taking a deep interest in how technology is shaping our society and seeking to make some fundamental changes to the operating system of our democracies so that we take a more active role in addressing the impacts that this technology has on us rather than succumbing to fatalism and treating the march of AI into all parts of our lives as an inevitability.

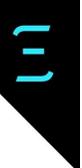

### Applying arms-control frameworks to autonomous weapons

[Original article by [Brookings](#)]

**What happened**: With the recent utilization of an autonomous weapon by Israel to assassinate the top nuclear scientist in Iran and last year the use of an automated drone by Turkey to target members of the Libyan National Army, discussions on autonomous weapons and their limitations and capabilities are gaining steam. The article makes the case that leaning on existing arms controls treaties as models can help regulate this field. Namely, it points out how the Ottawa Convention on Anti-personnel landmines provided a good starting point to bring together actors in the space for more fruitful discussions later. Building consensus and gaining momentum through targeted treaties that can help separate the concerns that militaries have in giving up on weapons vs. those that we want to absolutely stop the proliferation of will be a meaningful outcome from such an approach.

**Why it matters**: Autonomy in weapons systems can be something as simple as a sensor that is able to detect changes in the environment, some computing capability to act on changes signaled by the sensor, and then dispensing the payload of the weapon based on that computation. This spans the gamut from simple pressure-triggered landmines to the more sophisticated swarm drones that are being created by national militaries in their pursuit of dominance on the battlefield. The big concern raised by anyone participating in the domain comes down to how much autonomy and what meaningful human control looks like in these scenarios, and we don't yet have any concrete answers to these questions.

**Between the lines**: The problem with such approaches to regulation always come down to how strictly they can be enforced, and whether all countries who sign onto this will uphold the same high standards of robustness and verification that are required for safe operations. There are calls to completely ban such weapons but resistance emerges from some countries who claim that while they might halt such work, there are those who won't and their lackadaisical approach might cause more net harm. And this ultimately fuels the arms race where each pushes to develop the technology defensively but in the process furthers the state-of-the-art. Hopefully, those efforts while still being pursued are aimed towards making these systems safer rather than more lethal and unethical.

### Europe wants to champion human rights. So why doesn't it police biased AI in recruiting?

**[Original article by Sifted]**

**What happened**: Making the case for how Europe is in dire need of innovation and growth, something that diversity in hiring can enable, the author makes the case that at all levels of the regulations and legislations, the impact of biased hiring algorithms is being ignored, leaving job seekers at the mercy of systems that are highly problematic. For those unfamiliar with hiring practices in Europe, the CVs typically include pictures along with the name which can lead to implicit bias on race. The Digital Services Act in Europe is currently ill-equipped to handle this.

**Why it matters**: This goes against laws in several countries in Europe that prohibit the use of race in hiring decisions. Given that a lot of companies use automated systems to process incoming applications and fast-track the process that is time- and resource-intensive, illegality might be getting buried behind an opaque wall of black-box systems where it is difficult to point out what factors have been used to make a hiring decision.

**Between the lines**: Many examples have already demonstrated that hiring decisions made on the basis of algorithmic filtering tend to reproduce strong biases, especially along gender and race lines. This happens even when data related to these protected attributes is not collected and this manifests itself in the form of proxy variables that capture correlations between the protected attributes and non-protected attributes, negating the effectiveness of no data collection related to the protected attributes. Without stronger mandates in the form of law, firms may continue to exercise such biased systems severely impacting the livelihoods of people who become the subjects of algorithmic discrimination.

### Thousands of Geofence Warrants Appear to Be Missing from a California DOJ Transparency Database

**[Original article by The Markup]**

**What happened**: Investigation by the publication The Markup found discrepancies in the number of geofence warrants that were reported in the public database from the California DOJ and the number of requests that Google received for geofences. The article reports that such discrepancies might arise because of requests being revised during the warrant granting process, the lack of standards in filing, lack of entering this information in the database, sealed warrants where the information is not filed in the public database, and how the information is captured and reported when the requests are made by agencies that are outside of the state.

**Why it matters**: But, all of these pose significant challenges to those who might want to challenge unlawful warrants, especially civil society agencies that keep track of these from public databases. These challenges also defeat the efficacy of transparency requirements and laws like California Electronic Communications Privacy Act. Geofences by their very nature don't have a specific target individual, and hence can be quite invasive, especially scooping up data about a bunch of unrelated individuals who happen to be in the area that the geofence targets. This stands in contrast to wiretaps where the warrants are highly targeted.

**Between the lines**: This is a great demonstration of how even when we have laws and transparency requirements, the way they are enforced and the reporting standards can make or break whether we actually get the results that they set out to achieve. Standardization in reporting and more stringent requirements placed on the agencies seeking these geofence warrants can help alleviate some of the challenges identified in the article.

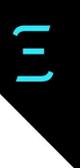

# 8. Trends

**Introduction** by Abhishek Gupta, Founder and Principal Researcher, Montreal AI Ethics Institute

Great to see that you've chosen to open up this chapter! It definitely is one of my favorites in this report because it is indicative of things that you can expect to see in the field based on current research. Before going ahead, I would caveat this by saying that in an accelerating field like AI, a crystal ball is nothing but an illusion. Change comes fast and it comes big. But, from observing the field since the publication of the last report, I can say that the ideas mentioned in this chapter and the sub-domains within which they are present will come to play somewhat of a significant role in 2022. If you're reading in the future, you can send us an email and let us know how we did!

In the first piece, "Machines as teammates: A research agenda on AI in team collaboration", we get an opportunity to explore what it would mean for us to operate within human-machine ensembles. My take has always been that AI is better as a tool for augmentation rather than replacement of human beings. It is complementary in many ways helping us bridge the gaps in our cognitive capabilities to ultimately achieve better results. But, this also means that we need to think about how we might need to adapt to a workspace where we have machines as constant collaborators. Building on this idea, in a world that will increasingly have AI systems all around us playing various roles, what skills will be most necessary in that space? In "Digital transformation and the renewal of social theory: Unpacking the new fraudulent myths and misplaced metaphors", we get a peek into some ways that we can cope with this.

In "AI Ethics: Enter the Dragon!", we look at how China has taken on a more serious interest in integrating ethics into the entire lifecycle of AI. In my view, this will certainly be interesting as we get to explore AI ethics from the lens of a different social system compared to the typical Western-centric formulations of AI ethics principles. The next piece on "Balancing Data Utility and Confidentiality in the 2020 US Census" covers a development that I had been watching with bated breath because it was one of the first mass implementations of the idea of differential privacy in practice. The piece details how the US Census Bureau rolled this out and what it could have done better to build public trust using the privacy-preserving concept of differential privacy. In an era where even small actors can get their hands on massive computational power, it is not surprising that we have state-level data collection authorities trying out newer mechanisms to uphold their mandates of protecting critical citizen data.

The rest of the chapter covers a plethora of interesting notions, notable amongst them how even experts are too quick to rely on outputs from AI systems, an exploration of deepfakes from an ethics standpoint and how in the future we might have more and more people "brought back from the dead" by creating deepfake likeness of their voice and visuals. The rise of foundation models and pre-trained models will pose even greater ethical challenges for the field as the provenance of the models and their data remains inscrutable at times to those using them downstream. It will also widen the chasm between those who have resources to create such systems and those who don't. Finally, a quick peek into the impacts that AI has on labor and how it will continue to impact labor, especially through "management by algorithm", we see how Alibaba tracks and dominates their delivery agents through the use of algorithmic monitoring and management. The final piece in this chapter paints a more optimistic note explaining how workers might still play a key role in warehouses and robots can only provide partial solutions (at least the way they operate currently).

I hope that you enjoy the eclectic mix of topics in this chapter and if you have other trends that you see coming down the pipe in 2022, please feel free to reach out to me at abhishek@montrealethics.ai and let's chat about them!

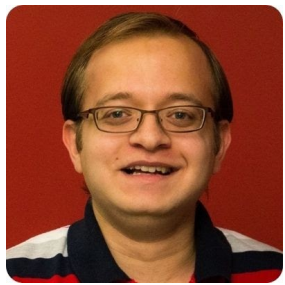

**Abhishek Gupta (@atg_abhishek)**
Founder, Director, & Principal Researcher
Montreal AI Ethics Institute

Abhishek Gupta is the Founder, Director, and Principal Researcher at the Montreal AI Ethics Institute. He is a Machine Learning Engineer at Microsoft, where he serves on the CSE Responsible AI Board. He also serves as the Chair of the Standards Working Group at the Green Software Foundation.

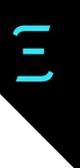

# Go Deep: Research Summaries

## Machines as teammates: A research agenda on AI in team collaboration

[Original paper by Isabella Seebera , Eva Bittnerb , Robert O. Briggsc , Triparna de Vreeded , Gert-Jan de Vreeded, *, Aaron Elkinsc , Ronald Maiera , Alexander B. Merza , Sarah Oeste-Reiße , Nils Randrupf , Gerhard Schwabeg , Matthias Söllner]
[Research Summary by Connor Wright]

**Overview**: The importance of collaboration in the AI space is not only between humans but also between humanity and AI. Imagining working with an AI teammate may no longer be imaginary in the future, and understanding how this will affect collaboration will be essential. For, understanding this will highlight the importance of the human cog to the human-AI machine.

**Introduction**
Have you ever imagined consulting an AI co-worker? What would you like them to be like? The implications of AI as a teammate are considered within this piece, stretching from how they look to how they could upset human team dynamics. While we must consider the benefits of this collaboration, the human element to the process must remain, especially in terms of human development itself.

**A different kind of teammate**
Whether the AI is in a physical robot or an algorithm, it cannot be compared to a regular human teammate. One key difference is its ability to assess millions of different alternatives and situations at a time, proving impossible for humans. While useful, the form in which the communication of this assessment arrives would need to be determined. It could be in speech or text, with or without facial expressions for visual feedback. Questions like these lead us to question what we prefer in an AI teammate over a human.

**What do we want in an AI team member?**
The paper holds the classic Alan Turing definition that "AI refers to the capability of a machine or computer to imitate intelligent human behavior or thought". In this sense, should our thinking about AI collaborators be centered in human terms? Like with chatbots, similar considerations are brought into play, such as whether the AI should have a gender, can it differentiate between serious and social chatter etc. Our decisions on these questions will then certainly affect how the team dynamic plays out.

**The effect on collaboration**

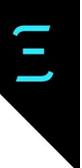

In this regard, it's essential to differentiate between AI as a teammate and AI as an assistant. Collaboration with AI as a tool is not as thought-provoking as holding it in the 'higher' regard as a counterpart.

In this way, collaborating with such an entity could enhance or negatively impact the team dynamic. The AI could become a leader in the group on specific issues it can handle best, yet depending too much on the machine could lose human competencies. Furthermore, the AI teammate could prove excellent at drawing insights from data, but the lack of out-the-box thinking could reinforce already present views. Hence, while collaboration is undoubtedly affected by introducing AI, the right balance still needs to be struck to make the most of it.

**Considerations when collaborating**

Given the novelty of this practice and AI in general, why an AI would suggest a particular course of action becomes a critical question. In addition, the extent to which we recognise the AI's involvement can also have far-reaching impacts. Should the AI become a leader on a topic, should it be credited with its work? Much of this stems from whether AI can be creative or not, which can be found in poetry, fashion and music.

**Between the lines**

While collaboration with AI teammates may be essential practice in the future, I would be cautious against throwing such collaboration into every problem possible. Sure, using AI's analytical capabilities will nearly always be helpful, but that pertains more to AI as an assistant rather than a counterpart. Hence, problems such as trying to solve world hunger, I believe, would not benefit from an AI as a teammate intervention, mainly due to how AI can never actually feel or understand what being hungry feels like. What's for sure is that while AI collaboration can reap benefits, human involvement remains paramount.

## Digital transformation and the renewal of social theory: Unpacking the new fraudulent myths and misplaced metaphors

[Original paper by Marinus Ossewaarde]
[Research Summary by Connor Wright]

**Overview**: With the emergence of technology, society has changed immeasurably. Questioning the status quo has become less of a pressing issue in favor of continuing to use a digital service. However, reflection is one of the most critical skills in preventing a digital future guided and dominated by the few.

**Introduction**
The lives of many have become densely linked with technology. The digital transformation is being led and developed by certain parties (labeled as the "googlization" of everything). Hence, social theory must adapt to the dominant economic and digital spheres, promoted and sustained through different technological "myths". To do so, acknowledging the status quo and advocating the importance of questioning will prove essential in both understanding and combating digital domination by the few. Up first is acknowledgement.

**Digital and physical life have become inseparable**
The reality we live in is becoming more and more recognised as inextricably linked with the digital space. The influence technology possesses often goes unnoticed until it is briefly taken away. Such influence is so strong that users readily accept the missions of businesses to continue using their services (especially when accepting the role that technology plays in our lives has led to the subsequent domination of digitalisation.

**The domination of digitalisation**
The information available through tech has ended up in its economisation. Less thinking and more simply accepting is the easiest way to drive profit, reducing the value of mental activities such as reflection. Whereas previously, society has been driven by the political and religious spheres of life, the economic sphere has overtaken them thanks to digital transformation. Through this, the few driving the transformation can dictate the game's rules as to what this digital transformation will look like. The changes undergone do not bode well for the academic sphere.

**The consequences for intellectual practices**
With digitalisation as the driver behind the dominant economic sphere, academic work becomes valued when it adopts the norms and language of the prevailing economic sphere. Such dominance then makes it difficult to imagine alternative scenarios to the reality in which we find ourselves. Instead of being encouraged to think, the mind is being used as an instrument of power rather than being critical. The ability of art and science to influence the mind gets weaker as this conditioning goes on while increasing the passive acceptance of the status quo. It is through this lack of questioning that "fake" myths can be developed.

**"Genuine" vs "fake" myths**
In contrast to a "fake" myth, a "genuine" myth leads to enlightenment through insights and a deeper understanding of the current state of affairs. The authors use Homer's myths in the Odyssey as some examples. On the other hand, a "fake" myth doesn't lead to enlightenment and is instead manufactured to reinforce the status quo. The fabrication is designed to blindfold the public so that they adhere to the status quo without question. One such example presented by the authors is Silicon Valley being lauded as the digital revolution heroes.

**The digital myth**
A reason Silicon Valley is seen in this way owes to how digitalisation is seen as bringing new enlightenment to the fore, making everyone more ready to accept whatever form it comes in. Promises of a new technological reality are made to condition how the public sees technology casting aside other potential realities to preserve the status quo. How this is done can be seen in the use of metaphors.

**The use of metaphors**
The digital reality desired by those at the helm of the technological reality is argued to utilize language to maintain the current state of affairs. Metaphors such as "data mining" and "the cloud" are employed but are inappropriate as they make data sound like a natural resource. Even metaphors such as "digital community" distracts from how communities are built on face-to-face interactions. Hence, again, the deep interaction between digital and physical reality comes through, subtly adjusting how we express ourselves and view technology itself.

**Between the lines**
I find how our use of language is also influenced by the technology we use very intriguing indeed. Similar occurrences can be seen in the anthropomorphic language used to describe AI at times, especially with self-driving cars being described as 'making decisions'. I also see how big corporations spin their own take on reality often undetected. As a result, the importance of building up civic competence shines even brighter. Should we choose to stop asking questions, those who dictate the space will stop giving answers and there remain many unanswered questions yet.

## AI Ethics: Enter the Dragon!

[Original document by Ministry of Science and Technology of the People's Republic of China]
[Research Summary by Angshuman Kaushik]

**Overview**: On September 25, 2021, the National New Generation Artificial Intelligence Governance Professional Committee issued the "New Generation of Artificial Intelligence Code of Ethics" (hereinafter "the Code"). According to the Code, its aim is to "integrate ethics into the entire life cycle of artificial intelligence, and to engage in artificial intelligence related activities".

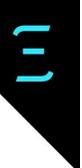

**Introduction**
It's quite mystifying to see a country as infamous as China globally for its AI ethics violations, come up with an Ethics Code for the world to sit up and take notice. Its violations list is endless, ranging from the use of Uighur-tracking facial recognition technology and the use of emotion detection software against them in its Xinjiang province, to its flouting of human rights norms and draconian manner of application of the social credit system. In fact, China's reputation as a country with an appalling human rights track record has gone from bad to worse in the past couple of years or so. To come up with an ethics code in such a setting and at this point in time, is quite surprising, to say the least. Travel back in time to 2017, and you have the "New Generation Artificial Intelligence Development Plan" which outlines China's policy to become the leading AI power by 2030. It is interesting to mention here that, according to the object part of the Code, a couple of its objectives includes, to thoroughly implement the "New Generation Artificial Intelligence Development Plan", and detailed implementation of the "New Generation Artificial Intelligence Governance Principles". Coming back to the Code, it contains 25 Articles divided into 6 chapters.

**High-level overview**
Below is a high-level overview of the Code:

**Chapter One ("One" as given in the Code) (General Provisions) (Articles 1- 4)**
This chapter talks about integrating ethics and morals into the full life cycle of AI, promoting fairness and avoiding problems such as discrimination, privacy etc. What is interesting to note here is that the chapter not only talks about incorporating ethics, but also morals. Therefore, clarity on the definition of morals for the purpose of this Code and how the integration will take place becomes imperative. Further, the chapter states that, apart from applying to natural and legal persons, the Code, also applies to 'other related institutions' engaged in related activities such as artificial intelligence management, research and development, supply, and use. There is ambiguity surrounding the meaning of the term 'other related institutions', and without any elucidation, the same can have disastrous consequences to the entities concerned in today's globalized world. Article 3 is one of the most important provisions of the Code, as it lays down some basic ethical norms to be followed by various AI related bodies, under six distinct heads.

They are as follows:

- **Enhance human well-being** – This first heading lists out several high-sounding ethical guidelines to be followed. Some of them include, follow the common values of mankind, respect human rights and fundamental interests of mankind, improve people's livelihood etc. One interesting norm is to promote harmony and friendship between man and machine. Whatever that means, it would be some task for the people associated with the field of AI to accomplish. Another ethical norm mentioned is "adhere to people-oriented", which is extremely arduous to comprehend.

- **Promote fairness and justice** – It talks about adherence to inclusiveness and inclusiveness, which again does not convey any meaning, whatsoever. The other ethical norms clubbed under this broad heading effectively protect the legitimate rights and interests of all relevant subjects, promote social fairness and justice and equal opportunities.
- **Protect Privacy and safety** – This head includes inter alia, fully respect the rights of personal information to know and consent, protect personal privacy and data security, information must not infringe on personal privacy etc. The above sounds more like clauses from a data protection statute. Although, incorporation of obligations concerning privacy and data protection seems like another layer of fortification for the rights-holders, but how far it will stay clear of not involving in an interpretation imbroglio with the recently passed Personal Information Protection Law (PIPL) will be one riveting duel to watch out for.
- **Ensure controllability and credibility** – It comprises ensuring that humans have full autonomous decision-making power, the right to choose whether to accept the services provided by artificial intelligence, the right to withdraw from the interaction with artificial intelligence at any time, and the right to suspend the operation of artificial intelligence systems at any time to ensure that artificial intelligence is always under human control. It is quite obvious that the above requirements (which entails some explanation), would prove extremely burdensome for the companies to follow.
- **Strengthen Responsibility** – Insist that human beings are the ultimate responsible subject, clarify the responsibilities of stakeholders, introspect and self-discipline in all links of the artificial intelligence life cycle etc. are some of the ethical norms included under this head. Explications required include 'ultimate responsible subject', 'introspect and self-discipline' etc.
- **Improve ethical literacy** – Actively learn and popularize artificial intelligence ethics knowledge, deeply promote the practice of artificial intelligence ethical governance etc., are some of the ethical norms contained under this head.

**Other chapters**

Without going into the comprehensibility and the interpretability issues, the other chapters containing the various articles are as follows:

**Chapter II (Management Standards) (Articles 5- 9)**
The management standards are contained in this chapter. Some of them include "stay true to reality and rush for quick success in the process of strategic decision-making, correctly exercise power and use power", etc.As it is obvious, the above requirements are extremely hard to comprehend and therefore, cannot be given effect to, in a meaningful manner.

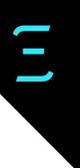

**Chapter III (R&D Specifications) (Articles 10- 13)**
Strengthen the awareness of self-discipline, improve data quality, strengthen security and transparency, and avoid prejudice and discrimination etc.

**Chapter IV (Supply Specifications) (Articles 14- 17)**
Respect market rules, strengthen quality control, protect the rights and interests of users, and strengthen emergency protection etc.

**Chapter 5 ("5" as given in the Code) (Specification) (Articles 18- 22)**
Promote good faith use, avoid misuse and abuse, forbid illegal use of artificial intelligence products and services etc.

**Chapter VI (Organization and Implementation) (Articles 23- 25)**
This chapter deals with the implementation aspect of the Code. It states that the specification is issued by the National New Generation Artificial Intelligence Governance Professional Committee, and is responsible for explaining and guiding its implementation. It further states that the management departments at all levels may formulate more specific ethical codes and related measures based on this code and combined with actual needs. Article 25 talks about coming into force of the specification on the date of promulgation, and its revision in due course according to the needs of economic and social development and the development of artificial intelligence.

**Between the lines**
Prima facie a proper drafting of the Code is conspicuous by its absence. In fact, it is very loosely drafted and seems not to have undergone any revision whatsoever, before publication. Further, it appears to have been passed in a hurry, the repercussions of which can be devastating. Apart from the syntactic and other grammatical gaffes, the Code brims with lofty ethical ideals, which are easy to prescribe but extremely difficult to implement in practice. Nevertheless, the burden now rests on the shoulders of the concerned authorities to provide more clarity, not only on the interpretation issues but also on the implementation aspects of the Code. Only time will tell as to whether the Chinese Government is able to deliver on the principles and standards enshrined in the Code. To sum up, China can draw inspiration from Robert Frost and his immortal lines, "And miles to go before I sleep", as far as implementing the Code is concerned.

**Balancing Data Utility and Confidentiality in the 2020 US Census**

[Original paper by danah boyd]
[Research Summary by Laird Gallaghar]

**Overview**: Due to advancements in computational power and the increased availability of commercial data, the traditional privacy protections used by the U.S. Census Bureau are no longer effective in preventing the mass reconstruction and reidentification of confidential data. In this paper, danah boyd explores the bureau's response for the 2020 Census: a new disclosure avoidance system called "differential privacy," which creates a mathematical trade-off between data utility and privacy. But the opaque manner in which the bureau has rolled out the changes has risked undermining trust between the bureau and the diverse stakeholders who use Census data in policymaking, research, and advocacy.

**Introduction**

Even before COVID-19 had taken hold in the United States, the 2020 Census was off to a rocky start, with a majority of U.S. adults mistakenly believing the form contained a citizenship question. Then, the pandemic upended the Census Bureau's normal operations and complicated efforts to ensure an accurate count.

However, barriers to enumeration are not the only challenges faced by the Census this year. Changes in the data and computing landscape over the past decade have made it much easier to reconstruct and reidentify confidential information out of Census data products. To respond to those threats, the Census has implemented an entirely new "disclosure avoidance system" (DAS). The system works by introducing noise–mathematical randomness–into the calculations used to generate data products.

But where and how much noise you inject matters. As danah boyd documents in Balancing Data Utility and Confidentiality in the 2020 US Census, the DAS requires a system-wide balance of privacy risk, which means that making certain statistical tables more accurate in turn requires others to include more noise. These trade-offs have widespread implications for the utility of data that stakeholders in government, academia, and the nonprofit and business sectors have come to rely on.

**How the Census constructs data products**

Since 1790, every ten years the U.S. government has conducted a census of all people living in the country. This decennial count determines the apportionment of legislative representation and the fair allocation of federal funding and resources. But the process also generates powerful data products used by policymakers, social science researchers, and others. In order to protect individual privacy, the Census does not release the full underlying data for 72 years. Instead, it releases aggregated and anonymized data products that ensure confidentiality while still providing valuable demographic information.

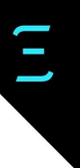

Through self-response and follow-up operations, the Census collects basic data about households: the type of housing unit; its ownership status; and the name, date of birth, sex, race, and Hispanic origin of everyone living there. After resolving addresses, this data becomes the "Census Unedited File" (CUF) which is used to calculate the population of each state and thus determine apportionment in the U.S. House of Representatives. Afterward, the Census resolves missing and conflicting demographic data, using statistical models to fill every cell with a value and produce the "Census Edited File" (CEF).

Then come the measures to avoid disclosure. Before this year, the Census would swap households from one location to another in an effort to scramble whether a record matches its real location. In addition, the Census would simply suppress certain information about subpopulations that would disclose too much detail. After recoding and quality assurance, these privacy-protected tabulations (the "Hundred-percent Detail File") would get released to the public as a series of data products. But swapping and suppressing is no longer enough.

**Why a new system to protect privacy?**
Due to increases in computing power, it is now much easier for attackers to rebuild individual records out of aggregate data. They do this by triangulating across statistical tables to determine which individuals likely contain which attributes, yielding a reconstructed list of individuals matched to attributes like race, sex, and Census block. From here, an attacker can then use external data sources, including widely-available commercial data, to link these anonymized yet reconstructed individual records and re-identify individuals by name and other characteristics.

boyd explains that while reconstruction, linkage, and reidentification attacks were once theoretical, they are no longer. "Using the published available statistical tables from only the 2010 decennial census, researchers at the bureau reconstructed a complete set of individual records that could effectively serve as a complete microdata file down to the block level," she writes. Due to swapping and other measures, the complete set did not fully match the unprotected, edited files–but fully 46 percent of individual records were perfect matches. And just by allowing the age variable to be +/- one year, fully 71 percent of individual records matched. From this reconstructed data, census researchers were able to re-identify (and confirm) 17 percent of individual records–tens of millions of U.S. residents.

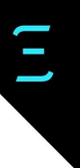

And that was with 2010 data. Since then, commercial data availability has increased nearly exponentially. It became clear to census researchers that far more than 17 percent of records could be exposed if they stuck to their standard practices of swapping and suppressing. They'd either have to release far less data products or implement a new system of privacy protection.

**Balancing confidentiality and accuracy**
For the 2020 census, the bureau has decided to implement a new "Disclosure Avoidance System" built on the principles of differential privacy. According to boyd:

*"Differential privacy works to prevent accurate reconstruction attacks while making certain that the data are still useful for statistical analyses. It does this by injecting a controlled amount of uncertainty in the form of mathematical randomness, also called noise, into the calculations that are used to produce data products. The range of noise can be shared publicly because an attacker cannot know exactly how much noise was introduced into any particular table. With differential privacy it is still possible to reconstruct a database, but the database that is reconstructed will include privacy-ensuring noise. In other words, the individual records become synthetic byproducts of the statistical system."*

The problem is that this system involves choices about where to introduce noise, and how much noise. In order to maintain a certain privacy-loss budget, designers must allocate noise levels throughout the data, prioritizing the accuracy of certain statistical tables over others. This is what makes the privacy differential. But as a consequence of how this top-down algorithmic approach works, it would create undesirable outcomes like geographic inconsistencies, partial people, and negative people, without additional processing. The need to perform a post-processing cleanup is primarily political, according to boyd. Laws around redistricting require the Census to prioritize making block-level data consistent and ensure the data consists only of non-negative integers (no negative or fractional counts of people). But this post-processing generates all sorts of statistical oddities entirely unrelated to privacy.

**Communication breakdown**
The Census' announcement of a new disclosure avoidance system in late 2018 caught many data users and advocates by surprise. The lack of education on how differential privacy works and why it is necessary left many stakeholders confused and frustrated. This new approach to protecting confidentiality required all data uses to be determined in advance so that the noise could be best allocated throughout the statistical tables, but most Census data users had never approached their work in this way. In addition, users didn't always understand why a new approach to privacy was even needed. And unlike the computer scientists who devised the disclosure avoidance system, they often lacked the skill set to analyze and comment on it. Data users became increasingly worried as they explored how the injection of noise could affect the reliability of their own scientific work.

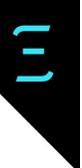

**Possible solutions**
How can the bureau best maximize data utility while minimizing privacy loss? boyd recommends several ways to relax the constraints on the data.

First is to reduce geographic precision. If the Census stopped publishing block-level data, more of the privacy budget could be spent elsewhere. Unfortunately, federal law dictates that the Census must produce redistricting files with block-level counts.

That fix is out of the question without an unlikely congressional intervention, so boyd suggests publishing "pre-post-processed data" so that users can get acclimated to negative counts, fractional people, and more. Doing so wouldn't jeopardize privacy.

In addition, the Census might also look to reduce the dimensions of certain variables and withhold publishing block-level data below a certain population threshold.

**Between the lines**
People like me–researchers for whom analyzing trends in Census data is a secondary aspect of our work–have by and large not even considered the effects of this sea-change in the Census approach to privacy protection. We didn't see the 2018 notice, didn't attend any meetings, and didn't look at the demonstration data. We haven't had the time. And now, we might not be able to use the Census like we did before. Luckily, differential privacy won't be applied to the American Community Survey until 2025, which buys us some time to understand this new reality. But the Census is in a challenging place. There is a major threat to public trust in Census data collection that requires these new privacy measures. If data collection suffers, the data products will suffer, too. But there's also a threat to the utility of the data, data which is important not just for advancing knowledge but also for public policy advocacy and more. Indeed, boyd is right in her premonition: "What's at stake is not simply the availability of the data; it is the legitimacy of the census."

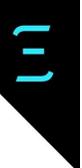

# Go Wide: Article Summaries (summarized by Abhishek Gupta)

## DeepMind AGI paper adds urgency to ethical AI

**[Original article by VentureBeat]**

**What happened**: Reinforcement learning (RL) is often much less discussed than other forms of machine learning. Usually, discussions around it in terms of its achievements in beating humans in games like Go. Adopting a different approach by creating a reward feedback loop with the environment, RL could be a pathway to achieving artificial general intelligence (AGI) with an accelerated timeline. The recent work from DeepMind has made some researchers revise their estimates of when AGI might become a reality, if at all.

**Why it matters**: The most frequently discussed ethical aspects in the context of RL include value alignment, reward hacking, safe exploration, and avoiding adverse side effects. In the current ecosystem of AI ethics research, these are severely under-discussed aspects, with most of the focus on issues like fairness and privacy.

**Between the lines**: Deployed ML systems will be a mixture of different approaches, and keeping an eye on developments like these and the implications they will have on ethics, safety, and inclusion is an integral part of working in the field. We need to broaden the scope of the discussion of concerns as they arise and relate to different ML methodologies so that our proposed approaches don't ignore essential facets of deployed ML systems.

## The Ethics of a Deepfake Anthony Bourdain Voice

**[Original article by The New Yorker]**

**What happened**: In the documentary titled "Roadrunner" about the life of Anthony Bourdain, there were segments of audio that were synthesized using previous audio data from his real voice. The words that were uttered in this synthetic voice were words he had actually written down. The use of synthetic media is rife with ethical troubles, as it became evident with the backlash that the producers of the documentary have faced since the release of their film. Notably, people have expressed concerns also in terms of disclosure that synthetic voice was used and the flippance with which the people involved in the making of the film dismissed some of the concerns when they were brought up the first time.

**Why it matters**: Synthetic media, notably deepfakes, is notorious for all the harm that it can cause. In some cases, there are positive uses for deepfakes as have outlined in a previous edition of this newsletter. But, when consent and intent is not clear, ethical qualms arise, especially for someone who passed recently and heavily emphasized authenticity as something he valued in his work.

**Between the lines**: We will see a rise in the use of synthetic media over time, especially as the technology becomes easier to use and the data available to train these systems becomes more widespread, as is the case with our increasing digital footprint. Building awareness around what constitutes appropriate and inappropriate use of synthetic media will make us more informed and nuanced in our discussions rather than lionizing or demonizing its use with a careful study of the underlying ideas of disclosure, consent, and context which are essential to discussing the ethics in the first place.

## Foundation models risk exacerbating ML's ethical challenges

[Original article by [VentureBeat](#)]

**What happened**: A massive report released recently from Stanford AI researchers titled "On the Opportunities and Risks of Foundation Models" has brought forth fervent discussion on the role that large-scale pretrained and other models are going to play in AI applications downstream that rely on them to build out their systems. An example of this is GPT-3 that now powers hundreds of apps processing billions of words every single day. Any bias in it gets amplified hundreds of times over in all its downstream uses. Such models also create risks of centralization of power in the hands of those who have the compute and data infrastructure to build such models.

**Why it matters**: Our penchant for larger AI systems has many impacts that exacerbate the problems in the domain of Responsible AI including bias and fairness, privacy, inclusion, accountability, and increasingly an environmental impact as well. Careful analysis needs to be performed and more research funded so that we can construct an in-depth understanding of the risks that such systems pose. The opportunities are quite clear in terms of being able to potentially democratize access to advanced AI capabilities and applying such advances to better humanity but as we've seen with most AI systems, there is always a cost that can have sinister consequences.

**Between the lines**: The newly formed Center for Research on Foundation Models at Stanford can become an example of encouraging cross-domain collaboration trying to answer fundamental questions about how we use AI and what the future holds. It would be interesting also to see how they choose to interact with groups like EleutherAI and HuggingFace which are more community-driven and are building such foundational models that will have an impact on our future.

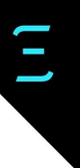

## Now That Machines Can Learn, Can They Unlearn?

**[Original article by [Wired](#)]**

**What happened**: The article covers the nascent area of "machine unlearning" which has the goal of effectively erasing personal information that is captured in parts by AI systems when they are trained and later on the user withdraws consent or wants to have their information erased. This is no easy task since millions of dollars might be spent in training up an AI system and asking to remove certain parts of the data from the training set means, at the moment, retraining the entire system and hence spending all that money again. This disincentivizes organizations from meeting these demands, especially when the financial burden is so high.

**Why it matters**: While there is a "right to be forgotten" in the EU, most of the current legislations focus on data erasure and consent withdrawal for data, but few talk about the need to also erase traces of the snippets of personal information that are incorporated in the learned representations in AI models. This will become a more essential consideration with more significant legislation coming up in the US and EU and will also be more meaningful as AI systems pervade more parts of our lives.

**Between the lines**: As pointed out in the article, the techniques of machine unlearning are still in the early days where their efficacy is quite limited. It's on the same journey as differential privacy where the technique is incredibly promising, tooling is being developed around it, and hopefully we will have more widespread utilization of the technique over time. What remains is for the efficacy to be proven along with it being practically viable, as we get more researchers and practitioners focussing on it, we will build up the tooling and related processes that will make this a more common practice.

## Even experts are too quick to rely on AI explanations, study finds

**[Original article by [VentureBeat](#)]**

**What happened**: The article covers a recently published research study that found discrepancies in the intention of features of AI systems as put together by designers and developers and those who use and interact with the systems in how they perceive them. Building on prior work from the domain of human-computer interaction (HCI), the researchers found that people both over-relied on the outputs from an AI system and misinterpreted what those outputs meant, even when they had knowledge about how AI systems work, when one might expect that to be the case only with those who don't know how such systems operate. The study evaluated these discrepancies through a game where a robot had to gather supplies for stranded humans in space and explain its actions as it navigated the terrain to get those supplies. Humans who were recruited to be a part of the experiment judged those robots more who provided numerical descriptions of their actions compared to those who provided natural language explanations.

**Why it matters**: This has direct implications for how we design explainability requirements, especially as those put forth by the EU, NIST, etc. in the sense that we need to know whether the perception of the provided explanations is the same as the ones that we intend. In particular, a mismatch between the two can lead to disastrous results and over- or underconfidence in situations where more human attention is warranted.

**Between the lines**: The results from the research study are not all that surprising. Perhaps the only novel element is that even those with a background in AI tended to fall for this trap and this only serves to underscore the problem more: we need to be more deliberate in how we design explanations for AI systems so that the gap between intended meaning and perceived meaning is minimized.

## Stopping Deepfake Voices

[Original article by USC Viterbi School of Engineering]

**What happened**: Researchers have discovered that voice assistants that use automatic speech recognition can be attacked using adversarial examples that can drop their performance accuracy from 94% to 0% in some cases. This work has also revealed strategies on how to add noise imperceptible to the human ear to surreptitiously attack such systems so that they behave in an unintended fashion aiding the malicious actor's goals. The paper also shares some research directions on defense strategies that can be used to protect against such attacks.

**Why it matters**: Given the rising proliferation of listening devices all around us, arguably waking up only on specific prompts, such vulnerabilities are important to analyze and defend against if they control important facets of our lives. Examples of this include things like the home's heating and cooling systems and security systems like door locks.

**Between the lines**: The field of adversarial machine learning and machine learning security are going to be foundational for the responsible deployment of AI technologies, and such research work helps to raise important questions and provide more research directions so that we can build more robust systems over time. While it is great to continue deploying AI systems and incorporating them into various products and services, without giving due consideration to how these systems might break down, we risk opening up new attack surfaces for malicious actors beyond the already vast vulnerabilities we face in the digital infrastructure that surrounds us.

## The Third Revolution in Warfare

[Original article by [The Atlantic](#)]

**What happened**: On the 20th anniversary of the 9/11 attacks, there has been a lot of reflection on warfare and terrorism. With AI pushing into all facets of our lives, it is natural to examine where we will end up with AI-enabled weapons systems. In this article, author and VC Kai-Fu Lee talks about some of the challenges, technical and moral, in the use of autonomous weapons systems. In particular, he highlights the clear moral dilemmas that arise when we don't have clear chains of accountability and a lack of transparency in terms of how the systems operate. He also points to potential solutions ranging from protocols of engagement to outright bans each of which have a different likelihood of success. There are some potential benefits in the use of AI in warfare, notably the potential to save lives and reduce collateral damage, but that comes at a cost.

**Why it matters**: The current state of the ecosystem is that we have an arms-race atmosphere where it appears that AI-enabled weapons are inevitable and countries are rushing to try out the technology to ensure that they don't get left behind. The article mentions the Harpy drone from Israel as an example. Of course, some hypothetical scenarios, like the Slaughterbots from a fictional short-film, point to a possible future where such capabilities are in the hands of malicious actors who don't need a lot of resources to execute fairly sophisticated and damaging attacks.

**Between the lines**: Ultimately, the biggest disruption that will arise from the use of AI is the degree of leverage it will create for non-state and small actors to utilize this technology, often using open-source designs and software, with cheap off-the-shelf hardware to assemble and deploy weapons that can wreak havoc, at least at a moderate scale, harming people and making it difficult to deter such attacks because of the nimbleness of such systems. At the moment, I don't believe that non-state and low-resourced actors will be able to use such systems to rival large militaries but it definitely gives them a leg up in small-scale combat due to lowering the costs and collateral that they have to put up to engage.

### Everyone will be able to clone their voice in the future

[Original article by The Verge]

**What happened**: The ability to clone voices has existed for some time now but the new crop of tools are faster, easier, and more realistic to boot. A simple web search yields results pointing to companies like Respeecher, Resemble.ai, Veritone, and Descript all of whom have product offerings that can create a clone of your voice that can be used for various purposes. One of the promising avenues advertised by firms like Veritone is that it allows creative talent, like influencers, to scale their impact by "loaning" out their likeness to advertisers without them needing to be present. The article points out though that the results still have a weird warble and lack the ability to charge the generated voice with emotion and intonation that a real actor can bring but the results are definitely realistic enough to be spooky.

**Why it matters**: The recent debacle with cloning Anthony Bourdain's voice showed that even potentially positive uses of such technologies can have an uncanny valley effect. In other cases, like the one where this technology was used to revive the voice of Val Kilmer who suffered from voice loss due to a tracheotomy, the results were perceived in a much more positive light. The technology can definitely be put to a positive use but this requires a careful consideration of pros and cons, as is the case with all dual-use technology.

**Between the lines**: Some interesting applications mentioned in the article include how voice clones could be used to make games more personalized by adding in the player's voice clone to deliver all the dialogues in-game from the protagonist making the game more immersive. Another one utilizing parents' voice clones to read bedtime stories to children when parents are away. As long as we can prevent stealing the likeness of our voices which can be used for automating fraud, such applications definitely have the potential to bring about some useful capabilities.

### The pandemic is testing the limits of face recognition

[Original article by MIT Technology Review]

**What happened**: The article dives into what happens when we have larger portions of our society's core operating infrastructure become automated, often run by private companies.

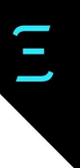

Facial recognition technology has already been shown to be deeply flawed, but the article documents the case of a transgender person who was put in a precarious financial position as the California government put in place facial recognition technology to verify identities and the change in appearance for this person wasn't correctly picked up by the system. In most places, automation like facial recognition technology is deployed as a force multiplier allowing low-resourced governments to provide more personalized services to a larger number of people. In other places, they are pitched as a health and safety option by providing contactless alternatives.

**Why it matters**: But, without an underlying supporting infrastructure composed of humans, such technology, when imperfect, and layered on top of an unjust society, can exacerbate injustices in society, making it particularly difficult for those who are already marginalized. When such technology for example works in 95% cases, the 5% who are left out need human intervention to still be able to access services. But, the current wave of automation often reduces human support down to the point where the 5% get permanently locked out of being able to access services and the help they need.

**Between the lines**: When thinking about deploying automation, design considerations are essential if they're going to achieve lofty goals of increasing access for everyone and improving the quality of service. While the technology might work in a large percentage of cases, those who are unable to be served by the technology, often those who were previously marginalized too, need to be provided alternatives that still meet their needs. Without that, we only risk making society worse than it is by promoting automation as a way forward when it might be one step forward and two steps backwards in reality.

### Three predictions for the future of responsible technology

[Original article by World Economic Forum]

**What happened**: Providing a quick overview of the work taking place at WEF on responsible technology, the article lays out three trends that they believe will come to pass in the field including investment efforts taking on responsible development as a pillar in assessing the quality of investments just as ESG became a criterion for assessments. They also believe that we are just at the beginning of targeted regulations and will only see them adopted in more countries in the world. Finally, they also see higher education making tech ethics a mandatory part of various curricula.

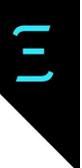

**Why it matters**: Broad adoption of responsible practices in technology design, development, and deployment is what is missing today and the three pillars identified by WEF do provide a good high-level roadmap for attacking this problem in a multi-pronged approach. What we need to think about in addition to this is what are the incentive structures that will actually enable these trends to come to pass, rather than requiring constant push to make them a reality, but ones that will evaporate without support.

**Between the lines**: Diving into just teaching AI ethics in higher education, there are many courses that are being developed and provided at universities with each pursuing this in their own manner. We have done a deep dive into this area through our series "Office Hours" that highlights how some educators are going about this. As for making this a criterion within investment assessments, I think we are still a long way away from that because there aren't yet enough market forces that call for such evaluations and the make-up of most investment shops continues to lean away from being diverse enough to acknowledge these problems in the first place. But, just as was the case with ESG criteria, I firmly believe that utilizing monetary incentives through investments will push the industry towards responsible technology practices faster than without it.

## A tiny tweak to Zomato's algorithm led to lost delivery riders, stolen bikes and missed wages

[Original article by Rest of World]

**What happened**: Zomato, a food delivery app popular in India, increased the delivery radius for workers from 10 km to 40 km which had an immediate impact on the number of deliveries they are able to complete in a day. The workers are forced to take on deliveries that push them progressively further from their "home zones." They tried things like switching off their GPS so that they would not receive orders taking them far away but that meant time off the app which reduced their earning potential. After significant protests by workers in Bengaluru, Zomato rolled the change back for workers who have been with the platform for more than 3 years, but not for the ones who are new.

**Why it matters**: The agents are incentivized based on the number of deliveries they are able to complete in a day and having to travel further diminishes the number of deliveries they are able to complete. Rejecting orders is also not an option since that directly affects their rating within the platform and the number of deliveries they get allocated based on that status. Finally, redressal mechanisms are mostly automated, fixed menu options that don't give them much agency with the company.

**Between the lines**: Workers have tried to organize and raise concerns with the Supreme Court of India to be classified as wage workers rather than contractors so that they have more rights and labor law protections but that has been unsuccessful so far. Similar to the gig economy issues elsewhere in the world, workers are disempowered and helpless, especially in a country like India where the wages they do receive are very low, barely helping them meet basic needs on a daily basis.

## Small Data Are Also Crucial for Machine Learning

[Original article by Scientific American]

**What happened**: The article makes the case that the technique of transfer learning whereby a small, highly domain-specific dataset can be used to leverage a pre-trained model to fine-tune performance on a task has a lot of promise and remains under-explored at the moment. It points out that there has been a lot of success in applying this to tasks in computer vision (CV) and natural language processing (NLP) such as the use of models pre-trained on ImageNet. But, it also points out if there is limited overlap in the domains of the new task and the dataset on which the model was pre-trained, performance can suffer.

**Why it matters**: Nonetheless, transfer learning is a promising area of research that deserves attention given that it can elevate the power of small data. Especially when there is a high financial cost to training large models, the ability to use pre-trained models fine-tuned using transfer learning can provide an avenue to resource-constrained researchers to harness the power of AI. This can also help us mitigate the environmental impact of AI systems by preventing the need to train really large models from scratch and operate well in small data regimes.

**Between the lines**: There are many benefits to having large, generalized models which can be taken off-the-shelf and fine-tuned for new tasks because they demonstrate the "generalizability" of such powerful models, one of the key things that any AI practitioner would love to have when developing AI systems. The more we are able to harness existing models where investments have already been made to bring them up to a baseline level of performance, the more we'll be able to democratize access to performant AI systems in novel domains to people who were previously limited in their ability to build and access such systems.

## How Big Tech Is Pitching Digital Elder Care to Families

[Original article by The Markup]

**What happened**: As the pandemic rolled on through all parts of the world, elder care facilities felt a particular twinge of isolation. In a woefully underprepared ecosystem, with constant understaffing, the elder population remained isolated and family members turned to consumer devices like Apple Watches and Alexas to step in partially in place of caregiver responsibilities, especially around monitoring and alerting in case of accidents. But, this comes with a slew of privacy and consent problems, given that elders are less likely to understand implications of the use of such technologies.

**Why it matters**: As an example, for elders with dementia, consent becomes problematic as their state of mind may not be such that they are fully able to grasp what it means to be monitored via an audio or visual device. In addition, their ability to withdraw consent also becomes limited if circumstances change. Then, the deployment of such technologies also have second-order effects, for example, the conversations of those around with such monitoring devices are also captured, not necessarily with their consent.

**Between the lines**: It is not surprising that such technology has taken off. There is an untapped market for technology in elder care and companies are trying to dive into this sector (as also covered in AI Ethics Brief #43). Also, caregivers tend to have a fair bit of power over elders and even through "benevolent coercion" nudge them into using technology that they might not otherwise be comfortable with. Finally, and most importantly, technology cannot serve as a replacement for human warmth and care. The rapid deployment of technology as a replacement for functions that are provided by human caregivers will only shift ecosystem investing away from what actually needs to be done (training, hiring, and paying well for human caregivers) towards technological solutions.

## How Alibaba tracks China's delivery drivers

[Original article by MIT Technology Review]

**What happened**: Getting meals delivered on time requires a coordinated effort across restaurants, service providers, and delivery drivers. With mounting pressure from consumers to get their deliveries on time, and a highly competitive landscape with many service providers trying to snatch up market share, innovation in tracking and estimating delivery times can offer an edge. In China, companies like Eleme, owned by Alibaba, with over 83 million monthly active users, have deployed more than 12,000 Bluetooth beacons to enable indoor tracking and figure out how long the driver waits for orders and when they enter and leave a restaurant.

**Why it matters**: The stated reason behind these deployments is that they will make the job of delivery drivers more efficient, since they won't have to pull out their phones every few minutes to "check-in" with the system on their status. Automatically doing so through Bluetooth beacons and proximity will alleviate this. But, more accurate location-data and using that to tighten up delivery times will also increase pressure on an already under-compensated and strenuous job. Gig workers with few rights will be forced to operate under even more draconian circumstances.

**Between the lines**: In addition to the many labor rights problems with such a technology, including the well-being of workers and stress concerns, having so many Bluetooth beacons, both virtual and physical, pose unexplored challenges when it comes to exchanging so much location-based information constantly throughout the day. Perhaps, tempering our expectations as consumers on delivery times and aiding workers in getting better rights is a more fruitful investment of resources than enabling more stringent technology from micromanaging every aspect of their job.

## Current AI Practices Could Be Enabling a New Generation of Copyright Trolls

[Original article by Unite]

**What happened**: In a study conducted by researchers from Huawei, they discovered that for the 6 most common datasets used by them in training their AI models, most of them posed significant legal challenges when it comes to commercial use. Specifically, challenges included things like what kind of licenses the models needed to be released under since they constituted derived work, whether commercialization was even possible, and the legal liabilities in case claims were made by anyone affected by adverse outcomes from the use of those models. In several of those datasets, there were challenges in tracing the lineages of the licenses that would be applicable given that they were curated and scraped datasets rather than original data gathering. In addition, most also come with auto-indemnification for the original authors of those datasets, placing the liability onto those who use them in building their models.

**Why it matters**: Given the push towards large models, which in the current paradigm of supervised learning mean the consumption of large datasets for training, the use of such datasets and their legal implications pose challenges if the current legal landscape evolves towards something stricter whereby such violations are pursued more stringently. The reasons identified in the article and paper that allow for such violations to pass unaddressed is that there is a laissez-faire and caveat-emptor approach, at least in the US. But, should that change or as is the case in other jurisdictions, where such activities are firmly disallowed, these violations will have to be tackled head-on rather than leaving them nebulously unaddressed.

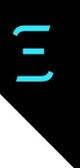

**Between the lines**: Large public datasets have been the bedrock upon which powerful models have been built in the modern era of AI. But, as has been showcased over the past 18-24 months, a lot of these datasets come with challenges in terms of not only biases, but also how that data was collected, often without consent. A deep-dive into the licensing lineage, as done by the paper cited in this article showcases that there are numerous challenges that are yet to be solved, especially as the regulatory regime stiffens with respect to the use of data in AI systems. This might also have implications for how AI systems are imported and exported if there are differences in the regulatory requirements across different jurisdictions.

### The Future of Digital Assistants is Queer

**[Original article by [Wired](#)]**

**What happened**: The article dives into building upon the case that was laid out in the UN report "I'd blush if I could" that highlighted how a lot of smart voice assistants have a feminized voice and are made to take on archaic, stereotypical feminine characteristics of obeisance emerging from the lack of diversity and other problems in the domain of technology. In particular, it showcases how the future for these assistants might be queer, not just in the formulation of the actual timbre of the voice, but more so in what being outside of traditional binaries mean when it comes to whether such an assistant should mimic humans in the first place.

**Why it matters**: Not only does such an approach eschew the problematic formulation of digital assistants today, it also enriches the discussion by providing alternate formulations for what digital assistants can look like. It helps us imagine an alternate future. One of the examples that they mention include an exploration of having multiple personalities that more accurately reflect the many versions of femininity, but even more on the point that such bots are not human. The example of Eno, the bot from Capital One stands out as an example where it talks about binary as 1s and 0s rather than gender when asked about its gender.

**Between the lines**: California in 2019 created the first legal precedent asking bots to identify themselves, something that is increasingly important as we have capabilities like Duplex from Google being capable of making appointments on our behalf sounding human. While the legal precedent is far from perfect, it lays down an imperative for us to start thinking differently about such technologies, especially as they inch into mimicking human assistants more and more. It will also shape the interactions between humans and machines much more.

## Robots Won't Close the Warehouse Worker Gap Anytime Soon

[Original article by Wired]

**What happened**: Anytime there is a conversation about the labor impacts of AI, the first thing that we hear about are the impacts that will take place on the factory floor. This article dives deeper into how that is actually manifesting and what it means for the future of work. Most of the robot deployments on the factory floor today are things that require limited intelligence and still rely heavily on human co-workers to complete jobs, where they only play a small part by taking over some tasks.

**Why it matters**: As we look for more nuance on the direct impacts from automation on factory floors and elsewhere, it helps to gain an understanding of which industries are deploying automation in what manners and to what extent. For example, when we look at Amazon putting out numbers saying they're hiring 150,000 more seasonal workers to meet the holiday demand, it helps to understand how they co-work in the warehouse environment, and given the capabilities of where robotics are headed, what can we reasonably expect to change in the future.

**Between the lines**: As is mentioned in the article by a lot of the robotics companies who supply places like FedEx and Amazon, there are a lot of unsolved and unanticipated edge cases which we can't design for just yet. What that means is that, at least in the near-future, we will continue to have both humans and machines working side-by-side. Or at least through isolated environments, given the current safety concerns where machines are housed in separate cages to prevent any accidents from taking place. The takeaway for me from this article is that as we think about upskilling and redeploying human labor capacities, keeping a keen eye on the edge cases that are still unsolved, and speaking with technical experts to gain an understanding of the timeline to solve them will be critical to better prepare for labor transitions as the need for those arise.

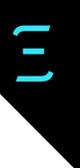

# 9. Outside the boxes

**Introduction** by Kathy Baxter, Principal Architect, Ethical AI Practice, Salesforce

[Recent US Congressional testimony](#) is another indication that the US Government continues to explore [regulating the tech industry](#). The debate itself touches on traditional issues such as antitrust, but also includes more novel issues like the impacts of artificial intelligence (AI). In particular, debates are focusing on [AI's carbon footprint](#), the [need for AI regulations](#), [if regulations actually impede innovation](#), and how to [identify and manage bias](#).

Since 2019, at least 17 US states have proposed 121 bills or resolutions with the goal of regulating AI applications. We saw the most regulation in 2019, with 10 out of 47 bills being adopted or enacted, and 2020 as the least prolific, with only one bill passing, five pending (all in New Jersey and all still pending at the end of 2021), and 36 failing to pass. In 2021, 32 bills or resolutions were proposed and six were enacted. For more information about the US AI bills or resolutions in each state, check out this [site](#).

The debate about regulating AI, though, is global. In February, [India published their approach for responsible AI](#), followed by the [EU's draft AI regulation](#) in April. In September, China published [Ethical Norms for the New Generation Artificial Intelligence](#), which "aims to integrate ethics into the entire life cycle of AI and provide ethical guidelines for natural persons, enterprises, and other related institutions engaged in AI-related activities." And, all of these discussions are happening as governments themselves increasingly [employ AI in public governance and decision-making](#).

However, some countries are still developing their approach to the global AI race (e.g., Vietnam). A [MAIEI op-ed](#) concluded that to be successful, these countries would have to secure significant investment in AI development, develop and retain technical talent, and cultivate a willingness to address ethical risks so AI benefits everyone equally in society.

With an increasing international ambition of achieving safe and responsible AI, organizations like the MAIEI have questioned whether the [world can unite under a global AI regulatory framework](#). Participants concluded that "global convergence could indeed help us overcome problems such as gaps in datasets." However, given the diversity of cultures, values, and AI capabilities around the world, "fragmentation of AI regulation is guaranteed, and the importance of local regulatory efforts is an essential consequence of that."

In light of this focus on AI development and regulation, it is not surprising that we have seen:
1. Many papers and presentations summarizing the various proposals (e.g. Stanford HAI's Policy Brief on the draft EU AI Act, visual guide to the EU AI Act by the founder of OpenEthics);
2. Wide-ranging analyses of their implications (e.g. Centre for European Policy Studies analysis of the costs of the EU AI Act); and
3. Remaining open questions about their implementation (e.g. IDC's list of unanswered questions in the EU AI Act).

Concerns have also been raised about "the dozens of separate AI ethics, policy, and technical working groups across the [US] federal government" and that "resulting policies may be incomplete, inconsistent or incompatible with each other." Many of those papers are featured in the current State of AI Ethics Report. I encourage you to take a few minutes to review them as it is critical to a deeper understanding of what is being proposed, the implications, and how we can improve these policy proposals.

Policymakers have varying levels of AI-literacy including knowledge of how the many different types of AI systems work, and how biases and harms emerge, as well as how to best mitigate them. In terms of bias mitigation, *we as a field of AI ethics researchers and practitioners still don't always know the optimal ways to do this,* depending on the type of AI application and the context of use. It is my hope that after reading this introduction and the papers linked here, you will be motivated to engage in discussions with policymakers on methodologies to identify bias and harms (e.g. balancing different measures of fairness), the downsides for each, realistic thresholds for bias (e.g. no dataset or model can ever be "bias-free"), and realistic mechanisms to monitor for and mitigate harms (e.g. human oversight, sandboxes, debiasing).

**Key issues to consider**

Many proposed regulations like the draft EU AI Act require developers of high-risk AI systems to perform both pre-deployment conformity assessments and post-market monitoring analyses to demonstrate that their systems are in compliance. Governments and companies alike need to invest more in their capacity to systematically measure and monitor the capabilities and impacts of AI systems.

Although AI governance tooling is one necessary component for creating and implementing AI responsibly, it is not sufficient. You can't know if your datasets or models are biased for or against some groups if you are unable to analyze measures like disparate impact or individual versus group fairness. However, identifying bias, analyzing by different definitions of fairness, and mitigating bias in datasets and models is [as much an art as a science](). Even small missteps can decrease the accuracy of your models and/or increase the risk for harmful outcomes without one realizing it.

This [amazing paper]() by EDRi discusses the need to go beyond debiasing as a solution to ensure AI is safe and fair. Different aspects of fairness and foundational assumptions about those that will be impacted are often left out of debiasing, providing creators the ability to game an audit and appear to comply with regulations when they do not.

Since a lack of representativeness in training and evaluation datasets is only one way that bias enters a system, automated debiasing won't address the root cause. Only by engaging with the populations underrepresented in the data can you improve your datasets, understand biased foundational assumptions, and know if your AI may result in unintended harm.

Additionally, AI can be unevenly applied to different groups (e.g. predictive policing or facial recognition surveillance applied only in predominantly black and brown neighborhoods). Debiasing mechanisms will do nothing to address those harms. The EDRi authors end their report with actionable recommendations for policymakers. Specifically, they recommend that:

- Policymakers adopting technocentric approaches to address the discriminatory impact of AI must define problems clearly, set criteria for solutions, develop guidance on known limitations, and support further interdisciplinary research.
- AI policies must limit the discretion of AI service providers in addressing discrimination and inequalities.
- AI regulation needs to go beyond ADMS [automated decision-making systems], data, and algorithms to include the spectrum of AI applications and the broader harms associated with the production and deployment of these systems.
- AI policies should empower individuals, communities, and organizations to contest AI-based systems and to demand redress.
- AI regulation cannot be divorced from the power of big tech companies to control computational infrastructures.
- AI regulation should protect, empower and hold accountable organizations and public institutions as they adopt AI-based systems.

A key component of responsible AI and often included in AI regulation is the requirement for explainability or interpretability -- making clear how a model works or why it is making a certain recommendation or prediction. It, too, is necessary but not sufficient for creating and implementing AI responsibly. [This paper]() by researchers at GA Tech, Cornell, and IBM found that "people both over-relied on the outputs from an AI system and misinterpreted what those outputs meant, even when they had knowledge about how AI systems work." Misunderstanding how AI systems work "can lead to disastrous results and over-or underconfidence in situations where more human attention is warranted." They conclude that "we need to be more deliberate in how we design explanations for AI systems so that the gap between intended meaning and perceived meaning is minimized."

AI governance tools typically require well-structured data that is labeled by demographic or sensitive variables (e.g. loan approval by race or gender). Some in the tech industry have relied on "ghost workers" -- low-paid, third-party workers primarily in the Global South or even [refugee camps](#) -- to label data to train image recognition systems, large language models, and even [self-driving cars](#). Increasingly, [ghost workers are demanding better conditions](#). Given the critical nature of their work in ensuring cars can accurately identify objects on the road or moderating toxic content and disinformation, regulations are needed to provide robust protection for these workers and society as a whole.

There are no silver bullets -- a multipronged, multistakeholder effort will be required that involves collaboration among governments, industry, academia, civil society, and consumers – especially the most underrepresented, historically marginalized, and vulnerable groups. If you are reading this report, you are a much-needed voice in this discussion!

---

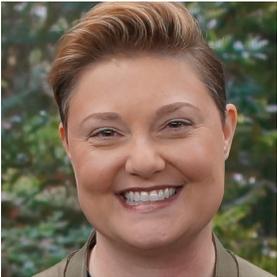

**Kathy Baxter**
Principal Architect, Ethical AI Practice
Salesforce

As a Principal Architect of Ethical AI Practice at Salesforce, Baxter develops research-informed best practices to educate Salesforce employees, customers, and the industry on the development of responsible AI. She collaborates and partners with external AI and ethics experts to continuously evolve Salesforce policies, practices, and products. Prior to Salesforce, she worked at Google, eBay, and Oracle in User Experience Research. She received her MS in Engineering Psychology and BS in Applied Psychology from the Georgia Institute of Technology. She is the coauthor of "Understanding Your Users: A Practical Guide to User Research Methodologies."

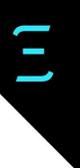

# From our events

## Top 10 Takeaways from our Conversation with Salesforce about Conversational AI

[Original article by Connor Wright]

**Overview**: Would you relate to a chatbot or voice assistant more if they were female? Would such conversational AI help you feel less lonely? Our event summary of our collaboration with Salesforce sets out to discuss just that.

**Introduction**
Would you feel less lonely if you had access to some conversational AI? Does the naming and gender of the chatbot matter? Some believe no, some believe yes, and some believe yes too much. Facilitated by Kathy Baxter, Yoav Schlesinger, Greg Bennett, Connor Wright and Abhishek Gupta, conversational AI as chatbots and voice assistants was deeply explored in our event with Salesforce. With so much potential for both positive and negative outcomes, it makes you start to wonder: can I have a good conversation with a chatbot?

**The key takeaways**
With our question prompts centering on the gender and name of different chatbots, the technology's effect on the vulnerable and the potential for bias that it brings with it, immediate reflection on chatbots itself is called into action. Specifically, how does it affect the basic notion of conversation itself?

**What makes a good conversation?**
When thinking of programming a chatbot, you may find yourself thinking about what actually makes a good conversation. Is it the speed at which you obtain an answer you were looking for? How did you feel afterwards? The information you learnt along the way? One thing's for sure, the context in which your chatbot is deployed plays a considerable role in determining what a 'good' conversation is.

**Context matters**
If your chatbot is to help customers with their banking, you're not going to prioritise making the customer feel good about themselves but rather achieve what they set out to do. From here, the distinction between 'narrow' and 'wide' chatbots comes to the fore. 'Narrow' chatbots are geared towards achieving a particular outcome within a very focussed context, such as a chatbot for a fashion brand helping you find the item of clothing you want. A 'wide' chatbot can be found in Alexa and Siri, tasked with a varied list of activities to do and accomplish in a wide range of contexts. For example, asking Alexa to both order something from Amazon and what the weather will be like tomorrow. However, whether 'narrow' or 'wide', does a chatbot's name and gender contribute to its overall success?

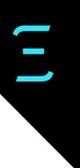

**Does the bot need a name and gender?**
It's curious as to why a majority of chatbots have been attributed a name and gender. Some say how, perhaps, the chatbot shouldn't have either as it's just a machine completing a task. For example, the streaming service Hulu's Hulubot in its help centre is an excellent example of a chatbot functioning without having a human name and a gender.

However, the norm is to assign a fixed gender and name. Doing so has a lot to do with the audience that the chatbot is being marketed towards. It is found that people are more likely to welcome into their home a female chatbot by finding the female voice more relatable and trustworthy. One problem this does cause is potentially reinforcing the gender stereotype of 'women assistants', so should you be allowed to choose whether you want your chatbot or voice assistant to be a particular gender?

**Should you be allowed to choose?**
ALongside avoiding any potential gender stereotypes, it may be that I feel like talking to different 'people' about various things, so deciding on gender and name should be left open. For example, having controls on the chatbot and voice assistant where I can play around with the pitch rather than feeling like I'm talking to the same person all the time.

However, if the choice is left open, you may run the risk of someone wanting to read in a potentially problematic persona (like a timid tone of voice to feel dominant over the chatbot). Furthermore, such customisation possibilities could lead to a severe attachment to the bot itself, making the line between humanity and machine even more blurred.

**Potentially getting too attached**
Although the human knowing whom they are talking to is a chatbot, it may still not be enough to prevent humans from getting attached and deceived about their other interlocutor, especially given how people still love anime characters despite knowing what they are. Such attachment could then be exploited by actors taking advantage of any vulnerability to use the human involved. A non-human name could potentially serve to combat this, but not all manipulation in itself could be a bad thing.

**Manipulation can have two sides to it**
Manipulation can be used to achieve more positive ends, such as a voice assistant reminding your sick Mother to take her medications and using persuasive language in doing so. Alternatively, the voice assistant could require a parent's voice authentication for certain products to be ordered off Alexa in order to dissuade children from abusing the service. In this sense, while chatbots can serve to manipulate, they can also serve to benefit human existence.

**Chatbots for good**
The chatbot evolution from just a machine to being a companion can substantially impact the darkest corners of some human lives. Chatbots can provide a 24/7 communication outlet to help combat loneliness and depression and serve as a digital companion in the dark depths of the pandemic last year. Fortunately, with such experiences not being shared by all, the importance of the chatbot process being inclusionary cannot be underestimated.

**Designing chatbots and voice assistants with all and not just for all**
A clear example can be found in differing opinions on voice recordings being done by voice assistants. Here, some are against voice assistants taking recordings of the daily happenings in the house. However, others believe that this can be a crucial step to combating gender violence, with voice recordings potentially proving key evidence of different incidents.

Making this kind of potential service accessible then proves paramount as well. Incorporating local dialects and different accents for optimum benefit to be guaranteed to all is one aspect of judging how good these conversational AI are. However, do we get too carried away with such technology?

**Seeing a chatbot for what it is**
Sometimes, without having any benchmarks, we may get over-excited about conversational AI in itself. This is not helped by any personal relationship developed through the voice assistant or chatbot having its own name and gender, which lead us to attribute more humanity to these AI than we actually should. For example, gender for voice assistants is instead just a pitch value to which we attribute our human interpretation, rather than a voice assistant or chatbot actually being on the gender spectrum.

It's important to note how chatbots and voice assistants are programmed to say things to you rather than to understand you. For example, Sirir may, at some point, be able to book you on a flight in your preferred window seat, but it would not know the reasoning behind it. Maybe this is, in fact, for the best, given the privacy concerns associated with chatbots themselves.

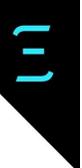

**Privacy issues**
Despite the California Chatbot Law and new EU AI law requirements regulating conversational AI, some thought-provoking questions still arise. For example, is it a privacy violation if we look at the chatbot conversations involving other people in our family? If we are caught on our neighbour's Amazon Ring committing a crime, does my neighbour have the right to share such information with the police? The more integrated into our lives conversational AI becomes, the more of these questions will surely surface.

**Between the lines**
Our event proved both inspiring and stimulating for me. The importance of involving all in the conversational AI design process is now abundantly clear, especially with the field of naming and assigning gender to your chatbot proving extremely rich with questions. I find attributing such aspects to the chatbot important given how it can affect how a conversation is conducted (such as feeling more trustworthy of a female voice assistant or chatbot). Although, what I caution against is attributing too much personality and humanity to such AI, which can only increase the likelihood of negative manipulation and harmful emotional attachment.

## Top 5 takeaways from our conversation with I2AI on AI in different national contexts

[Original article by Connor Wright]

**Overview**: Can the world unite under a global AI regulatory framework? Are different cultural interpretations of key terms a sticking point? These questions and more formed the basis of our top 5 takeaways from our meetup with I2AI. With such a variety of nations present, it shows that while we have different views on various issues, this is not a bad thing at all.

**Introduction**
Can the world unite under a global AI regulatory framework? Can problems with AI join together other nations in a common cause? These questions form the basis of the top 5 takeaways from our meetup with I2AI. Spanning topics like centralisation and the importance of localised AI regulations, our meetup showed how AI governance must be seen as a context-dependent phenomenon, starting with power relations.

**There are power relations at play**
Any talk about enacting localised regulations on AI must consider how uneven the playing field is in terms of decision-making and economic power. The extent to which local governments can instantiate local laws depends heavily on the resources available to each country. How this is conducted is then affected by the power relations in the international arena. For example, the attitude adopted to privacy laws could depend on the relationship a nation has with either China or the USA (two opposing views on privacy).

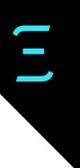

**Centralisation may be a moot point**
Given such global diversity, it may be difficult for all countries to follow one system to which there are even differences of interpretation within countries. Jamaica has dismissed digital ID cards in the Caribbean as unconstitutional, but Barbados is still trying to implement them. Furthermore, the interpretation of company data storage laws in India has a lot to do with cultural understanding. As a result, fragmentation of AI regulation may be inevitable, but is this a bad thing?

**Fragmentation isn't inherently undesirable**
Fragmentation doesn't mean that you have incoherent pieces. Peaceful coexistence between the AI regulation fragments can be moulded, primarily through a common thread. Setting a global target for all to reach can help direct all the different approaches towards the problem. Sure, there will be some inconsistencies in the approach, but arriving at the same point through different pathways is undoubtedly a viable option.

**The importance of local regulations**
To arrive at the same destination, local regulations and interpretations of the issues in AI are very important. They will serve to define what is meant by terms such as 'fair' and 'representative', as well as proving the most accurate expression of a country's views on issues within AI. If these were not in place, individual countries' values and concerns would be lost in the big-scale legislation conceived elsewhere. Without localised efforts, someone else ends up designing your AI for you.

**The language we speak and the language we use**
The importance of these regulations is most clearly seen in their relationship with language. Even reading the law in one language (say, German) can produce a completely different interpretation than reading in English. With our meetup spanning from South America to Europe, we found that some participants harboured different interpretations of the same legislation depending on the language used. The subtle meanings and context of each word changes throughout each language, emphasising the vital role of localisation even more.

It is not just the language in which we speak about AI that matters, but also how we talk about AI. At times, the AI vernacular tends to anthropomorphise the technology by saying "the AI decided" or "the AI is thinking". Furthermore, such ways of expressing AI renders countries like Brazil (with barely any initiatives towards AI) at risk of the buzzword effect that AI generates. For example, reading phrases like "the AI determined its course of action" and immediately thinking about terminators.

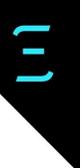

**Between the lines**
Technology is an excellent way of demonstrating how different countries treat their citizens and how difficult it is to find a common thread. Global convergence could indeed help us overcome problems such as gaps in datasets. With enough data in all the right places, researchers would no longer need to construct 'representative AI' where the data is most available but instead select the most relevant data. However, I believe that fragmentation of AI regulation is guaranteed, and the importance of local regulatory efforts is an essential consequence of that.

## Our Top-5 takeaways from our meetup "Protecting the Ecosystem: AI, Data and Algorithms"

**[Original article by Connor Wright]**

**Overview**: In our meetup with AI Policy Labs, we discussed AI's involvement with climate change. From the need for corporate buy-in to data centers, AI's role in the fight can often be confused. However, it starts with what is factual that will give us the best chance of using it.

**Introduction**
In partnership with AI Policy Labs, we discussed how AI is interconnected with the fight against climate change. The group quickly identified the role of misinformation; the group soon realized the need for a collective and not just individual effort. How this would be achieved then brought up questions regarding governance while the ever-present problem of tangibility continued to plague efforts to fight the crisis potentially. What is important to note is that knowing what's factual is the first step of many in confronting this challenge.

**Knowing what's factual**
Part of the problem of fighting climate change is combating those who deny there is any fight at all, with a worrying amount of counter-information on climate change in circulation. The role AI plays in this fight is resultantly confused, for example, AI being used to identify pollution hotspots and spread misinformation.

Therefore, part of the fight is understanding how to detect misinformation and how to know when something's factual. Demystifying climate change and knowing what is factual can help identify the actual problems, allowing us to focus on each issue one by one. The fight can seem overwhelming at the best of times, so different people concentrating their efforts can help to make great strides in the areas they choose.

*However, this can't be done alone.*

**Efforts at the individual level alone won't cut it**
Despite it being a global fight, only specific populations and sectors are buying in. Corporations are generally the most significant contributors to pollution. So, without their involvement in altering their habits, individual actions will become meaningless. The combination of personal and corporate action (whether a tech company or restaurant) will prove a potentially winning formula.

*However, while the corporate side has its challenges, so too does the individual.*

**The problem of data collection**
It must be acknowledged that even altering actions at the individual level is troublesome. Take, for example, the Sidewalk Labs' Smart City project in Toronto. Striving to try and create a revolutionized city, the data required to do so is deep and personal. Concerns about what this data would involve and how it would be stored were key in eventually stalling the project.

The kind of infrastructure needed for this project in the first place is also noteworthy, whether physical or regulatory. Data centers may provide the answer.

**Data centers**
Data centers could be a way to store and share data to facilitate a cooperative effort on the crisis, but this brings up governance problems. Any data that leaves a country's soil will involve relinquishing at least some control over what data is accessed and used. Different countries have different privacy laws, and the type of data that one country might want to collect may not be possible in another. Even then, 100% wifi reliability in both countries is needed to keep the data collected alive.

A theoretical approach and futuristic considerations are strongly present in discussing climate change. Yet, this sometimes generates a problem of tangibility.

**The tangibility problem**
At times, individuals tend to see climate change as a theoretical issue rather than seeing it for its effects on us. Here, mentioned in the meetup from a developer's view, the impacts of any non-climate-change-friendly policies are far removed. Helping to solve this could make carbon footprints of particular technologies, like washing machines, visible.

Although, the next question surrounds whether this would influence a consumer's decision. With so many choices in life to make, would a consumer want to be disposed to make another?

**Between the lines**

In answer to the previous question, individuals making choices are an essential component of the climate change fight. It provides an opportunity not to allow climate change compliance to be put on the back burner, especially when influencing what products companies are to produce. To facilitate this choice, AI needs to be seen as the right solution, not just another technological solution utilized just because. From my view, AI is still early enough to employ these kinds of considerations and with the correct factual information shared, these considerations can take a central role in the fight against climate change.

# Go Deep: Research Summaries

**Building Bridges: Generative Artworks to Explore AI Ethics**

[Original paper **by Ramya Srinivasan and Devi Parikh]**
[**Research Summary by Ramya Srinivasan]**

Overview: The paper outlines some ways in which generative artworks could aid in narrowing the communication gaps between different stakeholders in the AI pipeline. In particular, the authors argue that generative artworks could help surface different ethical perspectives, highlight mismatches in the AI pipeline, and aid in the visualization of counterfactual scenarios, and non-western ethical perspectives.

**Introduction**

*A picture is worth a thousand words!*

Indeed, visuals are extremely effective in conveying complex concepts in an accessible manner—they transcend language barriers, simulate engagement, trigger critical thinking, and leave lasting imprints in the minds of the observer. Backed by this understanding, the authors posit that generative artworks (i.e., artworks created by AI systems) could come handy in educating AI scientists with regards to potential pitfalls in the design, development, and deployment of the AI systems. To substantiate their argument, the authors lay out four potential pathways in which generative artworks could be leveraged in educating AI scientists about AI ethics, namely,—1) by visualizations of different ethical viewpoints, 2) by visualizations of mismatches in the AI pipeline, 3) by visualizations of counterfactual scenarios, and 4) by visualizations of non-western ethical perspectives.

**Key Insights**
Here, a brief description of each of the four aforementioned potential pathways (through which generative artworks could aid in enhancing AI ethics) is provided.

**Visualizations of different ethical perspectives**: Different ethical theories emphasize different principles in decision making, and can thus shed light on varying viewpoints relevant in a given context. For example, in utilitarian ethics, the emphasis is on maximizing the well-being of all stakeholders, which is not necessarily the case in deontological ethics, where the emphasis is on following the laws and regulations.  Thus, even within the context of a single problem setting, there can be diverse viewpoints about what is right, fair, just, or appropriate. In order to enhance AI ethics, it thus becomes important to educate AI researchers and developers about these diverse viewpoints and thereby aid in reflexive design. Generative artworks could serve as powerful visualization tools to surface such diverse perspectives. For example, through generative artworks, it may be possible to visualize the compounded adverse effects of an AI decision in an individual's life, a consequentialism ethical perspective.

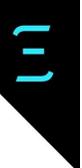

**Visualizations of mismatches in the AI pipeline**: Computational systems involve quantitatively modeling abstract concepts or constructs which may or may not be observable. Furthermore, there may be unobservable factors that affect the constructs themselves. Consider, for example, a construct such as "skill" or "ability", which is relevant across many applications such as hiring and admissions. These constructs can be influenced by both innate potential specific to the individual and other factors such as socio-economic status. Thus, a mismatch can be introduced even before measuring a construct. Generative artworks could aid in visualizing such mismatches. For example, it may be possible to highlight differences in measurement of similar constructs, thereby aiding AI researchers and developers in understanding system behavior. Consider an AI based hiring use case. Suppose one of the features in making the decision concerns measuring social skills of the candidate. In this regard, one might expect the constructs "self-esteem" and "confidence" to be related. Visualizations of AI system's behavior under different scenarios could reveal whether it treats these constructs similarly – whether it exhibits "convergent validity" , which refers to the degree to which two measures of constructs that theoretically should be related, are in fact related.

**Visualizations of counterfactuals**: Generative artworks could also aid in visualizing counterfactual situations which in turn can be beneficial in reflexive design via empathy fostering. Counterfactual thinking can help in engendering empathy by enabling one to visualize situations through another person's world. Thus, certain situations that may be irrelevant in one person's context, but relevant in another person's context, can be understood via such counterfactual visualizations. Generative artworks could be used as tools to visualize the consequences of AI decisions so AI researchers and developers (for instance), who may not necessarily be affected by the decision, can empathize with the impacted population, and thereby redesign their system for the better.

**Visualizations of non-western perspectives**: Generative artworks can serve as visualizations of social, cultural, and economic differences that exist across geographies. For example, through generative artworks it may be possible to highlight different viewpoints regarding fairness based on the local context such as social practices, religious beliefs, economic status, etc. By training generative models on data across cultures and looking at the latent visualizations, it might also be possible to view how different everyday practices (e.g. dress, food, etc.) and objects (e.g.

furniture, houses, etc.) can vary across cultures thereby shedding light on local contexts which can be valuable in AI system design.

**Between the lines**
The ideas postulated in the paper offer promise in that they can open up new ways of reflexive design and facilitate introspection. Generative artworks could be especially beneficial in highlighting counterfactual scenarios— given that such visualizations may not exist in the real world, and thereby could shed light on new and latent perspectives. That said, for surfacing non-western perspectives and viewpoints based on various ethical theories, existing artworks could also be used. Also, as the authors acknowledge, generative artworks could themselves be biased, so it is necessary to employ these tools mindfully. Ecological costs/environmental impacts of generative artworks are however not discussed in the paper. Given that generating artworks requires significant computational resources, there exists a tradeoff between the ecological cost and educational benefit, which calls for further analysis.

## Brave: what it means to be an AI Ethicist

[Original paper by Olivia Gambelin]
[Research Summary by Connor Wright]

**Overview**: The position of AI Ethicist is a recent arrival to the corporate scene, with one of its key novelties being the importance of bravery. Whether taken seriously or treated as a PR stunt, alongside the need to decipher right or wrong is the ability to be brave.

**Introduction**
The position of AI Ethicist is a recent arrival to the corporate scene. Tasked with ethical evaluations of AI systems, there may be times that the role feels lonely. Potentially being the only objector to the deployment of an AI product which could earn your company a healthy profit, no matter how sure you are, is a scary thought. Hence, it is important to note that the AI Ethicist's role requires bravery. Yet, the AI Ethicist is not the only agent operating in the Ethical AI space.

**AI Ethics is not just for the AI Ethicist**
An important distinction is how an AI Ethicist is not the only one who engages in AI Ethics. With AI stretching into multiple walks of life and business practices, a sole AI ethicist would not be able to capture the different perspectives needed to consider. Hence, technologists, data scientists, lawyers, and the public form part of the field's multidisciplinary nature. Different backgrounds are more suited to identifying different types of ethical risks. Be it a lawyer identifying a tricky definition used in describing an AI system, or a public member bringing up their view of how it would affect their lives.

To illustrate more clearly, an example involving autonomous vehicles fits. While an Ethicist can comment on the traditional Trolley Problem, data engineers must also understand how to incorporate its thinking into hard code. Not only that, but consultation with the broader public can help understand the broader requirements these vehicles are meant to fill, especially with the older population. All in all, just because the AI Ethicist's job title is closest semantically to AI Ethics doesn't mean it's the sole actor in the space.

**The role of an AI Ethicist**
Nevertheless, an AI Ethicist still has a role to fill within the field. The job includes potentially being the only member of a team to veto an AI product that could earn your company a healthy profit. Whilst other team members could be "silenced by a profit margin", an AI Ethicist is expected to draw on moral principles to help decipher what is right and wrong within an AI context before applying their deduction to concrete examples. The application then needs to be presented in an empathetic manner not to receive defensive responses.

It is also the AI Ethicist's responsibility to maintain objectivity in ethically charged situations within this process. As a result, the Ethicist may become the default General of assigning responsibility when consulted on the location of potential ethical faults in an AI product. To do this effectively, proficiency in the design, development and deployment of the AI system at hand is paramount. This does not mean that the ethicist must be fluent in every ethical system in existence, but how they must be fluent in their industrial context.

Part of understanding the context lies in recognising both the logical and illogical inputs present in making a decision. There is no point in simply appealing to logic when trying to explain an illogical decision made, making the quality of awareness of an AI Ethicist a vital tool. One such example could be how IBM released their facial recognition technology despite the bias problems that resulted. Here, it doesn't help to ask 'why did they release a harmful product?' but rather examine other factors in the decision. There could've been a lack of information about the potential for bias, or internal company pressure to release the product. It is not the AI Ethicist's job to excuse any form of industry behavior, but to be sensitive to non-logical factors.

All of this requires bravery.

**Why bravery is needed**
An AI Ethicist is to be prepared to walk into a room where they only disagree with an AI proposal. This also means that the AI Ethicist becomes the focal point of responsibility when discussing ethical decisions and may be used as a scapegoat should the product not be launched. Cases may arise where a moratorium results, placing the blame more on society 'not being ready' rather than an AI Ethicist being difficult.

However, policies that result from a moratorium aren't guaranteed to be water-tight. Some procedures could potentially only command the bare minimum for a compliant AI product yet still leave room for an AI Ethicist to give a red light. It could be that a company keeps the raw data for an AI system private to external parties in one national context (as mandated by the law) but doesn't do so in a different space. So, while technically being compliant, an AI Ethicist may still need to step in to encourage against damaging the company's reputation. To do so, requires bravery.

**Between the lines**
With the AI Ethicist position becoming more and more prominent, certain qualities are required to prevent it from becoming a marketing stunt. The paper claims that bravery is one of them, and I wholeheartedly agree. One thing that I believe can help is, as mentioned in my last research summary, being more than one AI Ethicist involved. Instead, boasting of AI Ethicists disseminated throughout the company will allow ethical problems to be picked up and talked about far quicker. Nevertheless, every one of these positions, no matter how many there are, will require bravery.

## You cannot have AI ethics without ethics

[Original paper by Dave Lauer]
[Research Summary by Connor Wright]

**Overview**: AI systems are often fixed by looking for the broken part, rather than the system that allowed the error to occur. The paper advocates for a more systematic examination of the AI process, which the more you think about it, the more sense it makes.

**Introduction**
Would Aristotle have bought into AI ethics? Or, does AI ethics sit as a separate entity to all that has gone before it? Given AI ethics' rise in popularity, it has often been held in its own regard, with special mentions of AI principles at big corporations like Facebook and Google. Nevertheless, the answer to the question 'can AI ethics exist in a vacuum?' is a resounding no. An examination of an 'unethical AI' problem needs to be systemic and aware of the incentives involved in the process, rather than just looking for the 'broken part'. Thus, let's first look at why AI ethics does not exist in a vacuum, with a comparison to medical ethics along the way.

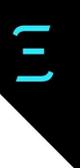

**AI Ethics does not exist in a vacuum**
The key notion that I found in this piece was how AI ethics could not come about without an ethical environment to surround it. As seen in medical ethics, the AI ethics space comes into contact with a whole host of issues also touched upon by other fields. Take, for example, the issues of autonomy and moral responsibility in AI ethics and for the past 500+ years of philosophy. Hence, without an all-encompassing ethical approach, the subfield of AI ethics quickly becomes isolated and ineffective.

In this sense, given AI ethics' ties to an overall ethical environment, we need to examine the system as a whole when something goes wrong with an AI system. Here, systems thinking is introduced to mention the relationships between parts of a process/product as being key, not just individual parts themselves. In other words, if an AI system fails, don't examine its features; examine its ecosystem.

**The broken part fallacy**
Tying into this last point, the "broken part fallacy" is introduced. About how humans examine problems, the fallacy lies in seeing that a system/product has malfunctioned and looking for the broken part with which to fix and resolve the issue. Such an approach deems the problem as something individualistic, which won't necessarily fix it if it's systemic. Looking for a broken part treats a systemic problem as too simple, given the complex interactional nature of an ecosystem.

Hence, looking for a malfunction in an AI system will not automatically fix its problem of being unethical. Instead, a thorough look at how that unethical behavior has surpassed the checks and balances is required, especially surrounding the product's deployment into social and cultural contexts.

**The importance of social and cultural sensitivity**
When examining the systemic nature of an AI's deployment, more abstract notions are discovered that require change than a simple 'broken part'. Listening to those closest to the problem and avoiding top-down legislation is an excellent first step. This offers a closer look at the situation from those who designed the AI product, cultivating a more trusting relationship.

**The question of incentives**
The next question is whether businesses can enact this kind of approach and whether they are incentivised to. The incentives created by law and policy can be a good starting point, examining whether there is a legislative push behind specific actions that can be deemed 'ethical'.

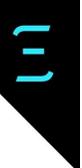

Such examinations can then expose the type of ownership within a business. To illustrate, Facebook operates on an Absentee Ownership model, whereby the "locus of control and locus of responsibility are different". In Facebook's case, they control what is allowed on their platform but do not have legal responsibility for the content that's eventually put on there. In this case, an AI ethics programme coming out of Facebook would not prosper without sharing in the center of responsibility. Instead, ethical frameworks are needed to be part of the company's ethos and not just something to be checked off the list. AI ethics can then be a branch of central ethical practices and frameworks instead of holding its own fort.

**Between the lines**
I very much share how AI ethics is not born in a vacuum. I liken it to conversations about bias in AI systems, whereby if the humans programming the AI product have their own biases, then we cannot expect some of these to turn up in the AI system. The aim is then to mitigate the harm that is produced from these biases taking hold. Applied to our present context, I would not be surprised if a company with a flawed ethical approach created an' unethical AI'. Without self-reflection on the AI process itself, the reason why an AI is producing the 'unethical' behavior that it does will remain an even darker black box. Hence, before looking for the broken part, we should ask ourselves how it got there.

## Implications of the use of artificial intelligence in public governance: A systematic literature review and a research agenda

[Original paper by Anneke Zuiderwijk, Yu-Chen Chen and Fadi Salem]
[Research Summary by Angshuman Kaushik]

**Overview**: The expanding use of Artificial Intelligence (AI) in public governance worldwide, has not only opened up new opportunities, but has also created challenges. This paper makes a systematic review of existing literature on the implications of the use of AI in public governance, and thereafter, develops a research agenda.

**Introduction**
There is no denying the fact that AI has been used for quite some time now, and its use has resulted in both positive and negative outcomes. Further, considering its scope, AI is a multidisciplinary area of research, rich with a vast number of papers pertaining to its myriad applications. Within that extensive gamut, the emphasis of this paper is on the literature that addresses the effects of the uses of AI in the public governance setting. This paper narrows down its focus on the articles that research on the implications of AI in the context of public administration, digital government, management, information science and public affairs. It deals with the issues relating to fairness, bias and governance questions pertaining to transparency, and regulatory frameworks. For instance, how does the implementation of specific AI technologies affect accountability of government institutions? As far as the research approach is concerned, the researchers conducted a systematic literature review of the relevant material. After an extensive review, the researchers found a number of potential benefits and challenges relating to the use of AI in public governance enumerated in them.

They identified the benefits in nine categories: 1) efficiency and performance benefits, 2) risk identification and monitoring benefits, 3) economic benefits, 4) data and information processing benefits, 5) service benefits, 6) benefits for society at large, 7) decision-making benefits, 8) engagement and interaction benefits, and 9) sustainability benefits. In addition to the potential benefits, they also identified eight challenges of the use of AI in their literature review, which are divided into eight categories: 1) data challenges 2) organizational and managerial challenges 3) skills challenges 4) interpretation challenges 5) ethical and legitimacy challenges 6) political, legal and policy challenges 7) social and societal challenges, and 8) economic challenges.

**Use of AI in public governance**
One example relating to the application of AI in the governance-setting is the use of SyRI ("System Risk Indication") by the Dutch Government to detect possible social welfare fraud. It had not only issues with transparency and a host of other factors, but the algorithm also turned out to be a 'black box'. Its operation was eventually brought to an end by the court for violating Article 8 of the European Convention on Human Rights (ECHR), which protects the right to respect for private and family life. The requirement of Article 8 is that, any legislation should strike a 'fair balance' between social interests and violation of the private life of the individuals. (The intention of citing this particular example is not to portray the deleterious effect of AI, but to show the application of AI in governance, in general.) There are numerous such other cases, with beneficial outcomes pertaining to the use of AI in various sectors of the government.

**Potential benefits**
The use of AI in governance has massive implications for society, in general and individuals, in particular. The reason being, the administration, and its various functionalities have to directly deal with the masses, within their respective spheres of jurisdiction. Through this paper, the researchers have uncovered their findings with respect to a comprehensive review of 26 articles pertaining to the use of AI in public governance, which were published in the last 3 years. After analyzing the content of the articles, the researchers found that they contained a number of potential benefits of the use of AI in public governance. It was found that efficiency is improved by automating processes and tasks or by simplifying processes using machine learning. Further, AI aids in increasing monitoring of urban areas, fraud detection, law enforcement and enhancing the 'smartness' of the cities. The researchers also noticed that AI for public governance leads to economic benefits, such as making e-government services and systems more economical. Moreover, data and information processing benefits also accrue due to processing of large amounts of data in limited time. Another area where AI has a potential positive impact with respect to public governance is that it leads to improvement in the quality of public services. It also leads to the creation of public value, decision-making and sustainability benefits.

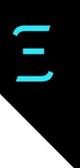

**Challenges**
Alongwith the potential benefits, the researchers also searched for the challenges of AI use in government, in the review of the 26 articles zeroed in on by them. They identified challenges relating to the availability and acquisition of data, organizational resistance to data sharing, limited in-house talent, complexity in interpreting AI results, ethical challenges, undermining the due process of law and effect on the labor market.

**Going Forward – The Research Agenda**
After an analysis of the various potential benefits and challenges, the researchers put forward a research agenda on the implications of the use of AI for public governance. It comprises eight process-related recommendations and seven content-related recommendations for researchers that examine the implications of AI use in public governance.

**Process-related research recommendations**
- Avoid applying AI-related terms superficially in public governance sources
- Move beyond the generic focus on AI in public governance sources
- Move to methodological diversity instead of dominant qualitative methods
- Expand conceptual and practice-driven research from the private to the public sector
- Increase empirical research on the implications of AI use for public governance
- Go beyond exploratory research and expand explanatory research
- Openly share the research data used for studies on the implications of the use of AI for public governance
- Learn from applicable pathways followed by digital government scholarship in its early phases

**Content-related research recommendations**
- Develop AI public governance scholarship from under-theorization into solid, multidisciplinary, theoretical foundations
- Investigate effective implementation plans and metrics for government strategies on AI use in the public sector
- Investigate best practices in managing the risks of AI use in the public sector

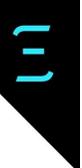

- Examine how governments can better engage with and communicate their AI strategic implementation plan to stakeholders
- Investigate a large diversity of possible governance modes for AI use in the public sector
- Research how the performance and impact of public sectors' AI solutions can be measured
- Examine the impact of scaling up AI usage in the public sector.

**Between the lines**

Although the paper has dealt with the subject very thoroughly, this is only the starting point of a 'journey of learning', which necessitates an iterative approach, involving relevant stakeholders. The findings matter, as it would enable the various actors involved in the process to have a better understanding of the issue in hand, and take appropriate steps, in the right direction, going forward. In my view, further deep dives into the application of AI in public governance at the grassroots level, through case studies, will yield specific insights. It deserves mention here that different cultures perceive AI and its outcomes in a different manner. Hence, more in-depth research, keeping in mind the cultural sensitivities, tastes and habits of different communities, would definitely, bring about a new flavor to the ever-growing field of AI, and its application, particularly, in the field of governance.

**Animism, Rinri, Modernization; the Base of Japanese Robotics**

[Original paper by Naho Kitano]
[Research Summary by Connor Wright]

**Overview**: Technology is not going anywhere anytime soon, so why not respect it for what it is? The approach adopted by Japanese culture is to recognize how natural and technological phenomena have a soul that intertwines with ours. The result: a beautiful sight of human-technological relations indeed.

**Introduction**

Have you ever felt your soul harmoniously intertwined with a technology you are using? As part of the Japanese government's aims for techno-integration, the Japanese tradition appeals for harmonious integration for the benefit of society. While the mystical element of this approach is notable, it creates a beautiful sight of what human-technology relationships can look like. Technology is now here to stay, so we may as well start off on a positive note.

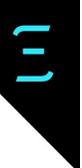

**The robotization of society is presented in a positive light**
To appropriately set the scene, the robotization and technologization of society are seen as a positive in Japanese culture. Interestingly, the change is seen as so much of a positive that the efforts placed on ethical considerations are not too deep. For, it is a cultural assumption that robots will be designed to keep the society's ethical values anyways. One way to explain the Japanese people's confidence is through exploring the role of spirit.

**The existence of spirits**
A strong Japanese cultural belief resides in how natural phenomena have spirits. Traditionally, this applied to the sun, moon and mountains, which had their own spirits and associated Gods. Subsequently, each God had a name and was assigned characteristics while having perceived control over natural events. Thanks to the technological revolution, the belief was expanded to artificial objects, which are believed to have souls in harmony with those of humans. Such belief consequently affects how the Japanese people interact with these objects, pleasantly seen when applied to tools.

**How this applies to tools**
Artificial tools made out of natural and unnatural phenomena possess anima (a soul). When in contact with humans, these objects are seen to work in tandem. Both the human and tool anima work harmoniously together. The relationship runs deep, seeing as tools are often companions for life, leading to them bearing names. Resultantly, the tools were traditionally inscribed with the owner's name and its date of first use when it took on its anima through coming into contact with humans.

The relationship between the human and their tool is respected even after it is no longer in use. Even today, tools that break are not thrown away but taken to a temple to be burned divinely. A sign of respect to how intertwined the instrument had become with its human. For example, in 2005, Tmsuk Co. Ltd. took their robot creation KIYOMORI to the shrine to pray for its success.

However, what is the benefit of all of this?

**Why talk about spirituality?**
It is important to acknowledge that the spirituality mentioned is not to advocate for a tool's subjectivity, but to show how it relates to its owner. It takes on and bears a spiritual connection with its human owner from the first minute it's used, forming the basis of the Japanese "Rinri" ("Ethics").

"Rinri" is the study of achieving harmony in human relationships, offering a guide on forming and maintaining lasting human relationships with the natural phenomena surrounding us. Each individual has a responsibility to the social wellbeing of their community. For example, in 2004, Mr. Koda apologized for disturbing the social peace and causing harm to the people of Japan due to the diplomatic storm created with his traveling to Iraq. Guided by the attitude of "social harmonization over the individual subjectivity", the spirituality of Japanese culture aims to foster a lasting relationship with technology. We see an interconnected reliability on one another; the tool on the human for its anima and the human on the tool for the task at hand.

This is a wonderful sight if you ask me.

**Between the lines**
Thanks to my links with Ubuntu philosophy (see previous research summary and panel discussion), I'm a fan of the interconnectedness and community-orientated approach offered. I think relating to technology in this way readjusts how we are to use it, namely for the benefit of the community we live in. While there are warranted concerns about what kind of relationship derives from this interconnectedness (such as sexual), grounding the action in respect is the way to go. By respecting the technology for what it is and can do for us, we can better learn how to develop this relationship with others.

## Who is afraid of black box algorithms? On the epistemological and ethical basis of trust in medical AI

[Original paper by Juan Manuel Durán & Karin Rolanda Jongsma]
[Research Summary by Marianna Ganapini]

**Overview**: The use of AI in medicine promises to advance this field and help practitioners make faster and more accurate diagnosis and reach more effective decisions about patient's care. Unfortunately, this technology has also come with a specific set of ethical and epistemological challenges. This paper aims at shedding some light on these issues and providing solutions to tackle the problems connected to using AI in clinical practice. We ultimately concur with the authors of the paper that medical AI cannot and should not replace physicians. We also add that a trustworthy AI will probably lead to more trust among humans and increase our reliance on experts. Thus, we propose that we start looking at the question: under what conditions is an AI system conducive to more human-to-human trust?

**Introduction**
The use of AI in medicine promises to advance this field and help practitioners make faster and more accurate diagnosis and reach more effective decisions about patient's care. Unfortunately, this technology has also come with a specific set of ethical and epistemological challenges. The epistemological challenges are specifically connected to the opacity of the so-called "black-box algorithms": "black boxes are algorithms that humans cannot survey, that is, they are epistemically opaque systems that no human or group of humans can closely examine in order to determine its inner states". The problem is that these algorithms make assessments in a way that is opaque to both their designers and the physicians using them because it seems impossible to know how the algorithms came to their conclusions.

The challenges that this epistemic opaqueness poses are both epistemic (are these algorithms in fact reliable?) and ethical (are these algorithms ethical, e.g. fair, respectful of human autonomy?). Both of these challenges touch on the issue of warranted trust in AI: if I can't check whether an algorithm is trustworthy (reliable & ethical), is trusting it ever permissible?

Even though this is not something the authors point out, it is worth noticing that 'trust' is already a loaded term: so let's unpack it a little bit. Say, an agent A can be said to trust B on some issue Y if A is willing to do at least one of the following: (i) A comes to believe what B says about Y and (ii) A uses what B says about Y as a sufficient reason for reaching a specific decision (e.g. making a certain diagnosis). Though we don't employ the same terminology, I believe the authors of the paper would agree that (i) and (ii) are not the same thing: (i) is what we can call 'doxastic trust' and (ii) is 'pragmatic trust' (note: the normative standards for doxastic trust might not be the same as for pragmatic trust).

We are now in a position to reformulate the question of the paper: When is it permissible for a physician to pragmatically trust a black box algorithm? The authors' answer is: even if the algorithm is reliable, what it says should rarely be used as a sufficient reason to make a diagnosis, prescribe a cure and so on. The algorithms' recommendations need to be interpreted by the physician's knowledge and understanding of the context and situation of the patient.

**Key Insights**
To answer the question above the authors of the paper look at the relationship between transparency and opacity in black box algorithms.

**Transparency** "refers to algorithmic procedures that make the inner workings of a black box algorithm interpretable to humans. To this end, an interpretable predictor is set out in the form of an exogenous algorithm capable of making visible the variables and relations acting within the black box algorithm and which are responsible for its outcome."

**Opacity** "focuses on the inherent impossibility of humans to survey an algorithm, both understood as a script as well as a computer process."

Relation between transparency and opacity:

*"designing and programming interpretable predictors that offer some form of insight into the inner workings of black box algorithms does not entail that the problems posed by opacity have been answered. To be more precise, transparency is a methodology that does not offer sufficient reasons to believe that we can reliably trust black box algorithms. At best, transparency contributes to building trust in the algorithms and their outcomes, but it would be a mistake to consider it as a solution to overcome opacity altogether."*

The authors are arguing here that transparency is not the solution to the problems of an opaque AI: it might be part of the solution, but it is not enough. What is the missing piece? Ensuring our blackbox AI is reliable.

**Solution**: as part of the solution the authors adopt computational reliabilism (CR). As the authors put it, "CR states that researchers are justified in believing the results of AI systems because there is a reliable process (ie, the algorithm) that yields, most of the time, [correct/accurate] results." They provide some insights on how reliability-assessments should be made in the context of blackbox algorithms by discussing some reliability-indicators (e.g. verification, expert knowledge, transparency). These reliability-indicators are still quite unclear, though.

However, the key point is that doxastically trusting AI might not be enough to justify acting on it, as we mentioned earlier. This is a contextual matter: what constituted enough reason for acting may vary given the context and what is at stake. This could mean two things: one has to do with the fact that epistemic standards for pragmatic trust may be more stringent than for doxastic trust. The second has to do with the fact that reliability is just one among the factors that make AI trustworthy: we need to make sure AI is also ethical (e.g. fair) before acting on its assessments and predictions. The authors explain that "if recommendations provided by the medical AI system are [doxastically] trusted because the algorithm itself is reliable, these should not be followed blindly without further assessment. Instead, we must keep humans in the loop of decision making by algorithms."

In other words, even if considered reliable, an algorithm should be rarely used as the only reason for reaching a decision in clinical practice. "It follows that it is unlikely and undesirable for algorithms to replace physicians altogether."

**Between The Lines**
The authors of this paper rightly argue that given what is at stake, (pragmatic) trust in medical blackbox algorithms is rarely justified. Practitioners and doctors still provide the necessary experience, reliability and commitment for patients to trust their decisions and diagnosis. That is, patients should trust doctors not algorithms. Doctors may trust algorithms to form beliefs but should not base their decisions only on what those algorithms say.

As a result, I believe we need to focus our attention on how AI can be trust-conducive: experts that rely on a robust, ethical and helpful AI are also themselves more trustworthy. Doctors that rely on a trustworthy AI system will themselves be and be perceived as more skillful, experienced and reliable. Hence, AI does not replace physicians: a trustworthy AI is conducive to more and better trust among humans and will probably make us rely on our experts even more. So from now let's ask the following question: under what conditions is an AI system conducive to human-to-human trust?

## Anthropomorphic interactions with a robot and robot-like agent

[Original paper by Sara Kiesler, Aaron Powers, Susan R. Fussell, and Cristen Torrey]
[Research Summary by Connor Wright]

**Overview**: Would you be more comfortable disclosing personal health information to a physical robot or a chatbot? In this study, anthropomorphism may be seen as the way to interact with humans, but most definitely not in acquiring personal information.

**Introduction**

Would you be more comfortable disclosing personal health information to a physical robot or a chatbot? Explored in this study is whether a humanlike robot solicits stronger anthropomorphic interactions than just a chatbot. With both physical presence and physical distance measured, the anthropomorphised robot wins the interaction race hands-down. However, when it comes to acquiring the medical information, the anthropomorphic strategy leaves much to be desired.

**Setting the scene**

The main actors of the study were a physically embodied robot, the same robot projected onto a screen, a software agent (like a chatbot) on a computer next to the participant and a software agent projected onto a big screen farther away. From there, four scenarios were set out (p.g. 172):

- The participant interacts with a physically present and embodied robot.
- The participant communicates with the same robot, but it is projected on a big screen.
- The participant engages with a software agent on a nearby laptop.
- The participant converses with the software agent on the further away big screen.

**Two hypotheses were proposed**
- The participants will interact and thus anthropomorphise the physically embodied robot more so than the software agent. However, they won't disclose as much personal information to the embodied agent.
- The participants will interact and thus anthropomorphise a software agent on a computer more than a robot projected onto a big screen.

The instructions for the discussion mentioned how the goal was to "have a discussion with this robot about basic health habits." (p.g. 173). Once carried out, the first conclusion drawn was on the importance of embodiment.

**Robot embodiment is key**
The participants interacted with the embodied robot a lot more than the social agent. Not only that, but it ranked top of all the robot trait ratings (such as trustworthiness and competency, see the table on p.g. 178).

In addition, the software agent was not seen as a "real" robot. The participants, of course, had their own preconceptions about how the robot was to look, with some being left disappointed when faced with a software agent.

**The embodied agent vs. the software agent**
Alongside the superior level of interaction, the first hypothesis was confirmed by how the participants did disclose less to the physical robot than the software agent. Instead, the software agent was viewed more as an administrative process that simply required personal information, which participants were more comfortable giving. While the software agent may have suffered in lacking human interaction, this proved beneficial in acquiring the desired medical information.

**The distance factor**
About the physical distance between the participant, the physical robot and the software agent did not differ. The variation in engagement time between having the robot and software agent projected and not projected was negligible. Hence, the study's second hypothesis was proved false.

**Between the lines**
While the physical robot was more anthropomorphised, it was still not seen as a fully human interlocutor. Participants mentioned how the robot, at times, wasn't flexible and interruptible enough for an entirely natural conversation to flow. Furthermore, the higher level of anthropomorphisation did not immediately lead to a sufficient level of trust to disclose personal health information. Hence, while anthropomorphisation does generate increased human interaction, it does not naturally follow that we trust the technology.

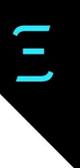

## Ubuntu's Implications for Philosophical Ethics

[Original talk by Thaddeus Metz]
[Research Summary by Connor Wright]

**Overview**: Philosophers have been puzzled over searching for an underlying principle expounded by a moral theory for over 400 years. Through his talk, Thaddeus Metz demonstrates how Ubuntu is also worth considering in the journey to solving this puzzle.

**Introduction**
Thaddeus Metz aims to demonstrate how Ubuntu looks when construed as a moral theory. The goal is not to show Ubuntu being 'better' when compared to other moral theories but rather as a perspective worthy of consideration. With the slogan of "a person is a person through other persons", we shall explore what Ubuntu construed as such entails and how this is applied to different situations. The Utilitarian and Kantian views are explored as comparisons, with the path that Ubuntu utilizes to arrive at similar conclusions proving particularly interesting.

**Key Insights**
Ubuntu is first represented as a moral theory. To be the case, it must offer the following:

- A comprehensive account of right and wrong.
- A specification of what all immoral actions have in common.
- A reduction of various duties down to just one.

Interpreting Ubuntu as such has the following benefits:

- Having a fundamental principle in philosophy would be super interesting.
- Having an underlying ethical principle can also be used to solve controversial issues (like abortion and the death penalty).
- Hence, the question becomes how we might draw on indigenous African thought to construct a moral theory?

Figures such as Archbishop Desmond Tutu, Professor Gessler Muxe Nkondo and Justice Yvonne Mokgoro have commented on Ubuntu. Here, they mention Ubuntu's emphasis on being

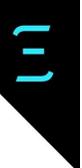

generous, hospitable, holding a commitment to the community and towards sympathetic social relations as basic tenets of the moral theory. As a result, Metz suggests the following two guidelines:

A real person becomes so through respecting others' capacity to relate harmoniously.
An act is wrong if and only if it fails to honor those that commune or be communed with.
However, what is a communal relationship, and how do you relate communally? A communal (harmonious) relationship includes two different strands: Identity and Solidarity. Identity is a sense of togetherness and coordination. Solidarity (caring for someone else's quality of life) includes sympathetic altruism.

Three corollaries of Ubuntu as a moral principle follow to pursue the harmonious relationship:

- You must avoid treating people in the opposite way to harmony; there is no us vs them.
- You must go out of your way to relate communally (exhibit identity and solidarity) and emphasize another person's dignity by allowing them to identify communally.
- Prioritize maintaining ties with people you already have a relation with, rather than strangers.

Following these steps leads to a communal relation. To fully manifest this, we can expect to see actions like those listed below:

- **Appealing to consensus** – everyone sits together until a solution is reached – necessary conditions for a just way of going forward. COnsensus = no split between majority and minority.
- **Collective labor** – everyone gathers to help one another harvest from plot to plot. Mutual aid for one another's sake.
- **Reconciliation** – rather than punishment that seeks to confine, punishment that aims to reconcile differences is pursued, like with the South Africa Truth and Reconciliation Commission.
- Moral value attributed to tradition, ritual and custom.

To show Ubuntu as a moral theory in action, Metz draws on examples from two different forms of considerations. Here, a basic intuition is explored through the Utilitarian, Kantian and Ubuntu views, allowing us to see how each differs. In this sense, Ubuntu entails the same kind of intuition as other Western theories, but for different reasons. To demonstrate, I have selected the most personally interesting examples from each section and listed how Ubuntu differs from the other two views explored.

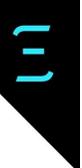

The first comparison: whether to fight poverty:

- **The intuition**: it is unjust for the extremely wealthy not to help out those who are poor due to circumstances out of their control.
- **Ubuntu**: poverty is unjust because the poor now have nothing to give to others, rather than the harm it does to the individual (Utilitarianism) or because the poor are less free to choose (Kantianism).

The second sort of comparison: whom to rescue from death:

- **The Intuition**: when having to choose between a young adult stranger and your mother, you should save your mother instead of a stranger.
- **Ubuntu**: the long-standing communal tie with your Mum means you should save her, rather than the stranger. A utilitarian would advocate for saving the stranger as they probably take up less resources and the Kantian would advocate for randomizing on who to save, seeing as their dignity is equal.

**Between the lines**

While it certainly proves controversial at times to say that one moral theory is 'outrightly' better than another, it is certainly less so to say one is worthy of consideration. I think Metz does exceptionally well not to force Ubuntu down our throats but to succinctly demonstrate why it ought to be considered. At times, I find that discussions in the West are susceptible to being stuck in the conventional ways of thinking about problems, a well-worn path, if you will. Ubuntu, in this sense, provides a welcomed new perspective on the issues at hand. An Ubuntu perspective is not only worth considering, but it's also beneficial.

## Risk and Trust Perceptions of the Public of Artificial Intelligence Applications

[Original paper by Keeley Crockett, Matt Garratt, Annabel Latham, Edwin Colyer, Sean Goltz]
[Research Summary by Connor Wright]

**Overview**: Does the general public trust AI more than those studying a higher education programme in computer science? The report aims to answer this very question, emphasising the importance of civic competence in AI.

**Introduction**

Is the opinion of the general public on AI different to those studying computer science in higher education? With a survey titled "You, me and "AI": What's the risk in giving AI more control?"

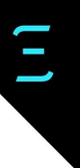

The paper aims to compare responses on the level of trust and risk that people of the general public and students studying computer science in higher education give. What is for sure is that civic competency in AI is crucial in creating representative technology, something we hold dear to our hearts here at MAIEI.

**Civic competency in AI**
One of the main slogans of this paper, my TEDxYouth talk and what we do at MAIEI is the importance of civic competency in the AI field. By improving public understanding of AI, we better equip them to fight any misinformation on the subject. One way to do this is to develop online courses, following in the footsteps of the University of Helsinki. By allowing non-experts to become involved in the debate, we enrich and make the AI space more representative.

Nevertheless, the paper points out how some may feel intimidated by courses offered by universities, for they don't feel they have the right qualifications. Hence, a future focus can be in creating courses specifically designed for the common person.

One of my core beliefs is that everyone can bring something to the AI table, no matter the level of expertise. Such value is clearly demonstrated in the data collated in the paper's surveys.

**The Results**
One of the main driving forces behind the survey is how previous studies conducted on the general public show varying degrees of knowledge about AI, but they all lack a robust description of the general public. Hence, the paper takes the general public to be those who have no specific knowledge in AI.

The groups of participants were split into Group 1 (the general public) and Group 2 (students of a higher education computer science programme). The groups were then asked questions on 3 different themes: trust, risk and questions on a scale of 0-10. A bird's eye view of the results are as follows:

**Trust**
- The groups were found to agree on questions such as not trusting an automated message from their boss, but differed on questions as to whether to trust a driverless car that had passed a "digital MOT" (p.g. 4).
- In this case, university students were more trusting of the AI involved.

**Risk**
The students always associated the same if not more risk to different AI applications, especially in terms of following instructions from a recognisable digitised voice.

The State of AI Ethics Report, Volume 6 (January 2022)                                                     283

**On a scale from 0-10**
There was general parity between the two groups on statements such as "I believe the minority of AI systems are biased". The only difference came in how students placed less emphasis on AI system decisions being explainable.

**Between the lines**
While the general public is defined as being without deep knowledge in the field, it is crucial that they are deemed to be a key stakeholder. In this way, their interactions with AI systems must be considered when evaluating an AI model's performance. As the paper rightly mentions, risk can occur at different points of the AI lifecycle, making system monitoring a vital aspect of a successful AI system. I hold that we cannot view AI systems as able to generalise over the whole population, meaning the practice is critical in ensuring that the system accurately tends to what it is designed to do and the problems this could bring.

## The Ethics of Sustainability for Artificial Intelligence

[Original paper by Andrea Owe and Seth D. Baum]
[Research Summary by Andrea Owe]

**Overview**: AI can have significant effects on domains associated with sustainability, such as aspects of the natural environment. However, sustainability work to date, including work on AI and sustainability, lacks clarity on the ethical details, such as what is to be sustained, why, and for how long. Differences in these details have important implications for what should be done, including for AI. This paper provides a foundational ethical analysis of sustainability for AI and calls for work on AI to adopt a concept of sustainability that is non-anthropocentric, long-term oriented, and morally ambitious.

**Introduction**
Sustainability is widely considered a good thing, especially a good thing related to environment-society interactions. It is in this spirit that recent initiatives on AI and sustainability have emerged, such as the conference AI for People: Towards Sustainable AI, of which this paper is part. But what exactly should be sustained, and why? Should, for example, the natural environment be sustained only to the extent that it supports the sustaining of human populations, or should natural ecosystems and nonhuman populations be sustained for their own sake? Is it enough to sustain something for a few generations or should sustainability endure into the distant future? Is sustainability even enough, or should we strive toward loftier aspirations? These are important ethical questions whose answers carry diverging implications for AI.

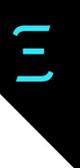

This paper surveys existing work on AI sustainability, finding that it lacks clarity on its ethical dimensions. This is shown through quantitative analysis of AI ethics principles and research on AI and sustainability. The paper then makes a case for a concept of sustainability for AI that is long-term oriented, including time scales in the astronomically distant future, and non-anthropocentric, meaning that humans should not be the only entities sustained for their own sake. The paper additionally suggests the more ambitious goal of optimization rather than sustainability.

**The ethical dimensions of sustainability**
To understand the ethics of sustainability for AI, it is essential to first understand the ethics of sustainability. In its essence, "sustainability" simply refers to the ability of something to continue over time; the thing to be sustained can be good, bad, or neutral. However, common usage of the term assumes that the thing to be sustained is some combination of social and ecological systems, with the most prominent definition being that of the 1987 Brundtland Report, defining sustainable development as "meeting the needs of the present without compromising the ability of future generations to meet their own needs." Since then, "sustainability" has been widely applied, often in ways that are imprecise or inconsistent with the basic idea of the ability to sustain something. This paper argues that usage of the term should be sharpened, and specifically that it should address three ethics questions:

- What should be able to be sustained, and why? For example, common conceptions of sustainability are anthropocentric in that they only aim to sustain humans for their own sake, with the natural environment or other nonhumans sustained only for the benefit of humans. In contrast, a wide range of moral philosophy calls for non-anthropocentric ethics that value both humans and nonhumans for their own sake.
- For how long should it be able to be sustained? There is a big difference between sustaining something for a few days or indefinitely into the distant future. For example, the Brundtland Report's emphasis on future generations implies a time scale of at least decades, but how many future generations? The limits of known physics suggest that it may be possible to sustain morally valuable entities for millions, billions, or trillions of years into the future, or even longer.
- How much effort should be made for sustainability? Should a person or an organization give "everything they've got" to advance sustainability or is just a little effort enough? How much should sustainability be emphasized relative to other competing values? The Brundtland definition was specifically crafted to acknowledge the competing values of present and future generations.

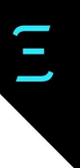

The paper additionally compares sustainability to the ethics concept of optimization. Sustainability means enabling something to be sustained in at least some minimal form, whereas optimization means making something be the best that it can be. For example, the Brundtland Report calls for the present generation to act "without compromising the ability of future generations to meet their needs". Arguably, the present generation should act to enable future generations to do much better than meeting their basic needs. Likewise, if human civilization has to focus on sustaining itself rather than loftier goals like optimization, then it is in a very bad situation.

**Empirical findings: AI and Sustainability**
Based on these ethical dimensions, the paper presents a quantitative analysis of published sets of AI ethics principles and academic research on AI and sustainability. The paper finds that most work on AI and sustainability focuses on common conceptions of socio-environmental sustainability, with smaller amounts of work on the sustainability of AI systems and other miscellaneous things. Further, most work is oriented toward sustaining human populations, with AI and the environment having value insofar as they support human populations. Most work does not specify the timescales of sustainability, nor the degree of effort to be taken, and overall lack clarity on the ethical dimensions presented above.

**The case for long-term, non-anthropocentric sustainability**
Following these findings, the paper gives its own answers on the ethical dimensions. First, sustainability should be non-anthropocentric, meaning that both humans and nonhumans should be sustained for their own sake. This is motivated by the scientific observation that humans are members of the animal kingdom and part of nature, and that nonhumans often possess attributes that are considered to be morally significant, such as the ability to experience pleasure and pain or have a life worth living. Second, sustainability should focus on long timescales, including the astronomically distant future. This is motivated by a principle of equality across time: everything should be valued equally regardless of what time period it exists in. Third, a large amount of effort should be made toward sustainability, and optimization should be emphasized over sustainability where the two diverge. Long-term sustainability of any Earth-originating entities will eventually require expansion into space, making it necessary to first handle any major threats on Earth, such as global warming and nuclear warfare. Additionally, the astronomically distant future offers astronomically large opportunities for advancing moral value, making an objective to optimize moral value diverge significantly from an objective of sustaining moral value only.

**Implications for AI**
Finally, the paper presents implications of the above for AI.

- First, AI should be used to improve long-term sustainability and optimization. For current and near-term forms of AI, this includes addressing immediate threats to the sustainability of global civilization, such as global warming and pandemics.
- Second, attention should be paid to long-term forms of AI, which could be particularly consequential for long-term sustainability and optimization due to its potential. Long-term AI is seldom discussed in relation to sustainability, but the paper argues that these topics are a more appropriate focus for work on AI and sustainability. Long-term AI could bolster efforts to address threats such as global warming, and it could also pose threats of its own, especially for runaway AI scenarios. Furthermore, it could play an important role in space expansion, which is central to the long-term sustainability and optimization of moral value.

**Between the lines**

In sum, this paper calls for work on AI and sustainability to be specific about its ethical basis and to adopt non-anthropocentric, long-term oriented concepts of sustainability or optimization. In practice, that entails focusing on applying AI to address major global threats and improving the design of long-term AI, in order to ensure the long-term sustainability of civilization and to pursue opportunities to expand civilization into outer space. Actions involving AI are among the most significant ways to affect the distant future. The field of AI therefore has special opportunities to make an astronomically large positive difference.

# Go Wide: Article Summaries (summarized by Abhishek Gupta)

**The hacker who spent a year reclaiming his face from Clearview AI**

[Original article by [Coda Story](#)]

**What happened**: A person living in Germany on reading the Clearview AI story in the NYT in 2020 wanted to check if Clearview AI had any data about him given his concern about his privacy and how rarely he shared pictures of himself online. He was shocked to discover that Clearview AI had found two images of him that he didn't even know existed. He raised a complaint in the Hamburg Data Protection Authority which after a 12-month back and forth with the company finally ordered them to remove the mathematical hash that characterized his biometric data, his face.

**Why it matters**: With the way that Clearview AI has gathered data from public sources on people's faces, the person from Germany rightly claims in the interview that the company has made it impossible to remain anonymous now. It is not like a regular search engine process in that on inputting faces it digs up specific matches to your face thus making it perhaps impossible to participate in a protest for the fear of being identified, even when it is legal to do so. More so, it has implications for what happens to all the data that is captured from CCTVs and other surveillance mechanisms that capture our data without our consent all the time, thus potentially limiting freedom of movement of people in the built environment.

**Between the lines**: Finally, the thing that caught my attention was the fact that the person from Germany mentioned that there were erroneous matches that were returned to him as a part of his data request. This is to be expected because no algorithm can be perfect but there is a severe consequence: if there are authorities that rely on this data to make determinations about the movements of people, they might draw false conclusions. Also, it might still be OK in a perfectly functioning democracy (which rarely if ever exists or will exist) but what would happen to this technology and its capabilities if the regime changes to something more authoritarian?

### An Artificial Intelligence Helped Write This Play. It May Contain Racism

**[Original article by Time]**

**What happened**: Human-machine teaming is always an interesting domain to surface unexpected results. In a play staged in the YoungVic in London, playwrights have joined forces with GPT-3 to generate a script on the spot which is enacted by a troupe of actors without any rehearsals giving unique plays every night that they are on stage. Taking on an uncensored and unfiltered approach, the harsh stage lighting of the YoungVic will also lay bare the biases that pervade the outputs of GPT-3 mirroring the realities of the world outside the stage. Jennifer Tang, the director, sees this as an exciting foray into the future of what AI can do for the creative field.

**Why it matters**: While we have seen a lot of debates around the role that AI will play in the creative fields, something that we've covered in Volume 5 of the State of AI Ethics Report as well, using a very powerful model like the GPT-3 to work side-by-side with humans is novel in generating creative outputs. While scholars interviewed in the article caution against attributing creativity to the outputs from the system, it might be worth considering if we can say that such a tool helps to boost creativity for artists by expanding the solution space that the artists can then explore.

**Between the lines**: Biases in the outputs from GPT-3 are very problematic - with stereotypical dialogue allocation based on the religion of the actors to outright homophobia and racism, the issues with such large-scale models are numerous. How such problems are tackled and if they can be brought to the stage where they become trusted tools in the creative process remains to be seen. The first step in that process is highlighting the problems and beginning to build tools that can address those issues before this becomes a common practice in the creative industries.

### The Stealthy iPhone Hacks That Apple Still Can't Stop

**[Original article by Wired]**

**What happened**: In a not-so-surprising revelation, high-profile individuals were targeted by the Bahraini government using zero-click attacks that targeted vulnerabilities in the iMessage app on the iPhone. Dubbed "Megalodon" and "Forced Entry" by Amnesty International and Citizen Lab respectively, the attacks bypass critical protections created by Apple, called BlastDoor, to guard against these kinds of attacks. The zero-click attacks don't require any interaction from the user and that's what gives them potency and effectiveness.

**Why it matters**: Though these attacks cost millions of dollars to develop and they often have short shelf-lives because of security patches and updates issued by manufacturers, they still pose an immense threat to the targeted individuals. The issue is exacerbated because Apple refuses to allow users to disable apps that they provide natively on the iPhone. Past releases have shown that the attack surface for such kinds of apps is quite large and protecting against such threats is increasingly difficult and requires significant overhauls of the core architecture, which will require tons of resources from Apple to make the necessary changes, something that they are unlikely to do in the short-run.

**Between the lines**: iMessage is not the only app that faces such zero-click attacks; there are other apps like Whatsapp that are also susceptible to different attacks that follow similar patterns. With the utilization of AI to discover vulnerabilities, we are perhaps entering an era whereby the detection of vulnerabilities is greatly accelerated allowing malicious actors to craft even more sophisticated attacks by directing their energies towards developments of those exploits more so than having to discover the vulnerabilities first before crafting the attacks. This places an additional burden on the manufacturers to ensure that they have more robust development practices guaranteeing security of the end user's devices and apps.

## The limitations of AI safety tools

[Original article by VentureBeat]

**What happened**: With the inclusion of the word "safety" in various trustworthy AI proposals from the EU HLEG and NIST, this article talks about the role something like the Safety Gym from OpenAI can play in achieving safety in AI systems. It provides an environment to test reinforcement learning systems in a constrained setting to evaluate their performance and assess them for various safety concerns. The article features interviews with some researchers in the field who mention how such a gym might be inadequate since it still depends on specifying rules to qualify what constitutes safe behavior. And such rules will have to constantly grow and adapt as the systems "explore" new ways of achieving the specified tasks.

**Why it matters**: Environments like the Safety Gym help to provide a sandbox to test digital twins of systems that will be deployed in production, especially when the costs of such testing might be prohibitive in the real-world or too risky. This applies to cases of autonomous driving, industrial robots working alongside other humans, and other use-cases with humans and machines operating in a shared environment.

**Between the lines**: A single environment with a specific modality of operation can never comprehensively help to identify all the places where alignment problems might arise for an AI system, but they do provide a diagnostic test to at least identify how the system can misbehave or deviate from expectations. Using that information to iterate is a useful outcome from the use of such an environment. More so, such environments offer a much more practical way to go about safety testing rather than just theoretical formulations which to a certain extent are limited by the ingenuity of the testers and developers' to imagine how an AI system might behave.

## AI fake-face generators can be rewound to reveal the real faces they trained on

[Original article by MIT Technology Review]

**What happened**: The article covers a recent paper that utilized membership inference attacks to determine what face images might have been used in training a facial recognition technology system. There are many websites like This Person Does Not Exist that offer AI-generated faces by utilizing GANs, but some of them resemble real people too closely. The paper sought to demonstrate that by generating faces from the GAN and then using a separate facial recognition system to see if any of them were a match.

**Why it matters**: Such a technique has the potential to allow people to check if their image has been used in training an AI system. But, it also showcases latent vulnerabilities in such systems when they can leak what kind of data was used to train them. Especially when you have pre-trained models that are re-used downstream by other developers. Other techniques like model inversion and model stealing, falling in the broad category of machine learning security demonstrate such weaknesses in AI systems today and provide us with a pathway towards building more robust AI systems.

**Between the lines**: The area of machine learning security today is highly under-explored with most of the focus on issues like fairness and privacy, which while important don't cover the full gamut of ethical issues with AI systems. We need to ensure that the systems are robust as well and today we are in the early stages of machine learning security attacks and defenses as was the case with cybersecurity for more traditional software systems a few years ago.

## What Apple's New Repair Program Means for You (And Your iPhone)

**[Original article by [NYTimes](NYTimes)]**

**What happened**: Apple has announced a new program under which they are making replacement parts available to a wider set of repair services providers, including to consumers so that they can make minor repairs either themselves, or take it to other repair shops to extend the life of their devices. This has direct implications in terms of increasing accessibility of these devices, since consumers who were charged a lot of money at Apple stores or authorized services can now get a cheaper pathway to continue using their devices. And most importantly, extending the lifespan of the device means that we will reduce the impact on the planet, given that the embodied carbon emissions constitute a major chunk of the environmental impacts of technology, this is a great step forward.

**Why it matters**: This is a huge win for advocates of the "Right to Repair" movement, and as mentioned in the article, a huge company like Apple making such a move can act as a trendsetter for other companies to follow suit and offer similar services. Given that we cycle through our devices fairly quickly, extending the lifespan of these devices can have an indirect impact also on the kind of software that is developed which can continue to leverage older hardware rather than constantly creating backward-incompatible updates that necessitate moving to newer devices.

**Between the lines**: The concerns that are usually flagged for not offering such programs has traditionally been that unauthorized repair centers might pose security and privacy risks to the data of the consumers on those devices. The current move might be coming on the heels of hints from the FTC that they might make more stringent regulations that mandate providing options to consumers to be able to repair their devices either on their own or get access to replacement parts so that they can pick replacement service providers outside of those authorized by the manufacturer.

# Closing Remarks

Congratulations for making it all the way to the end! This was our longest edition of the State of AI Ethics Report thus far!

Every few months as we embark on capturing the latest in research and reporting in the domain of AI ethics, we are surprised by the richness of the domain, the indefatigable efforts of activists, researchers, and practitioners around the world, and above all the amazing repertoire of ideas permeating the domain. We count ourselves lucky to have all of you as our supporters who encourage us to continue our exploration and help us realize our mission to **Democratize AI Ethics Literacy**.

Many pieces in this edition of the report resonated deeply with me and made me pause and wonder what it is that we can be doing to elevate the level of conversation in the field and provide more meaningful ways for the community to engage with each other and support each other as we all strive to make society better, using technology to solve problems rather than create new ones.

The report is one amongst many different ways that you can stay connected with us. We also publish The AI Ethics Brief that is read by technical leaders and policymakers from around the world. We invite you to stay in touch with us between reports through that. Above all, if you're working on interesting problems and are looking for sounding boards, we're around!

Until the next report, hope you stay safe and healthy, and let's all work together to Make Responsible AI the Norm rather than the Exception!

---

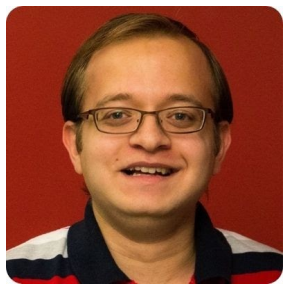

**Abhishek Gupta (@atg_abhishek)**
Founder, Director, & Principal Researcher,
Montreal AI Ethics Institute

Abhishek Gupta is the founder, director, and principal researcher at the Montreal AI Ethics Institute. He is also a machine learning engineer at Microsoft, where he serves on the CSE Responsible AI Board. He also serves as the Chair of the Standards Working Group at the Green Software Foundation.

# Support Our Work

The Montreal AI Ethics Institute is committed to democratizing AI Ethics literacy. But we can't do it alone.

Every dollar you donate helps us pay for our staff and tech stack, which make everything we do possible.

With your support, we'll be able to:

- Run more events and create more content
- Use software that respects our readers' data privacy
- Build the most engaged AI Ethics community in the world

Please make a donation today at **montrealethics.ai/donate**.

We also encourage you to sign up for our weekly newsletter *The AI Ethics Brief* at **brief.montrealethics.ai** to keep up with our latest work, including summaries of the latest research & reporting, as well as our upcoming events.

If you want to revisit previous editions of the report to catch up, head over to **montrealethics.ai/state**.

Please also reach out to **Masa Sweidan** **masa@montrealethics.ai** for providing your organizational support for upcoming quarterly editions of the *State of AI Ethics Report.*

**Note:** All donations made to the Montreal AI Ethics Institute (MAIEI) are subject to our **Contributions Policy**.